\documentclass[twoside,fleqn]{ActaStyle}
\usepackage[colorlinks=true, pdfstartview=FitV, linkcolor=black,citecolor=black, urlcolor=black]{hyperref}
\usepackage{times,cite}
\usepackage[all,matrix,frame,arrow]{xy}
\usepackage[active]{srcltx}
\usepackage{fancyhdr}
\renewcommand{\subsectionmark}[1]{}
\renewcommand{\thesection}{\arabic{section}}
\renewcommand{\thesubsection}{\thesection.\arabic{subsection}}

\renewcommand{\thefigure}{\thesection.\arabic{figure}}
\renewcommand{\thetable}{\thesection.\arabic{table}}

\include{diagrams}

\usepackage{epsfig,array}
\usepackage{amsbsy,latexsym}
\usepackage{amsmath}
\usepackage{amssymb, mathrsfs}
\usepackage[mathscr]{eucal}

\usepackage[english]{babel}
\usepackage[latin1]{inputenc}
\usepackage[T1]{fontenc}
\usepackage{graphicx,color,setspace,amsmath}
\usepackage{amssymb}
\usepackage{mathrsfs}
\usepackage{young} 

\usepackage{lscape}
\usepackage{rotating}
\usepackage{verbatim}
\usepackage{amsfonts}  
\usepackage{amsbsy}   

\input{qcircuit}

\newcommand{\bra}[1]{\left\langle{#1}\right\vert}
\newcommand{\ket}[1]{\left\vert{#1}\right\rangle}
\def\spc#1{\mathcal{#1}}

\def\d{\operatorname{d}\!}
\def\<{\langle}
\def\>{\rangle}
\def\Span{\operatorname{Span}}
\def\Tr{\operatorname{Tr}}
\def\rank{\operatorname{rank}}
\def\Rng{\operatorname{Rng}}

\def\:{\hbox{\bf :}}

\def\vec#1{{\boldsymbol{#1}}}
\def\set#1{{\sf #1}}
\def\dag{\dagger}
\def\geq{\geqslant}
\def\leq{\leqslant}
\def\map#1{\mathcal #1}
\def\sH{\mathcal{H}}

\def\dim{\operatorname{dim}}
\def\Ker{\operatorname{Ker}}

\def\supp{\set{Supp}}
\def\qed{$\,\blacksquare$\par}
\def\kk{\rangle\!\rangle}
\def\bb{\langle\!\langle}

\newcommand{\Ket}[1]{| #1 \rangle \! \rangle}
\newcommand{\Bra}[1]{\langle \! \langle #1 |}
\newcommand{\BraKet}[2]{\langle \! \langle #1 | #2 \rangle  \! \rangle}
\newcommand{\KetBra}[2]{\Ket{#1} \Bra{#2}}
\newcommand{\braket}[2]{\langle #1 | #2 \rangle}
\newcommand{\ketbra}[2]{\ket{#1} \bra{#2}}

\newcommand{\hilb}[1]{\mathcal{#1}}
\newcommand{\defset}[1]{{\sf #1}}

\newcommand{\group}[1]{\mathbf{#1}}

\newcommand{\Rel}[1]{{C^{#1}_d}}

\newtheorem{Def}{Definition}[section]
\newtheorem{lemma}{Lemma}[section]

\newtheorem{corollary}{Corollary}[section]
\newtheorem{theorem}{Theorem}[section]
\newtheorem{remark}{Remark}[section]

\def\authorheading{A. Bisio, G. Chiribella, G. M. D'Ariano, P. Perinotti }
\def\shorttitle{Quantum Networks: General Theory and Applications}

\numberwithin{equation}{section}

\begin{document}
\pagerange{273}{390}   

\title{QUANTUM NETWORKS: GENERAL THEORY AND
  APPLICATIONS\footnote{The material in this article was presented as a
    PhD thesis of A. Bisio at the University of Pavia. The work was
    conducted under the supervision of professor G. M. D'Ariano}}

\author{A.~Bisio\email{alessandro.bisio@unipv.it}$^*$, G.~Chiribella$^\dag$,
 G.~M.~D'Ariano$^*$, P.~Perinotti$^*$}
 {$^*$Quit group, Dipartimento di Fisica ``A. Volta'', via Bassi 6, 27100 Pavia, Italy
 \\ $^\dag$Perimeter Institute for Theoretical Physics, 31 Caroline St. North, Waterloo, Ontario N2L 2Y5, Canada}

\datumy{29 August 2011}{30 August 2011}

\abstract{In this work we present a general mathematical framework to
 deal with \emph{Quantum Networks}, i.e.  networks resulting from the interconnection of elementary quantum circuits.  The cornerstone of our approach is a generalization of the Choi isomorphism that allows one to efficiently represent any given
 Quantum Network in terms of a single positive operator. Our formalism allows one to face and solve many quantum
 information processing problems that would be hardly manageable otherwise,   the most relevant of which are reviewed in this work: quantum process tomography, quantum cloning and learning of transformations, inversion of a unitary gate, information-disturbance tradeoff in estimating a unitary transformation, cloning and learning of a measurement device.}

\quad{\small {\sf DOI:}}\ 10.2478/v10155-011-0003-9

\vspace{0.3cm}

\begin{minipage}{2.5cm}
\quad{\small {\sf KEYWORDS:}}
\end{minipage}
\begin{minipage}{10cm}
{\bfseries\itshape Quantum Information Processing, Quantum Circuits,
  Quantum Networks, Quantum Tomography, Quantum Cloning, Quantum
  Learning, Unitary Channels, Group theory in Quantum Mechanics}
\end{minipage}

\tableofcontents
\newpage

\setcounter{equation}{0} \setcounter{figure}{0} \setcounter{table}{0}\newpage
\section{Introduction}

In standard textbook Quantum Mechanics every physical system
corresponds to a Hilbert space. The states of a system are unit rays
in the corresponding Hilbert space, transformations of closed systems
are described by unitary operators acting on the states, and
measurements correspond to complete sets of orthogonal projectors,
each projector corresponding to a measurement outcome. Born's
statistical formula provides the outcome probability as the
expectation of the corresponding projector in the state of the system.
This formalism can be generalized to the case of open systems by
including the environment in the dynamical description. It is then
possible to describe any phenomenon in quantum mechanics in terms of
unitary transformations and von Neumann or L\"uders measurements.
Despite this fact, a convenient formalism for the field of Quantum
Information \cite{holevo, nielsenchuang} is rather provided by the
notions of statistical operator, channel, and positive operator valued
measure (POVM). One of the advantages in using such tools is that they
provide an effective description of physical devices avoiding a
detailed account of their implementation in terms of unitary
interactions and von Neumann measurements. 
This concise description is extremely useful when dealing with
optimization problems, like state estimation \cite{helstrom}, where
one can looks for the optimal measurement among all those allowed by
quantum mechanics.

A recent trend in Quantum Information is to consider transformations,
rather than states, as carriers of information e.g.  in gate
discrimination \cite{acin, leaderdiscr, maxdiscrimi, memoryeffects,
  watrousdiscrimi}, programming \cite{noprogramming}, teleportation
\cite{teleport1, teleport2, teleport3} and tomography \cite{tomoexp1,
  tomoexp2}, along with multi-round quantum games \cite{watrousgame},
standard quantum algorithms \cite{deutschjoz,Grover,shor} and
cryptographic protocols \cite{bitcommitment1, bitcommitment2,pironio}.
This new perspective requires an appropriate description not only of
state processing, but more generally of transformation processing.
Such processing is obtained through more general physical
devices---what we call \emph{Quantum Networks}---that are made of
composition of elementary circuits. A Quantum Network can be used to
perform a huge variety of different tasks like transformations of
states into channels, channels into channels, and even sequences of
states/channels into channels. However, describing a large quantum
network in terms of channels and POVMs is very inefficient. Indeed, if
one needs to optimize a quantum network for some task, one is forced
to carry out a cumbersome elementwise optimization. For this reason,
having new notions that generalize those of channels and POVMs is
crucial. Luckily enough, a general treatment of Quantum Networks on
the same footing as states, channels, and POVMs is possible, both for
the deterministic and the probabilistic case. \par
In this paper we review the aforementioned unified framework along
with some of its most relevant applications. Our approach is based on
a generalization of the Choi isomorphism that allows us to represent
any Quantum Network in terms of a suitably normalized positive
operator. This general theory is reviewed in Chapter
\ref{chapter:gentheory}, where we also provide some basic results of
linear algebra that are needed in order to prove most results of the
general theory. Following the exposition of Refs.
\cite{QCA, comblong}, we will introduce the notion of Quantum Network from
a constructive point of view that consists in looking at networks as a
result of composition of elementary circuits. We will begin by
considering deterministic Quantum Networks, and then we will extend
the results to the probabilistic case.\par
The Chapters from \ref{chaptertomo} to \ref{chapter:observables} are
devoted to the applications of the developed formalism. The first
application that we consider is the optimization of Quantum
Tomography, where we will derive the optimal networks for tomographing
states, transformations and measurements.  The material of this
Chapter was published in Refs \cite{optimaltomo, tomoieee}. \par
In Chapter \ref{chapter:cloning}, based on Ref. \cite{cloningunit}, we
discuss the concept of \emph{quantum cloning of a transformation}.
While cloning of quantum states has been subject of many works,
cloning of a transformation was never treated before. In particular, a
general no-cloning theorem for transformations and the derivation of
the optimal cloning network for a unitary transformation are
shown.\par
\emph{Quantum Learning} of a transformation is another task that is
possible to analyse within the new theory of Quantum Networks.
Suppose that a user is provided with $N$ uses of an undisclosed
transformation $\mathcal{T}$ today, and he needs to reproduce the same
transformation on an unknown state provided tomorrow.  The most
general strategy the user can follow, is to exploit the $N$ uses of
$\mathcal{T}$ into a Quantum Network today, in order to store the
transformation on a quantum memory. Tomorrow, the user will use the
quantum memory to program a retrieving channel that reproduces
$\mathcal{T}$. In Chapter \ref{chapter:learning} we will review Ref.
\cite{optimallearning}, in which the optimal learning network for a
unitary transformation is derived. The most relevant result here is
that the optimal storing of a unitary can be achieved by making use
only of a classical memory.\par
The optimal inversion of a unitary transformation is the subject of
Chapter \ref{chapter:inversion}.  We will derive the optimal network
that realizes this task considering two different scenarios and we
will prove that the ultimate performances in the inversion of a
unitary are achieved by an estimate and prepare strategy.  These
results were published in Refs \cite{optimallearning, algorithm}.\par
In Chapter \ref{chapter:infotradeoff}, based on
Ref. \cite{unitradeoff},
 we consider the tradeoff between 
information and disturbance in estimating a unitary transformation.
We suppose that we have 
 a black box implementing an unknown
unitary transformation, with the restriction that
 the
On the one hand, we may try
 to identify the unknown unitary
on the other hand, we may want to use the black box on a variable
input state.
 Since the two tasks are in general incompatible,
 there is a trade-off between the amount of information that 
can be extracted about a black box and the disturbance caused on its
action:  we cannot estimate an unknown quantum dynamics 
without perturbing it. In Chapter \ref{chapter:infotradeoff} we find
the optimal
 scheme that introduces the minimum amount of disturbance for any given amount of extracted information. 
\par
The last application we consider regards Quantum Networks that
replicate measurements.  We will study the problem of optimal learning
and cloning of von Neumann measurements.  In particular we will show
how the optimal learning from $3$ uses can be achieved only by a
sequential strategy.  These results are the subject of
Refs.~\cite{learnobs,clonobs}, and are presented in Chapter
\ref{chapter:observables}.\par
Two appendices close this work: in the first one we introduce the
notion of channel fidelity \cite{raginskyfide}, which is frequently
used in the applications and in the second one we review some basic
results from group representation theory, with special emphasis on the
decomposition of tensor product representations.


\setcounter{equation}{0} \setcounter{figure}{0} \setcounter{table}{0}\newpage
\section{Quantum Networks: general theory}\label{chapter:gentheory}

In this chapter we expose the general theory of Quantum Networks that
was developed in \cite{QCA, comblong, algorithm}.  We will start the
presentation with some preliminary results of linear algebra, with
emphasis on the Choi isomorphism.  This theorem will allow us to
represent quantum networks in terms of positive operators which are
subject to a normalization constraint. A key point of the formalism is
the notion of link product of operators that translates the physical
link between quantum networks into the mathematical language.

After this fully mathematical section we recall some basic notions of
ordinary quantum mechanics (states, quantum operations and POVMs) that
we will use as a testbed for the mathematical tools previously
introduced.

Section~\ref{section:construct} is 
 is focused on the definition of Quantum Network as a set of linear maps
linked together; in the following sections
the Choi representation  of Quantum Networks is introduced
first for the deterministic case and then for the probabilistic case.

In the final section the link product of operators will be used to express
the link of quantum Networks.

\subsection{Linear maps and linear operators}
Let us start with some notational remarks: we denote as
$\mathcal{L}(\hilb{H})$ the set of linear operators $A$ on $\hilb{H}$
while $\mathcal{L}(\hilb{H}_a,\hilb{H}_b)$ denotes linear
transformation from $\hilb{H}_a$ to $\hilb{H}_b$. The dimension of
space $\hilb{H}_a$ is denoted by $d_a$. We denote as
$\mathcal{L}(\mathcal{L}(\hilb{H}_a),\mathcal{L}(\hilb{H}_b))$ the set
of linear maps $\mathcal{M}$ from $\mathcal{L}(\hilb{H}_a)$ to
$\mathcal{L}(\hilb{H}_b)$.  Given a map $\mathcal{M} \in
\mathcal{L}(\mathcal{L}(\hilb{H}_a),\mathcal{L}(\hilb{H}_b))$ we refer
to $\mathcal{L}(\hilb{H}_a)$ as the \emph{input space} of
$\mathcal{M}$ while $\mathcal{L}(\hilb{H}_b)$ is called the
\emph{output space}.  We make use of the following notation:
\begin{itemize}
\item $\supp(A)$ denotes the support of $A$ and $\Rng(A)$ denotes the range of $A$;
\item $^T$ denotes transposition and $^*$ denotes complex conjugation;\footnote{Both
 transposition and complex conjugation are meant with respect to a fixed orthonormal basis.}
\item $A^{-1}$ denotes the inverse of an operator  $A \in \mathcal{L}(\hilb{H})$;
 if $\supp(A)$ is not the whole $\hilb{H}$, then
$A^{-1}$ will denote the inverse on its support.\footnote{More precisely
 $A^{-1}$ denotes the Moore-Penrose generalized inverse.} 
\end{itemize}
 
Within this presentation (unless explicitly mentioned)
the Hilbert spaces are assumed to be finite dimensional.
In order to avoid confusion when the number of Hilbert spaces proliferates
we adopt this convention:
\begin{itemize}
\item $\hilb{H}_{ab\dots n} := \hilb{H}_{a}\otimes \hilb{H}_{b} \otimes \dots \otimes \hilb{H}_n$
where $a,b,\dots,n$ are integer numbers;
\item $A_{ab \dots n}$ means $A \in \mathcal{L}(\hilb{H}_{ab \dots n})$;
\item $\ket{n}_a$ means $\ket{n} \in \hilb{H}_{a}$;
\item $\Tr_a$ denotes partial trace over $\hilb{H}_{a}$;
\item  $^{T_a}$ denotes partial transposition over $\hilb{H}_{a}$.
\end{itemize}

Given an operator $A \in \mathcal{L}(\hilb{H}_a)$ 
 and a Hilbert space
 $\hilb{H}_{a'}$  isomorphic
to $\hilb{H}_{a}$ $\hilb{H}_{a'} \cong \hilb{H}_{a}$,
it is possible to  define 
\begin{align}
\label{eq:relabeled}
 A_{a'} :=  T_{a \rightarrow a'} A_a  T_{a \rightarrow a'}
\end{align}
where $T_{a \rightarrow a'} = \sum_k \ket{k}_{a'}\bra{k}_{a}$ and
$\{ \ket{k}_{a} \}$, $\{ \ket{k}_{a'} \}$  are orthonormal bases for 
$\hilb{H}_{a}$ and $\hilb{H}_{a'}$ respectively.
The above procedure is  implicit
whenever
we  make a change of label $A_{ab\dots n} \rightarrow A_{a'b'\dots n'}$

It is possible to define the following isomorphism between
$\mathcal{L}(\hilb{H}_b,\hilb{H}_a)$ and $\hilb{H}_a\otimes \hilb{H}_b$
\begin{align}
  A = \sum_{n,m}\bra{n}A\ket{m}\ket{n}\bra{m} \leftrightarrow
\Ket{A} = \sum_{n,m}\bra{n}A\ket{m}\ket{n}\ket{m} \label{eq:doubleket}
\end{align}
where $\{\ket{m}\} (\{\ket{n}\})$ is a fixed orthonormal basis in
$\hilb{H}_b (\hilb{H}_a)$.  In the following we implicitly choose such
a basis in every Hilbert space.  The double-ket notation $\Ket{A}$ is
used to stress that the vector lives in a tensor product of Hilbert
spaces (from a quantum mechanical perspective, $\Ket{A}$ is
proportional to a pure bipartite state) We will use the notation
$\Ket{A}_{ab}$ with the meaning $\Ket{A} \in \hilb{H}_{ab} =
\hilb{H}_{a}\otimes \hilb{H}_b$.

By making use of  Eq. (\ref{eq:doubleket}) it is  possible to prove that the following identities hold
\begin{align}
  &A \otimes B \Ket{C} = \Ket{ACB^T} \label{eq:dketid1}
 \\
&A \in \mathcal{L}(\hilb{H}_a,\hilb{H}_c ), 
\;\;
B \in \mathcal{L}(\hilb{H}_b,\hilb{H}_d ),
\;\;
C \in \mathcal{L}(\hilb{H}_b,\hilb{H}_a ). \nonumber\\
& \Tr_b[\KetBra{A}{A}_{ab}] = AA^\dagger \qquad \Tr_a[\KetBra{A}{A}_{ab}] = A^TA^* 
\label{eq:dketid1bis}\\
&\Tr_a[A_{ab}(\KetBra{I}{I}_{ac})] = A_{bc}^{T_c} \label{eq:dketid2}\\
&\Bra{I}_{ac}A_{ab}\Ket{I}_{ac} = \Tr_a[A_{ab}] \label{eq:dketid3}\\
& \mbox{for}\  d_a \leq d_c, \quad  \Ket{I}_{ac} = \sum_{n=1}^{d_a}\ket{n}_a \ket{n}_c
\nonumber\\
&\Ket{I}_{abcd}= \Ket{I}_{ac}\Ket{I}_{bd} \label{eq:dketid4}\\
&(\Bra{I}_{ac}\otimes I_{bd})\Ket{A}_{abcd}
=(\Bra{I}_{ac}\otimes I_{bd})(A_{ab}\otimes I_{cd})\Ket{I}_{abcd} =\Ket{\Tr_{a}[A]}_{bd} \label{eq:dketid5}
\end{align}
Through this isomorphism it is possible to translate the inner product in $\hilb{H} \otimes \hilb{H}$
into the Hilbert-Schmidt product in $\mathcal{L}(\hilb{H})$
\begin{align}
  \BraKet{A}{B}= \Tr[A^\dagger B]
\end{align}

\subsubsection{Choi isomorphism}
The following theorem, which is a generalization of the one in Refs.
\cite{depillis, choiisom, jamioisom}, introduces an isomorphism
between linear maps and linear operators which is a a foundation stone
of the theory of Quantum Networks.
\begin{theorem}[Choi isomorphism]\label{th:Choijamiso}
Consider the map $\mathfrak{C}:\mathcal{L}(\mathcal{L}(\hilb{H}_0),\mathcal{L}(\hilb{H}_1))
\to \mathcal{L}(\hilb{H}_0 \otimes  \hilb{H}_1)  $ defined as
\begin{align}
  \mathfrak{C}:\mathcal{M}\mapsto M_{10} \qquad M_{10}:= \mathcal{M} \otimes \mathcal{I}_{0} (\KetBra{I}{I}_{00})
\end{align}
where $\mathcal{I}_{0}$ is the identity map on $\mathcal{L}(\hilb{H}_0)$.
Then $\mathfrak{C}$ defines an isomorphism between $\mathcal{L}(\mathcal{L}(\hilb{H}_0),$ $\mathcal{L}(\hilb{H}_1))$
and $\mathcal{L}(\hilb{H}_0 \otimes  \hilb{H}_1)$.
The operator $M = \mathfrak{C}(\mathcal{M})$ is called
the Choi operator of $\mathcal{M}$.
\end{theorem}
\begin{Proof}
  To prove the thesis we will provide an explicit expression for 
the inverse map $\mathfrak{C}^{-1}:
 \mathcal{L}(\hilb{H}_0 \otimes  \hilb{H}_1)
\rightarrow
\mathcal{L}(\mathcal{L}(\hilb{H}_0),\mathcal{L}(\hilb{H}_1))$.
Let us define 
\begin{align}
  [\mathfrak{C}^{-1}(M)](X) =  \Tr_0[(I_1 \otimes (X_0)^T )M_{10}] \label{eq:Choi-1};
\end{align}
it is easy to verify that $[\mathfrak{C}^{-1}(M)](X) = \mathcal{M}(X)$.

Suppose $X_0 = \ketbra{i}{j}_0$; we have
\begin{align}
  \Tr_0[(I_1 \otimes (\ketbra{i}{j}_0)^T )M_{10}] &=
\Tr_0[(I_1 \otimes (\ketbra{j}{i}_0 )\mathcal{M}\otimes \mathcal{I}_0(\KetBra{I}{I}_{00})] = \nonumber \\
&= \bra{i}_0(\mathcal{M}\otimes \mathcal{I}_0(\KetBra{I}{I}_{00}))\ket{j}_0 = \nonumber \\
&= \bra{i}_0(\sum_{m,n}\mathcal{M}(\ketbra{n}{m}_0)) \otimes \ketbra{n}{m}_0)\ket{j}_0= \nonumber \\
&=
\mathcal{M}(\ketbra{i}{j}_0).
\end{align}
From $[\mathfrak{C}^{-1}(M)](\ketbra{i}{j}) = \mathcal{M}(\ketbra{i}{j})$ for any $\ketbra{i}{j}$
 it follows $[\mathfrak{C}^{-1}(M)](X) = \mathcal{M}(X)$  for any $X$ by linearity. \qed
\end{Proof}
\begin{corollary}[Operator-sum representation]
Let $\mathcal{M}$ be in \\ $ \mathcal{L}(\mathcal{L}(\hilb{H}_0),\mathcal{L}(\hilb{H}_1))$;
then there exist $\{A_i|A_i \in  \mathcal{L}(\hilb{H}_0, \hilb{H}_1)\}$
and $\{B_i|B_i \in  \mathcal{L}(\hilb{H}_0, \hilb{H}_1)\}$ such that
\begin{align}
  \mathcal{M}(X) = \sum_i A_i X B_i^\dagger \qquad\
 \Tr[A_i^\dagger  A_j] = \lambda_i \delta_{ij}
 \qquad
  \Tr[B_i^\dagger  B_j] = \mu_i \delta_{ij}.
\end{align}
$\lambda_i, \mu_i \in \mathbb{R}$ and $ \delta_{ij}$ is the Kronecker delta.
\end{corollary}
\begin{Proof}
  Exploiting Th. \ref{th:Choijamiso}
we can write the action of $\mathcal{M}$ as
$\mathcal{M}(X) = \Tr_0[(I_1 \otimes (X_0)^T )M_{10}]$
where $M_{10}$ is the Choi operator of
$\mathcal{M}$.
Now consider the singular value decomposition of $M_{10}$,
$M = \sum_i\KetBra{A_i}{B_i}_{01}$, $\BraKet{A_i}{A_j}=\Tr[A_i^\dagger
A_j] = \lambda_i \delta_{ij}$,
$\BraKet{B_i}{B_j}=\Tr[B_i^\dagger B_j] = \mu_i \delta_{ij}$;
If we insert this decomposition into Eq. (\ref{eq:Choi-1})
we have
\begin{align}
 \mathcal{M}(X)& = \Tr_0[(I_1 \otimes (X_0)^T )\sum_i\KetBra{A_i}{B_i}] =  \nonumber \\
 &=\sum_i\Tr_0[(I_1 \otimes (X_0)^T )(I_1 \otimes (A_i)^T)\KetBra{I}{B_i}] = \nonumber \\
&= \sum_i\Tr_0[(I_1 \otimes (A_iX)^T_0)\KetBra{I}{B_i}]= \sum_i\Tr_0[\KetBra{I}{B_i} I_1 \otimes (A_iX)^T_0] = \nonumber  \\
&=\sum_i\Tr_0[\KetBra{I}{B_iX^\dagger A_i^\dagger}] =\sum_i\Tr_0[\KetBra{I}{I}(A_iXB^\dagger_i)_1 \otimes I_0] \nonumber \\
& = \sum_iA_iXB^\dagger_i
\end{align}
where we used Eq. (\ref{eq:doubleket}) and the cyclic property of the trace. \qed
\end{Proof}
Th. \ref{th:Choijamiso} provides an extremely powerful representation of linear maps
between operator spaces in terms of just one linear operator
acting on a bigger Hilbert space.
The following results will show how the properties of a linear map $\mathcal{M}$
translates into the properties of its Choi operator.

\begin{lemma}[Trace preserving condition] \label{lem:tpcond} 
Let $\mathcal{M}$ be in   $ \mathcal{L}(\mathcal{L}(\hilb{H}_0),\mathcal{L}(\hilb{H}_1))$
and $M \in \mathcal{L}(\hilb{H}_0 \otimes \hilb{H}_1)$ be its Choi
operator.
Then we have
\begin{align}
\Tr[\mathcal{M}(X)] = \Tr[X] \quad \forall X \in \mathcal{L}(\hilb{H}_0)
\;
\Leftrightarrow
\;
 \Tr_1[M_{01}] = I_0 \label{eq:tracepreserv}
\end{align}
\end{lemma}
\begin{Proof}
If we insert Eq. (\ref{eq:Choi-1}) into Eq.(\ref{eq:tracepreserv})
we get
\begin{align}
  \Tr[\mathcal{M}(X)] 
&= \Tr[(I_1 \otimes (X)^T )M_{10}] = 
 \Tr_0[(X)^T\Tr_1[M_{01}]] =
\nonumber \\
&= \Tr[X] 
= \Tr[X^T] \qquad \forall X \in \mathcal{L}(\hilb{H}_0) \label{eq:tracepreserv2}.
\end{align}
Since $\Tr[AB] = \Tr[A] \;\;\forall A \Leftrightarrow B = I$
we have that Eq. (\ref{eq:tracepreserv2}) holds if and only if $\Tr_1[M_{01}] = I_0$. \qed
\end{Proof}

\begin{lemma}[Hermitian preserving condition]
Let $\mathcal{M}$ be in \\ $ \mathcal{L}(\mathcal{L}(\hilb{H}_0),\mathcal{L}(\hilb{H}_1))$
and $M \in\mathcal{L}(\hilb{H}_1 \otimes \hilb{H}_1)$ be its Choi
operator.
Then we have
\begin{align}
\mathcal{M}(X)^\dagger =  \mathcal{M}(X^\dagger)
\;
\Leftrightarrow
\;
 M_{01}^\dagger = M_{01} \label{eq:hermitpreserv}
\end{align}
\end{lemma}
\begin{Proof}
If we take the adjoint in  Eq. (\ref{eq:Choi-1}) 
we have
\begin{align}
  \mathcal{M}(X)^\dagger= 
 \Tr_0[(X)^{\dagger T}M^\dagger_{01}] = 
\end{align}
If $M^\dagger= M $ clearly we have $\mathcal{M}(X)^\dagger = \mathcal{M}(X^\dagger)$.
On the other hand if \\
$ \Tr_0[(X)^{\dagger T}M^\dagger_{01}] = \Tr_0[(X)^{\dagger T}M_{01}] $ for all $X$
then $[\mathfrak{C}^{-1}(M^\dagger_{01})](X) = [\mathfrak{C}^{-1}(M_{01})](X)$
for all $X$ and so $\mathfrak{C}^{-1}(M_{01})=\mathfrak{C}^{-1}(M^\dagger_{01})$
that implies $M = M^\dagger$ \qed
\end{Proof}

\begin{lemma}[Completely-positive condition] \label{lem:cpcond}
Let $\mathcal{M}$ be in \\ $ \mathcal{L}(\mathcal{L}(\hilb{H}_0),\mathcal{L}(\hilb{H}_1))$
and $M \in \mathcal{L}(\hilb{H}_1 \otimes \hilb{H}_1)$ be its Choi
operator.
Then we have
\begin{align}
\mathcal{M} \otimes \mathcal{I}_2 (X) \geq 0 
\quad \forall X   \in \mathcal{L}(\hilb{H}_0\otimes \hilb{H}_2)
\;
\Leftrightarrow
\;
 M_{01} \geq 0  \label{eq:cppreserv}
\end{align}
Where  $\hilb{H}_2$  is  an Hilbert space
 of arbitrary dimension.
A linear map that satisfies condition (\ref{eq:cppreserv})
is called completely positive (CP).
\end{lemma}
\begin{Proof}
If $\mathcal{M} \otimes \mathcal{I}_2 (X) \geq 0 $
for all $X\in \mathcal{L}(\hilb{H}_0\otimes \hilb{H}_2)$
then clearly
$\mathcal{M} \otimes \mathcal{I}_0 (\KetBra{I}{I}) = M \geq 0$.
On the other hand, suppose that $M \geq 0$.
Then $M$ can be diagonalized in this way  $M= \sum_i\KetBra{A_i}{A_i}$
and  the operator-sum representation of $\mathcal{M}$
becomes
\begin{align}
 \mathcal{M}(X)= \sum_i A_i X A_i^\dagger \label{eq:krausform}.
\end{align}
If we introduce an auxiliary Hilbert space $\hilb{H}_2$
we have
\begin{align}
   \mathcal{M}\otimes \mathcal{I}_2 (X)= \sum_i (A_i \otimes I_2) X (A_i^\dagger  \otimes I_2) \geq0
\Leftrightarrow X \geq0
\end{align}
The operator-sum decomposition in Eq. (\ref{eq:krausform}) 
is called  canonical Kraus form. \qed
\end{Proof}

\subsubsection{The link product}

Given two linear maps $\mathcal{M} \in \mathcal{L}(\mathcal{L}(\hilb{H}_0),\mathcal{L}(\hilb{H}_1))$
and $\mathcal{N} \in \mathcal{L}(\mathcal{L}(\hilb{H}_1),\mathcal{L}(\hilb{H}_2))$
it is possible to consider the composition
\begin{align}
  \mathcal{C}:=\mathcal{N} \circ  \mathcal{M} : \mathcal{L}(\hilb{H}_0) \rightarrow \mathcal{L}(\hilb{H}_2)
\qquad
\mathcal{N} \circ  \mathcal{M}(X) := \mathcal{N}(\mathcal{M}(X))
 \quad \forall X \in \mathcal{L}(\hilb{H}_0).
\nonumber
\end{align}

Since we can represent $\mathcal{M}$ and $\mathcal{N}$
with the corresponding Choi operators $M$ and $N$,
it is reasonable to ask how the Choi operator $C$ of the composition 
$\mathcal{C}$ can be expressed in terms of $M$ and $N$.
Consider the action of $\mathcal{C}$ on an operator $X \in \mathcal{L}(\hilb{H}_0)$
\begin{align}
 \mathcal{C}(X) &= \Tr_1[(I_2 \otimes \Tr_0[(I_1 \otimes X_0^T)M]^T )N] = \nonumber \\
& = \Tr_0[(I_2 \otimes X_0^T) \Tr_1[(I_2 \otimes M_{01}^{T_1})(I_0 \otimes N_{12})]]; \label{eq:choicomp}
\end{align}
if we compare Eq. \ref{eq:choicomp} with Eq. \ref{eq:Choi-1}
we get
\begin{align}
  \mathfrak{C}(\mathcal{C})= \Tr_1[(I_2 \otimes M_{01}^{T_1})(I_0 \otimes N_{12})]=
N*M. \label{eq:choicomp2}
\end{align}
where we introduced the notation $N*M$ for the expression 
$\Tr_1[(I_2 \otimes M_{01}^{T_1})(I_0 \otimes N_{12})]$.

 If we consider maps such that their input and output spaces
are tensor product of Hilbert spaces it is possible to
compose these maps only through  some of these spaces.
For example if we have $\mathcal{M} \in 
\mathcal{L}(\mathcal{L}(\hilb{H}_0\otimes \hilb{H}_2),\mathcal{L}(\hilb{H}_1\otimes \hilb{H}_3))$
and
$\mathcal{N} \in 
\mathcal{L}(\mathcal{L}(\hilb{H}_3\otimes 
\hilb{H}_5),
\mathcal{L}(\hilb{H}_4\otimes \hilb{H}_6))$
it is possible to define the composition
\begin{align}
  \mathcal{N} \star \mathcal{M} := 
(\mathcal{N}\otimes \mathcal{I}_1) \circ (\mathcal{M} \otimes \mathcal{I}_5) \label{eq:compositionoftwomap}.
\end{align}
Following the same steps  as before we have that 
\begin{align}
  \mathcal{M} \in 
\mathcal{L}(\mathcal{L}(\hilb{H}_0\otimes \hilb{H}_2),\mathcal{L}(\hilb{H}_1\otimes \hilb{H}_3))
\leftrightarrow
 M \in \mathcal{L}(\hilb{H}_0\otimes \hilb{H}_2 \otimes \hilb{H}_1\otimes \hilb{H}_3)
\nonumber \\
\mathcal{N} \in 
\mathcal{L}(\mathcal{L}(\hilb{H}_3\otimes 
\hilb{H}_5),
\mathcal{L}(\hilb{H}_4\otimes \hilb{H}_6))
\leftrightarrow
N \in 
\mathcal{L}(\mathcal{L}(\hilb{H}_3\otimes 
\hilb{H}_5 \otimes \hilb{H}_4\otimes \hilb{H}_6))
\nonumber \\
\mathcal{N} \star \mathcal{M} \leftrightarrow N*M = \Tr_{3}[(I_{456} \otimes M_{0123}^{T_3})(I_{012} \otimes N)].
\end{align}
The above results  suggest us the following definition
\begin{Def}[Link product]\label{def:link}
  Let $M$ be an operator in $\mathcal{L}(\bigotimes_{i\in \defset{I}}\hilb{H}_i)$
and $N$ be an operator in $\mathcal{L}(\bigotimes_{j\in \defset{J}}\hilb{H}_j)$
where $\defset{I}$ and $\defset{J}$ are two finite set of indexes.
Then the \emph{link product} $N*M$ is an operator in
$\mathcal{L}(\hilb{H}_{\defset{I}\setminus \defset{J}}\otimes \hilb{H}_{\defset{J}\setminus \defset{I}} )$
defined as
\begin{align}
  N*M := \Tr_{\defset{I} \cap \defset{J}}[(I_{\defset{J}\setminus \defset{I}} \otimes M^{T_{\defset{I} \cap \defset{J}}})
(I_{\defset{I}\setminus \defset{J}} \otimes N) ]
\end{align}
where $\defset{A}\setminus \defset{B}:= \{i \in \defset{A} | i \notin  \defset{B}  \}$
and $\hilb{H}_{\defset{A}}:= \bigotimes_{i\in \defset{A}}\hilb{H}_i$
\end{Def}
\begin{remark}
It is worth noting that if $\defset{I} \cap \defset{J} = \emptyset$
we have $N*M = N \otimes M$ while if $\defset{I} = \defset{J}$
$N*M = \Tr[M^TN]$; 
\end{remark}

The previous discussion is summarized by the following theorem.
\begin{theorem}[Composition of linear maps]
Let $\defset{in}_{\mathcal{M}}, \defset{out}_{\mathcal{M}}, \defset{in}_{\mathcal{N}}, \defset{out}_{\mathcal{N}}$
be four sets of indeces such that $\defset{in}_{\mathcal{M}} \cap \; \defset{out}_{\mathcal{N}} = \emptyset$.\\
  Let $\mathcal{M}$ be map in 
$\mathcal{L}(\mathcal{L}(\bigotimes_{i\in \defset{in}_{\mathcal{M}}}\hilb{H}_i),
\mathcal{L}(\bigotimes_{j\in \defset{out}_{\mathcal{M}}}\hilb{H}_j)$,
$\mathcal{N}$ be map in \\
$\mathcal{L}(\mathcal{L}(\bigotimes_{i\in \defset{in}_{\mathcal{N}}}\hilb{H}_i),
\mathcal{L}(\bigotimes_{j\in \defset{out}_{\mathcal{N}}}\hilb{H}_j)$
and $M \in \mathcal{L}(\bigotimes_{n\in \defset{in}_{\mathcal{M}} \cup \; \defset{out}_{\mathcal{M}}}\hilb{H}_m)$,\\
$N \in \mathcal{L}(\bigotimes_{n\in \defset{in}_{\mathcal{N}} \cup \; \defset{out}_{\mathcal{N}}}\hilb{H}_n)$ 
be their respective Choi operators.
Then the Choi operator of the composition
\begin{align}
\mathcal{M} \star \mathcal{N} := 
(\mathcal{I}_{\defset{in}_{\mathcal{N}}\setminus (\defset{in}_{\mathcal{N}} \cap\; \defset{out}_{\mathcal{M}})}
\otimes
\mathcal{M})
\star
(\mathcal{I}_{\defset{out}_{\mathcal{M}}\setminus (\defset{out}_{\mathcal{M}} \cap\; \defset{in}_{\mathcal{N}})}
\otimes
\mathcal{N})
\end{align}
is given by
\begin{align}
\mathfrak{C}(\mathcal{M} \star \mathcal{N}) = M*N
\end{align}
\end{theorem}

We conclude this section with some properties of the link product

\begin{lemma}[Properties of link product]\label{lem:propertiesoflink}
    Let $M_1,M_2, M_3$  $M_4$
 be  operators in $\mathcal{L}(\bigotimes_{i \in \defset{I}_1}\hilb{H}_i)$,
 $\mathcal{L}(\bigotimes_{i\in \defset{I}_2}\hilb{H}_i)$,
 $\mathcal{L}(\bigotimes_{i\in \defset{I}_3}\hilb{H}_i)$
and $\mathcal{L}(\bigotimes_{i\in \defset{I}_4}\hilb{H}_i)$
respectively.
Then we have
\begin{itemize}
\item
If $N_{1}$ is an operator on $\mathcal{L}(\bigotimes_{i \in \defset{I}_1}\hilb{H}_i)$
 $(\alpha N_1 + \beta  M_1)*M_3 = \alpha (N_1*M_3) + \beta (M_1*M_3)$
for any $\alpha,\beta \in \mathbb{C}$
\item $M_1*M_2 = M_2*M_1$.
\item If $M_1^\dagger = M_1$ and $M_2^\dagger = M_2 $ then 
$(M_1*M_2)^\dagger = M_1*M_2$.
\item If $\defset{I}_1 \cap \defset{I}_2 \cap \defset{I}_3 = \emptyset$
then $(M_1*M_2)*M_3 = M_1*(M_2*M_3)$
\item 
If $M \geq 0 $ and $N \geq 0$ then 
$ M*N \geq 0 $.
\end{itemize}
\end{lemma}
\begin{Proof}
  The first four properties  trivially follow from the definition. 
To prove the last property consider the maps $\mathfrak{C}^{-1}(M)$
and  $\mathfrak{C}^{-1}(N)$. Since $M$ and $N$ are positive 
  $\mathfrak{C}^{-1}(M)$
and  $\mathfrak{C}^{-1}(N)$ are completely positive and also
$\mathfrak{C}^{-1}(M) \star \mathfrak{C}^{-1}(N)$ is completely positive. Then 
$\mathfrak{C}(\mathfrak{C}^{-1}(M) \star \mathfrak{C}^{-1}(N)) = M*N$ is positive.
\qed
\end{Proof}

\subsection{Diagrammatic representation of linear maps}

\begin{figure}[tb]
  \begin{center}
    \includegraphics[width=2cm ]{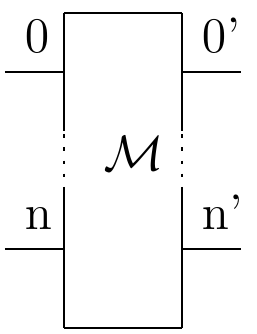}
    \caption{\label{fig:tesi-box} Pictorial representation of a linear
      map $\mathcal{M}: \mathcal{L}(\hilb{H}_0\otimes \cdots \otimes
      \hilb{H}_n) \rightarrow
      \mathcal{L}(\hilb{H}_{0'}\otimes\cdots \otimes \hilb{H}_{n'})$;
      the input wires are labelled according to the labeling of the
      Hilbert spaces.}
  \end{center}
\end{figure}

It is  useful to provide a pictorial representation of linear maps
and their composition.
We will sketch a linear map $\mathcal{M}:
\mathcal{L}(\hilb{H}_0\otimes \cdots \otimes \hilb{H}_n) \rightarrow
\mathcal{L}(\hilb{H}_{0'}\otimes\cdots \otimes \hilb{H}_{n'})$
 as box  with $n$ \emph{input wires} on the left $m$ \emph{output wires} on the right
as in Fig. \ref{fig:tesi-box}.

Using this representation the composition in Eq. (\ref{eq:compositionoftwomap})
can be sketched as follows
\begin{equation}
  \begin{aligned}
    \Qcircuit @C=2em @R=1.5em { \ustick{0} & \multigate{1}{\mathcal{M}} & \qw  & \ustick{1} \qw  & \qw   \\
      \ustick{2} & \ghost{\mathcal{M}} & \ustick{3}\qw &\multigate{1}{\mathcal{N}} & \ustick{4} \qw\\
      & \ustick{5}\qw & \qw&\ghost{\mathcal{N}} & \ustick{6} \qw}
  \end{aligned}\quad.
\end{equation}
or equivalently
\begin{equation}
  \begin{aligned}
    \Qcircuit @C=2em @R=1.5em {
      \ustick{0} & \multigate{1}{\mathcal{M}} & \ustick{1} \qw  & \ustick{5} &  \multigate{1}{\mathcal{N}} & \ustick{4} \qw   \\
      \ustick{2} & \ghost{\mathcal{M}} & \ustick{3} \qw &  \qw &\ghost{\mathcal{N}} & \ustick{6} \qw\\
    }
  \end{aligned}\quad.
\end{equation}

We do not draw wires corresponding to one dimensional Hilbert spaces.
We will sketch a map  $\mathcal{M}:\mathbb{C}\rightarrow \mathcal{L}(\mathcal{H}_0)$
 with a one dimensional input 
 as follows
\begin{equation}
\begin{aligned}
\Qcircuit @C=2em @R=1em {
  \prepareC{M} &\ustick{0}\qw
}
\end{aligned},
\end{equation}
where we use the label $M$ instead of $\mathcal{M}$.
In a similar way we represent  maps 
$\mathcal{M}: \mathcal{L}(\mathcal{H}_0)  \rightarrow \mathbb{C}$
that have one dimensional output space
\begin{equation}
\begin{aligned}
\Qcircuit @C=2em @R=2em 
{
\ustick{0}&\measureD{M}  \\
}
\end{aligned}\quad
\end{equation}


\subsection{States, Channels and POVMs}\label{sec:statechanpovm}
In the ordinary description of Quantum mechanics each physical system is
associated with a Hilbert space $\hilb{H}$ (that we will assume to be finite-dimensional)
and the states of the system are represented by positive operators with unit trace
 $ \rho \in \mathcal{L}(\hilb{H}) 
\rho \geq 0, \Tr[\rho]=1$.
Deterministic transformations of states are described by linear maps
 $\mathcal{C}: \mathcal{L}(\hilb{H}_0) \rightarrow \mathcal{L}(\hilb{H}_1)$
that have to be
\begin{itemize}
\item completely positive $\mathcal{C} \otimes \mathcal{I}_2(\rho) \geq 0  $
 for all $ \rho \in \mathcal{L}(\hilb{H}_0 \otimes \hilb{H}_2)$;
\item trace preserving $\Tr[\mathcal{C}(\rho)] = 1$ for all  $ \rho \in \mathcal{L}(\hilb{H}_0), \Tr[\rho]=1 $
\end{itemize}
 Deterministic transformations of states are called \emph{quantum channels}.
Thanks to Th. \ref{th:Choijamiso}
and lemmas \ref{lem:tpcond} and \ref{lem:cpcond} of the previous section 
we know that a quantum channel $\mathcal{C}  \in \mathcal{L}(\mathcal{L}(\hilb{H}_0),\mathcal{L}(\hilb{H}_1)) $
 can be represented
by its Choi operator $C \in \mathcal{L}(\hilb{H}_0 \otimes \hilb{H}_1)$
that  satisfies $C\geq0$ and $\Tr_0[C]= I_1 $
It is worth noting that the action $\mathcal{C}(\rho) = \Tr_0[(I_1 \otimes \rho^T)C]$
can be rewritten in terms of the link product as
\begin{align}
\mathcal{C}(\rho) = C *\rho \qquad
\begin{aligned}
\Qcircuit @C=2em @R=2em 
{
&\prepareC{\rho}& \gate{\mathcal{C}} & \qw \\
}
\end{aligned}\quad. \label{eq:applicationchoi}
\end{align}
where the state $\rho$ is interpreted as the Choi operator of a preparation device,
that is a channel 
$\tilde{\rho}$ from a one dimensional
Hilbert space to $\mathcal{L}(\hilb{H}_0)$
\begin{align}
&  \tilde{\rho}:\mathbb{C} \rightarrow \mathcal{L}(\hilb{H}_0) \qquad \tilde{\rho}(\lambda) = \lambda \rho \quad \forall
\lambda \in \mathbb{C} 
\qquad
\nonumber \\
&\mathfrak{C}(\tilde{\rho}) = (\tilde{\rho} \otimes \mathcal{I}_{\mathbb{C}}) (1 \otimes 1)=
\tilde{\rho}(1)= \rho  
\end{align}

Another relevant case is the one in which  the output space is one dimensional.
In this case we have a channel $\mathcal{C}:\mathcal{L}(\hilb{H}_0) \rightarrow  \mathbb{C}$
that receives a state $\rho$ as an input and outputs the normalization $\mathcal{C}(\rho)=\Tr[\rho]$;
it is easy to verify that its Choi operator is $\mathfrak{C}(\mathcal{C}) = I$
and so we have
\begin{align}
  \Tr[\rho] = \rho *I \qquad
 \begin{aligned}
\Qcircuit @C=2em @R=2em 
{
\prepareC{\rho}& \measureD{I} \\
}
\end{aligned}\quad. \label{eq:applicationchoi2}
\end{align}
We can then rewrite
the normalization condition  $\Tr_1[C_{01}] = I_0$
of the Choi operator of a quantum channel 
$\mathcal{C} \in \mathcal{L}(\mathcal{L}(\hilb{H}_0),\mathcal{L}(\hilb{H}_1))$
in the following way
\begin{align}\label{eq:traceaslink}
  C_{01}*I_1 = I_1
\end{align}

A relevant class of channels are the \emph{isometric channels}, that are defined as follows
\begin{align}
  \mathcal{V}:\mathcal{L}(\hilb{H}_0) \rightarrow \mathcal{L}(\hilb{H}_1)
\qquad 
\mathcal{V}(\rho) := V \rho V^\dagger \\
V \in \mathcal{L}(\hilb{H}_0,\hilb{H}_1), \quad V^\dagger V = I_0 .
\end{align}
The following theorem \cite{stinespring, JMP} states that every quantum channel can be realized
as an isometric channel on a larger system

\begin{theorem}[Stinespring dilation theorem]\label{th:stinedilation}
  Let $\mathcal{C} : \mathcal{L}(\hilb{H}_0) \rightarrow \mathcal{L}(\hilb{H}_1) $
be completely positive trace preserving linear map.
Then there exist an ancillary Hilbert space $\hilb{H}_{A}$ and an isometry
$V: \mathcal{L}(\hilb{H}_0) \rightarrow \mathcal{L}(\hilb{H}_1 \otimes \hilb{H}_{A}) $, $V^\dagger V= I_0$
such that
\begin{align}
\mathcal{C}(\rho) = \Tr_A[\mathcal{V}(\rho)] = \Tr_{A}[V \rho V^\dagger] \qquad 
\begin{aligned}
\Qcircuit @C=0.7em @R=1em 
{
\prepareC{\rho}& \gate{\mathcal{C}}& \qw \\
}
\end{aligned}
\; = \;
\begin{aligned}
\Qcircuit @C=0.7em @R=1em 
{
\prepareC{\rho}& \multigate{1}{\mathcal{V}}& \qw \\
&\pureghost{\mathcal{V}} &  \measureD{I}
}
\end{aligned}
\end{align}
$V$is called \emph{Stinespring dilation} of the channel $\mathcal{C}$
\end{theorem}
\begin{Proof}
Let $C$ be the Choi Jamio\l kowsky operator of $\mathcal{C}$ and
 define $\hilb{H}_{A} = \supp(C^*_{0'1'})$
(we introduced two auxiliary Hilbert spaces $\hilb{H}_{0'}$ and $\hilb{H}_{1'}$
and defined $C_{0'1'}$ according to Eq. (\ref{eq:relabeled})).
Now consider the operator
\begin{align}
  V:  \hilb{H}_0 \rightarrow \hilb{H}_1 \otimes \hilb{H}_{A}
\qquad V := I_1 \otimes C_{0'1'}^{\frac12*} \Ket{I}_{11'} T_{0 \rightarrow 0'} \label{eq:minimalstine}
\end{align}
where $T_{0 \rightarrow 0'} = \sum_k\ket{k}_{0'}\ket{k}_0$;
By using Lemmas \ref{lem:tpcond} and 
\ref{lem:cpcond} together with Eqs. (\ref{eq:dketid1}, \ref{eq:dketid2}, \ref{eq:dketid3})
 it is easy to  verify that $V$ is an isometry
\begin{align}
  V^\dagger V &=
 T_{0' \rightarrow 0}   \Bra{I}_{11'}
   (I_1 \otimes  C_{0'1'}^{\frac12 T}) 
   (I_1 \otimes C_{0'1'}^{\frac12*} 
\Ket{I}_{11'} T_{0 \rightarrow 0'}= \nonumber\\ 
&= T_{0' \rightarrow 0} \Tr_{1'}[C^T_{0'1'}] T_{0 \rightarrow 0'} = I_0,
\end{align}
and that
\begin{align}
  \Tr_{A}[V \rho V^\dagger] &= 
  \Tr_{A}[(I_1 \otimes C_{0'1'}^{\frac12*}) (\Ket{I}_{11'} T_{0 \rightarrow 0'} )
\rho
 (T_{0' \rightarrow 0}   \Bra{I}_{11'})
   (I_1 \otimes  C_{0'1'}^{\frac12 T}) ] = \nonumber \\
&=\Tr_{0'1'}[(I_1 \otimes C_{0'1'}^{T})(I_0' \otimes \KetBra{I}{I}_{11'} ) ( I_{11'}\otimes \rho_0') ]= \nonumber \\
&=
\Tr_{0'}[\Tr_{1'}[(I_1 \otimes C_{0'1'}^{T})(I_0' \otimes \KetBra{I}{I}_{11'} )] ( I_{1}\otimes \rho_0')] =
\nonumber \\ 
&=\Tr_{0'}[C_{0'1}^{T_{0'}} ( I_{1}\otimes \rho_0')] = C*\rho = \mathcal{C}(\rho) & \mbox{\qed} \nonumber
\end{align}
\end{Proof}
\begin{remark}
  The Stinespring dilation of a channel is generally non unique.  We now
  prove that the isometric dilation given by Eq.
  (\ref{eq:minimalstine}) has minimum ancilla dimension.  Suppose that
  there exists an isometric dilation $W: \hilb{H}_0 \rightarrow
  \hilb{H}_1 \otimes \hilb{H}_{B}$ such that $d_B=\dim(\hilb{H}_B) <
  \dim(\hilb{H}_A) = d_A$.  Each isometric dilation of a channel
  $\mathcal{C}$ provides operator-sum representation of
  $\mathcal{C}$ as follows $\mathcal{C}(\rho) = \Tr_A[W \rho W^\dagger] =
  \sum_{n=1}^{d_B} \bra{n}W \rho W^\dagger\ket{n} := \sum_{n=1}^{d_B}
  K_n \rho K^\dagger_n$ where $\ket{n}$ is an orthonormal basis in
  $\hilb{H}_B$.  From the operator sum representation it is possible
  to recover the Choi operator $C$ of $\mathcal{C}$ as follows
  $\mathfrak{C}(\mathcal{C}) = \sum_{n=1}^{d_B}(K_n \otimes I)
  \KetBra{I}{I}(K^\dagger_n \otimes I)= \sum_{n=1}^{d_B}
  \KetBra{K_n}{K_n}$.  Since $\BraKet{K_n}{K_m} = \bra{n}V^\dagger V
  \ket{m}=\braket{n}{m} = \delta_{nm} $, the vectors $\Ket{K_n}$ are
  linearly independent and this leads to the contradiction $\dim(\supp
  (C)) = \dim(\Span\{ \Ket{K_n} \}) = d_B < d_A =\dim(\supp(C^*)) =
  \dim(\supp(C))$.  The isometric dilation defined in Eq.
  (\ref{eq:minimalstine}) is called \emph{minimal Stinespring
    dilation}
\end{remark}

The probabilistic counterpart of a quantum channel is the \emph{quantum operation}.
A quantum operation  is a completely positive linear map  
$\mathcal{E} \in \mathcal{L}(\mathcal{L}(\hilb{H}_0),\mathcal{L}(\hilb{H}_1)) $ which is 
\emph{trace non-increasing} $\Tr[\mathcal{E}(\rho)] \leq 1$ for any state $\rho$.
The Choi Jamio\l kowski  operator $E$ of a quantum operation $\mathcal{E}$ satisfies the condition
$ E \leq \overline{E}$ where $\overline{E}$ is the Choi operator of a 
quantum channel.
A set  of quantum operation  $\{ \mathcal{E}_i \}$ that sum up to a channel $\mathcal{C}$
 is called a \emph{Quantum Instrument}\footnote{For simplicity we restricted ourselves
to the case of a finite number of outcomes. The generalization to an arbitrary
outcome space $\Omega$ can be obtained by defining a measure $\mathcal{E}_{B}$ that
associate to any event $B \subseteq \Omega$ a quantum operation $\mathcal{E}_{B}$
such that $\mathcal{E}_{\Omega}$ is a Quantum channel.}
and it is represented by a  set of positive operator $E_i $ such that $\sum_i E_i= C $;
the  index $i$ labels the possible classical outcomes of the instrument.
The action of a Quantum Instrument  is written as
\begin{align}
  \sum_i\mathcal{E}_i(\rho) = \sum_i\rho *E_i \qquad \qquad
 \begin{aligned}
\Qcircuit @C=2em @R=2em 
{
\prepareC{\rho}& \gate{\mathcal{E}_i}& \qw \\
}
\end{aligned}\quad. \label{eq:applicationinst}
\end{align}
and the probability that the Quantum Operation $\mathcal{E}_i$ takes place
is $p_i = \Tr[(I \otimes \rho)E_i ].$
A Quantum Instrument with one-dimensional output space is called \emph{POVM}
and is represented by a set of positive operator $P_i$ such that $\sum_i P_i = I$;
the elements $P_i$ of a POVM are called \emph{effects}.
The link product
\begin{align}
\Tr[\rho P_i^T] = \rho * P_i \qquad \qquad
 \begin{aligned}
\Qcircuit @C=2em @R=2em 
{
\prepareC{\rho}& \measureD{P_i}\\
}
\end{aligned}\quad. \label{eq:applicationinst2}
\end{align}
gives the probability $p_i$ of the outcome $i$ 
and coincides with the 
 usual Born rule
\begin{align}
  p_i = \Tr[\rho P_i]
\end{align}
if we make the substitution $P_i \leftrightarrow P^T_i$.
We conclude this section with a  theorem \cite{ozawadilationtheorem, JMP}
  that provides a realization scheme for  Quantum Instruments
in terms of a deterministic evolution on a bigger system followed by a  measurement
on the ancilla.
\begin{theorem}[Realization of Quantum Instruments]\label{th:realinst}
Let $\{ \mathcal{E}_i \}$, \\
$\mathcal{E}_i \in \mathcal{L}(\mathcal{L}(\hilb{H}_0),\mathcal{L}(\hilb{H}_1))$
 be a Quantum instrument.
 Then there exist an Hilbert space  $\hilb{H}_A$,  a channel $\mathcal{C} \in \mathcal{L}(\mathcal{L}(\hilb{H}_0),
\mathcal{L}(\hilb{H}_1 \otimes \hilb{H}_A))$ and a POVM $\{ P_i \}$, $P_i \in \mathcal{L}( \hilb{H}_A)$
such that
\begin{align}
  \mathcal{E}_i(\rho) = \Tr_A[\mathcal{C}(\rho)(I_1 \otimes P_i) ]
\qquad
\begin{aligned}
\Qcircuit @C=0.7em @R=1em 
{
\prepareC{\rho}& \gate{\mathcal{E}_i}& \qw \\
}
\end{aligned}
\; = \;
\begin{aligned}
\Qcircuit @C=0.7em @R=1em 
{
\prepareC{\rho}& \multigate{1}{\mathcal{C}}& \qw \\
&\pureghost{\mathcal{C}} &  \measureD{P_i}
}
\end{aligned}
\end{align}
\end{theorem}
\begin{Proof}
  Let us define $\mathcal{C} := \sum_i \mathcal{E}_i $ and 
let $C $ be the Choi operator of $\mathcal{C}$
and $E_i$ be the Choi operator of $\mathcal{E}_i$.
Since $\mathcal{C}$ is a quantum channel, we can consider its minimal Stinespring
dilation   $V:\hilb{H}_0 \rightarrow \hilb{H}_1 \otimes \hilb{H}_A$,\\
$\hilb{H}_A = \supp(C)$.
Now we introduce the POVM $\{P_i \in \mathcal{L}(\hilb{H}_A) \}$
$P_i= C^{-\frac12 T } E_i^{T} C^{-\frac12 T}$
(clearly $\sum_i E_i = I_A $ and $P_i^\dagger = P_i$).
It is easy to verify that
\begin{align}
\Tr_A[\mathcal{C}(\rho) (I_1 \otimes P_i)] &=   \Tr_A[V \rho V^\dagger (I_1 \otimes P_i) ] =  \nonumber \\ 
&=\Tr_A[(I_1 \otimes C_{0'1'}^{\frac12*})(\rho_{0'} \KetBra{I}{I}_{11'})
(I_1 \otimes C_{0'1'}^{\frac12 T})(I_i \otimes P_i)]  =\nonumber  \\
&=\Tr_A[(\rho_{0'} \KetBra{I}{I}_{11'})(I_1 \otimes  E_i^T )] = \mathcal{E}_i(\rho)
\end{align}
\end{Proof}

\subsection{Quantum Networks: constructive approach}
\label{section:construct}
In this section we introduce the formal definition of Quantum Network.
Within our approach a Quantum Network
is obtained by assembling elementary circuits linking outputs
of a circuit
to inputs of another circuit;
we consider ``elementary circuits'' channels, quantum operations, effects or state preparations
each of them represented with the corresponding linear map.
The restriction that we can connect only outputs with inputs and that
we cannot have closed loops ensures causality (see Remark \ref{rem:loop}) and motivates the following
definition


\begin{Def}[Quantum Network]
A \emph{quantum network} $\mathcal{R}$
is a linear map corresponding to directed acyclic graph (DAG)
in which
\begin{itemize}
\item each arrow is labeled with a non negative integer number $n$ (two different arrows cannot have the same label);
\item an arrow with label $n$ represents an Hilbert space $\hilb{H}_n$;
\item each vertex is labelled with a non negative integer number $i$ (two different vertexes cannot have the same label);
 \item each vertex $i$ represents a completely-positive trace non-increasing map
$\mathcal{C}_i \in
\mathcal{L}(\hilb{H}_{\defset{in}_i}\otimes \hilb{H}_{\defset{out}_i} )$ 
($\hilb{H}_{\defset{A}}= \bigotimes_{k \in \defset{A}}\hilb{H}_k$)
where $\defset{in}_i$ is the set of incoming arrows at vertex $i$
and $\defset{out}_i$ is the set of outgoing arrows at vertex $i$;
\item an arrow between two vertices's $i$ and $j$ corresponds to  the composition $\mathcal{C}_j\star\mathcal{C}_i$
of the  linear maps $\mathcal{C}_i$ and $\mathcal{C}_j$
\item we remove some vertices's with no incoming arrows (sources) and
  some vertices's
with no outgoing arrows (sink). The free incoming arrows remaining represent
input systems entering the network while the free outgoing arrows carry the output systems.
\end{itemize}
If $\mathcal{C}_i$ is a channel for each vertex $i$
$\mathcal{R}$ is called a \emph{deterministic quantum network}.
If $\mathcal{C}_i$ is a trace decreasing for some vertex $i$,
$\mathcal{R}$ is called a \emph{probabilistic quantum network}.
\end{Def}

\begin{figure}[tb]
\begin{center}
\includegraphics[width=12cm ]{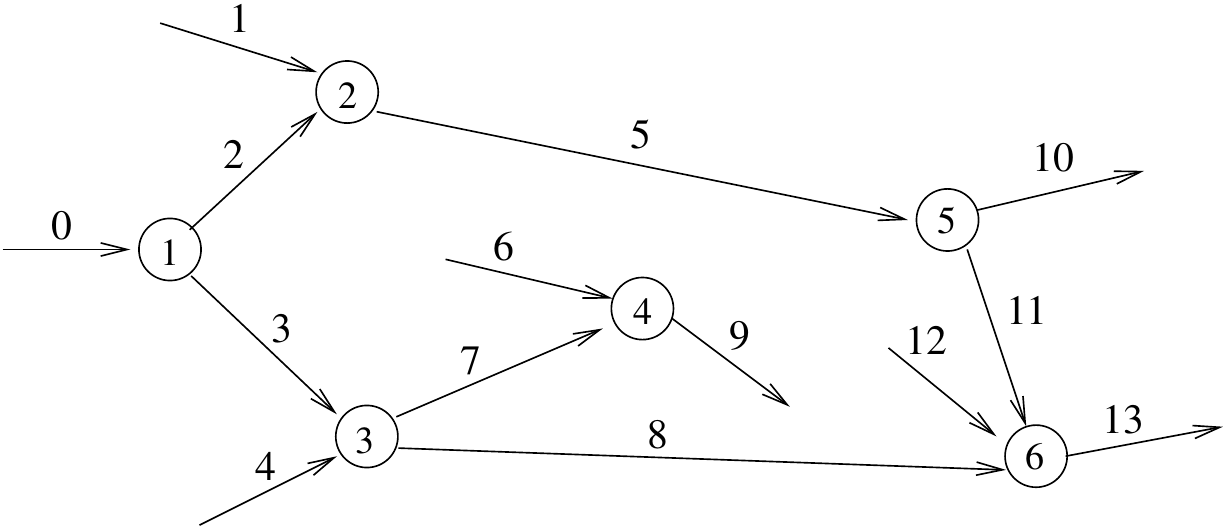}
  \caption{\label{fig:DAG} Graphical representation of a quantum network. The directions of the arrows represent
the flow of quantum information in the network, that is quantum systems travelling
from a vertex to another. Free incoming arrows represent input systems entering the network
while free outgoing arrows represent output systems of the network.}
\end{center}
\end{figure}

Fig. \ref{fig:DAG} provides a typical example of a  quantum network.

\begin{remark}\label{rem:equivalenceclasses}
It is worth noting that the same Quantum Network 
can be realized in different ways as a sequence of  maps
\begin{align}\label{eq:equivalenceofnetwork}
    \mathcal{R}^{(N)} = \mathcal{C}_1 \star \mathcal{C}_2 \star \cdots \star \mathcal{C}_N= 
\mathcal{C}'_1 \star \mathcal{C}'_2 \star \cdots \star \mathcal{C}'_N
\end{align}
and this fact reflects different possible physical implementation of the same network.
In this work we are not interested  in the inner structure of a network
but only in its properties as a linear map from input spaces to output spaces.
Because of this, whenever we introduce a Quantum Network $\mathcal{R}^{(N)}$,
we actually  mean an equivalence class of sequence of maps
that give the same overall operator $\mathcal{R}^{(N)}$, i.e. we consider the two sequences of maps
$\mathcal{C}_1 \star \mathcal{C}_2 \star \cdots \star \mathcal{C}_N$
and
$\mathcal{C}'_1 \star \mathcal{C}'_2 \star \cdots \star \mathcal{C}'_N$
in Eq. \ref{eq:equivalenceofnetwork} as the same object.
\end{remark}

\begin{remark}\label{rem:loop}
  The condition that the graph is acyclic means that no closed path is allowed.
This requirement ensures that causality is preserved, since the flow of quantum information
induces a causal order inside the network and a closed path would correspond to a time-loop.
It is worth stressing that in our representation
a  physical closed loop in the lab, that is taking the output of a device and then 
sending it as  an input to the same device,
 corresponds to many uses of the same transformation
\begin{equation}
\begin{aligned}
\Qcircuit @C=1.5em @R=1em {
&\qw &\qw \\
\qwx& \gate{\mathcal{C}} & \qw \qwx\\
}
\end{aligned}
\quad \rightarrow \quad
\begin{aligned}
\Qcircuit @C=1.5em @R=1em {
 & \gate{\mathcal{C}} & \gate{\mathcal{C}}& \qw &\dots & & \gate{\mathcal{C}}&\qw \\
}
\end{aligned}.
\end{equation}
In this work we use the convention that a vertex in a network or a box in a circuit represents
a single use of a physical device.
\end{remark}

\begin{figure}[tb]
\begin{center}
\includegraphics[width=12cm ]{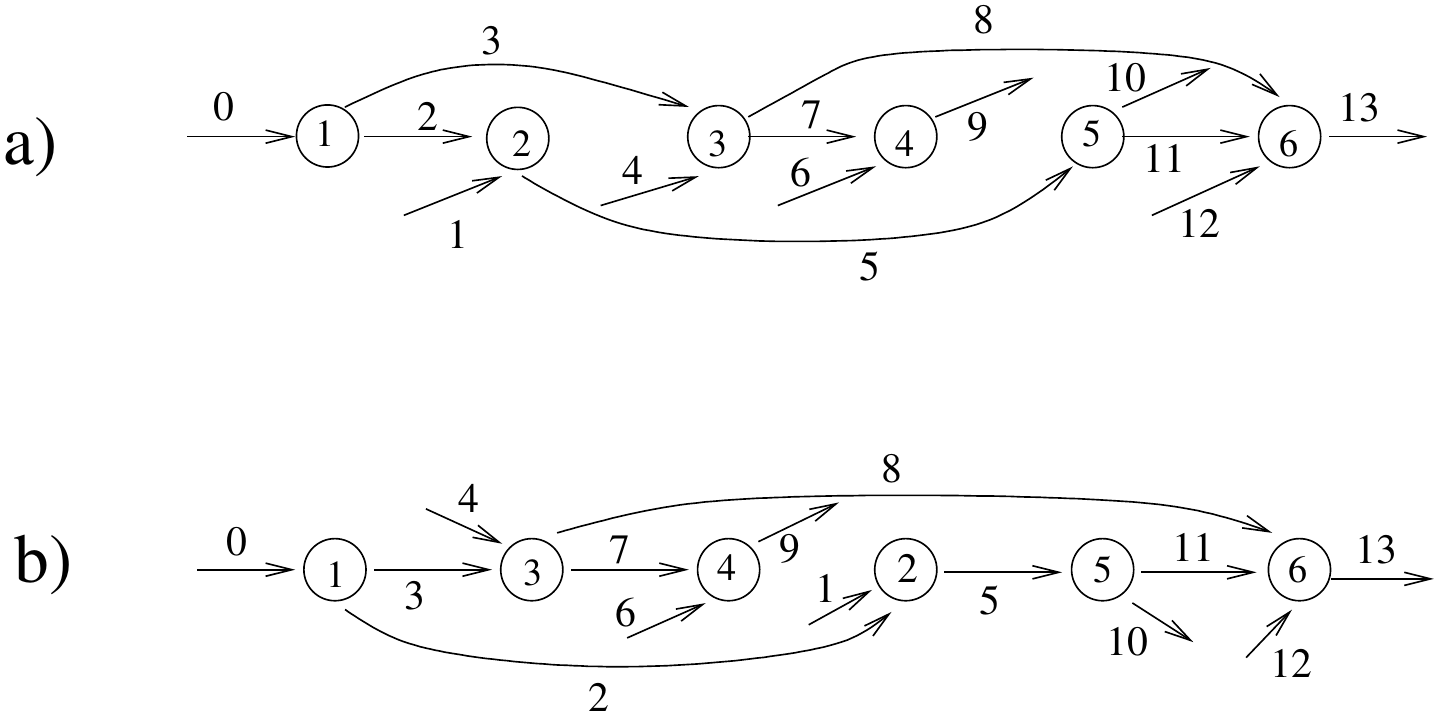}
  \caption{\label{fig:DAG2} Two possible total orderings of the network in Fig. \ref{fig:DAG}.}
\end{center}
\end{figure}

Any direct acyclic graph is naturally endowed with a partial ordering $\preceq$
among the vertices's, which is the causal ordering induced by the flow of quantum information
(see Remark \ref{rem:loop});
 we say that vertex $i$ causally precedes vertex $j$ ($i \preceq j$ )
if there exists a directed path 
from $i$ to $j$.
It is possible to prove that for a directed acyclic graph the partial ordering
$\preceq$ can be extended, in a generally
 non unique way, to a total ordering $\leq$ (See Fig. \ref{fig:DAG2}).

Each vertex in the network corresponds to a step of a computation and
the relation $i \preceq j$
means that step $j$ cannot be performed before step $i$.
If two vertexes are incomparable this means that the two steps can be run in parallel;
extending the partial ordering to a total ordering consists in
arbitrarily fixing
an ordering among parallel computational steps that is compatible with
the partial ordering $\preceq$.

Since each vertex $i$ in a quantum network corresponds to a linear map $\mathcal{C}_i$  and 
any arrow between two vertexes corresponds to a composition, we can
 exploit the diagrammatic representation that
 we introduced before and represent a quantum network in a circuit form
\begin{equation}
\begin{aligned} \label{circ:example}
\Qcircuit @C=1.5em @R=1em {
\ustick{0}&\multigate{2}{\mathcal{C}_1}&\ustick{1}& \multigate{1}{\mathcal{C}_2}
&\ustick{4}&\multigate{2}{\mathcal{C}_3} &\ustick{6}& \multigate{1}{\mathcal{C}_4} & \ustick{9}\qw & \pureghost{\mathcal{C}_5} & \ustick{10}\qw &\ustick{12} 
&\multigate{2}{\mathcal{C}_6}&\ustick{13} \qw\\
&\pureghost{\mathcal{C}_1}&\ustick{2} \qw& \ghost{\mathcal{C}_2}&\qw& \pureghost{\mathcal{C}_3}&\ustick{7}\qw& \ghost{\mathcal{C}_4}& 
&\multigate{-1}{\mathcal{C}_5}  & \ustick{11}\qw &  \qw &\ghost{\mathcal{C}_6}\\
&\pureghost{\mathcal{C}_1}&\qw& \ustick{3} \qw&\qw \qwx &  \ghost{\mathcal{C}_3} &\qw& \ustick{8} \qw  &  \qw \qwx& \qw& \qw&\qw &\ghost{\mathcal{C}_6}\\
&&&&\qwx&\qw&\ustick{5}\qw& \qw& \qw \qwx\\
}
\end{aligned}\quad;
\end{equation}
where the free incoming/outgoing arrows are now  substituted
by free input/output wires;
The flow of quantum information is from left  to right and the numbering of the boxes
is chosen accordingly.

To avoid drawing crossing  wires, it is possible to enlarge each box
by tensoring with the identity map i.e.
\begin{equation}
\begin{aligned} \label{circ:enlargeboxes}
\Qcircuit @C=1.5em @R=1em {
\ustick{n}&\gate{\mathcal{C}_i}& \ustick{m}\qw\\
\ustick{k} &  \qw& \qw\\
}
\end{aligned}
\quad=\quad
\begin{aligned}
\Qcircuit @C=1.5em @R=1em {
\ustick{n}&\multigate{1}{\mathcal{C}_i}& \ustick{m}\qw\\
\ustick{k} & \ghost{\mathcal{C}_i} & \ustick{k} \qw\\
}
\end{aligned}
\qquad \qquad
\mathcal{C}_i \rightarrow \mathcal{C}_i \otimes \mathcal{I}_k
\end{equation}
 in this way
the network takes the shape of a chain
\begin{equation}
\begin{aligned} \label{circ:example2}
\Qcircuit @C=1.5em @R=1em {
\ustick{0} &\multigate{2}{\mathcal{C}_1}& \ustick{1}  &  \multigate{2}{\mathcal{C}_2}& \ustick{4}   & \multigate{3}{\mathcal{C}_3}&
\ustick{6} & \multigate{3}{\mathcal{C}_4} &\ustick{9} \qw & &\pureghost{\mathcal{C}_5}& \ustick{10} \qw &
\ustick{12} & \multigate{2}{\mathcal{C}_6}& \ustick{13}  \qw\\
&\pureghost{\mathcal{C}_1}& \ustick{2} \qw& \ghost{\mathcal{C}_2}  & \ustick{5} \qw
&  \ghost{\mathcal{C}_3}&\ustick{5} \qw& \ghost{\mathcal{C}_4} &\qw &\ustick{5} \qw&\ghost{\mathcal{C}_5} &\qw&\ustick{11} \qw&\ghost{\mathcal{C}_6} \\
&\pureghost{\mathcal{C}_1}&\ustick{3}  \qw&  
\ghost{\mathcal{C}_2} & \ustick{3} \qw&   \ghost{\mathcal{C}_3}& \ustick{8}  \qw&  \ghost{\mathcal{C}_4}&  \qw  
&\ustick{8} \qw&\multigate{-2}{\mathcal{C}_5}&\qw&\ustick{8} \qw&\ghost{\mathcal{C}_6}\\
&&& &  & \pureghost{\mathcal{C}_3} &\ustick{7}\qw&   \ghost{\mathcal{C}_4}& &  &&\\
}
\end{aligned}\quad;
\end{equation}
we can further lighten the diagram by merging the internal wires connecting two boxes
\begin{align}
\begin{aligned} \label{circ:mergingwires}
\Qcircuit @C=1em @R=0.7em {
&\multigate{3}{\mathcal{C}_i}&\qw&  &  &  \multigate{3}{\mathcal{C}_{i+1}} & \qw\\
& \ghost{\mathcal{C}_i}& \qw& \ustick{n} \qw &\qw & \ghost{\mathcal{C}_{i+1}}&\qw \\
& \ghost{\mathcal{C}_i}& \qw&\ustick{m} \qw  & \qw& \ghost{\mathcal{C}_{i+1}}&\qw \\
& \ghost{\mathcal{C}_i}& \qw& \ustick{l}\qw & \qw&  \ghost{\mathcal{C}_{i+1}}&\qw \\
}
\end{aligned}
\quad=\quad
\begin{aligned} 
\Qcircuit @C=1em @R=0.7em {
&\multigate{3}{\mathcal{C}_i}&\qw&  &  &  \multigate{3}{\mathcal{C}_{i+1}} & \qw\\
& \ghost{\mathcal{C}_i}& &  & & \pureghost{\mathcal{C}_{i+1}}&\qw \\
& \ghost{\mathcal{C}_i}& \qw& \ustick{A_i} \qw  & \qw& \ghost{\mathcal{C}_{i+1}}&\qw \\
& \ghost{\mathcal{C}_i}&&  & &  \pureghost{\mathcal{C}_{i+1}}&\qw \\
}
\end{aligned}
\end{align}
\begin{align}
\hilb{H}_{A_i} =  \hilb{H}_n \otimes \hilb{H}_m \otimes \hilb{H}_l.
  \nonumber
\end{align}
In this way the circuit \ref{circ:example2}
becomes

\begin{equation}
\begin{aligned} \label{circ:example3}
\Qcircuit @C=1.5em @R=1.5em {
\ustick{0}&\multigate{1}{\mathcal{C}_1}& \ustick{1}&  
 \multigate{1}{\mathcal{C}_2} & \ustick{4}&   \multigate{1}{\mathcal{C}_3} &  &\ustick{6}&
\multigate{1}{\mathcal{C}_4}&\ustick{9}\qw&& \pureghost{\mathcal{C}_5}&\ustick{10} \qw &\ustick{12}&\multigate{1}{\mathcal{C}_6}&\ustick{13}\qw\\
& \pureghost{\mathcal{C}_1}& \ustick{A_1} \qw& \ghost{\mathcal{C}_1}  &
\ustick{A_2}\qw& \ghost{\mathcal{C}_1}& \ustick{A_3}\qw&\qw& \ghost{\mathcal{C}_4}&\ustick{A_4}\qw&\qw &\multigate{-1}{\mathcal{C}_5} & \ustick{A_5}\qw&\qw&
\ghost{\mathcal{C}_6}\\
\\
}
\end{aligned}\quad.
\end{equation}
The previous considerations can be summarized in the following
\begin{lemma} [Circuit form for Quantum Networks] \label{lem:netcircuit}
 Any quantum network $\mathcal{R}$ with $N$ vertexes
is equivalent to a concatenation of $N$
 completely positive trace non increasing linear maps
\begin{align}\label{eq:networkascircuit}
&  \mathcal{R} = \mathcal{C}_1 \star \mathcal{C}_2 \star \cdots \star \mathcal{C}_N
\end{align}
\begin{align}
&\begin{aligned}
\Qcircuit @C=1.5em @R=1em {
\ustick{0}&\multigate{1}{\mathcal{C}_1}& \ustick{1} \qw & \ustick{2}&  
 \multigate{1}{\mathcal{C}_2} &  \ustick{3} \qw \\
& \pureghost{\mathcal{C}_1}& \ustick{A_1} \qw&\qw& \ghost{\mathcal{C}_1}  &
\ustick{A_2}\qw\\
}
\end{aligned}
\qquad\cdots\qquad
\begin{aligned}
\Qcircuit @C=1.5em @R=1em {
\ustick{2N-2}&   \multigate{1}{\mathcal{C}_N} &  \ustick{2N-1} \qw \\
 \ustick{A_{N-1}} & \ghost{\mathcal{C}_N}  &
\\
}
\end{aligned}
\nonumber
\end{align}
where $\mathcal{C}_i: \mathcal{L}(\hilb{H}_a\otimes \hilb{H}_{A_{i-1}}) 
\rightarrow \mathcal{L}(\hilb{H}_b\otimes \hilb{H}_{A_{i}})$.
\end{lemma}
\begin{remark} \label{rem:netlabeling}
  In Eq. (\ref{eq:networkascircuit})
we  chose to attach one free incoming and  one free outgoing wire 
to each map $\mathcal{C}_i$.
This is our standard representation of quantum network; we can without loss of generality
sketch any quantum network in this way, since network in which
some input/output wires are missing (like in  \ref{circ:example3})
are just special cases.
We can stress, if present, a tensor product structure $\mathcal{H}_a = \otimes_j\mathcal{H}_{a_{j}}$
of the Hilbert space carried by a free input/output wire $a$,
by  drawing as many wires as the number of factors in the tensor product,
 for example
\begin{equation}
\begin{aligned}  
\Qcircuit @C=1.5em @R=1em {
\ustick{0}&\multigate{1}{\mathcal{C}_1}& \ustick{1} \qw & \ustick{2}&  
 \multigate{1}{\mathcal{C}_2} &  \ustick{3} \qw \\
 & \pureghost{\mathcal{C}_1}& \ustick{A_1} \qw&\qw& \ghost{\mathcal{C}_1}  &
}   
\end{aligned}
\quad = \quad 
 \begin{aligned}
\Qcircuit @C=1.5em @R=1em {
\ustick{0_1}&\multigate{1}{\mathcal{C}_1}& \ustick{1} \qw & \ustick{2}&  
 \multigate{1}{\mathcal{C}_2} &  \ustick{3_1} \qw \\
\ustick{0_2} & \ghost{\mathcal{C}_1}& \ustick{A_1} \qw&\qw& \ghost{\mathcal{C}_1}  &
\ustick{3_2}\qw
} 
\end{aligned}
\nonumber \quad;
\end{equation}
where $\hilb{H}_0 = \hilb{H}_{0_1} \otimes \hilb{H}_{0_2}$
and
$\hilb{H}_3 = \hilb{H}_{3_1} \otimes \hilb{H}_{3_2}$.
We also
choosed to label the free input/output wires
with increasing integer numbers;
in this way the Hilbert spaces of the input wires are labeled with even numbers
while the output ones correspond to odd numbers.
We can define the overall input space of the network as $\hilb{H}_{{\rm in}} = \bigotimes_{i=1}^{N} \hilb{H}_{2i-2}$
and $\hilb{H}_{{\rm out}} = \bigotimes_{j=1}^N \hilb{H}_{2i-1}$
\end{remark}

Lemma \ref{lem:netcircuit} reveals the equivalence between a Quantum Network
and a sequence of $N$ channels with memory; if we
 stretch and rearrange the input 
and the output wires 

\begin{align}
\begin{aligned}
\Qcircuit @C=1em @R=1em 
{
\ustick{0}&\multigate{1}{\mathcal{C}_1}& \ustick{1} \qw & \ustick{2}&  
 \multigate{1}{\mathcal{C}_2} &  \ustick{3} \qw & \ustick{4} & \multigate{1}{\mathcal{C}_3} &  \ustick{5} \qw \\
 & \pureghost{\mathcal{C}_1}& \ustick{A_1} \qw&\qw& \ghost{\mathcal{C}_2}  &  \ustick{A_2}\qw
& \qw &  \ghost{\mathcal{C}_n} &   &
}
\end{aligned}
\quad = \quad
\begin{aligned}
\Qcircuit @C=1em @R=1em 
{
&\pureghost{\mathcal{C}_1}&  \qw &\qw &\qw & \qw & \ustick{1}\qw \\
\ustick{0} & \multigate{-1}{\mathcal{C}_1 } & \ustick{A_1} \qw  & \multigate{1}{\mathcal{C}_2} & \qw &\qw &\ustick{3} \qw\\
 \ustick{2} &\qw& \qw  & \ghost{\mathcal{C}_2} & \ustick{A_2} \qw & \multigate{1}{\mathcal{C}_3} & \ustick{5} \qw \\
\ustick{4} &\qw& \qw & \qw &\qw & \ghost{\mathcal{C}_3} \\
}
\end{aligned}
\nonumber
\end{align}
from a Quantum Network we get a sequence of memory channels from the
left side to the right side.  Since a Quantum Network is a sequence of
linear maps, it can be considered as a linear map from
$\mathcal{L}(\hilb{H}_{{\rm in}})$ to $\mathcal{L}(\hilb{H}_{{\rm
    out}})$.  It is then possible to define the Choi operator of a
Quantum Network
\begin{align}
  \mathfrak{C}(\mathcal{R}^{(N)})   = R^{(N)}, \label{eq:choiofnetwork}
\end{align}
where we add the superscript $^{(N)}$ to record the number of vertex in the network i.e.
$\mathcal{R}^{(N)}$ denotes a quantum network with $N$ vertexes.

\subsection{Deterministic Quantum Networks}
The main aim of the following sections will be to inspect the structure of the Choi operator
of a Quantum Network.
In this section we consider the deterministic case while the probabilistic case will be discussed in the next section.
Specializing Lemma \ref{lem:netcircuit} a deterministic Quantum Network $\mathcal{R}^{(N)}$  can be
presented as a concatenation of $N$ quantum channels $\mathcal{C}_i$; then
the Choi operator of $\mathcal{R}^{(N)}$ is given by the link product
of the $C_i$'s. This structure leads to a peculiar normalization constraint for
$R^{N}$.
\begin{theorem}[Normalization Condition]\label{th:normcond}
  Let $\mathcal{R}^{(N)}$ be a deterministic Quantum Network and $R^{N} \in \mathcal{L}(\bigotimes_{i=0}^{2N-1} \hilb{H}_i)$
(we use the labeling introduced in Lemma \ref{lem:netcircuit} and Remark \ref{rem:netlabeling})
be its Choi operator.
Then $R^{(N)} \geq 0$ and satisfies the following  condition
\begin{align}\label{eq:recnorm1}
  \Tr_{2N-1}\left[ R^{(N)} \right] = I_{2N} \otimes R^{(N-1)}
\end{align}
\begin{align}
\begin{aligned}
\Qcircuit @C=0.6em @R=0.5em {
&\multigate{1}{\mathcal{C}_1}& \qw & &  
 \multigate{1}{\mathcal{C}_2} & \qw \\
& \pureghost{\mathcal{C}_1}&  \qw&\qw& \ghost{\mathcal{C}_1}  &
\qw\\
}
\end{aligned}
\cdots
\begin{aligned}
  \Qcircuit @C=0.6em @R=0.5em {
&   \multigate{1}{\mathcal{C}_N} &\qw&   \measureD{I} \\
  & \ghost{\mathcal{C}_N}  &
\\
}
\end{aligned}
= 
\begin{aligned}
\Qcircuit @C=0.6em @R=0.5em {
&\multigate{1}{\mathcal{C}_1}&  \qw & &  
 \multigate{1}{\mathcal{C}_2} & \qw \\
& \pureghost{\mathcal{C}_1}&  \qw&\qw& \ghost{\mathcal{C}_1}  &
\qw\\
}
\end{aligned}
\;\; \cdots \;\;
\begin{aligned}
  \Qcircuit @C=0.6em @R=0.5em {
&   \multigate{1}{\overline{\mathcal{C}_{N-1}}} &   \qw  \\
  & \ghost{\overline{\mathcal{C}_{N-1}}}  &
\\
}
\end{aligned}
\otimes
\begin{aligned}
  \Qcircuit @C=0.6em @R=0.5em {
&\measureD{I} \\
}
\end{aligned}
\nonumber
\end{align}
where $R^{(N-1)} \in \mathcal{L}(\bigotimes_{i=0}^{2N-1} \hilb{H}_i)$ is the 
Choi operator of the reduced Quantum Network with $N-1$ vertexes
and $\overline{\mathcal{C}_{N-1}}$ is a quantum channel
such that $\mathfrak{C}(\overline{\mathcal{C}_{N-1}}):= \overline{C_{N-1}} = \Tr_{A_{N-1}}[C_{N-1}]$.
\end{theorem}
\begin{Proof}
Since $\mathcal{R}^{(N)}$  is a quantum Network with $N$ vertexes, we can express it
in terms of a concatenation of $N$ channels
\begin{align}
  &\mathcal{R}^{(N)} = \mathcal{C}_1 \star \mathcal{C}_2 \star \cdots \star\mathcal{C}_N \\
&\mathcal{C}_i : \mathcal{L}(\hilb{H}_{2i-2} \otimes \hilb{H}_{A_{i-1}} )
\rightarrow
\mathcal{L} (\hilb{H}_{2i-1} \otimes \hilb{H}_{A_{i}}) 
\qquad \hilb{H}_{A_{0}} \cong \hilb{H}_{A_{N}} \cong \mathbb{C} \nonumber.
\end{align}
Let $C_i \in \mathcal{L}(\bigotimes_{k \in \defset{I}_i}\hilb{H}_{k}$
be the Choi of $\mathcal{C}_i$
where we introduced the set $\defset{I}_i := \{  2i-2, A_{i-1}, 2i-1, A_{i} \}$;
we notice that $\defset{I}_i \cap \defset{I}_j \cap \defset{I}_k = \emptyset$
for all $i,j,k = 1, \dots, N$ and so, exploiting Lemma \ref{lem:propertiesoflink},
we have
\begin{align}
  R^{(N)} = C_1*C_2*\cdots*C_N.
\end{align}
Since $\mathcal{C}_N$ is channel in $\mathcal{L}(\mathcal{L}(\hilb{H}_{2N-2}\otimes \hilb{H}_{A_{N-1}}),
\mathcal{L}(\hilb{H}_{2N-1}))$  its Choi-Jamio\l kowsky operator satisfies
$\Tr_{2N-1}[C_N] = I_{2N-2} \otimes I_{A_{N-1}}$ then we have
\begin{align}
  \Tr_{2N-1}[R^{(N)}] &= C_1*C_2*\cdots* C_{N-1}*\Tr_{2N-1}[C_N] =
\nonumber \\ 
&=C_1*C_2*\cdots*(C_{N-1}*I_{2N-2} \otimes I_{A_{N-1}} ) =\nonumber \\ 
&=
C_1*C_2*\cdots* \Tr_{A_{N-1}}[C_{N-1}] \otimes I_{2N-2}  =\nonumber \\ 
&=
C_1*C_2*\cdots* \overline{C_{N-1}} \otimes I_{2N-2}  =\nonumber \\ 
&= R^{N-1}\otimes I_{2N-2}
\end{align}

\begin{align*}
  \begin{aligned}
\Qcircuit @C=0.6em @R=0.5em {
&\multigate{1}{\mathcal{C}_1}&  \qw  \\
& \pureghost{\mathcal{C}_1}&  \qw \\
}
\end{aligned}
\cdots
\begin{aligned}
  \Qcircuit @C=0.6em @R=0.5em {
&\multigate{1}{\mathcal{C}_{N-1}}&\qw&&   \multigate{1}{\mathcal{C}_N} &   \measureD{I} \\
 &\ghost{\mathcal{C}_{N-1}}&\qw&\qw& \ghost{\mathcal{C}_N}  &
\\
}
\end{aligned}
&=
  \begin{aligned}
\Qcircuit @C=0.6em @R=0.5em {
&\multigate{1}{\mathcal{C}_1}&  \qw \\
& \pureghost{\mathcal{C}_1}&  \qw\\
}
\end{aligned}
\cdots
\begin{aligned}
  \Qcircuit @C=0.6em @R=0.5em {
&\multigate{1}{\mathcal{C}_{N-1}}&\qw&&&  \measureD{I} \\
 &\ghost{\mathcal{C}_{N-1}}&\measureD{I}  &
\\
}
\end{aligned}
= \nonumber \\
&=
  \begin{aligned}
\Qcircuit @C=0.6em @R=0.5em {
&\multigate{1}{\mathcal{C}_1}&  \qw \\
& \pureghost{\mathcal{C}_1}&  \qw\\
}
\end{aligned}
\cdots
\begin{aligned}
  \Qcircuit @C=0.6em @R=0.5em {
&\multigate{1}{\overline{\mathcal{C}_{N-1}}}&\qw&& & \measureD{I} \\
 &\ghost{\overline{\mathcal{C}_{N-1}}}&&
\\
}
\end{aligned}
\end{align*}
\qed
\end{Proof}
\begin{corollary}
  Let
 $R^{N} \in \mathcal{L}(\hilb{H}_{{\rm out}} \otimes \hilb{H}_{{\rm in}})$ 
( $\hilb{H}_{{\rm in}} = \bigotimes_{i=1}^{N} \hilb{H}_{2i-2}$
and $\hilb{H}_{{\rm out}} = \bigotimes_{j=1}^N \hilb{H}_{2i-1}$)
be the Choi operator of
  a deterministic Quantum Network $\mathcal{R}^{(N)}$ .
Then $R^{(N)}$ satisfies
\begin{align}\label{eq:recnorm2}
  \Tr_{2k-1}[R^{(k)}] = I_{2k-2} \otimes R^{(k-1)}, \qquad  1 \leq k \leq N  
\end{align}
where $R^{(0)}=1$, $R^{(k)} \in  \mathcal{L}(\hilb{H}_{{\rm out}_k} \otimes  \hilb{H}_{{\rm in}_k})$,
$\hilb{H}_{{\rm in}_k}= \bigotimes_{i=0}^{k-1} \hilb{H}_{2i}$, 
$\hilb{H}_{{\rm out}_k}= \bigotimes_{i=0}^{k-1} \hilb{H}_{2i+1}$. 
\end{corollary}
\begin{Proof}
Eq.  (\ref{eq:recnorm2}) can be obtained by recursively applying Eq. (\ref{eq:recnorm1}). \qed
\end{Proof}
\begin{remark}\label{rem:causalordervsnorm}
  We want to stress that Eq. (\ref{eq:recnorm2}) reflects the causal ordering of the Quantum Network.
This property translates the fact that information can be transmitted from system $i$ to a system $j$
if $i<j$ but not to a system $j'<i$.
Consider the Network 
$\mathcal{R}^{(2)}
 \in 
\mathcal{L}(\mathcal{L}(\hilb{H}_0 \otimes \hilb{H}_2),
\mathcal{L}(\hilb{H}_0 \otimes \hilb{H}_2))$
\begin{align*}
  \begin{aligned}
\Qcircuit @C=1.5em @R=1em {
\ustick{0}&\multigate{1}{\mathcal{C}_1}& \ustick{1} \qw & \ustick{2}&  
 \multigate{1}{\mathcal{C}_2} &  \ustick{3} \qw \\
 & \pureghost{\mathcal{C}_1}&  \qw&\qw& \ghost{\mathcal{C}_1}  &
}       
  \end{aligned}
\end{align*}
We will now prove that the condition that no information flows from
$2$ to $1$ is equivalent to $\Tr_{3}[R^{(2)}_{0123}] = I_2 \otimes
R_{01}^{(1)}$.  The condition that there is no flow of information
from $2$ to $1$ can be expressed by saying that upon application of
the memory channel represented by $R^{(2)}$ to a general input state
$\rho_{02}$, the partial state in $1$ does not depend on the local
state in $0$ i.e. $\Tr_3[\sigma_{13}] =
\Tr_{3}[\Tr_{02}[(\rho^{T}_{02}\otimes I_{13})R^{(2)}_{0123} ]]
=\mathcal{A}(\Tr_{0}[\rho_{02}])$ for a fixed channel $\mathcal{A}$.
If $\Tr_{3}[R^{(2)}_{0123}] = I_2 \otimes R_{01}^{(1)}$ we have
$\Tr_3[\sigma_{13}] = \Tr_{3}[\Tr_{02}[(\rho^{T}_{02}\otimes
I_{13})R^{(2)}_{0123} ]] =\Tr_{02}[(\rho^{T}_{02}\otimes
I_{13})R^{(1)}_{013} \otimes I_2] = \Tr_0[(\Tr_{2}[\rho^T_{02}]\otimes
I_1) R^{(1)}_{013}] = \mathcal{A}(\Tr_{0}[\rho_{02}])$ if we define
$\mathcal{A}:= \mathfrak{C}^{-1}(R^{(1)})$.

On the other hand let us suppose that 
$\Tr_{3}[\Tr_{02}[(\rho^{T}_{02}\otimes I_{13})R^{(2)}_{0123} ]]
=\mathcal{A}(\Tr_{0}[\rho_{02}])$
for a fixed $\mathcal{A}$ . In particular if $\rho_{02} = \tau_0 \otimes \omega_2$  we have
$\Tr_{3}[\Tr_{02}[(\tau^{T}_{0} \otimes \omega^T_2) \otimes I_{13})R^{(2)}_{0123} ]]=
\Tr_{0}[(\tau^{T}_{0} \otimes I_1)\Tr_2[ ( \omega^T_2 \otimes I_{10}) \Tr_3[R^{(2)}_{0123}] ]]=
\Tr_{0}[(\tau^{T}_{0} \otimes I_1)A_{01}]
=\mathcal{A}(\Tr_{0}[\tau_{0}])$,
where $ \mathfrak{C}(\mathcal{A}) = A_{01}=\mathcal{S}(\omega_2))$
and $\mathcal{S} = \mathfrak{C}^{-1}(\Tr_3[R^{(2)}_{0123}]) $.
Since $\mathcal{A}$ is a constant we have 
$\mathcal{S}(\omega_2) = A_{01}$ for all $\omega$, that implies
 $\mathfrak{C}(\mathcal{S}) = I_2 \otimes A_{01}$.
The Quantum Network $\mathcal{R}^{(2)}$
when considered as channel from $\mathcal{L}(\mathcal{L}(\hilb{H}_{02}))$
to  $\mathcal{L}(\mathcal{L}(\hilb{H}_{13}))$
has the properties of a \emph{semicausal channel}
as discussed in Refs. \cite{semicausal1, semicausal2}
\end{remark}
The recursive normalization condition (\ref{eq:recnorm2}) and the positivity constraint
characterize the Choi Operator of a deterministic Quantum Network.
The following theorem tells us that a positive operator satisfying Eq. (\ref{eq:recnorm2})
is the Choi operator  of a deterministic Quantum Network.

\begin{theorem}[Realization of deterministic Quantum Networks]\label{th:realinet}
  Let
 $R^{N} \in \mathcal{L}(\hilb{H}_{{\rm out}} \otimes \hilb{H}_{{\rm in}})$ 
($\hilb{H}_{{\rm in}} = \bigotimes_{i=1}^{N} \hilb{H}_{2i-2}$
and $\hilb{H}_{{\rm out}} = \bigotimes_{j=1}^N \hilb{H}_{2i-1}$) be
a positive operator satisfying Eq. (\ref{eq:recnorm2}).
Then $R^{(N)}$ is the Choi operator
of a deterministic Quantum Network $\mathcal{R}^{(N)}$ given by the concatenation
of $N$ isometries followed by a trace on an ancillary space:
for every state $\rho \in \mathcal{L}(\hilb{H}_{{\rm in}})$ one has
\begin{align}\label{eq:concatenation}
 & \mathcal{R}^{(N)}(\rho) = \Tr_{A_N}[V^{(N)} \cdots V^{(1)} \rho V^{(1)\dagger} \cdots V^{(N)\dagger}]\\
\nonumber \\
&\begin{aligned}
\Qcircuit @C=1.5em @R=1em {
\ustick{0}&\multigate{1}{\mathcal{V}_1}& \ustick{1} \qw & \ustick{2}&  
 \multigate{1}{\mathcal{V}_2} &  \ustick{3} \qw \\
& \pureghost{\mathcal{V}_1}& \ustick{A_1} \qw&\qw& \ghost{\mathcal{V}_1}  &
\ustick{A_2}\qw\\
}
\end{aligned}
\qquad\cdots\qquad
\begin{aligned}
\Qcircuit @C=1.5em @R=1em {
\ustick{2N-2}&   \multigate{1}{\mathcal{V}_N} &  \ustick{2N-1} \qw \\
 \ustick{A_{N-1}} & \ghost{\mathcal{V}_N}  &\ustick{A_N}\qw &\measureD{I}
\nonumber \\
}    
  \end{aligned}
\end{align}
where $V^{i} \in \mathcal{L}(\mathcal{L}(\hilb{H}_{2k-2} \otimes \hilb{H}_{A_{k-1}})
,\mathcal{L}(\hilb{H}_{2k-1} \otimes \hilb{H}_{A_k}))$
and $ \hilb{H}_{A_k}$ is an ancillary space, $ \hilb{H}_{A_0} = \mathbb{C}$
(in Eq. (\ref{eq:concatenation}) we omitted the identity operators on the Hilbert spaces
where the isometries do not act).
\end{theorem}
\begin{Proof}
  Define $\hilb{H}_{A_k} = \supp(R^{(k)*}) $ and
  \begin{align}\label{eq:realisometries}
V^{(k)} = I_{2k-1} \otimes R^{(k)\frac12*}R^{(k-1)-\frac12 *} \,\Ket{I}_{(2k-1)(2k-1)'}T_{(2k-2)\rightarrow(2k-2)'}    
  \end{align}
  where $T_{n \rightarrow m} = \sum_i\ket{i}_m \bra{i}_n$.\\
Using Eq. (\ref{eq:recnorm2}) one has
$V^{(k)\dagger}V^{(k)}=\left( R^{(k-1)*}\right)^{-\frac12}
\Tr_{2k-1}[R^{(k)*}] \left( R^{(k-1)*}\right)^{-\frac12} = I_{2k-2} \otimes I_{A_{k-1}}$
that is $V^{(k)}$ is an isometry.
Now consider $W^{(N)}=V^{(N)} \cdots V^{(1)}$, which goes from
$\hilb{H}_{{\rm in }}$ to $\hilb{H}_{{\rm out }} \otimes \hilb{H}_{A_N}$;
From Eq. \ref{eq:realisometries} we have $W^{(N)}=(I_{{\rm out}} \otimes (R^{(N)*})^{\frac12})\Ket{I}_{({\rm out})({\rm out})'}
\otimes T_{{\rm in}\rightarrow{\rm in}'}$ and Theorem \ref{th:stinedilation}
tells us that $W^{(N)}$ is an isometric dilation of $\mathcal{R}^{N}$ and
so
\begin{align}
 \mathcal{R}^{N}(\rho) = \Tr_{A_N}[W^{(N)}\rho W^{(N)\dagger}] = 
\Tr_{A_N}[V^{(N)} \cdots V^{(1)} \rho   V^{(1)\dagger} \cdots V^{(N)\dagger}].  
\end{align}
 \qed
\end{Proof}

\begin{corollary}
  The minimal dimension of the ancilla space $\hilb{H}_{A_k}$
is $\dim(\supp(R^{(k)}))$ in Theorem \ref{th:realinet}
\end{corollary}
\begin{Proof}
  Consider the isometries $W^{(k)}=V^{(k)} \cdots V^{(1)}$ where $V^{(i)}$ are defined according to Eq. \ref{eq:realisometries}.
Theorem \ref{th:stinedilation} tells us that $W^{(k)}$ is an isometric dilation of $\mathcal{R}^{k}$
with minimal ancillary space; then it is not possible to choose an ancillary space $\hilb{H}_{B_k}$
with $\dim(\hilb{H}_{B_k}) < \dim(\hilb{H}_{A_k}) = \dim(\supp(R^{(k)}))$. \qed
\end{Proof}
\begin{remark}
  The maximum $d_{{\rm max}} := \max_{1 \leq k \leq N} d_{A_k}$
provides an upper bound on the complexity of the Network in terms of quantum memory.
Indeed, the Stinespring  dilation theorem preserves coherence up to the last step; 
for example it can happen
that some ancillary degrees of freedom are used only up to a step $k<N$
and then the isometries $V^{(k+1)},\dots, V^{(N)}$ act only trivially on them.
In this case one can trace out some degrees of freedom before the last step.
This deeper analysis of  resources  can be performed only by inspecting
the structure of the isometries $V^{(k)}$.
\end{remark}
\begin{remark}
  We stress that the set of the Choi operators
is a convex set; indeed, imposing linear constraints (like the one in Eq. \ref{eq:recnorm2})
on a given convex set (like the set of positive operators)
does not spoils the convexity.
\end{remark}
Theorems \ref{th:normcond}  and \ref{th:realinet}
provide a one to one correspondence between the set of deterministic Quantum Networks
(considered as  equivalence classes of different implementations as pointed out in Remark 
\ref{rem:equivalenceclasses}) 
and the set of positive operators satisfying the normalization (\ref{eq:recnorm2})
\begin{align}
  \begin{array}{c}
  \mathcal{R}^{(N)} \\  
\\
   \begin{aligned}
\Qcircuit @C=0.6em @R=0.5em {
&\multigate{1}{\mathcal{C}_1}&  \qw & &  
 \multigate{1}{\mathcal{C}_2} & \qw \\
& \pureghost{\mathcal{C}_1}&  \qw&\qw& \ghost{\mathcal{C}_1}  &
\qw\\
}
\end{aligned}
\cdots
\begin{aligned}
  \Qcircuit @C=0.6em @R=0.5em {
    &   \multigate{1}{\mathcal{C}_N} &   \qw  \\
  & \ghost{\mathcal{C}_N}  &
\\
}
\end{aligned}
  \end{array}
\;\;\;  
\leftrightarrow
\;\;\;
\begin{array}{c}
R^{(N)} \geq 0\\
\mbox{such that} \\
\Tr_{2k-1}[R^{(k)}] = I_{2k-2} \otimes R^{(k-1)}
\end{array}
\quad ;
\nonumber
\end{align}
following the same terminology introduced in Refs. \cite{QCA, comblong}
we call the Choi operators of a Quantum Network
\emph{Quantum Combs}\footnote{Whenever we want to stress the distinction between deterministic  and probabilistic case
we use the terms \emph{deterministic Quantum Combs} and  \emph{probabilistic Quantum Combs}
respectively.
 }.
This result (and its generalization to the probabilistic case)
allows to represent every Quantum Networks  in terms of 
a single positive operator subjected to linear constraints.
This is extremely relevant for applications.
Indeed, optimizing a Quantum Network by  separately optimizing each device 
is extremely demanding. Thanks to this representation 
the optimization problem is reduced to an optimization problem over
 a convex  set  of suitably normalized positive operators.
Moreover we notice that through Eq. (\ref{eq:realisometries})
we are provided with an explicit expression
of a Quantum Network that is represented by a given quantum comb $R^{(N)}$.

This allows us to formulate an algorithm for designing
optimal Quantum Networks for a given task (e.g. cloning, discrimination, 
estimation):
\begin{enumerate}
\item Choose a suitable figure of merit $F$ for the task of interest.
\item  Find the positive operator $R^{(N)}$ satisfying constraint
  in Eq.~(\ref{eq:recnorm2}) and optimizing $F$.
\item Set $R^{(0)}=1$ and $I_{A_0} = 1$.
\item For $k=1$ to $k=N$ do the following:
  \begin{enumerate}
  \item Calculate $I_{\overline{\rm{in}_k}}\otimes R^{(k)} = \Tr_{ \overline{\rm{out}_{k} }}[C]$, where $I_{\overline \sH}$  ($\Tr_{\overline \sH})$ denotes the identity (partial trace) over all Hilbert spaces but $\sH$;  
  \item define $V^{(k)}$ as in Eq. (\ref{eq:realisometries}).
  \end{enumerate}
\item The optimal network is given by the concatenation of the
    $V^{(k)}$'s in Eq. (\ref{eq:concatenation})
\end{enumerate}
 
\subsection{Probabilistic Quantum Network}
The aim of this section is to provide the equivalents
of Theorems \ref{th:normcond} and \ref{th:realinet}
for the case in which probabilistic Quantum Network are considered.
We remind that a probabilistic Quantum Network $\mathcal{R}^{(N)}$
is equivalent to a concatenation of $N$ completely 
positive trace non increasing linear maps\footnote{This definition 
includes deterministic networks as a special case.}
\begin{align}
  \mathcal{R}^{(N)} = \mathcal{C}_1 \star \mathcal{C}_N \star \cdots \star \mathcal{C}_N. \nonumber
\end{align}
\begin{theorem}[Sub-normalization condition]\label{th:subnormcond}
  Let $\mathcal{R}^{(N)}$ be a probabilistic Quantum Network.
and  $R^{(N)} \in \mathcal{L}(\bigotimes_{i=0}^{2N-1}\hilb{H}_i)$ be
 its Choi-Jamiolkowski operator;
then there exists
a Choi operator $S^{(N)}$ of a deterministic Quantum Network such that
\begin{align}\label{eq:subnormcond}
  0 \leq R^{(N)} \leq S^{(N)},
\end{align}
\end{theorem}
\begin{Proof}
The proof is by induction.
For $N=1$ the probabilistic quantum Network
is just a quantum operation and we know
that its Choi operator $E^{(1)}$
is upper bounded by the Choi operator
of  a Quantum Channel, i.e. of a deterministic Quantum Network
with $1$ vertex.
Now suppose that the statement holds for $N-1$.
Since   $\mathcal{R}^{(N)} = \mathcal{C}_1 \star \mathcal{C}_N \star \cdots \star \mathcal{C}_N$
we have $R^N = C_1 * C_2 * \cdots *C_N$ where $C_i \leq \overline{C_i}$
for some  $\overline{C_i}$ which is the Choi operator of a quantum channel.
Exploiting the induction hypothesis we have that
$C_1 * C_2 * \cdots *C_{N-1} := D \leq \overline{D}$
where is the Choi of a deterministic Quantum Network.
Exploiting Lemma~\ref{lem:propertiesoflink} we have
\begin{align}
R^{(N)} = D*C_N \leq \overline{D}*C_N \leq \overline{D}*\overline{C_N} := S^{(N)}
\end{align}
that proves the statement. \qed
\end{Proof}
\begin{theorem}[Realization of probabilistic Quantum Networks]\label{th:realiprobnet}
  Let 
$R^{(N)} \in \mathcal{L}(\bigotimes_{i=0}^{2N-1}\hilb{H}_i)$ be
 a positive operator satisfying Eq. (\ref{eq:subnormcond}).
Than this is the Choi-Jamio\l kowsky operator
of a probabilistic Quantum Network $\mathcal{R}^{(N)}$, consisting of $N$
isometric channels followed by an effect on an ancillary space.
For any $\rho \in \mathcal{L}(\hilb{H}_{{\rm in}})$
we have
\begin{align}\label{eq:concatenationprob}
 & \mathcal{R}^{(N)}(\rho) = \Tr_{A_N}[(V^{(N)} \cdots V^{(1)}) \rho (V^{(1)\dagger} \cdots V^{(N)\dagger})E]\\
\nonumber \\
&\begin{aligned}
\Qcircuit @C=1.5em @R=1em {
\ustick{0}&\multigate{1}{\mathcal{V}_1}& \ustick{1} \qw & \ustick{2}&  
 \multigate{1}{\mathcal{V}_2} &  \ustick{3} \qw \\
& \pureghost{\mathcal{V}_1}& \ustick{A_1} \qw&\qw& \ghost{\mathcal{V}_1}  &
\ustick{A_2}\qw\\
}
\end{aligned}
\qquad\cdots\qquad
\begin{aligned}
\Qcircuit @C=1.5em @R=1em {
\ustick{2N-2}&   \multigate{1}{\mathcal{V}_N} &  \ustick{2N-1} \qw \\
 \ustick{A_{N-1}} & \ghost{\mathcal{V}_N}  &\ustick{A_N}\qw &\measureD{E}
\nonumber \\
}    
  \end{aligned}
\end{align}
where $V^{i} \in \mathcal{L}(\mathcal{L}(\hilb{H}_{2k-2} \otimes \hilb{H}_{A_{k-1}})
,\mathcal{L}(\hilb{H}_{2k-1} \otimes \hilb{H}_{A_k}))$
and $ \hilb{H}_{A_k}$ is an ancillary space, $ \hilb{H}_{A_0} = \mathbb{C}$
(in Eq. (\ref{eq:concatenationprob}) we omitted the identity operators on the Hilbert spaces
where the $V^{(k)}$'s and $E$ do not act).
\end{theorem}
\begin{Proof}
  Let $S^{(N)}$ be the Choi  operator of a deterministic Quantum Network such that
$R^{(N)} \leq S^{(N)}$.
Now we define
$\hilb{H}_{A_{k}}$ and $V^{(K)}$ for $S^{(N)}$ as in Eq. (\ref{eq:realisometries})
and $E = S^{(N)*-\frac12}R^{(N)*}S^{(N)*-\frac12}$;
It is easy to verify that
\begin{align}
&  \Tr_{A_N}[(V^{(N)} \cdots V^{(1)}) \rho (V^{(1)\dagger} \cdots V^{(N)\dagger})E] = \nonumber \\
&= 
\Tr_{A_N}[(I_{{\rm out}} \otimes (S^{(N)*})^{\frac12})( \rho_{{\rm in}'}\otimes  \KetBra{I}{I}_{({\rm out})({\rm out})'})
(I_{{\rm out}} \otimes (S^{(N)*})^{\frac12})\cdot \nonumber \\
&\cdot
(I_{{\rm out}} \otimes S^{(N)*-\frac12}R^{(N)*}S^{(N)*-\frac12})]= \nonumber  \\
&=
\Tr_{A_N}[(I_{{\rm out}} \otimes  \rho_{{\rm in}}^T)R^{(N)}] = \mathcal{R}^{(N)}(\rho) \nonumber
\end{align}
\end{Proof}
\begin{remark}
  Theorem \ref{th:realiprobnet}
says that any probabilistic Quantum Network can be split
into a coherent part (sequence of isometries)
and a final effect on an ancillary space.
\end{remark}
Thanks to theorems \ref{th:subnormcond} and \ref{th:realiprobnet}
we can represent  any probabilistic Quantum Network in terms of a positive operator
i.e. its probabilistic Quantum Comb.
We now introduce the Quantum Network analogue of Quantum Instruments and POVMs;
both of them will be exploited in the applications.
\begin{Def}[Generalized Instrument]\label{def:geninst}\ 
  A \emph{Generalized Instrument} is a set of probabilistic Quantum Networks
$\{ \mathcal{R}^{(N)}_i\}$ whose sum is a deterministic Quantum Network $\mathcal{R}^{(N)}_{\Omega} = \sum_i\mathcal{R}^{(N)}_i$.
The index $i$ represents the classical outcome of the Network\footnote{As we did
when we introduced the concept of Quantum Instrument,  we restrict ourselves
to the case of finite number of outcomes. The generalization to an arbitrary
outcome space $\Omega$ can be obtained by defining a measure $\mathcal{R}_{B}$ that
associates to any event $B \subseteq \Omega$ a  probabilistic Quantum Network $\mathcal{R}_{B}$
such that $\mathcal{R}_{\Omega}$ is a deterministic Quantum Network.}.
\end{Def}
For Generalized Instruments  the following analogue of Th. \ref{th:realinst} holds:
\begin{theorem}[realization of Generalized Instruments]\label{th:realgeninst}
  Let \\ $\{ \mathcal{R}^{(N)}_i,  \mathcal{R}^{(N)}_i \in \mathcal{L}(\mathcal{L}(\hilb{H}_{{\rm in}}),
\hilb{H}_{{\rm out}}  )  \}$, $\mathcal{R}^{(N)}_{\Omega} = \sum_i  \mathcal{R}^{(N)}_i$ be a Generalized Instrument.
Then there exist an Hilbert space $\hilb{H}_{A_N}$,
 a deterministic Quantum Network 
 $\mathcal{S}^{(N)} \in \mathcal{L}(\mathcal{L}(\hilb{H}_{{\rm in}}), \mathcal{L}(\hilb{H}_{{\rm out}}\otimes \hilb{H}_{A_N}))$
and a POVM $\{P_i, P_i \in  \mathcal{L}(\hilb{H}_{A_N}) \}$
such that for any $\rho \in \mathcal{L}(\hilb{H}_{{\rm in}})$ we have
\begin{align}\label{eq:concatenationprob2}
 & \mathcal{R}_i^{(N)}(\rho) = \Tr_{A_N}[(\mathcal{S}^{(N)}(\rho)) P_i]]\\
&\begin{aligned}
\Qcircuit @C=1.5em @R=1em {
\ustick{0}&\multigate{1}{\mathcal{V}_1}& \ustick{1} \qw & \ustick{2}&  
 \multigate{1}{\mathcal{V}_2} &  \ustick{3} \qw \\
& \pureghost{\mathcal{V}_1}& \ustick{A_1} \qw&\qw& \ghost{\mathcal{V}_1}  &
\ustick{A_2}\qw\\
}
\end{aligned}
\qquad\cdots\qquad
\begin{aligned}
\Qcircuit @C=1.5em @R=1em {
\ustick{2N-2}&   \multigate{1}{\mathcal{V}_N} &  \ustick{2N-1} \qw \\
 \ustick{A_{N-1}} & \ghost{\mathcal{V}_N}  &\ustick{A_N}\qw &\measureD{P_i}
\nonumber \\
}    
  \end{aligned}
\nonumber 
\end{align}
\end{theorem}
\begin{Proof}
  The Proof is the same as in Th. \ref{th:realiprobnet};
we just define $\mathcal{S}^{(N)} = \mathcal{V}_1 \star \cdots \star \mathcal{V}_N$,
where the $\hilb{H}_{A_k}$'s and $V^{(i)}$'s are defined as in Eq. (\ref{eq:realisometries})
(now $R_\Omega^{(N)}$ plays the role of $R^{(N)}$) and 
\begin{align}\label{eq:realpovminst}
  P_i = R^{(N)-\frac12*}_\Omega R^{(N)}_i  R^{(N)-\frac12*}_\Omega.
\end{align}
It is easy to verify that
\begin{align}
& \Tr_{A_N}[(\mathcal{S}^{(N)}(\rho))P_i] = 
  \Tr_{A_N}[(V^{(N)} \cdots V^{(1)}) \rho (V^{(1)\dagger} \cdots V^{(N)\dagger})P_i] = \nonumber \\
&= 
\Tr_{A_N}[(I_{{\rm out}} \otimes (R_\Omega^{(N)*})^{\frac12})( \rho_{{\rm in}')}\otimes  \KetBra{I}{I}_{({\rm out})({\rm out})'})
(I_{{\rm out}} \otimes (R_\Omega^{(N)*})^{\frac12})\cdot \nonumber \\
&\cdot
(I_{{\rm out}} \otimes R_\Omega^{(N)*-\frac12}R_i^{(N)*}R_\Omega^{(N)*-\frac12})]= \nonumber  \\
&=
\Tr_{A_N}[(I_{{\rm out}} \otimes  \rho_{{\rm in}}^T)R_i^{(N)}] = \mathcal{R}_i^{(N)}(\rho) \nonumber  \\
\end{align}
\qed
\end{Proof}
A relevant  class of Generalized Instrument is the the following
\begin{Def}[Quantum Tester]
  A \emph{Quantum Tester} is a Generalized Instrument $\{ \mathcal{R}_i^{N}\}$
such that $\dim(\hilb{H}_0)= \dim(\hilb{H}_{2N-1})=1$.

\end{Def}
\begin{theorem}[normalization of Quantum Tester]
  Let $\{ \mathcal{R}_i^{(N)} \}$ be a quantum Tester.
Then 
\begin{align}\label{eq:normtester}
&\sum_i R_i^{(N)} := R_\Omega^{(N)} = R^{(N-1)}_\Omega \otimes I_{2N-2}   \nonumber\\
&\Tr_{2k-1}[R_\Omega^{(k)}] = I_{2k-2} \otimes R_\Omega^{(k-1)}, \quad 2 \leq k \leq N-1 \nonumber\\
&\Tr_{1}[R_\Omega^{(1)}] = 1
\end{align}

\end{theorem}
\begin{Proof}
  Since $\dim(\hilb{H}_{2N-1})=1$ and applying Theorem \ref{th:normcond} to $R^{(N)}_\Omega$ we have
$R^{(N)} = \Tr_{2N-1}$ $[R^{(N)}] = I_{2N-1} \otimes R^{(N-1)}_\Omega$.
Clearly $\Tr_{1}[R_\Omega^{(1)}] = I_0 = 1$
since $\dim(\hilb{H}_{0})=1$.
\end{Proof}
\begin{theorem}[realization of Quantum Tester]
\hspace*{-0.2cm}    Let $\{ \mathcal{R}_i^{(N)} \}$ be a quantum Tester.
Then $\{ \mathcal{R}_i^{(N)} \}$ 
can be realized by a deterministic Quantum Network 
$\{ \mathcal{S}^{(N)} \}$ with $\dim(\mathcal{H}_0)=1$
followed by a POVM on $\hilb{H}_{2N-1}$
\begin{align}\label{eq:realitester}
\begin{aligned}
\Qcircuit @C=1.5em @R=1em {
&\multiprepareC{1}{\Ket{\Psi}}& \ustick{1} \qw & \ustick{2}&  
 \multigate{1}{\mathcal{V}_2}& \ustick{3} \qw\\
& \pureghost{\Ket{\Psi}}& \ustick{A_1} \qw&\qw& \ghost{\mathcal{V}_1}  &
\ustick{A_2}\qw\\
}
\end{aligned}
\qquad\cdots\qquad
\begin{aligned}
\Qcircuit @C=1.5em @R=1em {
\ustick{2N-2}&   \multigate{1}{\mathcal{V}_N} & \measureD{P_i} \\
 \ustick{A_{N-1}} & \ghost{\mathcal{V}_N}  &
\\
}    
  \end{aligned}  
\end{align}
\end{theorem}
\begin{Proof}
This result comes immediately from  theorem \ref{th:realgeninst}
by relabeling $\hilb{H}_{A_N} = \hilb{H}_{2N-1}$.
Since $\dim(\mathcal{H}_0)=1$
the first isometry is just the preparation of the   pure state
$\Ket{\Psi} := (I_1 \otimes R^{(1)*\frac12}_{\Omega}) \Ket{I}_{11'} = \Ket{R^{(1)*\frac12}_{\Omega}}$.
\qed
\end{Proof}
\begin{remark}
  By making the substitution
\begin{align*}
  \begin{aligned}
    \Qcircuit @C=1.5em @R=1em {
\ustick{2N-2}&   \multigate{1}{\mathcal{V}_N} & \measureD{P_i} \\
 \ustick{A_{N}} & \ghost{\mathcal{V}_N}  &
}
  \end{aligned}
\quad \rightarrow \quad
  \begin{aligned}
    \Qcircuit @C=1.5em @R=1em {
 \ustick{2N-2}&  
 \multimeasureD{1}{\widetilde{P_i}} \\
 \ustick{A_N}& \ghost{\widetilde{P_i}} \\
}
  \end{aligned}
\end{align*}
the realization scheme
\ref{eq:realitester}
can be rewritten as
\begin{align}\label{eq:realitester2}
\begin{aligned}
\Qcircuit @C=1.5em @R=1em {
&\multiprepareC{1}{\Ket{\Psi}}& \ustick{1} \qw & \ustick{2}&  
 \multigate{1}{\mathcal{V}_2}& \ustick{3} \qw\\
& \pureghost{\Ket{\Psi}}& \ustick{A_1} \qw&\qw& \ghost{\mathcal{V}_1}  &
\ustick{A_2}\qw\\
}
\end{aligned}
\qquad\cdots\qquad
\begin{aligned}
\Qcircuit @C=1.5em @R=1em {
\ustick{2N-2}&   \multimeasureD{1}{\widetilde{P_i}} \\
 \ustick{A_{N-1}} & \ghost{\widetilde{P_i}}  &
\\
}    
  \end{aligned}  
\end{align}
\end{remark}

A special class of Quantum Testers is the one in which $N=2$; this class
 has been independently introduced 
in Ref.  \cite{PPOVMziman}
under the name Process-POVM.
\begin{corollary}[characterization of Quantum $2$-Testers]\label{cor:2tester}
\hspace*{-0.1cm} Let $\{ \mathcal{R}^{(2)}_i\hspace*{-0.05cm},\mathcal{R}^{(2)}_i \hspace*{-0.05cm}
\in\hspace*{-0.05cm} \mathcal{L}(\mathcal{L}(\hilb{H}_1),\mathcal{L}(\hilb{H}_2))  \}$
 be a Quantum Tester with two vertexes.
Then we have  
\begin{align}\label{eq:normppovm}
  \sum_iR_i^{(2)} = \rho_1 \otimes I_2
\end{align}
  where $\rho$ is a state in $\mathcal{L}(\mathcal{H}_1)$.
$\{ \mathcal{R}^{(2)}_i\}$
can be split into a preparation of a pure state $\Ket{\sqrt{\rho}} \in \hilb{H}_1 \otimes \hilb{H}_{1'}$
and a POVM $\{P_i\}$ on the space $\hilb{H}_2 \otimes \hilb{H}_{1'}$
($\hilb{H}_{1'} = \supp(\rho)$)
\begin{align}\label{eq:realippovm}
  \begin{aligned}
    \Qcircuit @C=1.5em @R=1em {
&\multiprepareC{1}{\Ket{\sqrt{\rho}}}& \ustick{1} \qw & \ustick{2}&  
 \multimeasureD{1}{P_i} \\
& \pureghost{\Ket{\rho}}& \ustick{1'} \qw&\qw& \ghost{P_i} \\
}
  \end{aligned}\quad.
\end{align}
\end{corollary}
\begin{Proof}
  Eq. (\ref{eq:normppovm})
comes from from Eq. (\ref{eq:recnorm2}) and Eq. (\ref{eq:normtester}).
The realization (\ref{eq:realippovm}) is just a special case of 
(\ref{eq:realitester2})
with $A_1= 1'$.
\end{Proof}
\subsection{Connection of  Quantum Networks}
\begin{figure}[tb]
\begin{center}
\includegraphics[width=8cm ]{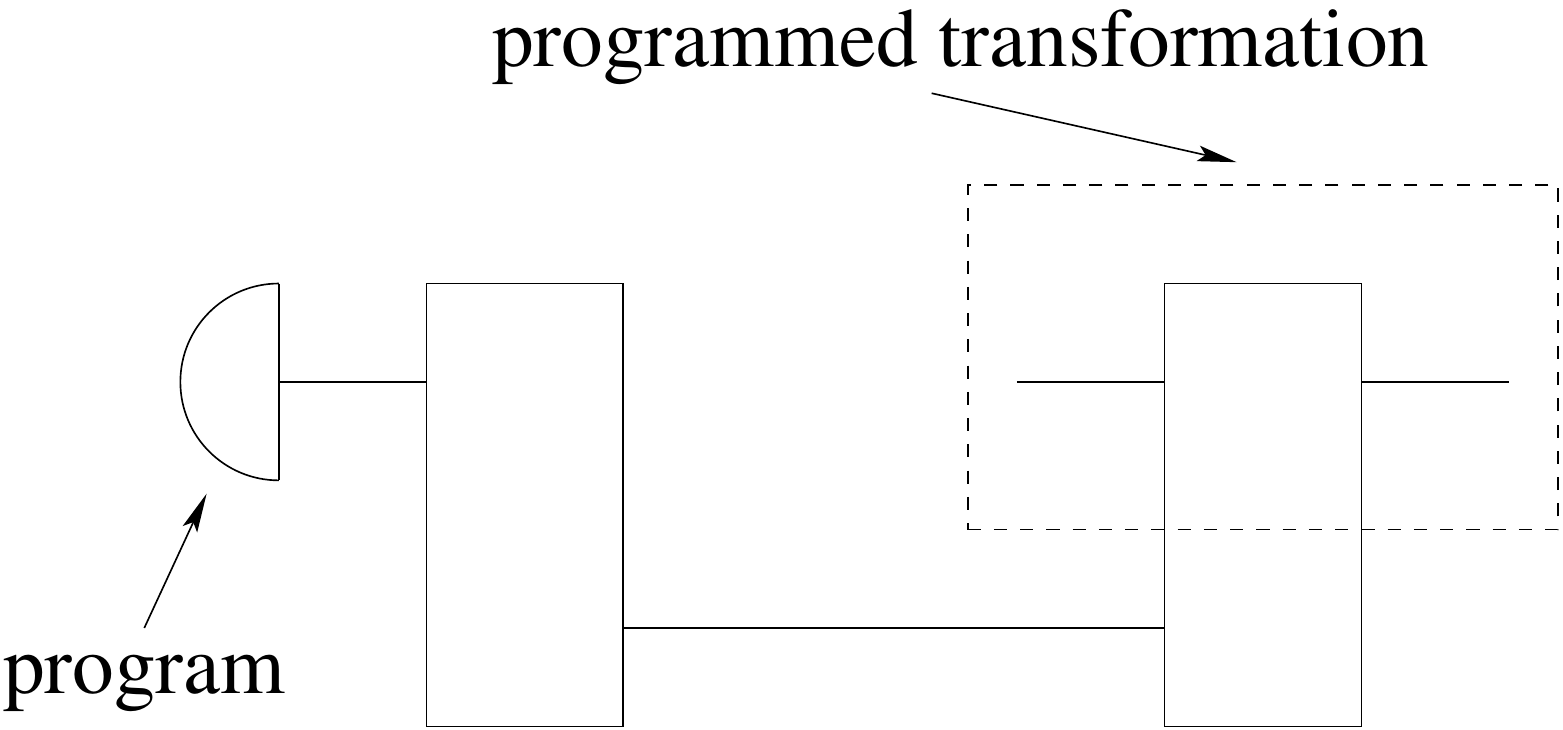}
  \caption{\label{fig:programtrans}
A Quantum Network with two vertexes used as a programmable device.}
\end{center}
\end{figure}

\begin{figure}[tb]
\begin{center}
\includegraphics[width=9cm ]{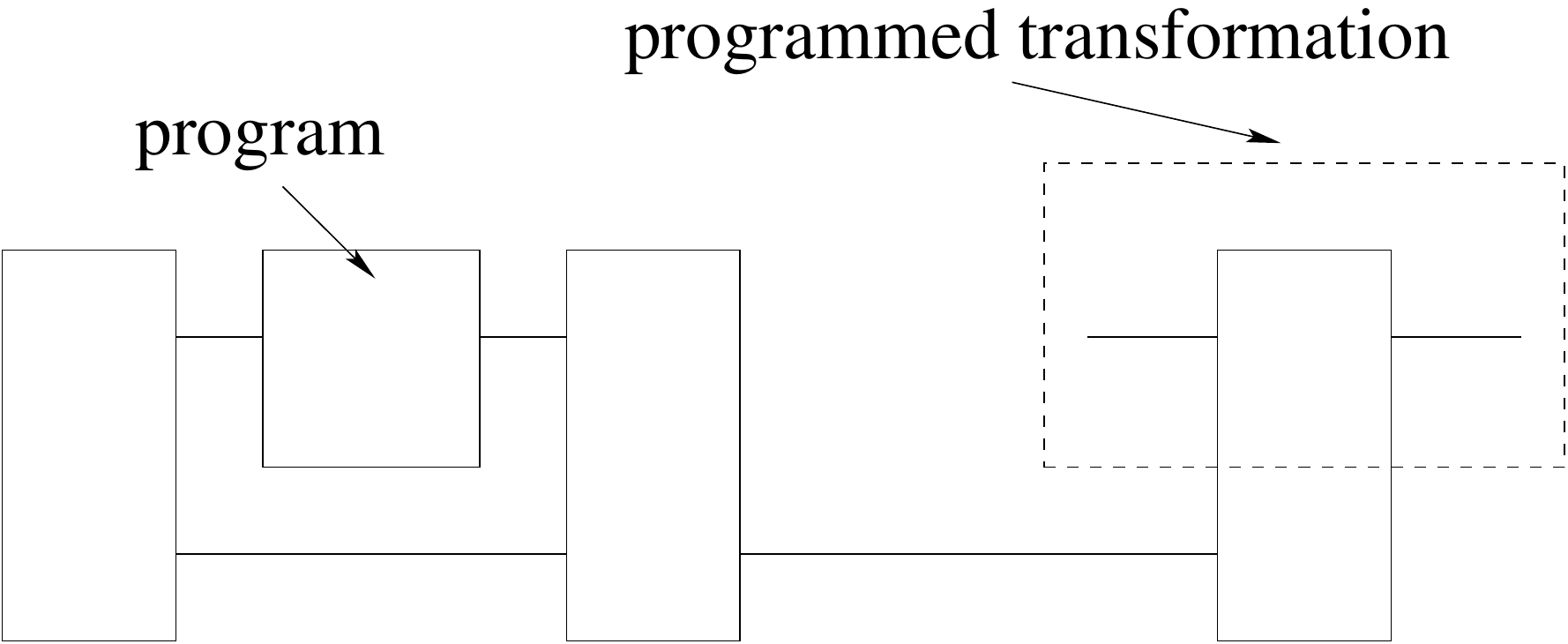}
  \caption{\label{fig:programtrans2}
A Quantum Network calls a quantum channel
as subroutine.
}
\end{center}
\end{figure}

\begin{figure}[tb]
\begin{center}
\includegraphics[width=10cm ]{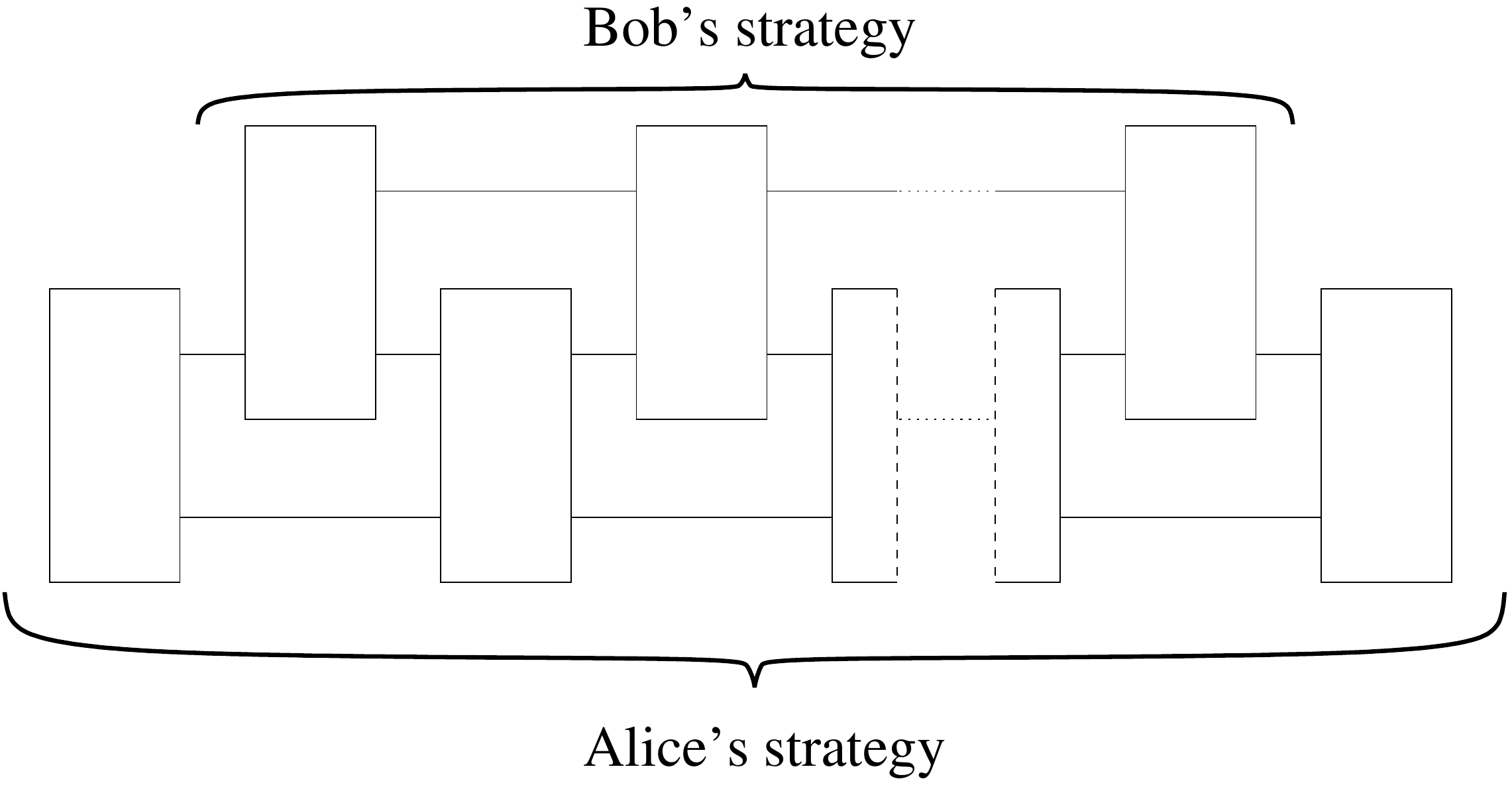}
  \caption{\label{fig:quantumgame}
A multi-round two party game: Alice's strategy is represented by the Quantum Network $\mathcal{A}$
and Bob's strategy is represented by the Quantum Network $\mathcal{B}$.
The outcome of the game can be seen as th interlinking of the two networks.
}
\end{center}
\end{figure}

A Network of Quantum transformations
can be used to achieve many different tasks.
We can imagine to use it as a programmable device
which implements different transformations
on some inputs depending on the quantum state of the program (see Fig. \ref{fig:programtrans}).
Moreover, the program itself of the Quantum Network  can be a quantum channel
rather then a state (Fig. \ref{fig:programtrans2}): during the computation the network call a variable
channel as a subroutine.
More generally a Quantum Network can call several different 
channels at different times and even another Quantum Network.
These kind of situation occur for example when multiple round Quantum games are considered;
in this scenario the overall outcome of the game depends on the strategies 
chosen by the players that can be modeled as Quantum Networks (Fig. \ref{fig:quantumgame}).

\begin{figure}[tb]
\begin{center}
\includegraphics[width=11cm ]{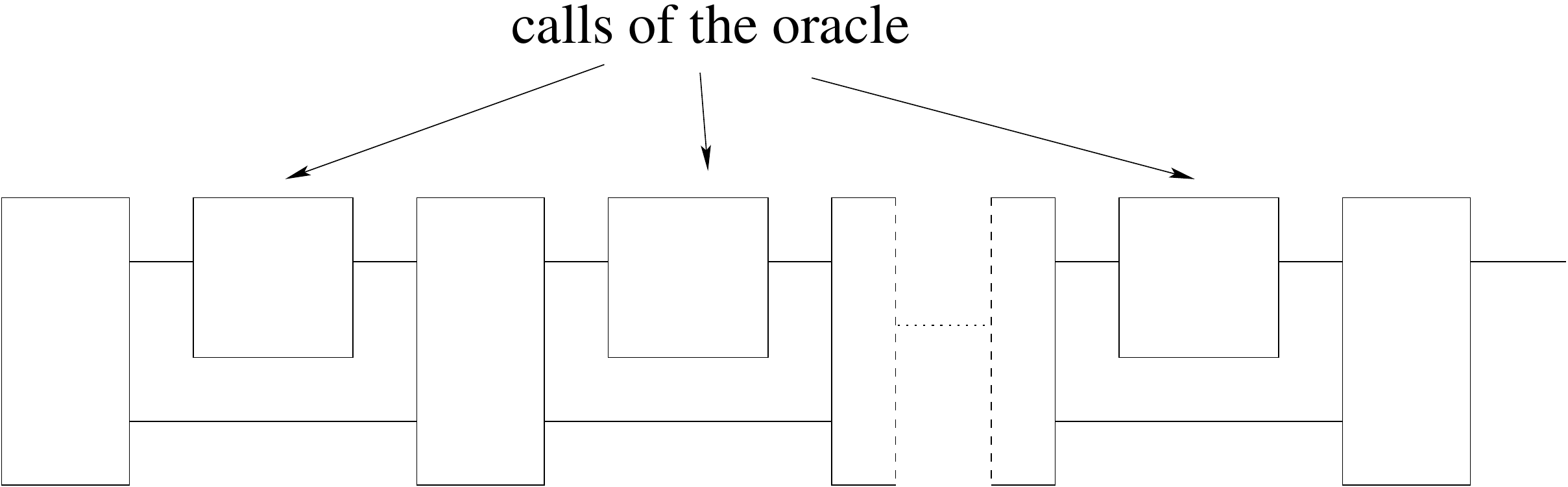}
  \caption{\label{fig:quantumalgorithm}
A Quantum algorithm realized by a Quantum Network in which
$N$ uses of the oracle are inserted.
}
\end{center}
\end{figure}

\begin{figure}[!b]
\begin{center}
\includegraphics[width=11cm ]{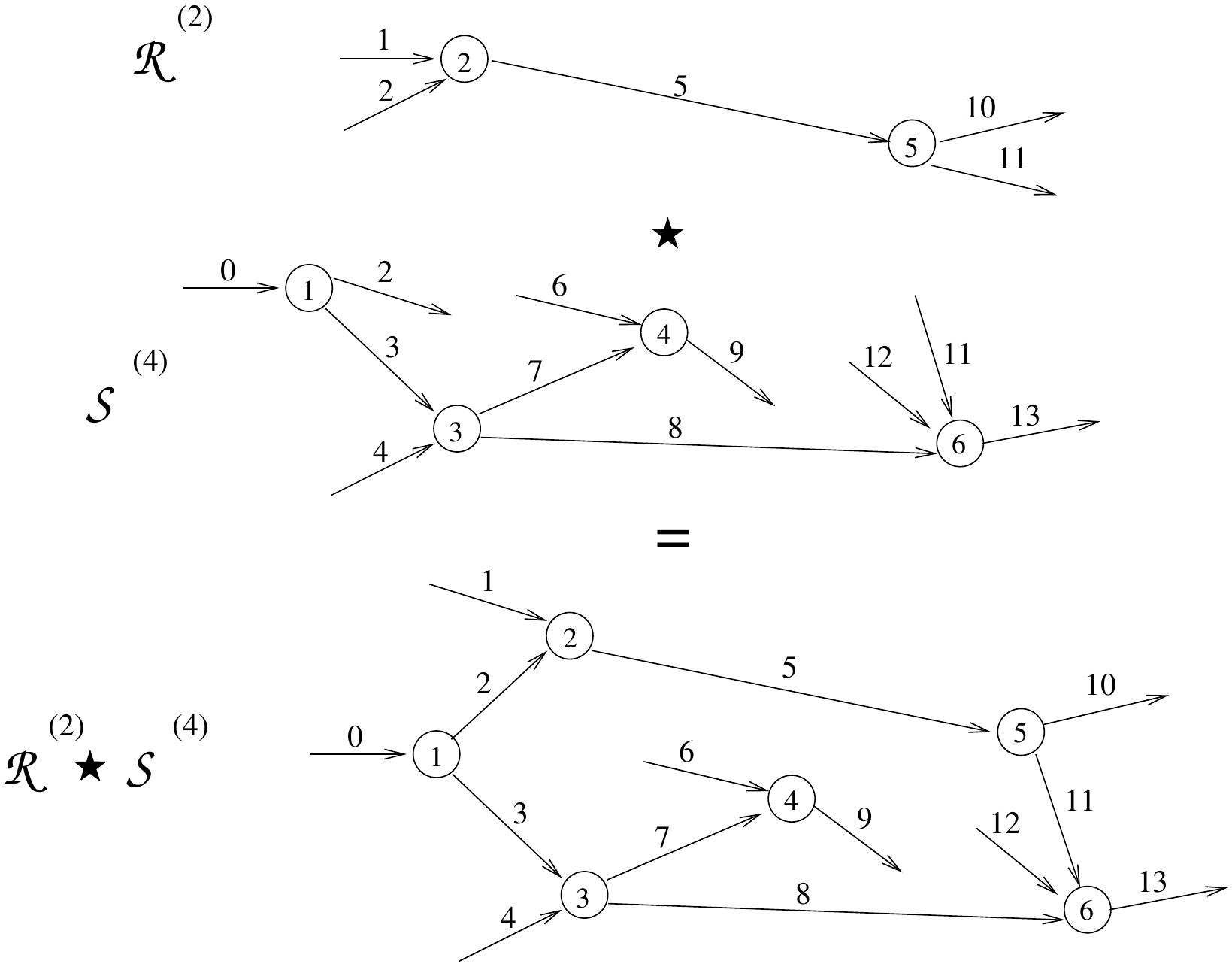}
  \caption{\label{fig:connectednetworks}
The scheme represents the connection of two quantum networks;
the arrows that we are going to connect have the same labels.
}
\end{center}
\end{figure}

Another relevant case are Quantum Algorithms:  they can be thought of as
Quantum Networks calling $N$ uses of a quantum oracle (Fig. \ref{fig:quantumalgorithm}).
All the possible uses of a Quantum Network  are then equivalent to
the connection of the network to another quantum network.
Connecting two network $\mathcal{R}^{(N)}$ and $\mathcal{S}^{(M)}$
means composing the corresponding graphs by joining some of the free
outgoing arrows of a network with free incoming arrows of the other in such a way that
the final network 
$\mathcal{R}^{(N)} \star \mathcal{S}^{(M)}$
 is still a directed acyclic graph\footnote{As  pointed out in Remark \ref{rem:loop}
this condition is necessary in order
to avoid time  loops };
we adopt the convention that if two vertexes $i \in \mathcal{R}^{(N)}$ and $j\in \mathcal{S}^{(M)}$
are connected by joining two arrows, the two arrows are identified with the same label
(see Fig.  \ref{fig:connectednetworks}).
As we said in Section \ref{section:construct},
a directed acyclic graph is endowed with a partial ordering 
among the vertexes that can be extended to a total ordering.
Given two quantum networks $\mathcal{R}^{N}$
and $\mathcal{S}^{M}$
there is a priori no relative ordering between the vertexes of 
$\mathcal{R}^{N}$ and the vertexes of $\mathcal{S}^{M}$.
However, since we require that
the final network is still a directed acyclic graph, it is possible to define
a total ordering among the vertexes in the union set 
$\mathcal{R}^{N} \cup \mathcal{S}^{M}$.
This allows us to sketch  the composition of two quantum networks
in the circuit form

\begin{align}
&  \begin{aligned}
     \Qcircuit @C=1em @R=0.7em {
\ustick{2}&\multigate{2}{\mathcal{C}_2}&  &  
 &    &  
\pureghost{\mathcal{C}_6}&\ustick{11}\qw&\\
\ustick{1}&\ghost{\mathcal{C}_2}&  &  
 &    &  
\pureghost{\mathcal{C}_6}& \ustick{10} \qw&\\
& \pureghost{\mathcal{C}_2}&  \qw & \qw&\qw& \multigate{-2}{\mathcal{C}_6}&&\\
}
  \end{aligned}
\star
  \begin{aligned}
     \Qcircuit @C=1em @R=0.7em {
\ustick{0}&\multigate{2}{\mathcal{C}_1}& \ustick{2}\qw &  
 & \ustick{4}&   \multigate{2}{\mathcal{C}_3} &  &\ustick{6}&
\multigate{2}{\mathcal{C}_4}&\ustick{9}\qw&&  & &\ustick{11}&\multigate{2}{\mathcal{C}_6}&\ustick{13}\qw\\
&\pureghost{\mathcal{C}_1}&  &  
 & &   \pureghost{\mathcal{C}_3} &  &&
\pureghost{\mathcal{C}_4}&&&  & &\ustick{12}&\ghost{\mathcal{C}_6}&\\
& \pureghost{\mathcal{C}_1}&  \qw& \qw  &
\qw& \ghost{\mathcal{C}_3}& \qw&\qw& \ghost{\mathcal{C}_4}&\qw&\qw &\qw & \qw&\qw&
\ghost{\mathcal{C}_6}\\
}
  \end{aligned}
\hspace{-0.1cm} = \nonumber
\end{align}
\begin{align}
  \nonumber\\
&=
  \begin{aligned}
     \Qcircuit @C=1em @R=0.7em {
&&& \pureghost{\mathcal{C}_2} &  \qw
 & \qw& \qw   & \qw &\qw&\qw
&\qw& \qw &\multigate{2}{\mathcal{C}_5}  &&&&\\
&&\ustick{1}& \ghost{\mathcal{C}_2} &  
 & &    &  &&
&&&  \pureghost{\mathcal{C}_5} &\ustick{10}\qw &&&\\
\ustick{0}&\multigate{2}{\mathcal{C}_1}& \ustick{2}  \qw & \multigate{-2}{\mathcal{C}_2} &  
 & \ustick{4}&   \multigate{2}{\mathcal{C}_3} &  &\ustick{6}&
\multigate{2}{\mathcal{C}_4}&\ustick{9}\qw&&
\pureghost{\mathcal{C}_5}   & \qw &\ustick{11}\qw&\multigate{2}{\mathcal{C}_6}&\ustick{13}\qw\\
&\pureghost{\mathcal{C}_1}&  &  
 & & &   \pureghost{\mathcal{C}_3} &  &&
\pureghost{\mathcal{C}_4}&&&  & &\ustick{12}&\ghost{\mathcal{C}_6}&\\
& \pureghost{\mathcal{C}_1}&  \qw& \qw  & \qw&
\qw& \ghost{\mathcal{C}_3}& \qw&\qw& \ghost{\mathcal{C}_4}&\qw&\qw &\qw & \qw&\qw&
\ghost{\mathcal{C}_6}\\
\\} 
  \end{aligned}
=\nonumber \\ \nonumber\\
&=
 \begin{aligned}
     \Qcircuit @C=1em @R=0.7em {
\ustick{0}&\multigate{1}{\mathcal{C}_1}& \ustick{1}&  
 \multigate{1}{\mathcal{C}_2} & \ustick{4}&   \multigate{1}{\mathcal{C}_3} &  &\ustick{6}&
\multigate{1}{\mathcal{C}_4}&\ustick{9}\qw&& \pureghost{\mathcal{C}_5}&\ustick{10} \qw &\ustick{12}&\multigate{1}{\mathcal{C}_6}&\ustick{13}\qw\\
& \pureghost{\mathcal{C}_1}&  \qw& \ghost{\mathcal{C}_1}  &
\qw& \ghost{\mathcal{C}_1}& \qw&\qw& \ghost{\mathcal{C}_4}&\qw&\qw &\multigate{-1}{\mathcal{C}_5} & \qw&\qw&
\ghost{\mathcal{C}_6}\\}
  \end{aligned}
\label{eq:circcomposition} \quad .\\
\nonumber
\end{align}

We now want to determine the Choi operator
of the  composite network $\mathcal{R}^{(N)} \star \mathcal{S}^{(M)}$ 
in terms of the Choi
operators $R^{(N)}$ and $S^{(M)}$  of the networks
$\mathcal{R}^{(N)}$ and $ \mathcal{S}^{(M)}$.
From Eq. (\ref{eq:circcomposition})
it is clear that the combined network can be  obtained by  combining
the linear maps $\mathcal{C}_i$,
then its Choi operator
will be the link product of all the $C_i$.
We have then the following
\begin{theorem}[Link of two Quantum Networks]\label{th:linknet}
  Let $\mathcal{R}^{(N)}$ and $ \mathcal{S}^{(M)}$ be
two Quantum Networks
and $R^{(N)} \in \mathcal{L}(\bigotimes_{i\in \defset{R}} \hilb{H}_{i}),
S^{(M)} \in \mathcal{L}(\bigotimes_{j\in \defset{S}} \hilb{H}_{j})$
be their Choi
operators
where we defined
$\defset{R}$ and $\defset{S}$
 the set of the free  arrows of 
$\mathcal{R}^{(N)}$ and $ \mathcal{S}^{(M)}$ respectively.
If $\defset{R} \cap \defset{S}$ is the set of connected arrows
then
\begin{align}
  \mathfrak{C}(\mathcal{R}^{(N)} \star  \mathcal{S}^{(M)}) = R^{(N)}*S^{(M)}
\end{align}
\end{theorem}
\begin{Proof}
  This result is an immediate consequence of Lemma \ref{lem:propertiesoflink}.
\end{Proof}
\begin{remark}\label{rem:tester}
  A relevant case of composition is the one in which we connect
a quantum network $\mathcal{R}^{(N)}$
with a quantum tester $\{ \mathcal{T}^{(N+1)}_i\}$ in this way:
\begin{eqnarray}
&\mathcal{T}^{(N+1)}_i& \nonumber\\
&\overbrace{\underbrace{
  \begin{aligned}
\Qcircuit @C=1em @R=0.7em {
\multiprepareC{1}{\Ket{\Psi}}&\qw&  \multigate{1}{\mathcal{D}_2}& \qw
 &   \qw \\
\pureghost{\Ket{\Psi}}&\multigate{1}{\mathcal{C}_1}& \ghost{\mathcal{D}_2} &   
 \multigate{1}{\mathcal{C}_2} &   \qw \\
& \pureghost{\mathcal{C}_1}&  \qw& \ghost{\mathcal{C}_1}  &
\qw\\
}
\end{aligned}
\qquad\cdots\qquad
\begin{aligned}
\Qcircuit @C=1.5em @R=1em {
&\qw    &  \multimeasureD{1}{P_i} \\
&   \multigate{1}{\mathcal{C}_N} &   \ghost{P_i}\\
& \ghost{\mathcal{C}_N}  &&
\\
}
\end{aligned}
}}& \\
&  \mathcal{R}^{(N)} &
\nonumber 
\end{eqnarray}
The composite network $\mathcal{R}^{(N)} \star \mathcal{T}^{(N+1)}_i$
has only a classical outcome, i.e. the the index $i$.
The link product $R^{(N)}*T^{(N+1)}_i$ gives the probability to obtain
output $i$:
\begin{align}\label{eq:newborn}
  p(i|\mathcal{R}^{(N)}) = R^{(N)}*T^{(N+1)}_i=\Tr[R^{(N)}T^{(N+1)T}_i]
\end{align}
Eq. (\ref{eq:newborn})  can be interpreted as a generalized version of the Born rule:
 $R^{(N)}$ plays the role of a quantum state
while the set $\{ T^{(N+1)T}_i  \} $ is the analogue of a POVM.
A quantum tester represents the most general measurement process we can perform
on a Quantum Network; Eq. (\ref{eq:newborn})
tells us that two Quantum Networks $\mathcal{R}^{(N)}$ and $\mathcal{S}^{(N)}$
 that have the same Choi Jamio\l kowski
operator, give the same probability distribution for all testers $\mathcal{T}^{(N+1)}$:
this means that $\mathcal{R}^{(N)}$ and $\mathcal{S}^{(N)}$
are experimentally indistinguishable.
\end{remark}

\setcounter{equation}{0} \setcounter{figure}{0} \setcounter{table}{0}\newpage
\section{Quantum Tomography}\label{chaptertomo}
Calibration of physical devices is  the  basis
of any experimental procedure, especially in quantum information, where the reliability
of  the processes involved in the computation is crucial.
\emph{Quantum Tomography} is the complete determination of physical devices
in a purely experimental manner
(by relying on some well established measurement instruments),
 without  using detailed theoretical knowledge
of its inner functioning.
Originally introduced for determine the quantum state
of radiation \cite{tomoptics, tomoptics2, tomoptics3},
 Quantum Tomography 
soon became the standard technique in the measuring the fine details of
any quantum  device.
In this chapter we will present a systematic theoretical approach to optimization
of Quantum Tomography of finite dimensional systems,
as it was introduced in \cite{optimaltomo, tomoieee}.
The optimization of a tomographic procedure involves two aspects:
i) optimization of the experimental setup and ii)
optimization of the data processing, that is the classical 
processing of the measurement outcomes.
Our approach is based on the notion of \emph{informationally complete measurement}
\cite{busch}.
The optimization of the data processing
\cite{optdataproc, scott1}
relies on the fact that the operators 
describing an informationally complete measurement are generally linearly dependent, thus allowing different expansions
coefficients.
For state tomography the  optimization of the setup consists in finding the best informationally
complete POVMs. However, when the more general scenario of quantum process tomography is considered,
the optimization problem involves the choice of the input state as well
(we are in the framework of the so called \emph{ancilla assisted process tomography} 
\cite{AAPT1, AAPT2});
for this reason we will take advantage of the general theory of Quantum Networks
that will allow us to optimize both the input state and final POVM at the same time.

We will begin by introducing Quantum  Tomography of states and the key concepts 
that are needed in order to  cope with the optimization.
Then, thanks to the tools developed 
in Chapter \ref{chapter:gentheory} we will generalize this setting from quantum states to Quantum Networks.
Finally, we will provide the optimal scheme for Quantum Tomography of states, channels and POVMs. 

 \subsection{State tomography}\label{sec:statetom}
Tomographing  an unknown state  $\rho$ of a quantum system means performing
a suitable POVM $\{ P_i \}$ in such a way that $\rho$
is completely  determined by the probability distribution
\begin{align}
  p_i = \Tr[\rho P_i].
\end{align}
 Completely determining a quantum state means being able to
predict the expectation value $ \< A \> = \Tr[\rho A]$
for any operator $A$, in terms of the probabilities $p_i$, i.e.
\begin{align}\label{eq:expectationvalue}
   \< A \> = \Tr[\rho A] = \sum_i p_i f(i,A) \qquad \forall A, \rho \in \mathcal{L}(\hilb{H})
\end{align}
where $f(i,A)$ denotes suitable expansion coefficients\footnote{we assumed
 a linear reconstruction of the expectation value, that is we are considering \emph{linear quantum tomography.} }.
The function $f:(i,A)\mapsto f(i,A)$ is called \emph{data processing}
since it represents  the processing of the outcomes $i$ of the measurement 
$\{P_i\}$ in order to recover $ \< A \>$

From Eq.  (\ref{eq:expectationvalue}) we get:
\begin{align}\label{eq:expansionofA}
&  \Tr[\rho A]& = \sum_i p_i f_i[A] = \sum_i \Tr[\rho P_i] f_i[A] = &\nonumber \\
&& = \Tr\left[\rho \sum_i f_i[A] P_i\right]\qquad \forall A, \rho \quad &\Leftrightarrow \quad
A = \sum_i f_i[A] P_i \quad \forall A
\end{align}
that is it is possible to expand any $A$  over the used POVM $\{ P_i \}$.
When the expansion (\ref{eq:expansionofA})
holds for for all the operators in $\mathcal{L}(\hilb{H})$,
we have that $\Span\{P_i\} = \mathcal{L}(\hilb{H})$ and 
we say that the POVM $\{ P_i \}$ is \emph{informationally complete}.
Informationally completeness of the POVM is equivalent to the
condition \cite{FRAMES1, FRAMES2}
\begin{align}\label{eq:frameop}
  \supp(F) = \hilb{H}\otimes\hilb{H} \qquad F = \sum_i \Ket{P_i}\Bra{P_i}
\end{align}
where we exploit the isomorphism (\ref{eq:doubleket}).
A set of vectors $\ket{v_i} \in  \hilb{H}$ such that 
   $\supp(F) = \hilb{H}$, $F = \sum_i \ketbra{v_i}{v_i}$
is called \emph{frame}\footnote{in this presentation
we are restricting ourselves to the finite dimensional case.}
 and the operator
$F$ is called \emph{frame operator}.
Given a frame $\{\ket{v_i}\}$ it is possible to introduce a set of vectors
$\{\ket{u_i}\}$, called \emph{dual frame}, such that
\begin{align}
  \sum_i \ketbra{v_i}{u_i} = I.
\end{align}
If the $\ket{v_i}$ are linearly dependent the dual frame $\{\ket{u_i}\}$ is not unique.
The expansion (\ref{eq:expansionofA})
can be rephrased in terms of the $\Ket{P_i}$ in the following way:
\begin{align}
  \Ket{A}= \sum_i f(i,A) \Ket{P_i}.
\end{align}
and if we introduce a dual frame 
$\Ket{D_i}$
($  \sum_i \Ket{P_i}\Bra{D_i} = I$)
we have 
\begin{align}\label{eq:expectA}
&   \Ket{A} = \left ( \sum_i \Ket{P_i} \Bra{D_i} \right ) \Ket{A} =
\sum_i \BraKet{D_i}{A}   \Ket{P_i} \quad \Rightarrow \quad  f_i[A] = \BraKet{D_i}{A} \nonumber \\
&\< A \> = \Tr[\rho A] = \BraKet{\rho}{A} = \sum_i \BraKet{D_i}{A}   \BraKet{\rho}{P_i}
\end{align}
 We requested that the POVM has to be informationally
complete because we have no prior information about the state $\rho$ of the system, i.e.
 $\rho$ can be an arbitrary normalized positive operator in $\mathcal{L}(\hilb{H})$.
However, we can suppose that the state $\rho$ belongs to a given subspace
$\hilb{A} \subseteq \mathcal{L}(\hilb{H})$;
in this case the  only operators we need to expand 
are the ones in $\hilb{A}$ since
$\Tr[A' \rho] = 0$ for all $A' \in \hilb{A}^\perp$.
Then the set $\{ P_i \}$ is required to span only $ \hilb{A}$.
Exploiting the isomorphism
(\ref{eq:doubleket}), if $\rho \in \mathcal{A}\subseteq \mathcal{L}(\hilb{H})$
and $\Span\{P_i\}= \mathcal{A}$, we have that
 $ \Ket{\rho} \in \mathcal{V}_{\mathcal{A}}$, where we defined
 $    \hilb{H} \otimes \hilb{H}\supseteq \mathcal{V}_{\mathcal{A}} := \Span\{\Ket{P_i}\}$.
If we denote with $Q_{\mathcal{A}}$ the projector on $\mathcal{V}_{\mathcal{A}}$
then Eq. (\ref{eq:expectA})
becomes
\begin{align}\label{eq:expectA2}
&   \Ket{A} = \left ( \sum_i \Ket{P_i} \Bra{D_i} \right ) \Ket{A} =
\sum_i \BraKet{D_i}{A}   \Ket{P_i} \quad \Rightarrow \quad  f_i[A] = \BraKet{D_i}{A} \nonumber \\
&\< A \> =  \sum_i \Bra{D_i} Q_{\mathcal{A}} \Ket{A}   \Bra{\rho}Q_{\mathcal{A}}\Ket{P_i}.
\end{align}
The condition that the POVM spans the subspace $\mathcal{A}$
can be rephrased in terms of the corresponding frame operator;
it is possible to prove that
\begin{align}\label{eq:tomspanesupp}
\Span\{P_i\}=\mathcal{A} \Leftrightarrow \supp(F) = \mathcal{V}_{\mathcal{A}}.
\end{align}
First we notice that Eq. (\ref{eq:tomspanesupp})
can be rephrased as
\begin{align}
\Span\{\Ket{P_i}\} = \supp(F);
\end{align}
we will verify both the inclusions
$\Span\{\Ket{P_i}\} \subseteq \supp(F)$
and
$\Span\{\Ket{P_i}\} \supseteq \supp(F)$.
 Since any vector $\Ket{X} \in \hilb{H}\otimes\hilb{H}$
can be decomposed as $\Ket{X} = \Ket{Y} + \Ket{Z}$
where $\Ket{Y} \in \mathcal{V}_{\mathcal{A}}$
and $\Ket{Z} \in \mathcal{V}_{\mathcal{A}}^{\perp}$
 ($\mathcal{V}_{\mathcal{A}} = \Span\{\Ket{P_i}\}$),
we have
\begin{align*}
&F\Ket{X} = \sum_i \KetBra{P_i}{P_i} (\Ket{Y} + \Ket{Z}) =
\sum_i\KetBra{P_i}{P_i}\Ket{Y} = 0 \Rightarrow \\
&
 \Rightarrow
\sum_i|\BraKet{P_i}{Y}|^2= 0  \Rightarrow
\BraKet{P_i}{Y}= 0\;\; \forall i  \Rightarrow \Ket{Y}=0  \Rightarrow
\\
&\Rightarrow \Ket{X} \in \mathcal{V}_{\mathcal{A}}^{\perp}
 \Rightarrow \Ker(F) \subseteq (\Span\{\Ket{P_i}\})^\perp \Rightarrow
\Span\{\Ket{P_i}\}  \subseteq \supp(F).
\end{align*}
On the other hand,
let $F^{-1}$ be the inverse of $F$
on its support;
since $F^\dagger = F$
we have $F^{-1}F = I_{\supp(F)}= FF^{-1}$;
then  it follows
\begin{align*}
&\Ket{X}\in \supp(F)
\Rightarrow \Ket{X} = F^{-1}F\Ket{X} = 
FF^{-1}\Ket{X} = \sum_i\Bra{P_i}F^{-1}\Ket{X} \Ket{P_i} = \\ &=\sum_i c_i \Ket{P_i}
\Rightarrow \Ket{X} \in \Span{\Ket{P_i}}
\Rightarrow \supp(F)\subseteq \Span\{\Ket{P_i}\} .
\end{align*}

We now need a criterion that quantifies how well our tomographic procedure
estimates the expectation $\< A \>$ of an observable $A$.
As we have previously shown, a tomographic procedure  involves  two steps:
\begin{itemize}
\item the  measurement process which is described by the infocomplete POVM $\{ P_i\}$
or equivalently by the frame $\Ket{P_i}$;
\item the processing of the outcomes which is described by the dual
$\Ket{D_i}$.
\end{itemize}
That being so, the optimization problem consists in finding the best
POVM $\{ P_i\}$ and the best dual $\Ket{D_i}$ according to a given figure of merit.
Suppose now that the POVM is fixed and that 
 every repetition of the experiment is independent;
if the experimental frequencies are $\nu_i := \frac{n_i}{N}$
($n_i$ is
the number of outcomes $i$ occurred, and $N$ is the total number
of repetitions),
the estimated expectation $\widetilde{\< A \>}$  is then
\begin{align}\label{eq:estimexpect}
  \widetilde{\< A \>} = \sum_i f(i,A)\nu_i \leadsto \< A \>
\end{align}
where the symbol $\leadsto$ means that, by the law of large numbers,
the left hand side converges in probability to the right hand side.
A good figure of merit for the data processing strategy 
 is the statistical error in the reconstruction of expectations,
i.e. the variance of the random variable $\widetilde{\< A \>} $.
Since the variance of the mean is proportional to the variance of the distribution
\cite{statistica},
 the statistical error
occurring when the processing in Eq. 
(\ref{eq:estimexpect})
is used,
can be written as:
\begin{equation}
  \delta(A):=\sum_i|f(i,A)-\<A\>|^2 \nu_i
\end{equation}
Averaging the statistical
error over all possible experimental outcomes  we have
\begin{align}\label{eq:tomfigmer}
  \overline{\delta(A)} &:= \sum_k \left(\sum_i|f(i,A)-\<A\>|^2 \nu_i^{(k)} \right) {\bf m}_k = \nonumber \\
 & =\sum_i|f(i,A)-\<A\>|^2 \left(\sum_k\nu_i^{(k)}{\bf m}_k\right) = \sum_i|f(i,A)-\<A\>|^2 p_i
\end{align}
where the index $k$ labels different experimental outcomes
(i.e. a possible set of frequencies)
and ${\bf m}_k$ is the multinomial distribution
\begin{align}
  {\bf m}_k = \frac{N!}{\prod_l n_l^{(k)}!}\prod_{l}p_l^{N \nu^{(k)}_l}
\end{align}
that gives the probability that the experiment gives
the frequencies $\{ \nu_l^{(k)} \}$ for each outcome $l$.
In terms of  $\rho$, $P_i$ and $D_i$ Eq. (\ref{eq:tomfigmer})
becomes
\begin{align}
\overline{\delta(A)} &=  
\sum_i|f(i,A)-\<A\>|^2 p_i = \sum_i|\BraKet{D_i}{A} - \BraKet{\rho}{A}|^2 \BraKet{\rho}{P_i} =
 \nonumber \\
& = \sum_i|\BraKet{D_i}{A}|^2\BraKet{\rho}{P_i} - |\BraKet{\rho}{A}|^2;
\end{align}
where we used Eq. (\ref{eq:expectA}) in the last equality.
In a Bayesian scheme the state $\rho$
is assumed to be randomly drawn from an ensemble
$\mathcal{S} = \{\rho_n, p_n  \}$
of state $\rho_n$ with prior probability
$p_n$.
If we average the quantity $\overline{\delta(A)}$ over $\mathcal{S}$
we get
\begin{align}
  \overline{\delta(A)}_{\mathcal S} &:=
  \sum_n\left (\sum_i|\BraKet{D_i}{A}|^2\BraKet{\rho_n}{P_i} - |\BraKet{\rho_n}{A}|^2\right )p_n = \nonumber \\
& = \sum_i|\BraKet{D_i}{A}|^2\BraKet{\rho_{\mathcal{S}}}{P_i} - \sum_n|\BraKet{\rho_n}{A}|^2p_n
  \label{eq:tomfigmer2}
\end{align}
where $\rho_{\mathcal{S}} = \sum_n p_n \rho_n$.
Moreover, a priori we can be interested in some observables
more than other ones, and this can be specified in terms of a weighted set  of observables
$\mathcal{G} = \{A_m, q_m \}$, with weight
$q_m >0$ for the observables $A_m$.
Averaging over $\mathcal{G}$ we have
\begin{align}\label{eq:tomfigmer3}
   \overline{\delta(A)}_{\mathcal S, \mathcal{G}} &:=
  \sum_m\left (\sum_i|\BraKet{D_i}{A_m}|^2\BraKet{\rho_{\mathcal{S}}}{P_i} - \sum_n|\BraKet{\rho_n}{A_m}|^2p_n\right )q_m =
\nonumber \\
&=\sum_i\Bra{D_i}G\Ket{D_i} \BraKet{\rho_{\mathcal{S}}}{P_i} - \sum_{n,m}|\BraKet{\rho_n}{A_m}|^2p_nq_m
\end{align}
where $G = \sum_m q_m \KetBra{A_m}{A_m}$.
Since only the first term of Eq. (\ref{eq:tomfigmer3})
depends on $P_i$ and $D_i$,  the figure of merit is finally given by:
\begin{align}\label{eq:tomfigmer4}
\eta :=  \sum_i\Bra{D_i}G\Ket{D_i} \BraKet{\rho_{\mathcal{S}}}{P_i} 
\end{align}
If $\rho_n \in \mathcal{A}$ for all $n$ then
$Q_{\mathcal{A}}\Ket{\rho_n} = \Ket{\rho_n}$ for all $n$;
then,  reminding Eq. (\ref{eq:expectA2}),    
Eq. (\ref{eq:tomfigmer3})
becomes
\begin{align}\label{eq:tomfigmer5}
\eta =  \sum_i\Bra{D_i}Q_{\mathcal{A}}GQ_{\mathcal{A}}\Ket{D_i} \Bra{\rho_{\mathcal{S}}}Q_{\mathcal{A}}\Ket{P_i}.
\end{align}
Then, the optimization problem consists in finding the POVM $P_i$ and the dual
$D_i$ that
minimize $\eta$.
In the following section we generalize this scenario from quantum states to quantum networks.

\setcounter{equation}{0} \setcounter{figure}{0} \setcounter{table}{0}\newpage
\section{Quantum Network Tomography}
At the beginning of this chapter we said that Quantum Tomography
consists in the determination of a physical device by means of experiments
that produce classical information.
If the physical device is a preparator of quantum system
 the experiments we can perform in order to determine its state are described by POVMs;
on the other hand, if the physical device is a Quantum Network,
the experiments are described by Quantum Testers, that are the generalization
of the POVMs (see Remark \ref{rem:tester}).
In analogy with what we did for the POVMs
in the previous section it is possible to introduce \emph{informationally complete tester},
that is a quantum tester $\{ \Pi_i, \Pi_i\in \mathcal{L}(\otimes_{k=1}^{2N}\hilb{H}_k) \}$
such that the probabilities $p_i = \Tr[\Pi_i^T R]$
are sufficient to completely characterize the (generally probabilistic) Quantum Network $\mathcal{R}$
(equivalently, to completely characterize its Choi 
operator $R \in \mathcal{L}(\otimes_{k=1}^{2N-2}\hilb{H}_k)$).
This condition can be rephrased by saying that the probabilities 
$p_i = \Tr[\Pi_i^T R]$ allow to evaluate $\Tr[TR]$
for all $T \in \mathcal{L}(\otimes_{k=1}^{2N-2}\hilb{H}_k)$:
\begin{align}\label{eq:tomtestexpand}
  \Tr[TR] = \sum_if(i,T)p_i =   \sum_if(i,T) \Tr[\Pi_i^T R].
\end{align}
Following the same line as in Eq. (\ref{eq:expansionofA}) we can
say that a tester $\{ \Pi_i \}$ is informationally complete when
\begin{align}\label{eq:infocomptester}
 \Span\{ \Pi_i^T \} = \mathcal{L}(\otimes_{k=1}^{2N-2}\hilb{H}_k) 
\end{align}

The following result proves that informationally complete testers actually exist
\begin{theorem}[informationally complete quantum testers]
  Let $\{ P_i, P_i \in \mathcal{L}(\otimes_{k=1}^{2N-2}\hilb{H}_k)\}$
be an informationally complete 
POVMs.
Then the tester $\Pi_i = (d_1d_2\cdots d_{2N-2}^{-1}P^{T}_i$
is informationally complete.
\end{theorem}
\begin{Proof}
  Since $P_i$ is informationally complete we have
$ \Span\{ P_i \} =  \Span\{ \Pi^T_i \} = \mathcal{L}(\otimes_{k=1}^{2N-2}\hilb{H}_k) $.
Then the set $\Pi_i$ is informationally complete.
Moreover $\sum_i\Pi_i = (d_1d_2\cdots d_{2N-2})^{-1}I$ and 
clearly $(d_1d_2\cdots d_{2N-2})^{-1}I$
satisfies Eq. (\ref{eq:recnorm2}). \qed
\end{Proof}

The condition that $\{ \Pi_i^T \}$ span the whole 
$\mathcal{L}(\otimes_{k=1}^{2N-2}\hilb{H}_k)$
can be relaxed if we know that the Quantum Network
$\mathcal{R}$ lies  in a subspace $\mathcal{A}$ of $\mathcal{L}(\otimes_{i=k}^{2n}\hilb{H}_k)$.
A relevant case is the one in which we know
that $\mathcal{R}$ is a deterministic network;
in this case 
the set $\{\Pi_i\}$
is required to span only the subspace $\mathcal{D}$
spanned by deterministic combs $\mathcal{D} := \Span\{R| R \mbox{ satisfies Eq. (\ref{eq:recnorm2})}   \}$.

If
$\{\Pi_i\}$ is an informationally complete tester the set $\{\Ket{\Pi_i}\}$
 is a frame
and we can write the expansion
\begin{align}
  \Ket{T} = \sum_i \BraKet{\Delta_i}{T}\Ket{\Pi_i}
\end{align}
where we introduced the dual $\Ket{\Delta_i}$.
It is then straightforward to generalize Eq. (\ref{eq:tomfigmer4})
\begin{align}\label{eq:tomfigmertes1}
\eta =  \sum_i\Bra{\Delta_i}G\Ket{\Delta_i} \BraKet{R_{\mathcal{S}}}{\Pi_i}.
\end{align}
where we introduced
an ensemble of quantum network
 $\mathcal{S}:= \{R_n,p_n  \}$
and a weighted set of observables
 $\mathcal{G}:= \{T_m, q_n  \}$,
and we defined $R_{\mathcal{S}}= \sum_np_nR_n$,
 $G= \sum_mq_m \KetBra{T_m}{T_m}$.

If $R_n \in \mathcal{A}$ for all $n$
it is possible to write an analogous of Eq. (\ref{eq:tomfigmer5})
\begin{align}\label{eq:tomfigmertes2}
\eta =  \sum_i\Bra{\Delta_i}Q_{\mathcal{A}}GQ_{\mathcal{A}}\Ket{\Delta_i} \Bra{R_{\mathcal{S}}}Q_{\mathcal{A}}\Ket{\Pi_i} 
\end{align}
where $Q_{\mathcal{A}}$ is the projector on 
$\mathcal{V}_{\mathcal{A}}$ ($\Ket{R_n} \in \mathcal{V}_{\mathcal{A}}$ for all $n$).

The analogy between 
Eqs. (\ref{eq:tomfigmertes1},\ref{eq:tomfigmertes2})
and Eqs. (\ref{eq:tomfigmer4},\ref{eq:tomfigmer5})
tells us that the optimization of  Quantum state tomography
and the optimization of  Quantum network tomography
consist in minimizing the same figure of merit;
the only difference is that
 $\{ \Pi_i \}$ 
is tester instead of a POVM and it must  satisfy the constraint 
(\ref{eq:normtester}).
\subsection{Optimal quantum tomography for states, effects and transformation}
In this section
 we will show how to perform the optimization of quantum
tomographic setups for (finite-dimensional) states, channels
and effects, according to the figure of merit defined in
Eqs. (\ref{eq:tomfigmertes1},\ref{eq:tomfigmertes2}).
As we pointed out in  Section \ref{sec:statetom},
 optimizing quantum tomography can be divided in two main steps;
the first optimization stage involves a fixed detector, and
only regards the data processing, namely the choice of the
dual $\Delta_i$ used to determine the expansion coefficients $f(i,T)$
for a fixed $T$. 
As we will prove in the following, the optimal dual $\Delta_i$ is
independent of $T$, and only depends on the ensemble $\mathcal{S}$.
The second stage consists in optimizing the detector, which is represented by a POVMs
for the case of state tomography and by a  Quantum $2$-tester
when the more general case of transformation is concerned.
\begin{remark}
It is worth noting that the  optimization of  the $2$-tester
covers  both the  choice of  the best input state for the transformation
and the choice of the best final measurement.
Even if at a first sight one could think to carry this two optimization separately,
thanks to 
the general theory developed in Chapter \ref{chapter:gentheory},
they can be rephrased as a single optimization problem
over a set of suitably normalized positive operators.  
\end{remark}
\subsubsection{Optimization of data processing}
In this section
we provide the optimization of the dual frames
(i.e. of the data processing)
for the general case of quantum networks;
this derivation is new and is a generalization of the one
used in \cite{scott1}. 

Let us fix the tomographing device, which is described by the frame $\Ket{\Pi_i}$,
and let us minimize Eq. (\ref{eq:tomfigmertes1}) over the possible data processing strategies, i.e.
over all the possible duals
$\{\Ket{\Delta_i}\}$. We notice that at this stage it is irrelevant 
whether $\Pi_i$ is a quantum tester or a POVM.
Let us introduce the operator
\begin{align}\label{eq:defineX}
\mathcal{X}  = \sum_i \frac{\KetBra{\Pi_i}{\Pi_i}}{\BraKet{R_{\mathcal{S}}}{\Pi_i}}
\end{align}
Since $\Ket{\Pi_i}$ is a frame, $F = \sum_i\KetBra{\Pi_i}{\Pi_i}$
is invertible and then also $\mathcal{X}$ is invertible.
We now introduce the set $\{\Ket{\widetilde{\Delta_i}}\}$ defined as follows:
\begin{align}\label{eq:defineoptdual}
  \Ket{\widetilde{\Delta_i}}:= \mathcal{X}^{-1}\frac{\Ket{\Pi_i}}{\BraKet{R_{\mathcal{S}}}{\Pi_i}}.
\end{align}
It is easy to verify that $\{\Ket{\widetilde{\Delta_i}}\}$ is a dual:
\begin{align}
  \sum_i \KetBra{\widetilde{\Delta_i}}{\Pi_i} = 
\mathcal{X}^{-1}\sum_i\left(\frac{\KetBra{\Pi_i}
{\Pi_i}}{\BraKet{R_{\mathcal{S}}}{\Pi_i}}\right) = \mathcal{X}^{-1}\mathcal{X} = I.
\end{align}
Before proving that $\{\Ket{\widetilde{\Delta_i}}\}$ is the optimal dual we need to prove the following lemma
 \begin{lemma}\label{lem:lemfordual}
Let   $\{ \Ket{\Pi_i} \}$ be a frame and $\{ \Ket{\widetilde{\Delta_i}} \}$
be defined as in Eq. (\ref{eq:defineoptdual}).
Then, for any dual $\{ \Ket{\Delta_i} \}$ we have 
\begin{align}
  \sum_i\BraKet{R_{\mathcal{S}}}{\Pi_i}\KetBra{\widetilde{\Delta_i}}{K_i} = 0
\end{align}
where $\Ket{K_i} = \Ket{\Delta_i}-\Ket{\widetilde{\Delta_i}}$.
 \end{lemma}
 \begin{Proof}
For any dual $\Ket{\Delta_i}$ we have  $\sum_i \KetBra{\Pi_i}{\Delta_i} = I$.
Then, using Eq. (\ref{eq:defineoptdual}) we have
  \begin{align}
      \sum_i\BraKet{R_{\mathcal{S}}}{\Pi_i}\KetBra{\widetilde{\Delta_i}}{K_i} = 
      \sum_i\BraKet{R_{\mathcal{S}}}{\Pi_i}\KetBra{\widetilde{\Delta_i}}{\Delta_i} -
      \sum_i\BraKet{R_{\mathcal{S}}}{\Pi_i}\KetBra{\widetilde{\Delta_i}}{\widetilde{\Delta_i}} =
\nonumber \\ 
=
\mathcal{X}^{-1}\sum_i\KetBra{\Pi_i}{\Delta_i} - 
\mathcal{X}^{-1}\sum_i\frac{\KetBra{\Pi_i}{\Pi_i}}{\BraKet{R_{\mathcal{S}}}{\Pi_i}}\mathcal{X}^{-1}=
\mathcal{X}^{-1} - \mathcal{X}^{-1}\mathcal{X}\mathcal{X}^{-1} = 0 \nonumber
  \end{align}
\qed
 \end{Proof}
 \begin{theorem}[Optimal dual]
    Let   $\{ \Ket{\Pi_i} \}$ be a frame and $\{ \Ket{\widetilde{\Delta_i}} \}$
be defined as in Eq. (\ref{eq:defineoptdual}).
Then, for any dual $\{ \Ket{\Delta} \}$ we have 
\begin{align}
   \sum_i\Bra{\Delta_i}G\Ket{\Delta_i} \BraKet{R_{\mathcal{S}}}{\Pi_i}
\geq \sum_i\Bra{\widetilde{\Delta_i}}G\Ket{\widetilde{\Delta_i}} \BraKet{R_{\mathcal{S}}}{\Pi_i}
\end{align}
i.e. the dual $\{ \Ket{\widetilde{\Delta_i}} \}$
minimizes Eq. (\ref{eq:tomfigmertes1})
\begin{Proof}
From Lemma \ref{lem:lemfordual}
  we have:
  \begin{align}
    0 &= \Tr \left[G  \left(\sum_i\BraKet{R_{\mathcal{S}}}{\Pi_i}\KetBra{\widetilde{\Delta_i}}{K_i}\right)\right]
= \sum_i \Bra{K_i}G\Ket{\widetilde{\Delta_i}} \BraKet{R_{\mathcal{S}}}{\Pi_i} = \nonumber \\
&= \sum_i \Bra{\widetilde{\Delta_i}}G\Ket{K_i} \BraKet{R_{\mathcal{S}}}{\Pi_i}.\nonumber
  \end{align}
It is now easy to verify that
\begin{align*}
   \sum_i\Bra{\Delta_i}G\Ket{\Delta_i} \BraKet{R_{\mathcal{S}}}{\Pi_i}
&= \sum_i(\Bra{\widetilde{\Delta_i}}+ \Bra{K_i})G(\Ket{\widetilde{\Delta_i}}+\Ket{K_i}) \BraKet{R_{\mathcal{S}}}{\Pi_i}= \\
&=\sum_i\Bra{\widetilde{\Delta_i}}G\Ket{\widetilde{\Delta_i}} \BraKet{R_{\mathcal{S}}}{\Pi_i}+
\sum_i\Bra{\widetilde{\Delta_i}}G\Ket{K_i} \BraKet{R_{\mathcal{S}}}{\Pi_i}+\\
&+\sum_i\Bra{K_i}G\Ket{\widetilde{\Delta_i}} \BraKet{R_{\mathcal{S}}}{\Pi_i}+
\sum_i\Bra{K_i}G\Ket{K_i} \BraKet{R_{\mathcal{S}}}{\Pi_i}=\\
&=\sum_i\Bra{\widetilde{\Delta_i}}G\Ket{\widetilde{\Delta_i}} \BraKet{R_{\mathcal{S}}}{\Pi_i}+
\sum_i\Bra{K_i}G\Ket{K_i} \BraKet{R_{\mathcal{S}}}{\Pi_i} \geq \\
&\geq\sum_i\Bra{\widetilde{\Delta_i}}G\Ket{\widetilde{\Delta_i}} \BraKet{R_{\mathcal{S}}}{\Pi_i}
\end{align*}
\qed
\end{Proof}
 \end{theorem}
 \begin{corollary}
   If $\Ket{\Delta_i}$ is the optimal dual
Eq. (\ref{eq:tomfigmertes1}) can be rewritten as:
\begin{align}\label{eq:tomfigmertes3}
\eta =   \sum_i\Bra{\Delta_i}G\Ket{\Delta_i} \BraKet{R_{\mathcal{S}}}{\Pi_i} = 
\Tr[\mathcal{X}^{-1}G]
\end{align}
where $\mathcal{X}$ was defined in Eq. (\ref{eq:defineX}).
 \end{corollary}
 \begin{Proof}
By making use of Eq. (\ref{eq:defineoptdual}) we have:
\begin{align*}
&\sum_i\Bra{\Delta_i}G\Ket{\Delta_i} \BraKet{R_{\mathcal{S}}}{\Pi_i}= 
\Tr \left[ \left(\sum_i\KetBra{\Delta_i}{\Delta_i} \BraKet{R_{\mathcal{S}}}{\Pi_i}\right)G 
 \right] = \\
&  \Tr \left[ \left(\mathcal{X}^{-1}\sum_i\frac{\KetBra{\Pi_i}{\Pi_i}}{\BraKet{R_{\mathcal{S}}}{\Pi_i}}
\mathcal{X}^{-1}\right)G 
 \right] = 
  \Tr \left[\mathcal{X}^{-1}G 
 \right]
\end{align*}
 \end{Proof}
 \begin{remark}
   It is worth noting that the optimal dual does not depend on the set of the observables
$\{T_m, q_m \}$.
On the other hand, the optimal dual  depends on the ensemble $\{R_n,p_n\}$
through $R_{\mathcal{S}}$ that appears in the definition
of $\mathcal{X}$.
 \end{remark}
 \begin{remark}
We derived the optimal dual for the case in which the ensemble $\{R_n\}$
spans the whole $\mathcal{L}(\otimes_{k=1}^{2N-2}\mathcal{H}_k)$.
When we consider the case $R_n \in \mathcal{A}$ for all $n$,
the inverse of $\mathcal{X}$
becomes the inverse on its support and Eq. (\ref{eq:tomfigmertes3})
becomes
\begin{align}\label{eq:tomfigmertes4}
\eta =   \Tr[\mathcal{X}^{-1}Q_{\mathcal{A}}  G Q_{\mathcal{A}}].
\end{align}
 \end{remark}

\subsubsection{Optimization of the setup}

In this section we address the problem of the optimization of the
tester $\{ \Pi_i \}$
that represents the experimental setup performing the measurement process
on the unknown device we want to tomograph.
We will analyze  the case in which the 
unknown device is a Quantum Operation
$\mathcal{R}:\mathcal{L}(\hilb{H}_0) \rightarrow \mathcal{L}(\hilb{H}_1)$
represented by its Choi operator
$R\in \mathcal{L}(\hilb{H}_0 \otimes \hilb{H}_1)$;
within this framework
the experimental setup is represented by a Quantum $2$-tester
$\Pi_i \in \mathcal{L}(\hilb{H}_0 \otimes \hilb{H}_1)$
\begin{eqnarray}
&\underbrace{
\begin{aligned}
    \Qcircuit @C=1.5em @R=1.5em {
\multiprepareC{1}{\Ket{\rho}}& \ustick{0} \qw &\gate{\mathcal{R}}& \ustick{1} \qw&  
 \multimeasureD{1}{P_i} \\
 \pureghost{\Ket{\rho}}&\qw &\ustick{A} \qw&\qw& \ghost{P_i}}
\end{aligned}}& \nonumber\\
&\Pi_i& \qquad .
\end{eqnarray}
We notice that  the special case $\dim(\hilb{H}_0)=1$ corresponds to 
tomography of states while $\dim(\hilb{H}_1)$ $=1$
corresponds to tomography of effects.
In order to avoid a cumbersome notation we will perform the optimization 
for the case $\dim(\hilb{H}_0)=\dim(\hilb{H}_1)=d$; however, the generalization
to the case $\dim(\hilb{H}_0) \neq \dim(\hilb{H}_1)$ is straightforward.
We now need to make two assumptions
about the ensemble of quantum operations $\{R_n,p_n\}$
and the weighted set of observables $\{T_n,q_n\}$:
\begin{itemize}
\item the average quantum operation is the maximally depolarizing channel
$\mathcal{R}_{\mathcal{S}}(\rho)= I$ for any $\rho$, whose Choi
operator is $R_{\mathcal{S}}= d^{-1}I_0 \otimes I_1$;
\item the weighted set $\mathcal{G}=\{T_m,q_m\}$
of observables is such that\\ $G = \sum_m q_m\KetBra{T_m}{T_m} = I_{01}$;
this happens for example when the set $\{ T_m \}$
is an orthonormal basis, whose elements are equally weighted.
\end{itemize}
With this assumption Eq. (\ref{eq:tomfigmertes3})
becomes
\begin{align}\label{eq:tomfigmertessym}
  \eta = \Tr[\mathcal{X}^{-1}]= \Tr\left[\left(\sum_i\frac{d\KetBra{\Pi_i}{\Pi_i}}{\Tr[\Pi_i]}\right)^{-1} \right]
\end{align}
We now prove that we can impose the covariance w.r.t. $\group{SU}(d) \times \group{SU}(d) $ 
on the tester.
Let $\Pi_i$ be the optimal quantum tester 
and $\Delta_i$ the corresponding optimal dual;
we  define
\begin{align}\label{eq:tomcovtest}
  \Pi_{i,U,V}:= (U_0 \otimes V_1)\Pi_i(U_0^\dagger \otimes V_1^\dagger) \nonumber\\
  \Delta_{i,U,V}:= (U_0 \otimes V_1)\Delta_i(U_0^\dagger \otimes V_1^\dagger)
\end{align}
where $U_0 \in \mathcal{L}(\hilb{H}_0)$,
$V_1 \in \mathcal{L}(\hilb{H}_1)$
are unitary matrices with determinant equal to $1$, i.e. they are two instances 
of the defining representation of $\group{SU}(d)$.
It is easy to check that $  \Delta_{i,U,V}$ is a dual of 
$  \Pi_{i,U,V}$; in fact we have
\begin{align}
     \sum_i 
\int\!\d U \d V \, |\Pi_{i,U,V}\kk\bb \Delta_{i,U,V}| &=  
\int\! \d g \d h \, W_{U,V} \left(\sum_i |\Pi_{i}\kk\bb \Delta_{i}|\right)W^\dag_{U,V} \nonumber \\
& = d^{-1} I \otimes I
\end{align}
where we defined $W_{U,V} \in \mathcal{L}(\hilb{H}_{010'1'})$, 
$W_{U,V}= U_0 \otimes V_1 \otimes U^*_{0'} \otimes V^*_{1'}$.
We now prove  that 
$\Pi_{i,U,V}$ and $\Delta_{i,U,V}$
give the same  value of $\eta$
as $\Pi_{i}$ and $\Delta_{i}$:
\begin{align*}
 &\int\!\!\! \d U \d V   
\sum_i d\BraKet{\Delta_{i,U,V}}{\Delta_{i,U,V}}\Tr[\Pi_{i,U,V}]= 
\sum_i d\BraKet{\Delta_{i}}{\Delta_{i}}\Tr[\Pi_{i}]
= \eta
\end{align*}

Because of this, we can w.l.o.g.
optimize over the set of covariant testers;
the condition that the covariant tester is informationally complete w.r.t.
the subspace of transformations to be tomographed will be verified
after the optimization.

Exploiting Theorem \ref{th:groupaverage}
we have
\begin{align}
  \int\!\!\!  \d U \d V   \Pi_{i,U,V} = 
\int\!\!\! \d U \d V   (U \otimes V)\Pi_i(U^\dagger \otimes V^\dagger) =
I_{01}\frac{\Tr[\Pi_i]}{d^2};
\end{align}
A generic covariant tester is then obtained by Eq.
(\ref{eq:tomcovtest}), with operators $\Pi_i$ 
becoming ``seeds'' of the covariant tester and now being required to satisfy only the
normalization condition\footnote{this is the analogous of covariant POVM normalization in 
\cite{holevo, covpovmnorm1}}
\begin{equation} \label{eq-norm} 
  \sum_i \Tr[\Pi_i]= d
\end{equation}
in such a way that 
\begin{align}\label{eq:tomnormcovtest}
\sum_i  \int\!\! \d U \d V   \Pi_{i,U,V} = d^{-1}I_{01}  
\end{align}
satisfies the normalization (\ref{eq:normppovm}).
Because of the normalization (\ref{eq:tomnormcovtest})
we have that $\Ket{\rho} = \frac{1}{\sqrt{d}}\Ket{I}$ in Eq. (\ref{eq:realippovm})
that is, $\frac{1}{\sqrt{d}}\Ket{I}$
is the optimal input state for the quantum operations $R_n$.

With the covariant tester  Eq.~(\ref{eq:tomfigmertessym}) becomes
\begin{equation}\label{eq:tomfigmertescov}
  \eta= \Tr[\widetilde{\mathcal{X}}^{-1}],
\end{equation}
where
\begin{align}
 \widetilde{\mathcal{X}}  = \sum_i \int \!\! \d  U \d V \; \frac{ d |\Pi_{i,U,V}\kk\bb \Pi_{i,U,V}| }{\Tr[\Pi_{i,U,V}]}=
\int \!\! \d U \d V \; W_{U,V}
 \mathcal{X} W_{U,V}^\dag.
\end{align}
Applying Theorem \ref{th:groupaverage}
and exploiting the decomposition of $U\otimes U^*$
(see Section \ref{subsec:uu*})
 we have
\begin{align}
&\widetilde{\mathcal{X}}=P^{pp} + A P^{qp} + B P^{pq} + C P^{qq},\label{eq:projcov}\\
&\!\!\begin{array}{ll}
P^{pp}=P^p_{00'} \otimes P^p_{11'} 
&P^{qp}=  P^q_{00'} \otimes P^p_{11'}\\
P^{pq} = P^p_{00'} \otimes P^q_{11'}
&P^{qq}= P^q_{00'} \otimes P^q_{11'}
\end{array}
\end{align}
having posed $P^{p}_{ab}= d^{-1}\KetBra{I}{I}_{ab}$, $P^{q}_{ab}= I_{ab}-P^{p}_{ab}$ and
\begin{align}
&A = \frac{\Tr[\mathcal{X}P^{qp}]}{\Tr[P^{qp}]}
 = \frac{1}{d^2-1} \left\{\sum_i\frac{\Tr[(\Tr_1[\Pi_i])^2]}{\Tr[\Pi_i]}-1\right\},\nonumber\\
&B = \frac{\Tr[\mathcal{X}P^{pq}]}{\Tr[P^{pq}]}
=\frac{1}{d^2-1} \left\{\sum_i\frac{\Tr[(\Tr_0[\Pi_i])^2]}{\Tr[\Pi_i]}-1\right\},
\label{eq:tomcoeff}\\
&C =\frac{\Tr[\mathcal{X}P^{qq}]}{\Tr[P^{qq}]}
= \frac{1}{(d^2-1)^2}\left\{\sum_i \frac{d\Tr[\Pi_i^2]}{\Tr[\Pi_i]}-(d^2-1)(A+B)-1\right\}\nonumber.
\end{align}
The identities in Eq. (\ref{eq:tomcoeff})
can be obtained by making use of the identities 
(\ref{eq:dketid4}) and (\ref{eq:dketid5})

We can now rewrite Eq. (\ref{eq:tomfigmertescov}) as
\begin{equation}\label{eq:tomtrx}
\Tr[\widetilde{\mathcal{X}}^{-1}]= 1 + (d^2-1) \left( \frac{1}{A}+\frac{1}{B} + \frac{(d^2-1)}{C}\right).
\end{equation}

Without loss of generality we can assume the operators $\{ \Pi_i \}$
to be rank one. In fact, suppose that $\Pi_i$ has rank higher than 1.
Then it is possible to decompose it as $\Pi=\sum_{j}\Pi_{i,j}$ with
$\Pi_{i,j}$ rank 1. The statistics of $\Pi_i$ can be completely
achieved by $\Pi_{i,j}$ through a suitable coarse graining. For the
purpose of optimization it is then not restrictive to consider rank
one $\Pi_i$, namely $\Pi_i=\alpha_i|\Psi_i\kk\bb\Psi_i|_{01}$, with
$\sum_i\alpha_i=d$ and $||\Psi_i\kk|^2=1$.  Notice that all multiple seeds of this form lead
to testers satisfying Eq. (\ref{eq:tomnormcovtest}). 
Since $\Pi_i=\alpha_i|\Psi_i\kk\bb\Psi_i|$, exploiting Eq. (\ref{eq:dketid1bis})
we have
\begin{align}
&\Tr[  (\Tr_{0}[\alpha_i |\Psi_i\kk\bb\Psi_i|_{01}])^2 ]=
\alpha_i^2\Tr[(\Psi_i\Psi_i^\dagger)^2] = 
\alpha_i^2\Tr[\Psi_i^\dagger\Psi_i\Psi_i^\dagger\Psi_i] = \nonumber \\ 
&\alpha_i^2\Tr[(\Psi_i^\dagger\Psi_i\Psi_i^\dagger\Psi_i)^T]=
\alpha_i^2\Tr[(\Psi_i^T\Psi_i^*)^2]= \Tr[(\Tr_1\alpha_i\KetBra{\Psi_i}{\Psi_i})^2]\quad
\Rightarrow \quad A=B \label{eq:tomA=B} \nonumber \\
& \Tr[(\alpha_i \KetBra{\Psi_i}{\Psi_i})^2]= \alpha_i^2\Tr[\KetBra{\Psi_i}{\Psi_i})^2]= \alpha_i^2
\quad \Rightarrow \quad C = \frac{d^2-1}{1-2A} 
\end{align}
Eq. (\ref{eq:tomtrx}) becomes then
\begin{align}\label{eq:tomtrx2}
  \eta=\Tr[\widetilde{\mathcal{X}}^{-1}]=1+(d^2-1)\left(\frac2A+\frac{(d^2-1)^2}{1-2A}\right) 
\end{align}
where 
\begin{align}
0\leq
A=\frac1{d^2-1}\left(\sum_i\alpha_i\Tr[(\Psi_i\Psi_i^\dag)^2]-1\right) \leq\frac1{d+1}<\frac12\quad.
\end{align}
Since $\eta $ is a differentiable function of $A$,
the  minimum can be determined  by deriving Eq. (\ref{eq:tomtrx2})
with respect to $A$, obtaining
\begin{align}
  A = \frac1{d^2+1};
\end{align}
the corresponding value of $\eta$ is
\begin{align}
  \eta = d^6+d^4-d^2.
\end{align}
This bound is achieved by a single seed
$\Pi_0=d|\Psi\kk\bb\Psi|$, with
\begin{equation}\label{eq:tompsi}
\Psi=[d^{-1}(1-\beta) I+\beta|\psi\>\<\psi|]^{\frac12}
\end{equation}
where $\beta=[(d+1)/(d^2+1)]^{1/2}$ and $\ket{\psi}$ is any pure state;
the optimal tester is then
\begin{align}\label{eq:tomopttester}
  \Pi_{0,U,V}= (U \otimes V) d|\Psi\kk\bb\Psi| (U^\dagger \otimes V^\dagger) \\
\nonumber \\
\begin{aligned}
    \Qcircuit @C=1.5em @R=1.5em {
\multiprepareC{1}{\frac{1}{\sqrt{d}}\Ket{I}}& \ustick{0} \qw &\gate{\mathcal{R}_n}& \ustick{1} \qw&  
 \multimeasureD{1}{d \Pi_{0,U,V}} \\
 \pureghost{\frac{1}{\sqrt{d}}\Ket{I}}&\qw &\ustick{A} \qw&\qw& \ghost{d\Pi_{0,U,V}}}
\end{aligned}\nonumber 
\end{align}
We now have to verify that the set 
$\Pi_{0,U,V}$ is informationally complete.
Exploiting Th. 
\ref{th:groupaverage}
we have that 
\begin{align}
F &=\int \!\! \d U \d V \, |\Pi_{0,U,V}\kk\bb\Pi_{0,U,V}| 
= \int \!\! \d U \d V W_{U,V} d^2 \Ket{\Psi}\Ket{\Psi}\Bra{\Psi}\Bra{\Psi} W_{U,V}^\dagger = \nonumber \\
 &= \bigoplus_{\mu,\nu \in \{p,q\}} I_{\mu\nu} \otimes \frac{\Tr[\Ket{\Psi}\Ket{\Psi}
\Bra{\Psi}\Bra{\Psi} P^{\mu\nu}]}{\Tr[P^{\mu\nu}]}.
\end{align}
From the definition 
of $\Psi$ given in Eq. (\ref{eq:tompsi}) 
we have that $\Tr[\Ket{\Psi}\Ket{\Psi}\Bra{\Psi}\Bra{\Psi} P^{\mu\nu}] \neq 0$
for all $\nu,\mu$
and thus $F$
is invertible.

We now consider two relevant cases in which
$R_n \in \mathcal{V} \subseteq \mathcal{L}(\hilb{H}_{01})$:
\begin{itemize}
\item channels:   $\mathcal{C} = \Span\{R \in \mathcal{L}(\hilb{H}_{01})|\Tr_1[R]=I_0\}$  $=$
 $\{R \in \mathcal{L}(\hilb{H}_{01})|\Tr_1[R]=\lambda I_0, \lambda = d^{-1}\Tr[R] \in \mathbb{C} \}$;
\item unital channels: $\mathcal{U} = \Span\{R \in \mathcal{C}|\Tr_0[R]=I_1\}$ 
$=$
 $\{R \in \mathcal{L}(\hilb{H}_{01})|\Tr_0[R] =\lambda I_1, \Tr_1[R]$ $=\lambda I_0, \lambda =d^{-1}\Tr[R] \in \mathbb{C} \}$.
\end{itemize}
It is easy to  prove that
\begin{align}\label{eq:tomspanchan}
\mathcal{V}_\mathcal{C} := \{\Ket{R}| R \in \mathcal{C}\} =\Ker(P^{qp})
\end{align}
exploiting Eq. (\ref{eq:dketid4}) and Eq. (\ref{eq:dketid5}):
\begin{align}
  &P^{qp}\Ket{R} = 
\left(I_{00'}-\frac{\KetBra{I}{I}_{00'}}{d}\right)\otimes \frac{\KetBra{I}{I}_{11'}}{d} \Ket{R}_{010'1'}= \nonumber \\
&\frac1d \Ket{\Tr_1[R]}_{00'}\Ket{I}_{11'} -\frac{\Tr[R]}{d^2}\Ket{I}_{00'}\Ket{I}_{11'}=0
\Leftrightarrow  \Tr_1[R]= \frac{\Tr[R]}{d} I_0.
\end{align}
In a similar way we have
\begin{align}
  &(P^{qp} + P^{pq})\Ket{R} = 
\frac1d \Ket{I}_{00'}\Ket{\Tr_0[R]}_{11'}
+\frac1d \Ket{\Tr_1[R]}_{00'}\Ket{I}_{11'} -2\frac{\Tr[R]}{d^2}\Ket{I}_{00'}\Ket{I}_{11'} \nonumber\\
&= \frac1d \left(\left( \Tr_1[R] - \frac{\Tr[R]}{d}I\right)_0 \otimes I_1 
  + I_0 \otimes \left( \Tr_0[R] - \frac{\Tr[R]}{d}I\right)_1   \right)\otimes I_{0'1'} 
\Ket{I}_{010'1'} = \nonumber\\
&= 0
\Leftrightarrow
\left( \Tr_1[R] - \frac{\Tr[R]}{d}I\right)_0 \otimes I_1 
  + I_0 \otimes \left( \Tr_0[R] - \frac{\Tr[R]}{d}I\right)_1 = 0  \Leftrightarrow \nonumber \\
&\Leftrightarrow \Tr_1[R] = \frac{\Tr[R]}{d}I_0, \quad \Tr_0[R] = \frac{\Tr[R]}{d}I_1
\end{align}
that is 
\begin{align}\label{eq:tomspanunitchan}
\mathcal{V}_\mathcal{U}  =\Ker(P^{qp}+P^{pq}).  
\end{align}
From Eq. (\ref{eq:tomspanchan}) and Eq. (\ref{eq:tomspanunitchan})
it follows
\begin{align}\label{eq:tomprojectorchan}
  Q_{\mathcal{C}} = P^{pp}+P^{pq}+P^{qq} \qquad   Q_{\mathcal{U}} = P^{pp}+P^{qq}.
\end{align}
Inserting Eq. (\ref{eq:tomprojectorchan})
into Eq. (\ref{eq:tomfigmertes4}) we have
\begin{align}
&\eta_\mathcal{C}=\Tr[\widetilde{\mathcal{X}}^{-1} Q_\mathcal{C}] 
=\Tr[P^{pp}+B^{-1}P^{qp}+C^{-1}P^{qq}] \nonumber \\
&\eta_\mathcal{U}=\Tr[\widetilde{\mathcal{X}}^{-1} Q_\mathcal{U}] 
=\Tr[P^{pp}+C^{-1}P^{qq}]
\end{align}
($\widetilde{\mathcal{X}}^{-1}$
is the inverse on the support of $\mathcal{X}$).
Reminding Eq. (\ref{eq:tomA=B})
the two figures of merit become
\begin{align}
&\eta_\mathcal{C} =1+(d^2-1)\left(\frac1A+\frac{(d^2-1)^2}{1-2A}\right) & \nonumber \\
&\eta_\mathcal{U} =1+(d^2-1)\left(\frac{(d^2-1)^2}{1-2A}\right) &.
\end{align}
and the minima can be determined by derivation with respect to $A$
thus leading to
\begin{equation}
  \begin{array}{ll}
\eta_\mathcal{C} = d^6+(2\sqrt2-3)d^4+(5-4\sqrt2)d^2+2(\sqrt2-1)  &\mbox{for }A=\frac{1}{\sqrt2(d^2-1)+2}\\
\eta_\mathcal{U} = (d^2-1)^3+1  &\mbox{for }A=0.    \nonumber
  \end{array}
\end{equation}
The same results for quantum operation and for unital channels have been obtained in 
\cite{scott2} in a different framework.
The optimal tester
for the two cases under examination
have the same structure as in Eq. (\ref{eq:tomopttester})
where now in Eq. (\ref{eq:tompsi}) we have
$\beta=[(d+1)/(2+\sqrt2(d^2-1))]^{1/2}$ for  channels and
$\beta=0$ for unital channels.
Since $\beta=[(d+1)/(2+\sqrt2(d^2-1))]^{1/2}$ implies
 $\Tr[\Ket{\Psi}\Ket{\Psi}\Bra{\Psi}\Bra{\Psi} P^{\mu\nu}] \neq 0$,
we have that  optimal tester for channel tomography
spans the whole $\mathcal{L}(\hilb{H}_{01})$
(i.e. $F$ is still invertible on the whole $\mathcal{H}_{010'1'}$).

In the case of unital channel
we have $\Ket{\Psi} = d^{-\frac12}\Ket{I}$ that leads to \\
 $\Tr[\Ket{I}\Ket{I}\Bra{I}\Bra{I} P^{\mu\nu}] = 0$
if $\nu \neq \mu$.
The frame operator becomes
 \begin{align}
F &=\int \!\! \d U \d V \, |\Pi_{0,U,V}\kk\bb\Pi_{0,U,V}| 
= \int \!\! \d U \d V W_{U,V} d^2 \Ket{\Psi}\Ket{\Psi}\Bra{\Psi}\Bra{\Psi} W_{U,V}^\dagger = \nonumber \\
&= \int \!\! \d U \d V W_{U,V}  \Ket{I}\Ket{I}\Bra{I}\Bra{I} W_{U,V}^\dagger = P^{pp}+P^{qq}.
\end{align} 
Since $\supp(F)=\supp(P^{pp}+P^{qq})= \mathcal{V}_{\mathcal{U}}$,
the optimal tester spans the whole $\mathcal{U}$ as required.

The same procedure can be carried on when the operator $G$
in Eqs. (\ref{eq:tomfigmertes3}) (\ref{eq:tomfigmertes4})  has the
more general form $G= g_1P^{pp} + g_2P^{qp} +  g_3P^{pq} +  g_4P^{qq}$, where
$P^{\nu\mu}$ are the projectors defined in   (\ref{eq:projcov}). In this case
Eq. (\ref{eq:tomtrx}) becomes
\begin{equation}
\Tr[\widetilde{\mathcal{X}}^{-1}G]= g_1 + (d^2-1) \left( \frac{g_2}{A}+\frac{g_3}{B} + \frac{(d^2-1)g_4}{C}\right),
\end{equation}
which can be minimized along the same lines previously followed. 
 $G$
has this form when optimizing measuring procedures of this kind:
\emph{i}) preparing an input state randomly drawn from the set $\{U
\rho U^{\dag}_g\} $; \emph{ii}) measuring an observable chosen from
the set $\{U_h A U^{\dag}_h\} $.

 With the same derivation, but keeping
$\dim(\sH_{0})\neq\dim(\sH_{1})$, one obtains the optimal
tomography for general quantum operations. The special case of
$\text{dim}(\sH_{0})=1$ (one has $P^{q}_{00'}=0,P^{p}_{00'}=1$ in Eq.
(\ref{eq:projcov})) corresponds to optimal tomography of states
and gives
\begin{align}
  \eta = \frac1d+ \frac{d^2-1}{A}
\end{align}
with $A= \frac1{d^2-1}\left( \sum_i\frac{d\Tr[P_i^2]}{\Tr[P_i]}\right)$.
If we assume w.l.o.g that $P_i$ is rank one
we get $A = d(d-1)$ 
and the optimal value of $\eta$ is
\begin{align}
  \eta = \frac1d\left(d^3-d^2+1\right)
\end{align}
(compare with Ref. \cite{scott1}).
This bound is simply achieved
by a covariant POVM
\begin{eqnarray}
 & P_{0,V}= Vd \ketbra{\psi}{\psi}V^\dagger \qquad \braket{\psi}{\psi}=1&
\\
\nonumber \\
&    \begin{aligned}
    \Qcircuit @C=1.5em @R=1.5em {
\prepareC{\rho_n}& \ustick{1} \qw &
 \measureD{P_{0,V}} }
\end{aligned}&\nonumber 
\end{eqnarray}
where $\ket{\psi}$
is any pure state.

On the other hand
the  case $\text{dim}(\sH_{1})=1$ ($P^{q}_{11'}=0,P^{p}_{11'}=1$) gives the optimal
tomography of effects.
The optimal value of $\eta$
turns out to be
\begin{align}
  \eta = \left(d^3-d^2+1\right).
\end{align}
and is achieved by a covariant tester $\{\Pi_{0,U}\}$ through following scheme
\begin{eqnarray}\label{eq:tomoptpovm}
 & \Pi_{0,U}= U\ketbra{\psi}{\psi}U^\dagger \qquad \braket{\psi}{\psi}=1&
\\
\nonumber \\
&    \begin{aligned}
    \Qcircuit @C=1.5em @R=1.5em {
\multiprepareC{1}{\frac1{\sqrt{d}}}& \ustick{1} \qw &
 \measureD{E} \\
\pureghost{\frac1{\sqrt{d}}}& \ustick{1'} \qw & \measureD{d\Pi_{0,U}}}
\end{aligned}&\nonumber 
\end{eqnarray}
where $\ket{\psi}$
is any pure state.
It is worth noting that both in the case of effects
and in the case of states the derivation of the optimal tester
is the same. The only difference is  
that for states we assume the average state $\rho_{\mathcal{S}}$
equal to $I/d$,while for effects we assumed
$E_\mathcal{S} = I$.
The assumption $E_\mathcal{S}= \sum_np_nE_n = I$
can be interpreted by saying that
we are considering
 a set of effects $\{\tilde{E}_n = p_nE_N\}$
that form  a POVM;
from this perspective
the scheme (\ref{eq:tomoptpovm})
represents the optimal tomography of a POVM.
\subsubsection{Realization scheme for the optimal tomography}
In this  section we  illustrate a possible realization scheme for the optimal tomography of transformation
in Eq. (\ref{eq:tomopttester})
that can be useful for an experimental realization.
The first step is to prove the equality
\begin{align}\label{eq:tommissile}
\begin{aligned}
    \Qcircuit @C=1.5em @R=1.5em {
\ustick{1}&\multimeasureD{1}{d\Pi_{0,U,V}}\\
\ustick{A}&\ghost{d\Pi_{0,U,V}}  }
\end{aligned}  
\quad = \quad
\begin{aligned}
    \Qcircuit @C=1.5em @R=1em {
&\multiprepareC{3}{\Ket{\Psi}} & \qw & \ustick{B_1}\qw &\qw&\multimeasureD{1}{d\KetBra{U}{U}}\\
&\pureghost{\Ket{\Psi}} & &&\ustick{1}&\ghost{d\KetBra{U}{U}}\\
&\pureghost{\Ket{\Psi}}  &&&\ustick{A}&\ghost{d\KetBra{V}{V}}\\
&\pureghost{\Ket{\Psi}}  &\qw&\ustick{B_2}\qw&\qw &\multimeasureD{-1}{d\KetBra{V}{V}}
 }
\end{aligned}  \quad:
\end{align}
we have
\begin{align}
\KetBra{\Psi}{\Psi}_{B_1B_2}*d\KetBra{U}{U}_{B_11}*d\KetBra{V}{V}_{B_2A}=
   U \otimes V d^2\KetBra{\Psi}{\Psi}U^{\dagger}\otimes V^{\dagger}=d\Pi_{0,U,V}
\end{align}
Exploiting a  result 
proved in \cite{contmeasurdiscrete}
we also have that the continuous measurement $\KetBra{U}{U}$
can be realized by 
applying a random unitary before 
a (discrete) Bell measurement, that is
\begin{align}\label{eq:tomcontmeasur}
\begin{aligned}
    \Qcircuit @C=1.5em @R=1.5em {
\ustick{1}&\multimeasureD{1}{d\KetBra{U}{U}}\\
\ustick{A}&\ghost{d\KetBra{U}{U}}  }
\end{aligned}  
\quad = \quad
\begin{aligned}
    \Qcircuit @C=1.5em @R=1.5em {
\ustick{1}&\gate{U}&\multimeasureD{1}{\mbox{Bell}}\\
&\ustick{A}&\ghost{\mbox{Bell}}  
 }
\end{aligned}  \quad:
\end{align}
Combining the scheme (\ref{eq:tomcontmeasur})
with the scheme (\ref{eq:tommissile})
we get:
\begin{align}\label{eq:missiletomografico}
  \begin{aligned}
    \Qcircuit @C=1.5em @R=1em {
&\multiprepareC{3}{\Ket{\Psi}} & \qw & \ustick{B_1}\qw &\qw&\gate{U}&\multimeasureD{1}{\mbox{Bell}}\\
&\pureghost{\Ket{\Psi}} & \multiprepareC{1}{\frac1d \Ket{I}}&\ustick{0}\qw&\gate{\mathcal{R}}&\ustick{1}\qw&\ghost{\mbox{Bell}}\\
&\pureghost{\Ket{\Psi}}  &\pureghost{\frac1d \Ket{I}}&\qw&\qw &\ustick{A}\qw&\ghost{\mbox{Bell}}\\
&\pureghost{\Ket{\Psi}}  &\qw&\ustick{B_2}\qw&\qw &\gate{V}&\multimeasureD{-1}{\mbox{Bell}}
 }
\end{aligned}  \quad.
\end{align}
Referring to Eq. (\ref{eq:missiletomografico})
 the bipartite system
carrying the Choi operator of the transformation is indicated with the
labels $1$ and $A$. We prepare a pair of ancillary systems $B_1$
and $B_2$ in the joint state $|\Psi\kk\bb\Psi|$, then we apply two
random unitary transformations $U$ and $V$ to $B_1$ and $B_2$,
finally we perform a Bell measurement on the pair $1\;\; B_1$ and
another Bell measurement on the pair $A\;\;B_2$.

 The scheme proposed is
feasible using e.~g. the Bell measurements experimentally realized in
\cite{zeilingerwalt}.

\setcounter{equation}{0} \setcounter{figure}{0} \setcounter{table}{0}\newpage
\section{Cloning a Unitary Transformation}\label{chapter:cloning}

The no-cloning theorem \cite{nocloning} is one of the
main results in  Quantum Information, and it is
the basis of the security of quantum cryptography.
Although the cloning of quantum states
has been extensively studied 
\cite{optclonvlado, optclonwerner, asymcloningfiura, cloningreview}
the cloning of transformation
is quite a new topic.
This chapter  reviews
Ref. \cite{cloningunit} where  the cloning of  a quantum transformation was introduced
and the optimal network that clones a single use of a unitary transformation was derived.
Cloning a single use of a transformation $\map T$ means exploiting a single use of
$\map T$ inside a quantum network,
in such a way that the overall transformation is as close as possible to two uses of $\mathcal{T}$

\begin{align} \label{eq:cloningscheme}
\begin{array}{ccc}
\underbrace{\begin{aligned}
    \Qcircuit @C=0.7em @R=1.5em
{
&\ustick{0}\qw &\multigate{1}{\mathcal{C}_1}
&\ustick{1}\qw&\gate{\;\map{T}\;}&\ustick{2}\qw&\multigate{1}{\mathcal{C}_2}&\ustick{3}\qw&\qw\\
&\ustick{4}\qw&\ghost{\map{T}_1}&\qw&\ustick{A_1}\qw&\qw&\ghost{\mathcal{C}_2}&\ustick{3'}\qw&\qw
}
  \end{aligned}}
&\quad \simeq \quad&
\begin{aligned}
    \Qcircuit @C=0.7em @R=1.5em
{
&\ustick{0}\qw&\gate{\;\map{T}\;}&\ustick{3}\qw&\qw \\
&\ustick{4}\qw&\gate{\;\map{T}\;}&\ustick{3'}\qw &\qw
}
\end{aligned}   \\
\mathcal{R}&&
\end{array}
\end{align}

\begin{figure}[tb]
\includegraphics[width=13cm,clip]{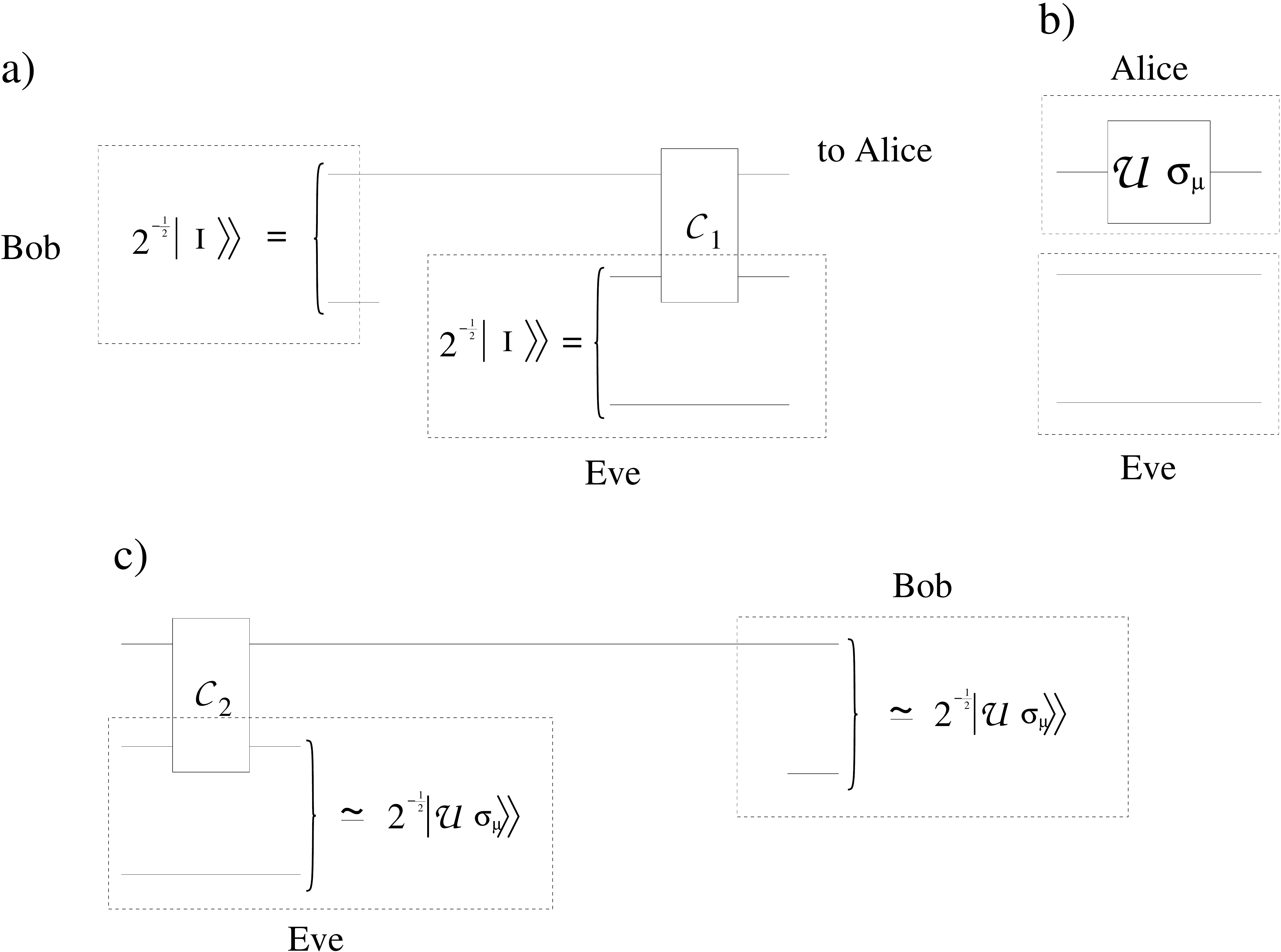}
\caption{Alternative version of the $BB84$ protocol with a possible eavesdropping.
a) Bob prepares the state  $2^{-\frac12}\Ket{I}$ and send one half of it to Alice.
Eve prepares the same state as Bob, intercepts the portion of system addressed to Alice
and performs the channel $\mathcal{C}_1$. Then Eve send one half of the outcome to Alice.
b) Alice applies $\mathcal{U}\sigma_\mu$ to her portion of system and send it back to Bob.
c) Eves intercepts the portion of system addressed to Bob and performs
the channel $\mathcal{C}_2$; if the quantum network
$\mathcal{C}_1 \star \mathcal{C}_2$
is a cloning network Eve obtains a state which is the same that Bob has
and that is as close as possible to $2^{-\frac12}\Ket{{U}\sigma_\mu}$.
 \label{fig:attackcloning}}
\end{figure}

Cloning quantum transformations can be used for copying quantum software with a limited number of
uses, and in other informational contexts, e.g.  in the security
analysis of multi-round cryptographic protocols with encoding on secret transformations.
We can consider for example this alternative version of the BB84 \cite{BB84} protocol.
Bob prepares the maximally entangled state $2^{-\frac12}\Ket{I}$ of two qubits
and sends one half of the system to Alice.
Alice perform either a unitary from the set
$ A_1 = \{ \sigma_\mu \} $ (where $\sigma_0 = I$ and 
$\sigma_{1,2,3}$ are the three Pauli matrices)
or a unitary from the rotated set
$A_2 =  \{ U\sigma_\mu \}$, where $U$ is a unitary in $\group{SU}(2)$.
Then Alice sends back  is portion of the system to Bob that finally
measures either the Bell basis $\{2^{-\frac12}\Ket{\sigma_\mu} \}$
or the rotated basis $\{2^{-\frac12}\Ket{U\sigma_\mu} \}$.
After they publicly announce their choice of basis
and discarded the cases in which they took different choices.
they use the values of $\mu$ as a secret key.
A natural attack to this protocol is  the quantum cloning
(see Fig. \ref{fig:attackcloning}).

Cloning an undisclosed transformation is a challenging
task not only from a quantum-theory perspective but even classically.
Indeed, the following result holds:
\begin{theorem}[no-cloning for transformations]
  Let $\map O_1$ and $\map O_2$
be two quantum or classical transformations and
let $p \leq 1/2$ denote the minimum of the  worst case error probability in discriminating between them.
Then $\map O_1$ and $\map O_2$
cannot be perfectly cloned by a single use unless $p =0$
  (perfect discrimination) or
$p = 1/2$ (i.e. $\map O_1 = \map O_2$)
\end{theorem}
\begin{Proof}
The proof is
simple: if perfect cloning is possible, we can get three copies, perform three times the minimum
error discrimination, and use majority voting to decide the most likely between $\map O_1$ and $\map
O_2$ with worst case error probability $p' = p^2 (3-2 p)$.  Since $p$ is the minimum error
probability, it must be $p \le p'$, whose acceptable solutions are only $p =0$ and $p=1/2$.
Viceversa, if 
$\map O_1$ and $\map O_2$
 can be perfectly distinguished (i.e. $p=0$), then they can be perfectly
cloned by a classical strategy based on discrimination and subsequent
re-preparation of the corresponding transformation.  
\end{Proof}
This result can be generalized to an arbitrary number of transformations:
\begin{corollary}
  Let $\{\map O_i\}, i=1,\dots,N$ a set of transformations.
Then perfect cloning is possible iff
either
$\mathcal{O}_i=\mathcal{O}_j$ for all $i,j$
or $\{\map O_i\}, i=1,\dots,N$ are perfectly discriminable by a single use
\end{corollary}

\begin{remark}
  It is worth noting that this result for $N >2$ is non trivial also for 
classical transformations.
Consider the following permutations of the set $\{1,2,3,4\}$
\footnote{
we use the following notation:
 the first row contains the elements $\{1,2,3,4\}$ ,
and the second row  contains the images under the permutation of the elements above.}
\begin{align}
\sigma_1=
\left(
  \begin{array}{cccc}
    1&2&3&4\\
    2&1&3&4
  \end{array}
\right),
\quad
\sigma_2=
\left(
  \begin{array}{cccc}
    1&2&3&4\\
    1&2&4&3
  \end{array}
\right),
\quad
\sigma_3=
\left(
  \begin{array}{cccc}
    1&2&3&4\\
    1&2&3&4
  \end{array}
\right);
\end{align}
there is no way to perfectly discriminate among them 
by evaluating the permutations on a single element.
\end{remark}

The existence of a no-cloning theorem immediately rises the problem of
finding the optimal  cloners:
In the following section we will derive 
the optimal network 
which produces two
approximate copies of a completely unknown unitary transformation $\mathcal{U} \in \group{SU}(d)$.

\subsection{Optimal cloning of a Unitary transformation}

Exploiting the general theory developed in chapter \ref{chapter:gentheory}
the cloning network $\mathcal{R}$ (see Eq. (\ref{eq:cloningscheme}) )
 can be represented by means of its Choi operator $R$
that has to satisfy the constraint (\ref{eq:recnorm2}), that is
\begin{align}\label{eq:clonnorm}
  \Tr_{35}[R]=I_2 \otimes R^{(1)} \qquad   \Tr_{1}[R^{(1)}]=I_{04}.
\end{align}
When we insert the unitary channel $\mathcal{U}$
in the network $\mathcal{R}$
we obtain the network
\begin{align}
  \mathcal{C}_U := \mathcal{R} \star \mathcal{U}
\end{align}
whose Choi operator is
\begin{align}
  C_U = R*\KetBra{U}{U}= \Bra{U^*}R\Ket{U^*}
\end{align}
As a figure of merit we use the channel fidelity
(see Appendix  \ref{chap:channelfidelity})
between   $\mathcal{C}_U$ and and the two uses
 $\mathcal{U} \otimes \mathcal{U}$
 of unitary channel, averaged over all the unitaries in $\group{SU}(d)$:
 \begin{align}\label{eq:clonfigmer1}
   F &:= \frac{1}{d^4}\int_{\group{SU}(d)} \d U \mathcal{F}(\mathcal{C}_U,\mathcal{U} \otimes \mathcal{U}) 
= \frac{1}{d^4}\int_{\group{SU}(d)} \d U \Bra{U}\Bra{U}C_U\Ket{U}\Ket{U} =  \\ \nonumber
& = \frac{1}{d^4}\int_{\group{SU}(d)} \d U \Bra{U}\Bra{U}\Bra{U^*}R\Ket{U^*}\Ket{U}\Ket{U}.
 \end{align}

The following Lemma exploits the symmetry of Eq. (\ref{eq:clonfigmer1})  and simplifies 
the structure of the optimal network:

\begin{lemma}\label{Lemma}
  The optimal cloning network maximizing the channel fidelity
  (\ref{eq:clonfigmer1})
 can be assumed without loss of generality 
to satisfy 
  the commutation relation
\begin{equation}\label{eq:cloncov}
[R, V_{04}^{\otimes 2} \otimes V_1^* \otimes W_2 \otimes W_{35}^{* \otimes 2}]=0 \quad \forall V,W \in \mathbb {SU} (d)~.
\end{equation}  
\end{lemma}
\begin{Proof}
 Let $R$ be optimal. Then consider the average
\begin{align}
\overline{R} = \int \d V \d W~
 (V_{04}^{\otimes 2} \otimes V_1^* \otimes W_2 \otimes W_{35}^{ *\otimes 2})
R
 (V_{04}^{\otimes 2} \otimes V_1^* \otimes W_2 \otimes  W_{35}^{*\otimes 2})^\dagger;
\end{align}
exploiting the properties of the Haar measures (see Definition \ref{def:haarmeasure})
we have
\begin{align}
F  & = \frac1{d^4}\int \d U \Bra{U}^{\otimes 2} \Bra{U^*} \overline{R}
\Ket{U^*} \Ket{U}^{\otimes 2}  = \nonumber \\
&=
\frac1{d^4}\int \d U \Bra{VUW^\dagger}^{\otimes 2} \Bra{V^*U^*W^T}
 \overline{R}
\Ket{V^*U^*W^T} \Ket{VUW^\dagger}^{\otimes 2}  = \nonumber \\
 & = \frac1{d^4}\int \d U \Bra{U}^{\otimes 2} \Bra{U^*}R
\Ket{U^*}\Ket{U}^{\otimes 2} 
\end{align}
that is $\overline{R}$ and $R$
give the same value of $F$.  
From Theorem \ref{th:groupaverage} we have that
$\overline{R}$ satisfies Eq. (\ref{eq:cloncov}).
Finally, we  verify that
$\overline {R}$ satisfies Eq. (\ref{eq:clonnorm}):
\begin{align}
&  \Tr_{35}[\overline{R}] = \nonumber \\
& = \Tr_{35}\left[  \int \d V \d W~
 (V_{04}^{\otimes 2} \otimes V_1^* \otimes W_2 \otimes W_{35}^{ *\otimes 2})
R
 (V_{04}^{\otimes 2} \otimes V_1^* \otimes W_2 \otimes  W_{35}^{*\otimes 2})^\dagger \right] = \nonumber \\
& = \int \d V \d W~
 (V_{04}^{\otimes 2} \otimes V_1^* \otimes W_2) 
\Tr_{35}\left[
R
\right]
(V_{04}^{\otimes 2} \otimes V_1^* \otimes W_2) = \nonumber \\
& = \int \d V \d W~
 (V_{04}^{\otimes 2} \otimes V_1^* \otimes W_2) 
I_2 \otimes R^{(1)}
(V_{04}^{\otimes 2} \otimes V_1^* \otimes W_2) =  \nonumber\\
& 
 = I_2 \otimes
\int \d V 
(V_{04}^{\otimes 2} \otimes V_1^*) 
 R^{(1)}
(V_{04}^{\otimes 2} \otimes V_1^*) = 
 I_2 \otimes \overline{R}^{(1)}\\
\nonumber \\
&\Tr_{1}[\overline{R}^{(1)}] = 
\Tr_{1}\left[ 
\int \d V 
(V_{04}^{\otimes 2} \otimes V_1^*) 
 R^{(1)}
(V_{04}^{\otimes 2} \otimes V_1^*) 
\right]= \nonumber\\
& =
\int \d V 
V_{04}^{\otimes 2}
\Tr_{1}\left[ V_1^*
 R^{(1)}
 V_1^* 
\right] 
V_{04}^{\otimes 2} =
\int \d V 
V_{04}^{\otimes 2}
\Tr_{1}\left[
 R^{(1)}
\right] 
V_{04}^{\otimes 2} =
\nonumber\\
&=
\int \d V 
V_{04}^{\otimes 2}
I_{04}
V_{04}^{\otimes 2} = I_{04}
\end{align}
\qed
\end{Proof}
Exploiting Eq.  (\ref{eq:cloncov})
the figure of merit (\ref{eq:clonfigmer1}) becomes:
\begin{align}\label{eq:clonfigmer2}
  F = \frac1{d^4}\Bra{I}  \Bra{I}  \Bra{I}R \Ket{I} \Ket{I} \Ket{I}.
\end{align}
Thanks to the commutation relation (\ref{eq:cloncov}),
we can apply the decomposition (\ref{decomposition}) to 
the Choi operator $R$:
\begin{align}\label{eq:clondecompor}
  R = \sum_{\nu, \mu \in \defset{S}} \sum_{i,j,k,l= \pm} T^{\nu,i,j} \otimes T^{\mu,k,l} \otimes r_{\nu\mu}^{ik,jl}
\end{align}
where we notice that 
$(r_{\nu\mu}^{ik,jl})$  is a non negative matrix for any $\nu, \mu$.

The decomposition (\ref{eq:UUU*decomposition}) induces the following identity
\begin{align}
  \hilb{H} \otimes \hilb{H} \otimes  \hilb{H} =
 \hilb{H}_{\alpha +}\oplus \hilb{H}_{\alpha -}   \oplus \hilb{H}_{\beta -} \oplus \hilb{H}_{\gamma -}
\end{align}
that leads to
\begin{align}\label{eq:clondecompoket}
&  \Ket{I}_{03}\Ket{I}_{45}  \Ket{I}_{12} = \Ket{I}_{(041)(352)} = \nonumber \\
& = ( T^{\alpha, +,+}\oplus T^{\alpha, -,-}   \oplus P^{\beta -} \oplus P^{\gamma -})\otimes I_{352} \Ket{I}_{(041)(352)} = \nonumber \\
&= \Ket{T^{\alpha, +,+}} + \Ket{T^{\alpha, -,-}}+\Ket{T^{\beta, +,+}}+\Ket{T^{\gamma,-,-}}
\end{align}
Inserting Eq.  (\ref{eq:clondecompoket}) into Eq. (\ref{eq:clonfigmer2})
and reminding the decomposition (\ref{eq:clondecompor})
we have
\begin{align}\label{eq:clonfigmerfinale}
  F& = \frac1{d^4}\Bra{I}  \Bra{I}  \Bra{I}R \Ket{I} \Ket{I} \Ket{I} = \nonumber \\
 & = \frac1{d^4} \sum_{i'}\sum_{\nu'} \Bra{T^{\nu', i',i'}}
\left(\sum_{\nu, \mu} \sum_{i,j,k,l = \pm} T^{\nu,i,j} \otimes T^{\mu,k,l} \otimes r_{\nu\mu}^{ik,jl}\right) 
\sum_{j'}\sum_{\mu'} \Ket{T^{\mu', j',j'}} = \nonumber \\
&= \frac1{d^4} \sum_\nu\sum_{i,j} d_\nu r_{\nu \nu}^{ii,jj}
\end{align}
where $d_\nu := \dim(\hilb{H}_\nu)$.

We now express the normalization constraint in terms of the 
$r_{\nu\mu}^{ik,jl}$.
Taking the trace over $\hilb{H}_{35}$ in Eq.
(\ref{eq:cloncov}) we get:
\begin{align}\label{eq:clondecompos}
&0 =   [\Tr_{35}[R],  V_{04}^{\otimes 2} \otimes V_1^* \otimes W_2 ] = 
[I_2 \otimes R^{(1)},V_{04}^{\otimes 2} \otimes V_1^* \otimes W_2] \Rightarrow \nonumber \\
&\Rightarrow [ R^{(1)},V_{04}^{\otimes 2} \otimes V_1^*] \Rightarrow
R^{(1)} := S = \sum_\nu\sum_{i,j}T^{\nu,i,j}s_{\nu}^{i,j}
\end{align}
Reminding the decomposition (\ref{eq:UUdecomposition})
we have:
\begin{align}\label{eq:clondecompotraces}
  [ \Tr_1[S], V_{04}^{\otimes 2}] = 0 \Rightarrow \Tr_1[S] = t_+P^+ \oplus t_-P^-
\end{align}
Comparing Eq. (\ref{eq:clondecompos}) and Eq. (\ref{eq:clondecompotraces}) we get:
\begin{align}\label{eq:clontraceofs}
 & \Tr_1[S]= \Tr_1[\sum_\nu\sum_{i,j}T^{\nu,i,j}s_{\nu}^{i,j}] = t_+P^+ \oplus t_-P^- \Rightarrow
\nonumber \\
&\Rightarrow 
t_i d_i= \sum_\nu d_\nu s_{\nu}^{i,i} \qquad i= \pm.
\end{align}
The normalization constraint $\Tr_1[S]= I_{04}$ becomes then
\begin{align}\label{eq:clonnormconstraint3}
  \Tr_1[S]= t_+P^+ \oplus t_-P^- =I_{04} \Rightarrow t_+=t_-=1 \qquad i= \pm.
\end{align}
Comparing now Eq. (\ref{eq:clondecompos}) with Eq. (\ref{eq:clondecompor})
we have
\begin{align}\label{eq:clontraceofr}
&  \Tr_{35}[R] = I_2 \otimes S \Rightarrow 
 \Tr_{35}\left[   R = \sum_{\nu, \mu} \sum_{i,j,k,l} T^{\nu,i,j} \otimes T^{\mu,k,l} \otimes r_{\nu\mu}^{ik,jl} \right] = \nonumber \\
&= I_2 \otimes \sum_\nu\sum_{i,j}T^{\nu,i,j}s_{\nu}^{i,j}  \Rightarrow s_{\nu}^{i,i} = \sum_k\sum_{\mu} \frac{d_\mu}{d} r_{\nu\mu}^{ik,ik}
\end{align}
Inserting Eq. (\ref{eq:clontraceofr}) into Eq. (\ref{eq:clontraceofs})
we obtain
\begin{align}\label{eq:clonquasinorm}
t_i d_i= \sum_{\nu \mu }\sum_k d_\nu \frac{d_\mu}{d} r_{\nu\mu}^{ik,ik}.
\end{align}
The normalization (\ref{eq:clonnormconstraint3}) becomes then
\begin{align}\label{eq:clonnormfinale}
dd_i= \sum_{\nu \mu }\sum_k d_\nu d_\mu r_{\nu\mu}^{ik,ik}
\end{align}

We are now ready to derive the optimal cloner:
\begin{theorem}[optimal cloning network]\label{th:optclon}
For the fidelity (\ref{eq:clonfigmerfinale}) the following bound holds:
\begin{align} \label{eq:clonmaxfide}
F \leq (d+ \sqrt
  {d^2-1})/d^3.  
\end{align}
The bound (\ref{eq:clonmaxfide}) can be achieved by a network as in
Eq. (\ref{eq:cloningscheme})
  with:
  \begin{itemize}
  \item $\mathcal{C}_1:  \mathcal{L}(\hilb{H}_{04})\rightarrow 
\mathcal{L}(\hilb{H}_{1A_1})$ is given by:
\begin{align}
\mathcal{C}_1 (\rho) = \sum_{i,j}  \Tr_{4}[P_i \rho P_j] \otimes |i\>\<j|  
\end{align}

  \item $\mathcal{C}_2: \mathcal{L}(\hilb{H}_{2 \, A_1})  \rightarrow  \mathcal{L}(\hilb{H}_{35})$
 is given by:
 \begin{align}
\mathcal{C}_2(\sigma) = \sum_{i,j}   \frac d {\sqrt{d_i d_j}} ~ P_i \left[\<i| \sigma |j\> \otimes I_{5}   \right] P_j~.   
 \end{align}
  \end{itemize}
  where $\hilb{H}_{A_1} = \mathbb{C}^2$ and $\{ \ket{+},\ket{-} \}$
is an orthonormal basis of $\hilb{H}_{A_1}$.

The resulting channel $\map C_U = \mathcal{R} \star \mathcal{U}$ is then given by
\begin{equation}\label{eq:clonoptimalcU}
\begin{split}
\map C'_U (\rho) &= \mathcal{C}_2 \star (\mathcal{U} \otimes \map I_{A_1}) \star \mathcal{C}_1(\rho)\\
 &=\sum_{i,j}  \frac d {\sqrt{d_i d_j}} ~ P_i\left[ U \Tr_{0E} [P_i \rho P_j] U^{\dag}  \otimes I  \right]P_j~.
\end{split}
\end{equation}
\end{theorem}
\begin{Proof}
Consider  the matrix $(a_{i,k}) := \left( \sum_\nu r_{\nu\nu}^{ik,ik} \right) $:
$(a_{i,k})$ is non negative and the bound $a_{i,k} \leq  \sqrt{a_{i,i}a_{k,k}}$  holds.
Then we have
\begin{align}\label{eq:clonfigmerbond1}
F & \leq \frac 1 {d^4} \left( \sum_{i} \sqrt{\sum_{\nu} d_{\nu}  r^{\nu\nu}_{ii,ii}} \right)^2  =
 \frac 1 {d^4} \left( \sum_{i} \sqrt{\sum_{\nu} \frac {d_{\nu}^2}{d_\nu}  r^{\nu\nu}_{ii,ii}} \right)^2 
\end{align}
where $\overline{\nu}$ labels the irreducible subspace of $U \otimes U \otimes U^*$
with minimum dimension, that is $\overline{\nu} = \alpha$.
Exploiting the constraint  (\ref{eq:clonnormfinale})
into Eq. (\ref{eq:clonfigmerbond1}) we get
\begin{align}
 F &\leq
 \frac1{d^4} \left( \sum_{i} \sqrt{ \frac1{d_{\alpha}} \sum_{\nu} d_{\nu}^2  r^{\nu\nu}_{ii,ii}} \right)^2  \leq \nonumber \\
&\leq \frac1{d^4} \left( \sum_{i} \sqrt{ \frac1{d_{\alpha}} d_id} \right)^2 = \frac{1}{d^4}(\sqrt{d_+}+\sqrt{d_-})^2
\end{align}
Direct computation of Eq. (\ref{eq:clonfigmer1}) with $\mathcal{C}_U$
as defined in Eq. (\ref{eq:clonoptimalcU}) proves the achievability. \qed
\end{Proof}

\subsection{The optimal cloning network}
In this section we discuss the inner structure of the optimal cloning network 
$\mathcal{R} = \mathcal{C}_2 \star \mathcal{C}_1$.
We can extend $\mathcal{C}_1$ 
 to a unitary interaction between the input
systems $\spc H_{0}, \spc H_{4}$ and the memory $\spc H_{a_1}$: $\map C_1
(\rho) = \Tr_{0E} [V (\rho \otimes |0\>\<0| ) V^\dag ]$, where $|0\> =
(|+\> +|-\>)/\sqrt{2} \in \spc M$, and $V$ is the controlled-swap $V=
I \otimes |+\>\<+| + S \otimes |-\>\<-|$, $S |\phi\>|\psi\>=
|\psi\>|\phi\>$.  Such an extension has a very intuitive meaning in
terms of quantum parallelism: for bipartite input $|\Psi\>_{04}$ the
single-system unitary $U$ is made to work on both $B$ and $E$ by
applying it to the superposition $|\Psi\>_{04}+S|\Psi\>_{04}$ and
discarding $E$. 

On the other hand the  channel $\map C_2$
can be interpreted as an extension of optimal
  cloning of pure states \cite{optclonwerner}: if 
$\map C_2$ receives the state $|\psi\>_2|+\>_{A_1}$
as an input, the output is
 $\map C_2 (|\psi\>\<\psi| \otimes |+\>\<+|) = d/d_+ \left[P_+
  (|\psi\>\<\psi| \otimes I) P_+\right]$, which are indeed two optimal
clones of $|\psi\>$.  This means that realizing the optimal cloning of
unitaries is a harder task than realizing the optimal cloning of
states: an eavesdropper that is able to optimally clone unitaries must
also be able to optimally clone pure states. This suggests that
cryptographic protocols based on gates (such the alternative $BB84$ protocol described
at the beginning of the chapter) might be harder to attack than protocols based on
states.

\begin{remark}
  It is worth notice that the optimal cloning network that we derived in the previous sections,
is not the optimal attack to the protocol in Fig.  \ref{fig:attackcloning}.
We derived the optimal cloning network  for an arbitrary unitary of $\group{SU}(d)$;
an optimization for the restricted set $A_1 \cap A_2$ of unitaries involved in the protocol
could in principle achieve better performances.
\end{remark}

\setcounter{equation}{0} \setcounter{figure}{0} \setcounter{table}{0}\newpage
\section{Quantum learning of a unitary transformation}\label{chapter:learning}

Quantum memory is a key resource for quantum information and computation and
great experimental efforts are in operation in order to realize it
\cite{quantmemory1, quantmemory2, quantmemory3}.
A quantum memory can be used  to store an unknown transformation;
in this way Alice can transmit the transformation to a distant Bob avoiding to send
the physical device; Bob retrieves the transformation from the quantum memory.

\emph{Quantum learning} is 
an example of storing and retrieving of a transformation.
Consider the scenario in which Alice puts at Bob's disposal $N$ uses
of a black box implementing an unknown unitary transformation $\mathcal{U}= U \cdot U^{\dagger}$.
Today Bob is allowed to exploit such uses at his convenience, running
an arbitrary quantum circuit that makes $N$ calls to Alice's black
box.  Tomorrow, however, Alice will withdraw the black box and ask Bob
to reproduce $\mathcal{U}$ on a new input state $|\psi\>$ unknown to him. Alice
will then test the output produced by Bob, and assign a score that is
as higher as the output is closer to $U|\psi\>$. 
More generally, Alice can
ask Bob to reproduce $\mathcal{U}$  more than once, i.e. to produce
$M \geq 1$ copies of $\mathcal{U}$.

Let us focus first on the case in which
a single use of the black box is available today ($N=1$)
and a single copy has to be produced tomorrow
($M=1$).
The only thing Bob can do today is to apply
the unknown unitary $\mathcal{U}$ to a known
(generally entangled state)
$\Ket{\Psi}$ thus producing the state
\begin{eqnarray*}
&\Ket{\Psi_\mathcal{U}}:= U \otimes I \Ket{\Psi}&  \\
\\
&\begin{aligned}
      \Qcircuit @C=1.5em @R=1em {
\multiprepareC{1}{\Ket{\Psi}}&\gate{\mathcal{U}}&\qw\\
\pureghost{\Ket{\Psi}}&\qw&\qw
}
\end{aligned}
\quad=\quad
\begin{aligned}
      \Qcircuit @C=1.5em @R=1em
 {
\multiprepareC{1}{\Ket{\Psi_\mathcal{U}}} & \qw\\
\pureghost{\Ket{\Psi_\mathcal{U}}} & \qw
}
\end{aligned}&
\end{eqnarray*}
after that Bob can store the state $\Ket{\Psi_\mathcal{U}}$
on a quantum memory.
Tomorrow, when Alice will provide the unknown state $\ket{\varphi}$,
Bob can send both $\ket{\varphi}$
and $\Ket{\Psi_\mathcal{U}}$ as input to a channel $\mathcal{C}$
whose output state has to be as close as possible to
$U\ket{\phi}$:
\begin{equation*}
    \begin{aligned}
      \Qcircuit @C=1.5em @R=1em {
\prepareC{\ket{\phi}}&\multigate{2}{\mathcal{C}}&\qw\\
\multiprepareC{1}{\Ket{\Psi_\mathcal{U}}}&\ghost{\mathcal{C}}\\
\pureghost{\Ket{\Psi_\mathcal{U}}}&\ghost{\mathcal{C}}
}
\end{aligned}
\quad \simeq \quad
\begin{aligned}
        \Qcircuit @C=1.5em @R=1em {
 \prepareC{\ket{\phi}}&\gate{\mathcal{U}}&\qw
}
\end{aligned}\qquad.
\end{equation*}
  When $N >1$ uses of the black box are available, Bob has several option
to encode the unknown unitary  into the state of the quantum memory:
 he can e.g.  opt for a {\em parallel
  strategy} where $\mathcal{U}$ is applied on $N$ different systems, yielding
\begin{eqnarray*}
&\Ket{\Psi_\mathcal{U}}=(U^{\otimes N } \otimes I)\Ket{\Psi} &\\
\\
&
\begin{aligned}
        \Qcircuit @C=1.5em @R=0.4em {
 \multiprepareC{4}{\Ket{\Psi}}&\qw&\gate{\mathcal{U}}&\qw\\
\pureghost{\Ket{\Psi}}&\qw &\gate{\mathcal{U}}&\qw\\
\pureghost{\Ket{\Psi}}&\rstick{\cdots}\\
\pureghost{\Ket{\Psi}}&\qw &\gate{\mathcal{U}}&\qw\\
\pureghost{\Ket{\Psi}}&\qw&\qw&\qw\\
}
\end{aligned} 
\quad = \quad
\begin{aligned}
        \Qcircuit @C=1.5em @R=0.4em {
 \multiprepareC{4}{\Ket{\Psi_\mathcal{U}}}&\qw&\qw\\
\pureghost{\Ket{\Psi_\mathcal{U}}}&\qw &\qw\\
\pureghost{\Ket{\Psi_\mathcal{U}}}&\rstick{\cdots}\\
\pureghost{\Ket{\Psi_\mathcal{U}}}&\qw &\qw\\
\pureghost{\Ket{\Psi_\mathcal{U}}}&\qw&\qw\\
}
\end{aligned} 
&
\end{eqnarray*}
 or for a {\em sequential
  strategy} where $\mathcal{U}$ is applied $N$ times on the same system,
generally alternated with other known unitaries, yielding
\begin{eqnarray*}
&\Ket{\Psi_\mathcal{U}}:=(UV_{N-1}\ldots V_2UV_1U \otimes I) \Ket{\Psi}&  \\
\\
&\begin{aligned}
  \Qcircuit @C=0.5em @R=1em
{
\multiprepareC{1}{\Ket{\Psi}}&\gate{\mathcal{U}}&\gate{{\mathcal{V}}_1}&\gate{\mathcal{U}}&\gate{{\mathcal{V}}_2}&\qw&\rstick{\cdots}&&&&&&\gate{{\mathcal{V}}_{N-1}}&\gate{\mathcal{U}}&\qw\\
\pureghost{\Ket{\Psi}}&\qw&\qw&\qw&\qw&\qw&\rstick{\cdots}&&&&&&\qw&\qw&\qw\\
}
\end{aligned}
\quad= \quad
\begin{aligned}
  \Qcircuit @C=0.5em @R=1em
{
\multiprepareC{1}{\Ket{\Psi_{\mathcal{U}}}}&\qw\\
\pureghost{\Ket{\Psi_{\mathcal{U}}}}&\qw
}
\end{aligned}
&\qquad.
\end{eqnarray*}
The most general
storing strategy is described by a quantum network in which the $N$
uses of the transformation $\mathcal{U}$
are inserted (see also Fig. \ref{fig:quantumalgorithm}):
\begin{eqnarray*}
&\Ket{\Psi_\mathcal{U}} := S*\KetBra{U}{U}*\cdots*\KetBra{U}{U}& \\
\\
&\underbrace{
\begin{aligned}
  \Qcircuit @C=0.5em @R=1em
{
\multiprepareC{1}{\Ket{\Psi}}
&\gate{\mathcal{U}}
&\multigate{1}{{\mathcal{C}}_1}
&\gate{\mathcal{U}}
&\multigate{1}{{\mathcal{C}}_2}
&\qw
&\rstick{\cdots}
&&&&&
&\multigate{1}{{\mathcal{C}}_{N-1}}
&\gate{\mathcal{U}}
&\multigate{1}{{\mathcal{C}}_{N}}
&\qw\\
\pureghost{\Ket{\Psi}}
&\qw
&\ghost{{\mathcal{C}}_1}
&\qw
&\ghost{{\mathcal{C}}_2}
&\qw
&\rstick{\cdots}
&&&&&
&\ghost{{\mathcal{C}}_{N-1}}
&\qw
&\ghost{{\mathcal{C}}_{N}}
&\qw
}
\end{aligned}}
\quad = \quad
\begin{aligned}
  \Qcircuit @C=0.5em @R=1em
{
\multiprepareC{1}{\Ket{\Psi_{\mathcal{U}}}}&\qw\\
\pureghost{\Ket{\Psi_{\mathcal{U}}}}&\qw
}
\end{aligned}
&
\qquad.\\
&
\!\!\!\!\!\!\!\!\!\!\!\!\!\!\!\!\!\!\!\!\!\!\!\!\!\!\!\!\!\!\!\!\!\!\!\!\!\!\!\!\!\!\!\!
\mathcal{S}
\end{eqnarray*}

Quantum learning of a transformation
can be seen as an instance of \emph{Quantum Programming}
\cite{noprogramming, programming1, programming2, programming3, programming4}:
the retrieving channel is indeed an example of a programmable device
that uses the state $\Ket{\Psi}_{\mathcal{U}}$
as a program.
The following result \cite{noprogramming} tells us that a universal programmable quantum
channel with a finite dimensional program state, does not exists.
\begin{theorem}[No Programming]\label{th:noprogramming}
There exists no universal programmable channel, that is a quantum channel
 $\mathcal{C}:\mathcal{L}(\hilb{H}_0\otimes \hilb{H}_{P})\rightarrow \mathcal{L}(\hilb{H}_1)$,
where
$\dim(\hilb{H}_0)= \dim(\hilb{H}_1)=d$
and $\dim(\hilb{H}_{P}) < \infty$,
with the following property:
\begin{align}
  \mathcal{C}(\rho \otimes \sigma_U) = U \rho U^\dagger 
\end{align}
for all state $\rho \in \mathcal{L}(\hilb{H}_0)$ and all
unitaries
$U \in \group{SU}(d)$.
\end{theorem}
\begin{Proof}
  Consider an isometric dilation $V$
of $\mathcal{C}$ and suppose that 
$\rho$ is a  pure state $\ket{\psi}$;
we have
\begin{align}
\Tr_{A}[V(\ketbra{\psi}{\psi} \otimes \sigma_U)V^\dagger] = U \ketbra{\psi}{\psi} U^\dagger  
\end{align}
Adding an auxiliary Hilbert space $\hilb{H}_{P'} \cong \hilb{H}_P$
we have the identity
\begin{align*}
&\Tr_{A'}[W\ketbra{\psi}{\psi} \otimes \KetBra{\sigma^{\frac12}_U}{\sigma^{\frac12}_U}W^\dagger ] 
=\Tr_{AP'}[V\otimes I_{P'}(\ketbra{\psi}{\psi} \otimes \KetBra{\sigma^{\frac12}_U}{\sigma^{\frac12}_U})V^\dagger\otimes I_{P'}] =\\
&=\Tr_A[V(\ketbra{\psi}{\psi} \otimes \Tr_{P'}\KetBra{\sigma^{\frac12}_U}{\sigma^{\frac12}_U} )V^\dagger]
=\Tr_{A}[V(\ketbra{\psi}{\psi} \otimes \sigma_U)V^\dagger]
\end{align*}
where we defined $\hilb{H}_{A'} = \hilb{H}_{P'}\otimes\hilb{H}_{A}$
and $W=V\otimes I_{P'}$;
then, w.l.o.g.
we can consider a pure program state $\ket{\tilde{\sigma}_U}$.
Since $U \ketbra{\psi}{\psi} U^\dagger$
is a pure state
we must have
$W(\ket{\psi} \otimes  \ket{\tilde{\sigma}_U})= U \ket{\psi} \otimes \ket{\tau_{{U}}}$ 
for some pure state $\ket{\tau_{U}}$.
First we prove that the state $\ket{\tau_{U}}$
does not depend on $\ket{\psi}$;
we have
\begin{align*}
& \braket{\tilde{\sigma}_U}{\tilde{\sigma}_U}\braket{\psi}{\psi'} = (\bra{\psi} \otimes  \bra{\tilde{\sigma}_U})
W^\dagger W(\ket{\psi'} \otimes  \ket{\tilde{\sigma}_U})=\\
&= \bra{\psi} \otimes \bra{\tau_{{U}}} U^\dagger U \ket{\psi'} \otimes \ket{\tau'_{{U}}}
= \braket{\tau_U}{\tau'_U}\braket{\psi}{\psi'}
\end{align*}
and so $\ket{\tau_U} = \ket{\tau'_U}$ if
if $\braket{\psi}{\psi'}\neq 0$.
On the other hand if $\braket{\psi}{\psi'}= 0$
we have
\begin{eqnarray*}
& U \frac1{\sqrt{2}}(\ket{\psi'}+\ket{\psi}) \otimes  \ket{\tau_{U}}
= W ( \frac1{\sqrt{2}}(\ket{\psi'}+\ket{\psi}) \otimes  \ket{\tilde{\sigma}_U})=\\
&=W (\frac1{\sqrt{2}}\ket{\psi'}\otimes  \ket{\tilde{\sigma}_U}) +
W (\frac1{\sqrt{2}}\ket{\psi}\otimes  \ket{\tilde{\sigma}_U}) = \\
&=U \frac1{\sqrt{2}}\ket{\psi'} \otimes  \ket{\tau'_{U}} +
U \frac1{\sqrt{2}}\ket{\psi} \otimes  \ket{\tau_{U}} 
 \Rightarrow  \ket{\tau'_{U}} = \ket{\tau_{U}}.
\end{eqnarray*}
Now let $U_1$ and $U_2$ be two  unitaries different up to a global phase;
for arbitrary $\ket{\psi}$ we have 
\begin{eqnarray*}
&W \ket{\psi} \otimes  \ket{\tilde{\sigma}_{U_1}} = U_1\ket{\psi} \otimes  \ket{\tau_{U_1}}\\
&W \ket{\psi} \otimes  \ket{\tilde{\sigma}_{U_2}} = U_2\ket{\psi} \otimes  \ket{\tau_{U_2}};
\end{eqnarray*}
if we take the scalar product of the previous two identities
we get
\begin{eqnarray*}
  \braket{\tilde{\sigma}_{U_1}}{\tilde{\sigma}_{U_2}} =\bra{\psi}U_1^\dagger U_2\ket{\psi}   \braket{\tau_{U_1}}{\tau_{U_2}}.
\end{eqnarray*}
If $\braket{\tau_{U_1}}{\tau_{U_2}} \neq 0$ we can write
\begin{eqnarray*}
  \frac{\braket{\tilde{\sigma}_{U_1}}{\tilde{\sigma}_{U_2}}}{\braket{\tau_{U_1}}{\tau_{U_2}}} 
= \bra{\psi}U_1^\dagger U_2\ket{\psi}
\end{eqnarray*}
and since the left hand side of the equation  does not depend on $\ket{\psi}$
we have that also $\bra{\psi}U_1^\dagger U_2\ket{\psi}$
must not depend on $\ket{\psi}$.
However, this is possible only if $U_1^\dagger U_2 = \lambda I$
for some $\lambda \in \mathbb{C}$ that is
$U_1$ is equal to $U_2$ up to a global phase which is contrary to the hypothesis.
Then it must be $\braket{\tau_{U_1}}{\tau_{U_2}} = 0$
that implies $\braket{\tilde{\sigma}_{U_1}}{\tilde{\sigma}_{U_2}}=0$ that is,
the programs of two distinct unitaries must be orthogonal states;
since there are infinite distinct unitaries in $\group{SU}(d)$
we cannot  have $\dim(\hilb{H}_P) < \infty$. \qed
\end{Proof}
The case in which the program state $\sigma_U$
is the output of a fixed quantum network 
in which $N$ uses of the unknown unitary $U$ are inserted,
corresponds to the learning scenario;
since Theorem \ref{th:noprogramming}
proved that perfect  programming is not possible,
quantum learning can be realized only approximately.\footnote{Whether the optimal  programming
of unitaries coincides with the optimal
 Quantum Learning is still an open question.}
That being so, the search for the 
optimal learning protocol deserves interest.

Moreover, we can think of quantum learning 
as an instance of quantum cloning of a transformation
as presented  in the previous chapter\footnote{
Clearly this interpretation make sense
only if the number of uses $N$ is greater than the number of replicas $M$}.
In the learning case we have the additional constraint
that the $N$ uses are provided before than the input states on which we want to apply the $M$ replicas.
Let us focus on the $N=1,M=2$ case;
the following two different scenarios are  possible:
\begin{align}
&
\underbrace{
  \begin{aligned}
    \Qcircuit @C=1em @R=1.5em
{
\ustick{0} &\multigate{1}{\mathcal{C}_1}&
\ustick{1}\qw&\gate{\;\mathcal{U}\;}&\ustick{2}\qw&\multigate{1}{\mathcal{C}_2}&\ustick{3}\qw\\
\ustick{0'}& \ghost{\mathcal{C}_1}&\qw&\ustick{A_1}\qw&\qw&\ghost{\mathcal{C}_2}&\ustick{3'}\qw
}
  \end{aligned}}
\label{eq:cloningunit12}\\
&\qquad\qquad\;\;\;\; \;\; \mathcal{E}\nonumber
\\
&\underbrace{
  \begin{aligned}
    \Qcircuit @C=1em @R=1.5em
{
&&&&&&&
\ustick{2}
&\multigate{2}{\mathcal{R}}&\ustick{3}\qw\\
\multiprepareC{1}{\Ket{\Psi}}&\ustick{0}\qw&\gate{\;\mathcal{U}\;}&\ustick{1}\qw&\multigate{1}{\mathcal{C}}&&&
\ustick{2'}
&\ghost{\mathcal{R}}&\ustick{3'}\qw\\
\pureghost{\Ket{\Psi}}&\qw&\ustick{A_1}\qw&\qw&\ghost{\mathcal{C}}&\qw&\ustick{M}\qw&\qw&\ghost{\mathcal{R}}\\
}
  \end{aligned}}
\label{eq:learnunit12} \\
&\qquad\qquad\qquad\qquad\; \mathcal{L}\nonumber
\end{align}

The  two networks
in Eqs. (\ref{eq:cloningunit12})  and (\ref{eq:learnunit12})
differ in their causal structure:
in the learning scheme
the input state $\ket{\varphi}$
cannot influence the  state 
$\Ket{\Psi}$ which the unitary is applied to;
on the other hand, the general cloning scheme
allows that the state
$\ket{\varphi}$ can affect the input state of $\mathcal{U}$.

As pointed out in Remark \ref{rem:causalordervsnorm},
different causal structures 
reflect into different normalization
of the Choi operators:
for the the network $\mathcal{E}$ in Eq. (\ref{eq:cloningunit12})
we have the constraint (see Eq. (\ref{eq:clonnorm}))
\begin{align}\label{eq:clon12const}
  \Tr_{33'}[E]= I_2 \otimes E^{(1)} \qquad \Tr_1[E^{(1)}] = I_{00'},
\end{align}
while for the learning scheme in Eq. (\ref{eq:learnunit12})
we have
\begin{align}\label{eq:learn12const}
  \Tr_{33'}[L]= I_{22'} \otimes I_1 \otimes \rho \qquad \Tr_0[\rho] = 1,
\end{align}
($\rho$ is the partial state of $\Ket{\Psi}$).
 
It is easy to prove that the constraint
(\ref{eq:learn12const}) is stronger than the constraint
(\ref{eq:clon12const}).
Suppose that the operator $E$ satisfies
Eq. (\ref{eq:learn12const});
then we have
\begin{align}
\Tr_{33'}[E] = I_1 \otimes (\rho_0 \otimes I_{22'} ) = I_1 \otimes E^{(1)}
\qquad
\Tr_0[E^{(1)}] = \Tr_0[\rho_0 \otimes I_{22'} ] = I_{22'}
\end{align}
that coincides with Eq. (\ref{eq:clon12const})
if we relabel $2 \rightarrow0$, $1 \rightarrow2$, $2' \rightarrow 0'$ and
$0 \rightarrow 1$.

This  proves  that the cloning scheme 
is more general than the learning scheme and contains
the latter as a special case.
As a consequence we will show that the performances of the learning network
are indeed worse than the performances of the cloning network.

\subsection{Optimization of quantum learning}
In this section, based on Ref. \cite{optimallearning} we derive the optimal quantum learning of an
unknown unitary randomly drawn  from a group representation.
The search of the optimal learning process can be divided into two steps:
\begin{itemize}
\item optimizing the \emph{storing} network $\mathcal{S}$, that is the
  device that encodes
the unknown transformation  $\mathcal{U}$ into the state
$\Ket{\Psi_\mathcal{U}}$ of a quantum memory;
\item finding the optimal  \emph{retrieving} channel $\mathcal{C}$, that receives
 $\Ket{\Psi_\mathcal{U}}$ and an unknown state
$\ket{\phi}$  as input and  emulates $\mathcal{U}$
applied to $\ket{\phi}$.
\end{itemize}

An alternative to coherent retrieval is to estimate $\mathcal{U}$, to store the
outcome in a classical memory, and to perform the estimated unitary on
the new input state.  This incoherent strategy has the double
advantage of avoiding the expensive use of a quantum memory, and of
allowing one to reproduce $\mathcal{U}$ an unlimited number of times with
constant quality.  However, incoherent strategies are typically
suboptimal for the similar task of quantum cloning, and
this would suggest that a coherent retrieval achieves better
performances.  Surprisingly enough, we find that the incoherent
strategies already achieve the ultimate performance of quantum
learning.  We analyze the case in which $U$ is a completely unknown
unitary in a group $\group{G}$, and we find that the performances of the
optimal retrieving machine are equal to those of the optimal estimation.

We will show that the solution has the following structure:
\begin{itemize}
\item apply the $N$ of the unknown unitary in parallel on a suitable entangled state;
\item estimate the unknown unitary by measuring the  state of the quantum memory
\item produce the estimated unitary $M$ times where $M$
is the number of replicas that are required.
\end{itemize}

\subsubsection{Considered scenario: $M=1$}
We tackle the optimization of learning starting from the case $M=1$.
\begin{figure}[tb]
\includegraphics[width=13cm,clip]{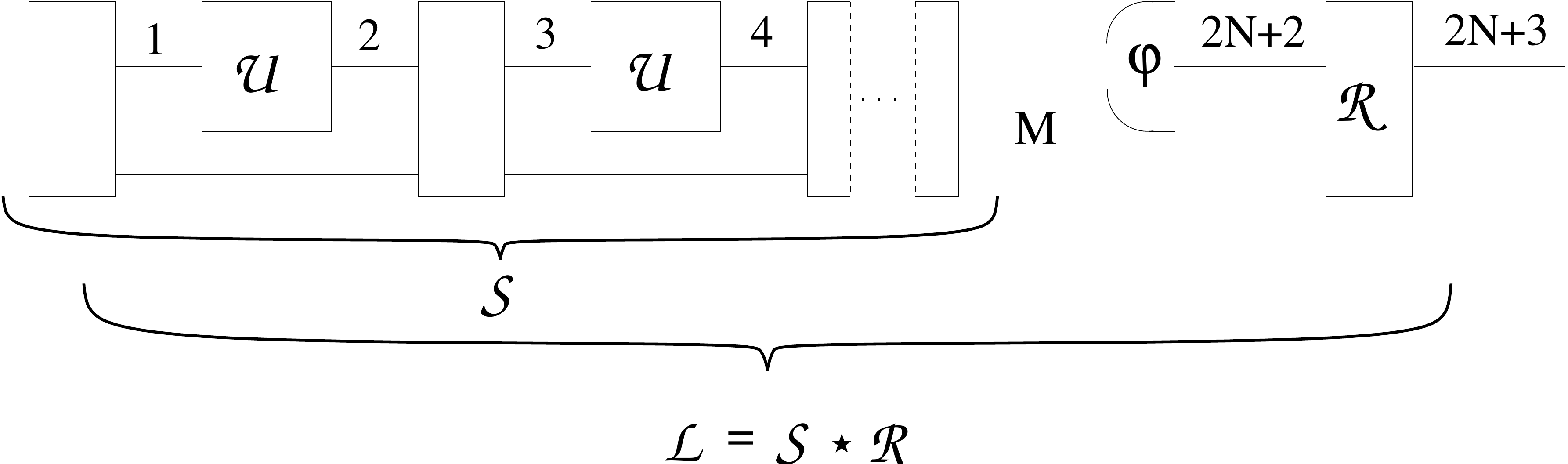}
\caption{The learning process is described by a quantum network $\mathcal{L}=\mathcal{S}\star\mathcal{R}$
 in which the $N$ uses of 
  $\mathcal{U}$ are plugged, along with the state $|\varphi\>$.  The wires
  represent the input-output Hilbert spaces.
The output of the network
$\mathcal{S}$ is stored in a
  quantum memory $M$, later used by the retrieving channel $\mathcal{R}$
  \label{fig:learning}}
\end{figure}

Referring to Fig.  \ref{fig:learning}, we label the Hilbert spaces of
quantum systems according to the following criterion:
$\hilb{H}_{2j-1}$ is the input of the $j$-th example
of $\mathcal{U}$, and $\hilb{H}_{2j}$ is the corresponding
output. We denote by $\hilb H_{\rm i} = \bigotimes_{j=1}^{N} \hilb H_{2j-1}$ the
Hilbert spaces of all inputs 
and by   $\hilb H_{\rm o} = \bigotimes_{j=1}^{N} \hilb H_{2j}$
the
Hilbert spaces of all outputs of the $N$ examples.  Alice's
input state $|\varphi\>$ belongs to $\hilb{H}_{2N+2}$, and the output
state finally produced by Bob belongs to $\hilb H_{2N+3}$.  All spaces
$\hilb H_j $ considered here are $d-$dimensional, except the spaces
$\hilb H_0$ and $\hilb H_{2 N +1}$ which are one-dimensional, and are
introduced just for notational convenience.

 The Choi operator
 $L \in \mathcal{L}(\hilb{H}_{\rm i} \otimes \hilb{H}_{\rm o} \otimes \hilb{H}_{2N+2}\otimes\hilb H_{2N+3})$ 
of the 
learning network $\mathcal{L}$ satisfies the normalization condition (\ref{eq:recnorm2}),
that becomes
\begin{equation}\label{eq:learnrecnorm}
\Tr_{2k-1} [ L^{(k)}] = I_{2k-2} \otimes L^{(k-1)} \qquad k=1,2, \dots, N+2~
\end{equation} 
where $L^{(N+2)}=L$, $L^{(0)} =1$, and $L^{(k)}\in\mathcal{L}(\hilb H_j)_{j=0}^{2k-1}$.

When we insert the $N$ example in the
learning board we obtain the network
\begin{align}
\mathcal{C}_U = \mathcal{L}\star \mathcal{U} \star \dots \star   \mathcal{U}
\end{align}
and then, according to Theorem \ref{th:linknet},
its Choi-Jamio\l kowsky operator is given by
\begin{align}\label{eq:choilearn}
C_U &= L * \KetBra{U}{U}_{12} * \dots *   \KetBra{U}{U}_{(2N-1)(2N)} = \nonumber \\
&= \Tr_{\rm i,o}
\left[ L \left( I_{2N+3} \otimes I_{2N+2}
    \otimes (|U\kk \bb U|^{ \otimes N})^T \right) \right]
=\Bra{U^*}^{ \otimes N}L\Ket{U^*}^{\otimes N}.
\end{align}

We now need to introduce a figure of merit that
quantifies how close the resulting channel
$\mathcal{C}_U$ is to the original unitary transformation $\mathcal{U}$.
A reasonable choice is to maximize the channel fidelity $\mathcal{F}$ (see Definition \ref{def:channelfidelity}
and the following lemmas)
between $\mathcal{C}_U$ and the target unitary 
averaged over $U$:
\begin{align}\label{eq:learfigmer3}
  F&:= \int_{\group{G}} \d U \mathcal{F}(\mathcal{C}_U, \mathcal{U})=
\frac1{d^2}\int_{\group{G}} \d U \Bra{U}C_U, \Ket{U})
\end{align}
Inserting Eq. (\ref{eq:choilearn})
into Eq. (\ref{eq:learfigmer3}) we have
\begin{align}\label{eq:learfigmer4}
    F=\frac1{d^2}\int_{\group{G}} \d U \Bra{U}_{(2N+3)(2N+2)} \Bra{U^*}^{ \otimes N}_{\defset{o}\;\defset{i}} L
  \Ket{U^*}^{\otimes N}_{\defset{o}\;\defset{i}}\Ket{U}_{(2N+3)(2N+2)} .
\end{align}

The following lemma 
simplifies the search for the optimal
learning network
\begin{lemma}\label{lem:learncov}
  The operator $L$
maximizing the fidelity (\ref{eq:learfigmer4})
can be assumed without loss of generality to satisfy the following commutation relation
\begin{equation} 
 [L,  
 U^{* \otimes N}_{\rm o} \otimes V^{ \otimes N}_{\rm i} \otimes U_{2N+3} \otimes V^*_{2N+2} ] = 0 
\qquad \forall U,V \in \group{G}. \label{eq:covlearn}
\end{equation}
\end{lemma}
\begin{Proof}
Let $L$ be the Choi operator of the optimal learning network;
if we define 
\begin{align}
  \overline{L} = 
\int \!\! \d Z \d W (Z^{* \otimes N}_{\rm o} \otimes W^{\otimes N}_{\rm i} \otimes Z_{2N+3} \otimes W^*_{2N+2})^\dagger  L  (Z^{* \otimes N}\otimes W^{\otimes N} \otimes Z \otimes W^*), \nonumber
\end{align}
exploiting the properties of the Haar measure (see Definition \ref{def:haarmeasure}),
we have
\begin{align}
  &\frac1{d^2}\int_{\group{G}} \d U \Bra{U}_{(2N+3)(2N+2)} \Bra{U^*}^{ \otimes N}_{\defset{o}\;\defset{i}} \overline{L}
\Ket{U^*}^{\otimes N}_{\defset{o}\;\defset{i}}  \Ket{U}_{(2N+3)(2N+2)}  = \nonumber \\
&=
\frac1{d^2}\int_{\group{G}} \d U \Bra{ZUW^\dagger}_{(2N+3)(2N+2)} \Bra{Z^*U^*W^T}^{ \otimes N}_{\defset{o}\;\defset{i}} \overline{L}
\Ket{Z^*U^*W^T}^{\otimes N}  \Ket{ZUW^\dagger}  = \nonumber \\
 & = \frac1{d^2}\int_{\group{G}} \d U \Bra{U}_{(2N+3)(2N+2)} \Bra{U^*}^{ \otimes N}_{\defset{o}\;\defset{i}} L
\Ket{U^*}^{\otimes N}_{\defset{o}\;\defset{i}}  \Ket{U}_{(2N+3)(2N+2)} 
\end{align}
that is $\overline{L}$ and $L$
give the same value of $F$.  
$\overline{L}$, thanks to Theorem \ref{th:groupaverage},
enjoys the property ~(\ref{eq:covlearn}).
Finally, it is easy to verify
that $\overline{L}$
satisfies the constraint
(\ref{eq:learnrecnorm}).
\end{Proof}

\subsubsection{Optimization of the storing strategy}
Lemma \ref{lem:learncov} allows us to look
for the optimal learning network among the ones that satisfy
Eq. (\ref{eq:covlearn}).

Using Eq. (\ref{eq:learnrecnorm}) with $k = N+2$
we have 
\begin{align} \label{eq:stornorm}
  \Tr_{2N+3}[L] = I_{2N+2} \otimes L^{(N+1)}.
\end{align}
The commutation (\ref{eq:covlearn})
can be rewritten as
\begin{align}\label{eq:covlearn2}
(  U^{* \otimes N}_{\rm o} \otimes V^{ \otimes N}_{\rm i} \otimes U_{2N+3} \otimes V^*_{2N+2})^\dagger L 
 U^{* \otimes N}_{\rm o} \otimes V^{ \otimes N}_{\rm i} \otimes U_{2N+3} \otimes V^*_{2N+2}  = L 
\end{align}
Taking the trace over $\hilb{H}_{2N+3}$ in Eq. (\ref{eq:covlearn2})
and using Eq. (\ref{eq:covlearn2})
we get
\begin{align}\label{eq:storcov}
&\Tr_{2N+3}[   U^{* \otimes N}_{\rm o} \otimes V^{ \otimes N}_{\rm i} \otimes U_{2N+3} \otimes V^*_{2N+2})^\dagger L 
 U^{* \otimes N}_{\rm o} \otimes V^{ \otimes N}_{\rm i} \otimes U_{2N+3} \otimes V^*_{2N+2} ] =  \nonumber\\
& = \Tr_{2N+3}[L ] \Rightarrow  U^{* \otimes N}_{\rm o} \otimes V^{ \otimes N}_{\rm i})^\dagger L^{(N+1)} 
 U^{* \otimes N}_{\rm o} \otimes V^{ \otimes N}_{\rm i} = L^{(N+1)} \Rightarrow \nonumber\\
&\Rightarrow [L^{(N+1)},  
 U^{* \otimes N}_{\rm o} \otimes V^{ \otimes N}_{\rm i} ] = 0 .
\end{align}

We now prove that the commutation
(\ref{eq:storcov}) implies that the parallel
storage is optimal.

\begin{lemma}[Optimality of parallel storage]\label{lem:parallelstor}
 The optimal storage of
  $U$ can be achieved by applying $U^{\otimes N}_{\defset{o}} \otimes I^{\otimes N}_{\defset{i}}$
 on a suitable input state $\Ket{\Psi} \in \hilb H_{\rm o} \otimes
  \hilb H_{\rm i}$.
\end{lemma}
\begin{Proof}
According to Th. \ref{th:realinet} the learning Network $\map L$ 
can be realized as a sequence of
isometries, followed by a measurement on  an ancillary space.
\begin{align}
\nonumber \\
&\begin{aligned}
\Qcircuit @C=1.5em @R=1em {
&\multiprepareC{1}{\Ket{L^{(1)*\frac12}}}& \ustick{1} \qw & \ustick{2}&  
 \multigate{1}{\mathcal{V}_2} &  \ustick{3} \qw \\
& \pureghost{\Ket{L^{(1)*\frac12}}}& \ustick{A_1} \qw&\qw& \ghost{\mathcal{V}_1}  &
\ustick{A_2}\qw\\
}
\end{aligned}
\quad\cdots\quad
\begin{aligned}
\Qcircuit @C=1.5em @R=1em {
\ustick{2N}  &\multigate{1}{\mathcal{V}_{N+1}}&
                 &\ustick{2N+2}            &   \multigate{1}{\mathcal{V}_{N+2}}  &  \ustick{2N+3} \qw \\
\ustick{A_N}  &\ghost{\mathcal{V}_{N+1}}& 
 \ustick{M} \qw &\qw                      & \ghost{\mathcal{V}_{N+2}}&\ustick{A_{N+2}}\qw &\measureD{I}
}    
  \end{aligned} \nonumber \\
&\;\;\quad \underbrace{\qquad \qquad \qquad\qquad\qquad\qquad\qquad \qquad\qquad\qquad\qquad}
 \;\;\; \underbrace{\qquad \qquad \qquad\qquad\qquad} \nonumber \\
&\qquad\qquad\qquad\qquad\qquad \qquad \;   \mathcal{S} 
\qquad\qquad\qquad \qquad  \qquad\qquad\qquad \qquad\;   \mathcal{R}
\nonumber
\end{align}
The storing network is then represented by the isometric channel 
$\mathcal{S}:=\mathcal{W}^{(N+1)}:=W^{(N+1)} \cdot  {W^{(N+1)}}^\dagger$
where
$W^{(N+1)}= V^{(N)}\cdots V^{(1)} = I_{\defset{o}}\otimes L^{(N+1)*\frac12}_{\defset{o}'\;\defset{i}'} \Ket{I}_{\defset{o}\;\defset{o}'}
\otimes T_{\defset{i} \rightarrow \defset{i}'}$
and $\mathcal{H}_M = \supp(L^{(N+1)*\frac12}_{\defset{o}'\;\defset{i}'})$.
The Choi Jamiolkowski operator of the storing network is then
 $S=W^{(N+1)}$ $\KetBra{I}{I}_{\defset{i} \; \defset{i}'}  {W^{(N+1)}}
^\dagger =
 \KetBra{L^{(N+1)*\frac12}}{L^{(N+1)*\frac12}}_{\defset{o}\,\defset{i} \;\defset{o}'\,\defset{i}'}$
When we connect the storing board with the $N$
copies of the unitary the final state on space $\mathcal{H}_M$  becomes
\begin{align*}
 &  \KetBra{\Psi_U}{\Psi_U} :=  S
*
\KetBra{U}{U}_{12}*\cdots*\KetBra{U}{U}_{(2N-1)(2N)} = \\
&=\KetBra{L^{(N+1)*\frac12}}{L^{(N+1)*\frac12}}_{\defset{o}\,\defset{i} \;\defset{o}'\,\defset{i}'}
*
\KetBra{U}{U}_{12}*\cdots*\KetBra{U}{U}_{(2N-1)(2N)} =
\end{align*}
and exploiting   Eq. (\ref{eq:storcov}) we have
\begin{align*}
  \Ket{\Psi_U} &= \Bra{U^*}^{\otimes N}_{\defset{o}\,\defset{i}}  \Ket{L^{(N+1)*\frac12}}_{\defset{o}\,\defset{i} \;\defset{o}'\,\defset{i}'} =
\Bra{I}^{\otimes N}_{\defset{o}\,\defset{i}}
( {U^{T}}^{\otimes N}_{\defset{o}} \otimes I_{\defset{i}})
 \Ket{L^{(N+1)*\frac12}}_{\defset{o}\,\defset{i} \;\defset{o}'\,\defset{i}'} 
= \\
&=
\Bra{I}^{\otimes N}_{\defset{o}\,\defset{i}}
 \Ket{({U^T}_{\defset{o}}^{\otimes N}\otimes I_{\defset{i}})L^{(N+1)*\frac12}}_{\defset{o}\,\defset{i} \;\defset{o}'\,\defset{i}'} =
\Bra{I}^{\otimes N}_{\defset{o}\,\defset{i}}
 \Ket{L^{(N+1)*\frac12}({U^T}_{\defset{o}}^{\otimes N}\otimes I_{\defset{i}})} = \\
&=\Bra{I}^{\otimes N}_{\defset{o}\,\defset{i}}
 ({U}_{\defset{o}'}^{\otimes N}\otimes I_{\defset{i}'})\Ket{L^{(N+1)*\frac12}}_{\defset{o}\,\defset{i} \;\defset{o}'\,\defset{i}'} =
(U_{\defset{o}'}^{\otimes N}\otimes I_{\defset{i}'})
\Bra{I}^{\otimes N}_{\defset{o}\,\defset{i}}\Ket{L^{(N+1)*\frac12}}_{\defset{o}\,\defset{i} \;\defset{o}'\,\defset{i}'} \\
&= (U_{\defset{o}'}^{\otimes N}\otimes I_{\defset{i}'})\Ket{\Psi}_{\defset{o}'\,\defset{i}'}.
\end{align*}
where we defined $\Ket{\Psi}_{\defset{o}\,\defset{i} \;\defset{o}'\,\defset{i}'}=
\Bra{I}^{\otimes N}_{\defset{o}\,\defset{i}}\Ket{L^{(N+1)*\frac12}}_{\defset{o}'\,\defset{i}'}$.
Then every storing board can be realized 
applying $(U_{\defset{o}'}^{\otimes N} \otimes I_{\defset{i}'})$
to a suitable input state $\Ket{\Psi} \in \hilb{H}_{\defset{o}' \,\defset{i}'}$.
\qed
\end{Proof}
\begin{remark}\label{parallelgruppi}
  It is possible to prove that the optimality of a parallel strategy
is a common feature of all the problems involving
estimation of group transformations.
However, the only covariance
(\ref{eq:storcov})
does not imply that the Quantum Network can be parallelized;
a crucial aspect of the problem
is that we have  access to the physical transformation
$U$ and that the scheme $(U^{\otimes N}\otimes I)\Ket{\Psi}$
is physically realizable.
We will see (see Chapter \ref{chapter:observables}) that there are cases in which
the quantum storing network is covariant 
but it cannot be parallelized
because the set transformations $R_U$ we want to learn,
even if they are orbit of a group representation
(e.g. $R_U=UR_IU^\dagger$),  do not form a group;
In this case, an analogous of Eq. (\ref{eq:storcov}) holds but since we do not have physical access
to the unitaries $U$, the optimal network cannot be assumed to be parallel.
\end{remark}

Optimization of learning is then reduced to finding the optimal input state
$|\Psi \>$ and the optimal retrieving channel $\map R$.  The
fidelity can be computed substituting $L = R * S$ in Eq.
(\ref{eq:learfigmer4}), and using the relation $\bb U| \bb U^* |^{\otimes N}
(R*S) |U\kk | U^* \kk^{\otimes N} = \bb U| R |U\kk * \bb U^*|^{\otimes
  N} S |U^*\kk^{\otimes N} = \bb U| R|U\kk *\KetBra{\Psi_U}{\Psi_U} $, which gives
\begin{align}\label{eq:learfigmer5}
  F = \frac 1 {d^2} \int_G \Bra{U} \< \Psi_U^* | R \Ket{U}
  |\Psi_U^* \> ~\d U.
\end{align}
The following lemma further simplifies the structure
of the optimal input state for storage

\begin{lemma}[Optimal states for storage]\label{lem:learnoptstate}
  The optimal input state for storage can be taken of the form
\begin{equation}\label{inpst}
  |\Psi \kk = \bigoplus_{j} \sqrt{\frac{p_j}{d_j}}  |I_j\kk  \in \widetilde{\hilb H}
  ~,\end{equation} 
where $p_j$ are probabilities, $\widetilde{\hilb H} = \bigoplus_j (\hilb H_j \otimes \hilb
H_j)$ is a subspace of $\hilb H_{\rm o} \otimes \hilb H_{\rm i}$ carrying the
representation $\widetilde{U} = \bigoplus_{j}( U_j \otimes I_{j})$,
$I_j$ being the identity in $\hilb H_j$, and the index $j$ labelling
the irreducible representations  $U_j$ contained in the decomposition of $U^{\otimes N}$.
\end{lemma}
\begin{Proof}
Let us consider the local state
\begin{align*}
&\rho := \Tr_{\rm i'}[|\Psi \kk\bb \Psi |]  =
 \Tr_{\rm i'}[\Bra{I}^{\otimes N}_{\defset{o}\,\defset{i}}
\KetBra{L^{(N+1)*\frac12}}{L^{(N+1)*\frac12}}_{\defset{o}\,\defset{i} \;\defset{o}'\,\defset{i}'}
\Ket{I}^{\otimes N}_{\defset{o}\,\defset{i}}
] = \\ 
&= \Tr_{\rm i'}[L^{(N+1)  \frac12}_{\defset{o}'\,\defset{i}'} \Bra{I}^{\otimes N}_{\defset{o}\,\defset{i}}
\KetBra{I}{I}_{\defset{o}\,\defset{i} \;\defset{o}'\,\defset{i}'}
\Ket{I}^{\otimes N}_{\defset{o}\,\defset{i}}L^{(N+1) \frac12}_{\defset{o}'\,\defset{i}'}
] = \\
&= \Tr_{\rm i'}[L^{(N+1)  \frac12}_{\defset{o}'\,\defset{i}'}
\KetBra{I}{I}_{\defset{o}'\,\defset{i}'}
L^{(N+1) \frac12}_{\defset{o}'\,\defset{i}'}]
\end{align*}
It is easy to prove that $\rho \in \mathcal{L}(\hilb{H}_{\defset{o}'})$
is invariant under $U^{\otimes N}$:
\begin{align*}
  &U^{\otimes N}\rho U^{\dagger \otimes N} = 
  U^{\otimes N} \Tr_{\rm i'}[L^{(N+1)  \frac12}_{\defset{o}'\,\defset{i}'}
\KetBra{I}{I}_{\defset{o}'\,\defset{i}'}
L^{(N+1) \frac12}_{\defset{o}'\,\defset{i}'} U^{\dagger \otimes N} = \\
&=\Tr_{\rm i'}[(U^{\otimes N}\otimes I_{\defset{i}'})L^{(N+1)  \frac12}_{\defset{o}'\,\defset{i}'}
\KetBra{I}{I}_{\defset{o}'\,\defset{i}'}
L^{(N+1) \frac12}_{\defset{o}'\,\defset{i}'}(U^{\dagger \otimes N}\otimes I_{\defset{i}'})] = \\
&=\Tr_{\rm i'}[L^{(N+1)  \frac12}_{\defset{o}'\,\defset{i}'}
(U^{\otimes N}\otimes I_{\defset{i}'})
\KetBra{I}{I}_{\defset{o}'\,\defset{i}'}
(U^{\dagger \otimes N}\otimes I_{\defset{i}'})
L^{(N+1) \frac12}_{\defset{o}'\,\defset{i}'}] =\\
&= \Tr_{\rm i'}[L^{(N+1)  \frac12}_{\defset{o}'\,\defset{i}'}
(I_{\defset{o}'}\otimes U^{T \otimes N})
\KetBra{I}{I}_{\defset{o}'\,\defset{i}'}
(U^{\dagger \otimes N}\otimes I_{\defset{o}'})
L^{(N+1) \frac12}_{\defset{o}'\,\defset{i}'}] =\\
&= \Tr_{\rm i'}[
(I_{\defset{o}'}\otimes U^{T \otimes N})
L^{(N+1)  \frac12}_{\defset{o}'\,\defset{i}'}
\KetBra{I}{I}_{\defset{o}'\,\defset{i}'}
L^{(N+1)\frac12}_{\defset{o}'\,\defset{i}'}
(U^{\dagger \otimes N}\otimes I_{\defset{o}'})
]=\\
&=
 \Tr_{\rm i'}[
L^{(N+1)  \frac12}_{\defset{o}'\,\defset{i}'}
\KetBra{I}{I}_{\defset{o}'\,\defset{i}'}
L^{(N+1)  \frac12}_{\defset{o}'\,\defset{i}'}
]= \rho
\end{align*}
 Decomposing $U^{\otimes N}$ into
irreducible representations (irreps) we have $U^{\otimes N } =
\bigoplus_j (U_j \otimes I_{m_j})$, where $I_{m_j}$ is the identity on
an $m_j$-dimensional multiplicity space $\mathbb C^{m_j}$.  Reminding theorem
\ref{th:characterizationcommutant},
$\rho$ must have the form $ \rho= \bigoplus_j p_j (I_j/ d_j \otimes
\rho_{j})$, where $\rho_{j}$ is an arbitrary state on the multiplicity
space $\mathbb C^{m_j}$.  Since $|\Psi \kk$ is a purification of
$\rho$, there exists a  basis in which  we have $|\Psi \kk=
|\rho^{\frac 1 2} \kk = \bigoplus_j \sqrt{p_j / d_j}~ |I_j \kk
|\rho_{j}^{\frac 1 2} \kk $, which after storage becomes
$|\Psi_U\kk= \bigoplus_j \sqrt{p_j/d_j}|U_j \kk |\rho_{j}^{\frac 1
  2} \kk$.  Hence, for every $U$ the state
$|\Psi_U\kk$ belongs to the subspace $\widetilde{\hilb{H}} =
\bigoplus_{j} ( \hilb H_j^{\otimes 2} \otimes |\rho^{\frac 1
  2}_{j}\kk) \simeq \bigoplus_j \hilb H_j^{\otimes 2}$.  \qed
\end{Proof}

\subsubsection{Optimization of the retrieving channel}
In this section we optimize the retrieving channel
$\mathcal{R}$; exploiting some symmetries  of 
$\mathcal{R}$ we can prove that
the optimal retrieval is achieved by a measure and re-prepare strategy.

Thanks to Lemma \ref{lem:learnoptstate}
we can restrict our attention to the subspace $\widetilde {\hilb
  H}$, and consider retrieving channels $\map R$ from $(\hilb H_{2N
  +2} \otimes \widetilde{\hilb{H}})$ to $\hilb H_{2N+3}$.
 The
normalization of the Choi operator is then 
\begin{equation}\label{normRetrieving}
\Tr_{2N+3} [R] = I_{2N+2}
\otimes I_{\widetilde {\hilb H}}~.
\end{equation}
The following lemma tells us that the optimal
retrieving channel can be chosen among the covariant ones:
\begin{lemma}
We can require without loss of generality that the operator $R$  maximizing the fidelity
~(\ref{eq:learfigmer5}) satisfies the commutation relation
\begin{align}\label{eq:learcovretr}
  \left[R, U_{2N+3}\otimes V^*_{2N+2} \otimes \left(\bigoplus_j (U^*_j \otimes V_j)   \right) \right]=0
\qquad\forall U,V \in \group{G}  .
\end{align}
where $\bigoplus_j (U^*_j \otimes V_j)$ acts on $\widetilde{\hilb{H}}$.
\end{lemma}
\begin{Proof}
  The proof consists in the same averaging argument that was used in the proof of Lemma \ref{lem:learncov}. \qed
\end{Proof}

According to Eq. (\ref{eq:completereducibility2}), 
the representation
$U \otimes U^*_j$
can be decomposed as
\begin{align}\label{eq:leardecompo1}
  U_{2N+3} \otimes U^*_j = \bigoplus_{K \in \defset{irrepS}(U \otimes U^*_j)}\left(U_K \otimes I_{m_K^{(j)}}  \right)
\end{align}
and in a similar way we have
\begin{align}\label{eq:leardecompo2}
  V_{2N+2}^* \otimes V_j = \bigoplus_{L \in \defset{irrepS}(V^* \otimes V_j)}\left(V^*_L \otimes I_{m_L^{(j)}}  \right).
\end{align}
Combining Eq. (\ref{eq:leardecompo1})
and Eq. (\ref{eq:leardecompo2})
we have
\begin{align}\label{eq:leardecompo3}
&U_{2N+3}\otimes V^*_{2N+2} \otimes \left(\bigoplus_j (U^*_j \otimes V_j)   \right) =
\bigoplus_j \left( (U \otimes U^*_j)\otimes (V^* \otimes V_j)    \right) =  \nonumber \\
&= \bigoplus_{K,L}\left( U_K \otimes V_L \otimes I_{m_{KL}}   \right)
\end{align}
where 
$I_{m_{KL}}$ is given by $I_{m_{KL}} = \bigoplus_{j \in
  \set{P}_{KL}}\left( I_{m_K^{(j)}} \otimes I_{m^{(j)}_L}\right)$,
where $\set P_{KL}$ is the set of values of $j$ such that the irrep
$U_K \otimes V^*_L$ is contained in the decomposition of $U_{2N+3} \otimes
V_{2N+2}^* \otimes U^*_j \otimes V_j$.

Inserting the decomposition (\ref{eq:leardecompo3})
into Eq. (\ref{eq:learcovretr})
we have
\begin{align}
  \left[R, \bigoplus_{K,L}\left( U_K \otimes V_L \otimes I_{m_{KL}}   \right)  \right]=0
\end{align}
that thanks to Theorem \ref{th:characterizationcommutant},
leads to the decomposition
\begin{align}\label{eq:learndecompoR}
  R = \bigoplus_{K,L}\left( I_K \otimes I_L \otimes R_{KL}   \right)
\end{align}
where $R_{KL}$ is a positive operator on the multiplicity space\\
$\mathbb C^{m_{KL}} = \bigoplus_{j \in \set P_{KL}} \left( \mathbb
  C^{m_K^{(j)}} \otimes \mathbb C^{m_{L}^{(j)}} \right)$

The decomposition (\ref{eq:leardecompo1})
induces the following  decomposition of Hilbert spaces
\begin{align}
  \hilb{H} \otimes \hilb{H}_j = 
\bigoplus_{K \in \defset{irrepS}(U \otimes U^*_j)}\left(\hilb{H}_K \otimes \mathbb{C}^{m_K^{(j)}}  \right)
\end{align}
that allows us to write
\begin{align}\label{eq:leardecompoI}
  I\otimes I_J  =\bigoplus_{K \in \defset{irrepS}(U \otimes U^*_j)}\left(I_K \otimes I_{m_K^{(j)}}  \right).
\end{align}
From Eq.  (\ref{eq:leardecompoI}) we have
\begin{align}\label{eq:leardecompoKet}
  \Ket{I}\otimes \Ket{I_J}  =\bigoplus_{K \in \defset{irrepS}(U \otimes U^*_j)}\left(\Ket{I_K} \otimes \Ket{I}_{m_K^{(j)}}  \right)
\end{align}
that leads to the following identity:
\begin{equation}\label{eq:learnsecompopsi}
\begin{split}
   \Ket{I} |\Psi^*\>&= \bigoplus_{j} \sqrt{\frac{p_j} {d_j}} \Ket{I} \Ket{I_j}  = \bigoplus_{j} \bigoplus_{K \in \defset{irrepS}(U\otimes U^*_j)}   \sqrt{\frac{p_j} {d_j}}  \Ket{I_K} \Ket{I_{m_K^{(j)}}} \\
  & = \bigoplus_{K} \bigoplus_{j \in \set P_{KK}}   \sqrt{\frac{p_j} {d_j}}  \Ket{I_K} \Ket{I_{m_K^{(j)}}} 
 = \bigoplus_{K}  |I_K \kk |\alpha_K \>~,
\end{split}
\end{equation}
where $|I_K\kk \in \hilb H_K^{\otimes 2}$ and $|\alpha_K\> \in \mathbb C^{m_{KK}} $ is given by
\begin{equation}\label{eq:alphadec}
|\alpha_K\> =\bigoplus_{j\in \set P_{KK}}\sqrt{p_j/d_j}\ |I_{m_K^{(j)}} \kk.
\end{equation}

Exploiting Eqs. (\ref{eq:learndecompoR}) and (\ref{eq:learnsecompopsi}) the fidelity (\ref{eq:learfigmer5})
can be rewritten as 
\begin{equation}\label{eq:learfigmer6}
F= \sum_K  \frac {d_K}{d^2}~ \<\alpha_K | R_{KK} |\alpha_K \>~.
\end{equation}

We now prove that 
the optimal retrieving  consists in a measure and re-prepare
channel;
we split the derivation into two parts.
\begin{lemma}\label{th:boundretr}
For the fidelity in Eq. (\ref{eq:learfigmer6})
the following bound holds
\begin{align}\label{eq:learnbound}
  F \leq  \sum_K \frac { \left(\sum_{j \in \defset{P_{KK}}} m_K^{(j)} \sqrt
  {p_j}\right)^2}{d^2}
\end{align}
where we remind that $m_K^{(j)}$ is the dimension of the multiplicity space  $\mathcal{C}^{m_K^{(j)}}$
and that where $\set P_{KK}$ is the set of values of $j$ such that the irreducible representation
$U_K \otimes V^*_K$ is contained in the decomposition of $U \otimes
V^* \otimes U^*_j \otimes V_j$.
\end{lemma}
\begin{Proof}
Taking the trace over $\hilb{H}_{2N+3}$
into Eq. (\ref{eq:learcovretr}) gives
\begin{align}
  \left[\Tr_{2N+3}[R],  V^*_{2N+2} \otimes \left(\bigoplus_j (U^*_j \otimes V_j)   \right) \right]=0 ;
\end{align}
reminding the decomposition (\ref{eq:leardecompo1})
and 
exploiting Theorem \ref{th:characterizationcommutant}
we can write
\begin{align}\label{eq:learpartrac2}
  \Tr_{2N+3}[R]= \bigoplus_j I_j \otimes \left(\bigoplus_L I_L \otimes   r_{L}^{(j)}\right)
\end{align}
where $r_{L}^{(j)}$ is a positive operator acting on $\mathbb{C}^{m_L^{(j)}}$.

Comparing Eq. (\ref{eq:learndecompoR}) traced over $\hilb{H}_{2N+3}$
with
Eq. (\ref{eq:learpartrac2})
we have
\begin{align}\label{eq:learcaos1}
& \bigoplus_L\left( \bigoplus_j I_j \otimes r_{L}^{(j)} \right) \otimes  I_L =
  \bigoplus_{L}   \left(\bigoplus_{K} \Tr_{2N+3}[I_K \otimes R_{KL}]  \right) \otimes I_L \Rightarrow \nonumber \\
&\Rightarrow \bigoplus_j I_j \otimes r_{L}^{(j)} = \bigoplus_{K} \Tr_{2N+3}[  I_K \otimes R_{KL}]
\end{align}
Let us now denote with
$P_j$ the projector on $\hilb{H}_j$
with $P_K$ the projector on $\hilb{H}_K$
and with $P_{K}^{(j)}$ the projector
on $\mathbb{C}^{m_K^{(j)}}$:
we can then rewrite  the decomposition
(\ref{eq:leardecompoI})
as $P_j \otimes I = \sum_K P_K \otimes P_{K}^{(j)}$ 
Projecting both the two sides of Eq. (\ref{eq:learcaos1})
on $\hilb{H}_j \otimes \mathbb{C}^{m_L^{(j)}}$ we get
\begin{align}\label{learcaos2}
 I_j \otimes r_{L}^{(j)} &= \nonumber
(P_j \otimes P_{L}^{(j)} )\bigoplus_{K} \Tr_{2N+3}[  I_K \otimes R_{KL}] (P_j \otimes P_{L}^{(j)}) = \nonumber\\
&=\bigoplus_{K} \Tr_{2N+3}[(P_j \otimes I_{2N+3} \otimes P_{L}^{(j)}) 
 I_K \otimes R_{KL} (P_j \otimes I_{2N+3} \otimes P_{L}^{(j)})]\nonumber\\
&= \bigoplus_{K} \Tr_{2N+3}\left[\sum_{Q}(P_{Q}\otimes P_{Q}^{(j)} \otimes  P_{L}^{(j)})  I_K \otimes R_{KL}
\sum_{G}(P_{G}\otimes P_{G}^{(j)} \otimes  P_{L}^{(j)})\right] \nonumber\\
& = \bigoplus_{K} \Tr_{2N+3}[  I_K \otimes R^j_{KL}].
\end{align}
where we used the notation $R^j_{KL} = (P_{K}^{(j)} \otimes  P_{L}^{(j)}) R_{KL} (P_{K}^{(j)} \otimes  P_{L}^{(j)})$.
Taking the trace over $\hilb{H}_j$ in Eq. (\ref{learcaos2}) 
leads to
\begin{align}
\label{learcaos3}
 &\Tr_j[I_j \otimes r_{L}^{(j)}] = \Tr_j\left[\bigoplus_{K} \Tr_{2N+3}[  I_K \otimes R^j_{KL}]\right] \Rightarrow \nonumber\\
&\Rightarrow d_jr_{L}^{(j)} = \Tr_{j \, 2N+3}\left[\bigoplus_{K} I_K \otimes R^j_{KL}\right]= \nonumber\\
&= \Tr_{(\bigoplus_{K} \!\!K \otimes\, m_K^{(j)})}\left[\bigoplus_{K} I_K \otimes R^j_{KL}\right] = 
\sum_Kd_K\Tr_{m_K^{(j)}}\left[R^j_{KL}\right]
\end{align}
where $\Tr_{(\bigoplus_{K} \!\!K \otimes\, m_K^{(j)})}$
denotes the trace over $\bigoplus_{K}\hilb{H}_K\otimes{C}^{m_K^{j}}= \hilb{H}_j\otimes \hilb{H}_{2N+3}$.

Exploiting Eq. (\ref{eq:learpartrac2})
into the normalization (\ref{normRetrieving}) we obtain
\begin{align}\label{eq:learnnorm2}
  \Tr_{2N+3}[R]= \bigoplus_j I_j \otimes \left(\bigoplus_L I_L \otimes   r_{L}^{(j)}\right) 
= I_{2N+2}\otimes I_{\widetilde{\hilb{H}}} \Rightarrow r_{L}^{(j)} = I_{m_L^{(j)}}
\end{align}
that together with  Eq. (\ref{learcaos3})
gives
\begin{align}
 I_{m_L^{(j)}} =  \sum_K\frac{d_K}{d_j}\Tr_{m_K^{(j)}}\left[R^j_{KL}\right]
\end{align}
that for $L=K$ implies the bound
\begin{align}\label{eq:learnormultimate}
  \Tr[R^j_{KL}] \leq \frac{d_j m_K^{(j)}}{d_K}
\end{align}
Reminding Eq. (\ref{eq:alphadec}),   for the fidelity (\ref{eq:learfigmer6}) we then have the  bound
\begin{align}
  F & = \sum_{K}  \frac{d_K} {d^2} \sum_{j, j' \in \set P_{KK}} \sqrt{\frac{p_j p_{j'}}{d_{j} d_{j'}}}  \bb I_{m^{(j)}_K} |R_{KK} |I_{m^{(j')}_K}\kk \leq\\
   & \le \sum_K \frac{d_K} {d^2} \left( \sum_{j \in \set P_{KK}} \sqrt{
      \frac{ p_j \bb I_{m_K^{(j)}}| R^{(j)}_{KK}
        |I_{m_K^{(j)}}\kk }{d_{j}}}\right)^2 
 \leq  \sum_K \frac { \left(\sum_{j \in \defset{P_{KK}}} m_K^{(j)} \sqrt
  {p_j}\right)^2}{d^2}\nonumber
\end{align}
having used the positivity of $R_{KK}$ for the first bound and
Eq. (\ref{eq:learnormultimate})
the second.
\qed
\end{Proof}

It is now easy to prove the following
\begin{theorem}[Optimal retrieving strategy]
  \label{th:OptRetr}
 The optimal retrieving of
  $U$ from the memory state $|\Psi_U \kk$ is
  achieved by measuring the ancilla with the  POVM $P_{\hat U} =
  |\eta_{\hat U} \kk \bb \eta_{\hat U}|$ given by $|\eta_{\hat U}\kk =
  \bigoplus_j \sqrt {d_j} |\hat U_j\kk$, and, conditionally on outcome $\hat U$,
  by performing the unitary $\hat U$ on the new input system
\begin{equation}
\begin{aligned}
\Qcircuit @C=0.6em @R=1.2em {
&\multigate{1}{\mathcal{R}}&\qw\\
\prepareC{|\Psi_U \kk}&\ghost{{\mathcal{R}}}
}
\end{aligned}\quad=\quad
\begin{aligned}
\Qcircuit @C=0.6em @R=1.2em {
&\gate{\mathcal{\hat U}}&\qw\\
\prepareC{|\Psi_U \kk}&\measureD{{P_{\hat U}}}
}
\end{aligned} \!\!\!\!\!\!\!\!\!\!\!\!\!\!
\uparrow \qquad\qquad
\end{equation}
(the arrow represents the  communication of  the classical outcome of the measurement).
\end{theorem}
\begin{Proof}
We now prove that the measure and prepare strategy described above
achieves the bound (\ref{eq:learnbound}).
First, the Choi operator of the measure-and-prepare strategy has the form
$R_{est} = \int_G |\hat U\kk\bb \hat U|_{(2N+3)(2N+2)} \otimes |\eta^*_{\hat U} \kk\bb
\eta^*_{\hat U} | \d \hat U$.
Using Eq.  (\ref{eq:learnsecompopsi}) with
$|\Psi^*\kk$ replaced by $|\eta_{I}^*\kk$
and applying theorem \ref{th:groupaverage} we have
\begin{align*}
  R_{est} &=
\int_G |\hat U\kk\bb \hat U| \otimes |\eta^*_{\hat U} \kk\bb
\eta^*_{\hat U} | \d \hat U= \\
&=\int_G |\hat UV\kk\bb \hat UV| \otimes |\eta^*_{\hat UV} \kk\bb
\eta^*_{\hat UV} | \d \hat U \d \hat V=\\
&= \int_G 
U \otimes V^T \otimes \widetilde{U^*} \otimes \widetilde{V^\dagger}
 | I\kk\bb  I| \otimes |\eta^*_{ I} \kk\bb\eta^*_{I} | 
U^\dagger \otimes V^* \otimes \widetilde{U^T} \otimes \widetilde{V}
\d \hat U \d \hat V = \\
&= \int_G 
U \otimes V^T \otimes \widetilde{U^*} \otimes \widetilde{V^\dagger}
\bigoplus_{KL}\KetBra{I_k}{I_L}\ketbra{\beta_K}{\beta_L}
U^\dagger \otimes V^* \otimes \widetilde{U^T} \otimes \widetilde{V}
\d \hat U \d \hat V = \\
&=\bigoplus_{K}I_K \otimes I_K \otimes \ketbra{\beta_K}{\beta_K}
\end{align*}
where
$\widetilde{U^*} \otimes \widetilde{V^\dagger} = \bigoplus_j U_j \otimes V_j $
and $\ket{\beta_K} = \bigoplus_{j \in \defset{P}_{KK}}\sqrt{d_j}\Ket{I_{m_k^{(j)}}}$.
Eq. (\ref{eq:learfigmer6}) then becomes
\begin{align}\label{eq:learnoptvalue}
  F_{est} = \sum_K \frac{|\braket{\alpha_K}{\beta_K}|^2}{d^2} = 
\frac { \left(\sum_{j \in \defset{P_{KK}}} m_K^{(j)} \sqrt
  {p_j}\right)^2}{d^2}.
\end{align}
\qed
\end{Proof}
By making use of the above result it is easy to optimize the input state for
storing. In fact, such a state is just the optimal state for the
estimation of the unknown unitary $U$ \cite{optimalestimunit1},
 whose expression
is known in most relevant cases.  For example, when $U$ is an unknown
qubit unitary in $SU(2)$, learning becomes equivalent to optimal
estimation of an unknown rotation in the Bloch sphere \cite{optimalestimunit2}.
For large number of copies, the optimal input state is given by
$\Ket{\Psi}\approx \sqrt{4/N}~\sum_{j= j_{\min}}^{N/2} \frac{\sin
  (2\pi j/N)}{\sqrt{2j+1}} ~ |I_{j}\kk$, with $j_{\min} = 0 (1/2)$ for
$N$ even (odd), and the fidelity is $F\approx 1-(2\pi^2)/N^2$.
Remarkably, this asymptotic scaling can be achieved without using
entanglement between the set of $N$ qubits that are rotated and an
auxiliary set of $N$ rotationally invariant qubits: the optimal
storing is achieved just by applying $U^{\otimes N}$ on a the optimal
$N$-qubit state \cite{optimalestimunit2}. Another example is that of an unknown
phase-shift $U= \exp[i \theta \sigma_z]$. In this case, for large
number of copies the optimal input state is $\Ket{\Psi} =
\sqrt{2/(N+1)}\sum_{m=-N/2}^{N/2} \sin[\pi(m+1/2)/(N+1)] |m\> $ and
the fidelity is $F\approx 1-2\pi^2/(N+1)^2$ \cite{optimalestmbuzek}. Again,
the optimal state can be prepared using only $N$ qubits.

\subsubsection{Generalization to the M > 1 case}

Our result can be extended to the case where the user must reproduce
$M>1$ copies of the unknown unitary $U$.   In this case, there are two different notions of optimality induced by two different figures of merit, namely the single-copy  and the global fidelity. In the following we will examine both cases.

\subsubsection{Optimal learning according to the single-copy fidelity}
Let $\map C_U$ be the
$M$-partite channel obtained by the user, and $\map C_{U,\Omega}^{(i)}$ be the
local channel $\map C_{U,\Omega}^{(i)} (\rho) = \Tr_{\bar i}[\map C_U (\rho
\otimes \Omega)]$, where $\rho$ is the state of the $i$-th 
system, $\Omega$ is the state of the remaining $M-1$ systems, and $\Tr_{\bar 1}$ denotes the trace over all
systems except the $i$-th. The local channel $\map C_{U,\Omega}^{(i)}$ describes
the evolution of the  $i$-th input of $\map C_U$ when the remaining
$(M-1)$ inputs are prepared in the state $\Omega$.
Since we can be interested in some replicas more than in other ones,
we can imagine to associate a weight $q_i$ ($\sum_i q_i =1$) to each of the $M$ copies;
in this way the figure of merit becomes:
\begin{align}\label{eq:learfigmersingleside}
F^{(s)} =  \int \d U \sum_i q_i \mathcal{F}(\map C^{(i)}_{U,\Omega_i}, \mathcal{U}).
\end{align}
  Of course, the
fidelity between $\map C_{U,\Omega_i}^{(i)}$ and the unitary $U$ cannot be larger
than the optimal fidelity of Eq.~(\ref{eq:learnbound});
moreover the optimal fidelity  depends neither on $q_i$ nor on $\Omega_i$.
Therefore, the
measure-and-prepare strategy presented in Theorem \ref{th:OptRetr} 
 is optimal also for the maximization of Eq. (\ref{eq:learfigmersingleside}), 
which do
not decrease with increasing $M$.

\subsubsection{Optimal learning according to the global fidelity} 
The optimization carried on for the case $M=1$ can be
extended to the maximization of the global fidelity between $\map C_U$
and $U^{\otimes M}$
\begin{align}\label{eq:learfigmerglobal}
  F^{(g)} =  \int \d U  \mathcal{F}(\map C_{U}, \mathcal{U}^{\otimes M}) = 
\frac1{d^{2M}}\int \d U  \Bra{U}^{\otimes M}\Bra{U^*}^{\otimes N}L\Ket{U}^{\otimes M}\Ket{U^*}^{\otimes N}
\end{align}
just by replacing $U$ with $U^{\otimes M}$ in all
derivations.  Indeed, the role of the target unitary $U$ in our derivations is completely generic: we never used the fact that the unitary emulated by the machine was equal to the unitaries provided in the examples.   
Therefore, following the same proofs for the case $M=1$ it is immediate to see that also for the case of $M>1$ copies with global fidelity the optimal strategy for storing consists in the parallel application of the examples on an input state of the form of Lemma \ref{lem:learnoptstate} and that the optimal strategy for retrieving consists in measuring
the optimal POVM $P_{\hat U} $ and in performing $\hat U^{\otimes M}$
conditionally on outcome $\hat U$.
  Note that in this case the coefficients $\{p_j\}$ in the optimal input state of Lemma \ref{lem:learnoptstate})
 generally depend on $M$. 

\begin{remark}
 Since we never used the fact that the $N$ examples are identical, all the previous results
   hold even when the input (output) uses are not identical copies
$U^{\otimes N}$ ($U^{\otimes M}$), but generally $N$ ($M$) different
unitaries, each of them belonging to a different representation of the
group $G$. For example, if $G={\rm SO}(3)$  the $N$ examples may correspond to rotations (of the same angle and around the same axis) of $N$ quantum particles with different  angular momenta.   Of course, the same remark also holds when the $M$ output copies.  
\end{remark}

\subsection{Comparison with the cloning}
Let us now focus on the optimal learning according to
 the global fidelity for the
$N=1$ and $M=2$ case
Specializing Eq. (\ref{inpst}) the optimal state for storage
becomes $\Psi = \frac1{\sqrt{d}}\Ket{I}$ 
and the optimal learning board is
\begin{align}\label{eq:learopt12}
  \begin{aligned}
    \Qcircuit @C=0.5em @R=1em
{
&&&&\gate{\hat{\mathcal{U}}}&\qw\\
&&&&\gate{\hat{\mathcal{U}}}&\qw\\
\multiprepareC{1}{\frac{1}{\sqrt{d}}\Ket{I}}&\qw&\gate{\mathcal{U}}&\qw&\multimeasureD{1}{d \KetBra{\hat{U}}{\hat{U}}}\\
\pureghost{\frac{1}{\sqrt{d}}\Ket{I}}&\qw&\qw &\qw&\ghost{d \KetBra{\hat{U}}{\hat{U}}}
\gategroup{1}{5}{4}{5}{1em}{--}
}
  \end{aligned}\quad.
\end{align}
The maximum value of the fidelity  
is given by replacing $\bigoplus_{j}(U^*_j\otimes V_j)$ 
with $U^*\otimes V$
and $U_{2N+3}\otimes V^*_{2N+2}$
with $U_{2N+3}\otimes V^*_{2N+2} \otimes U_{2N+5}\otimes V^*_{2N+4}$
in the previous derivation.
From the decomposition (\ref{eq:UUU*decomposition})
we have that $m_\alpha = 2,m_\beta = 1,m_\gamma = 1$,($m_\gamma = 0$ if $d=2$);
inserting these values into Eq.
(\ref{eq:learnoptvalue})
we get
\begin{align}
  F^{(g)}_{N=1,M=2} &= \frac{1}{d^4}\sum_\nu (m_\nu)^2  = \nonumber \\
&= \frac{6}{d^4} \mbox{ for $d>2$,} \quad\mbox{or}\quad \frac{5}{d^4} \mbox{ for  $d=2$}
\end{align}

 The learning with $N=1$ and $M=2$
can be compared with the optimal cloner $1 \rightarrow 2$
we derived in chapter \ref{chapter:cloning}.
The maximum value of the fidelity was (see Eq: (\ref{eq:clonmaxfide}))
\begin{align}
  F^{(clon)}  (d+ \sqrt {d^2-1})/d^3  
\end{align}
which is much higher than $F^{(g)}$.
This result stresses the difference between  cloning and learning:
since in the learning scenario we have to apply
the unitary to a fix input state, we cannot exploit the full computational
power of the unitary channel $\mathcal{U}$
and we cannot
 achieve the same performance of the optimal cloner.



\setcounter{equation}{0} \setcounter{figure}{0} \setcounter{table}{0}\newpage
\section{Inversion of a unitary transformation}\label{chapter:inversion}

In this chapter we consider the problem of finding the Quantum Network
that realizes the optimal inversion of a unitary transformation.
Let us suppose that we are provided with a single use of unitary transformation
$\mathcal{U} = U \cdot U^\dagger$
but we need to apply its inverse $\mathcal{U}^{-1} = U^\dagger \cdot U$
on an unknown state $\ket{\varphi}$\footnote{the generalization to the general case
with $N$ uses and and $M$ replicas of the inverse is work in progress.}.
The most general strategy we can follow in order to achieve this task
is to exploit the single use of $\mathcal{U}$
in a quantum network such that the resulting channel is as close as possible
to target unitary
$\mathcal{U}^{-1}$:
\begin{align}\label{eq:invsupermap}
&  \begin{aligned}
    \Qcircuit @C=0.7em @R=1.5em
{
\prepareC{\ket{\varphi}}&\ustick{0}\qw &\multigate{1}{\mathcal{C}_1}
&\ustick{1}\qw&\gate{\;\mathcal{U}\;}&\ustick{2}\qw&\multigate{1}{\mathcal{C}_2}&\ustick{3}\qw\\
&&\pureghost{\mathcal{C}_1}&\qw&\ustick{A_1}\qw&\qw&\ghost{\mathcal{C}_2}
}
  \end{aligned}
\quad \simeq \quad
\begin{aligned}
    \Qcircuit @C=0.7em @R=1.5em
{
\prepareC{\ket{\varphi}}&\ustick{0}\qw&\gate{\;\mathcal{U}^{-1}\;}&\ustick{3}\qw
}
\end{aligned} \\
&\qquad\quad\; \underbrace{\qquad\qquad \qquad \qquad \qquad} \nonumber\\
&\qquad\qquad\qquad \;\;\;\;\;\;\;\;  \mathcal{G}\nonumber
\end{align}

If  the use of the unitary $\mathcal{U}$
is available only today
while the state 
$\ket{\varphi}$
on which we need to apply 
the inverse $\mathcal{U}^{-1}$
will be provided tomorrow,
we cannot apply the scheme in Eq. (\ref{eq:invsupermap})
and the best we can do is to apply a learning strategy
(see Chapter \ref{chapter:learning}):
\begin{align}\label{eq:invlearn}
&\underbrace{
  \begin{aligned}
    \Qcircuit @C=0.6em @R=1.5em
{
&\multiprepareC{1}{\Ket{\Psi}}&\ustick{0}\qw&\gate{\;\mathcal{U}\;}&\ustick{1}\qw&\multigate{1}{\mathcal{C}}&&\prepareC{\ket{\varphi}}&\ustick{2}\qw&\multigate{1}{\mathcal{R}}&\ustick{3}\qw\\
&\pureghost{\Ket{\Psi}}&\qw&\ustick{A_1}\qw&\qw&\ghost{\mathcal{C}}&\qw&\ustick{M}\qw&\qw&\ghost{\mathcal{R}}\\
}
  \end{aligned}}
\quad \simeq \quad
\begin{aligned}
    \Qcircuit @C=0.6em @R=1.5em
{
\prepareC{\ket{\varphi}}&\ustick{0}\qw&\gate{\;\mathcal{U}^{-1}\;}&\ustick{3}\qw
}
\end{aligned}
 \\
&\qquad\qquad\qquad\qquad \mathcal{L}\nonumber
\end{align}

We encountered the same situation
when we compared the cloning and the learning of a unitary transformation.
The Choi-Jamio\l kowsky operators of
$\mathcal{G}$ and $\mathcal{L}$
satisfy the conditions:

\begin{align}\label{eq:invsupermapconst}
  &\Tr_3[G]= I_2 \otimes G^{(1)} \qquad \Tr_1[G^{(1)}] = I_0,
\\
\label{eq:invlearconst}
 & \Tr_3[L]= I_2 \otimes I_1 \otimes \rho \qquad \Tr_0[\rho] = 1,
\end{align}
that coincide with Eqs. (\ref{eq:clon12const}) and (\ref{eq:learn12const})
by defining $\hilb{H}_{0} \otimes \hilb{H}_{0'} := \hilb{H}_{0}$
and $\hilb{H}_{3} \otimes \hilb{H}_{3'} := \hilb{H}_{3}$.

As we noticed when we compared the learning
and the cloning strategies, the constraint
(\ref{eq:invlearconst}) is stronger than the constraint
(\ref{eq:invsupermapconst}), and this means that the learning scheme in  Eq. (\ref{eq:invlearn})
can be interpreted  as a special case of the scheme in Eq. (\ref{eq:invsupermap}).

In principle,
one could expect that the strategy (\ref{eq:invsupermap})
allows to achieve better performances that the learning scheme (\ref{eq:invlearn}).
However, as we will see in the next sections,
the optimal inversion is achieved by a measure and re-prepare strategy 
which is a special case of quantum learning.

\subsection{Learning scenario}\label{sec:inv}
In this section we show that it is possible to  extend the  results 
of chapter \ref{chapter:learning}
to the optimal learning  of the inverse of an unknown unitary $U$.
Then we  can consider the more general scenario
in which $N \geq 1$
uses of the unitary are available and $M\geq1$ replicas
have to be produced.
The figure of merit is than the averaged channel fidelity
between the inverses ${\mathcal{U}^{-1}}^{\otimes M}$
and the resulting replicas $\mathcal{L} \star \mathcal{U} \star \cdots \star \mathcal{U}$:
\begin{align}
F = \frac1{d^{2M}} \int_G \bb U^{\dag} |^{\otimes M} \bb
U^*|^{\otimes N} ~ L~|U^\dag\kk^{\otimes M} |U^*\kk^{\otimes N} \d U
\end{align}
 as obtained by substituting
$U$ with $U^{\dag \otimes M}$ in the target of Eq.~(\ref{eq:learfigmer4}).
  From this expression the commutation (\ref{eq:covlearn}) becomes
  \begin{align}
    [L, V^{\otimes M} \otimes U^{*\otimes M}\otimes U_o^{*\otimes N}
\otimes V_i^{\otimes N} ]=0
  \end{align}
Therefore, the optimal inversion is obtained from our  derivations by
simply substituting $U_{2N+3} \to V^{\otimes M}$ and $V_{2N+2} \to U^{\otimes M}$.  Accordingly, the
optimal inversion is achieved by measuring the optimal POVM $P_{\hat U}$ on the optimal state
$|\Psi_U\kk$ and by performing $\hat U^{\dag \otimes M}$ conditionally on outcome $\hat U$.

Focusing on the $N=1, M=1$ case the optimal network is:
\begin{align}\label{eq:invoptinverscheme}
  \begin{aligned}
    \Qcircuit @C=0.5em @R=1em
{
&&&&\gate{\hat{\mathcal{U}}^{-1}}&\qw\\
\multiprepareC{1}{\frac{1}{\sqrt{d}}\Ket{I}}&\qw&\gate{\mathcal{U}}&\qw&\multimeasureD{1}{d \KetBra{\hat{U}}{\hat{U}}}\\
\pureghost{\frac{1}{\sqrt{d}}\Ket{I}}&\qw&\qw &\qw&\ghost{d \KetBra{\hat{U}}{\hat{U}}}
\gategroup{1}{5}{3}{5}{1em}{--}
}
  \end{aligned}\quad.
\end{align}

The maximum value of $F$
for this case 
is obtained  by substituting
 $\bigoplus_{j}(U^*_j\otimes V_j)$ 
with $U^*\otimes V$
and $U_{2N+3}\otimes V^*_{2N+2}$
with $V_{2N+3}\otimes U^*_{2N+2}$
in the main derivation.
Reminding the decomposition ~(\ref{eq:UU*Youngdecomposition})
we have that $m_p = m_q = 1$
and thus
Eq. (\ref{eq:learnoptvalue})
gives 
\begin{align}
  F &= \frac{1}{d^4}\sum_\nu (m_\nu)^2  = \frac{2}{d^2}
\end{align}

\begin{remark}
The optimal ``learning of the inverse''
of a unitary transformation
provides the optimal approximate realignment of reference frames in the quantum communication
scenario  considered in Ref. \cite{spekkensframe}, 
proving the optimality of the ``measure-and-rotate" strategy conjectured therein. 
In that scenario, the storing state $|\Psi\kk$
 serves as a token of Alice's reference frame, and is sent to Bob along with
a quantum message $|\phi\> $.  Due to the mismatch of reference frames, Bob
receives the decohered state $\sigma_\phi = \int_G |\Psi_U\kk\bb\Psi_U| \otimes U
|\varphi\>\<\varphi|U^\dag \d U$, from which he tries to retrieve the message $|\varphi\>$ with maximum
fidelity $f = \int \d \varphi~ \<\varphi| \map R' (\sigma_\varphi)|\varphi\> \d \varphi$, where $\map R'$ is the
retrieving channel and $\d \varphi$ denotes the uniform probability measure over pure states.  The
maximization of $f$ is equivalent to the maximization of the channel fidelity $F' = \int_G \bb
U^\dag| \bb \Psi_U^*| R' |U^\dag\kk |\Psi_U^*\kk \d U$, which is the figure of merit for optimal
inversion.  It is worth stressing that the state $|\Psi\kk$ that maximizes the
fidelity is not the state $|\Psi_\mathrm{lik}\kk = \bigoplus_j \sqrt{d_j/L} |I_j\kk$, $L=\sum_j
d_j^2$ that maximizes the likelihood \cite{maxlikelihood}. For $M =1$ and $G = SU (2), U(1)$ the state
$|\Psi\kk$ gives an average fidelity that approaches 1 as $1/N^2$, while for
$|\Psi_\mathrm{lik}\kk$ the scaling is $1/N$.  On the other hand, Ref. \cite{spekkensframe} shows
that for $M=1$ $|\Psi_\mathrm{lik}\kk$ allows a perfect correction of the misalignment errors with
probability of success $p = 1- 3/(N+1)$, which is not possible for $|\Psi\kk$. The
determination of the best input state to maximize the probability of success, and the study of the
probability/fidelity trade-off remain open interesting problems for future research.
\end{remark}

\subsection{Supermap scenario}
In this section we will review the derivation
of Ref. \cite{algorithm} of the
optimal  inversion of a unitary transformation
according to the scheme (\ref{eq:invsupermap});
since the quantum network $\mathcal{G}$
can be interpreted as a \emph{supermap}
 $\mathcal{G}: \mathcal{L}(\hilb{H}_1,\hilb{H}_2)
 \rightarrow \mathcal{L}(\hilb{H}_0,\hilb{H}_3) $
that maps the unknown unitary transformation $\mathcal{U}\in \mathcal{L}(\hilb{H}_1,\hilb{H}_2)$
into another transformation $\mathcal{G}(\mathcal{U}):= \mathcal{G}\star \mathcal{U} \in \mathcal{L}(\hilb{H}_0,\hilb{H}_3)$
we call the scheme (\ref{eq:invsupermap}) the \emph{supemap scenario}
for the inversion of a unitary transformation.\footnote{Clearly, also the learning network can be thought as a supermap;
however, whenever in this chapter we use the term supermap, we refer to the scheme (\ref{eq:invsupermap}).}


In order to make a meaningful comparison,
we choose as figure of merit the averaged channel fidelity as we previously did in the  learning scenario:
\begin{align}\label{eq:invaverfidelity}
F &= \int_{SU(d)} \!\!\!  dU \mathcal{F}(\mathcal G,\mathcal U) & \nonumber \\
 &=\frac{1}{d^2}  \int_{SU(d)} \!\!\! dU \Bra{U^\dagger}_{30}\Bra{U^*}_{21} G \Ket{U^\dagger}_{30}\Ket{U^*}_{21}&
\end{align}

The following lemma holds:
\begin{lemma}
The operator $G$  maximizing the fidelity
  (\ref{eq:invaverfidelity}) can be assumed without loss of generality to satisfy
  the commutation relation
\begin{equation}
[G, U_3 \otimes W_2 \otimes U_1\otimes W_0]=0 \quad \forall V,W \in SU (d)~. \label{eq:invCovR}
\end{equation}  
\end{lemma}
\begin{Proof}
The proof consists in the standard averaging argument (see e.g. Lemma \ref{lem:learncov}):
Let $G$ be optimal. Then take its average
$\overline{G} = \int \d U \d W~  (U_3 \otimes
 W_2 \otimes  U_1 \otimes  W_0) G (U_3 \otimes
 W_2 \otimes  U_1 \otimes  W_0)^{\dagger}$: it is immediate to see that
$\overline {G}$ satisfies Eqs. (\ref{eq:invCovR}) and (\ref{eq:invsupermapconst})
 and has the same fidelity as $G$. \qed
\end{Proof}

Thanks to Theorem  \ref{th:groupaverage} and reminding the decomposition
(\ref{eq:UUdecomposition}) $G$ can be decomposed as
\begin{align}\label{eq:invdecomb}
C = \sum_{\mu, \nu \in \mathsf{S}}a^{\mu \nu}P^{\mu}_{31} \otimes P^{\nu}_{20},
\end{align}
where $\mathsf{S} = \{ +,- \}$, $P^{\pm}_{ij}$ is the projector onto the symmetric/antisymmetric subspace of
$\mathcal{H}_i\otimes \mathcal{H}_j$ ,  and $a^{\mu \nu} \geq 0$ $\forall \mu, \nu$.
Moreover, using Eq. (\ref{eq:invdecomb}) the fidelity (\ref{eq:invaverfidelity}) becomes
\begin{align}
F  &= \frac1{d^2} \Bra{I}_{30}\Bra{I}_{21} G \Ket{I}_{30}\Ket{I}_{21} & \nonumber \\
   &  =\frac{1}{d^2}\sum_{\nu \in \mathsf{S}} a^{\nu\nu }d_\nu , \; \quad d_\nu = \Tr[P^\nu],
\end{align}
while the normalization  (\ref{eq:invsupermapconst}) can be rewritten as
$\sum_{\mu \in \mathsf{S}}a^{\mu\nu}d_\mu = 1, \forall \nu\in \mathsf S$.
The last equality implies the  bound
$ F = \frac{1}{d^2}\sum_{\mu \in \mathsf{S}} a^{\mu\mu }d_\mu \leq  2/d^2$,
which is achieved if and only if $a^{\mu\nu }= \frac{\delta_{\mu \nu}}{d_\mu}$,  that is, if and only if
\begin{align}\label{eq:invoptimalinverter}
G&= \frac{P_{31}^+\otimes P_{20}^+}{d_+}+\frac {P_{31}^-\otimes P_{20}^-}{d_-} \nonumber\\
&  =\int_{SU(d)}  d \hat U  ~ |\hat U^\dag\kk\bb \hat U^\dag|_{30}  \otimes |\hat U^*\kk \bb \hat U^*|_{21}.
\end{align}

We have then proved that the learning 
and the supermap scenarios
achieves the same value of $F$.
Contrary to what one could expect
there is no coherent strategy
that achieves better performances than the measure and re-prepare learning scheme
in Eq. (\ref{eq:invoptinverscheme}).

\setcounter{equation}{0} \setcounter{figure}{0} \setcounter{table}{0}\newpage
\section{Information-disturbance tradeoff in estimating a unitary transformation}\label{chapter:infotradeoff}

One of the key features of Quantum Mechanics is the impossibility of extracting information from a system without producing
a disturbance on its state; this is the basis of the indeterminism of Quantum Mechanics 
 and of quantum cryptography.
However, a quantitative expression of the tradeoff between information and disturbance is generally a non trivial 
issue, and it has been the subject of numerous papers 
\cite{tradeoffscully, tradeofffuchs-peres, tradeoffbanaszek, tradeoffozawa,
 tradeoffmax, tradeoffdema, tradeoffmacca, tradeoffwerner, tradeoffbusc}
since Heisenberg's $\gamma$-ray microscope thought experiment  \cite{heisenberg}.

On the other hand, the case of extracting information from a black box without affecting
 the transformation it is expected to perform, has not been considered yet.
More precisely, we consider the problem of both applying the black box to an arbitrary input state
and estimating its transformation within the same use.
Similarly to the case of state estimation, the information-disturbance tradeoff for channels
is interesting for security analysis of two-way quantum cryptographic protocols 
\cite{bostromprotocol, lucamariniprotocol}.
An information-disturbance problem in the estimation of the state of a  quantum system
can be split into two parts;

\begin{itemize}
\item making a measurement which supplies information about the state of the system;
\item comparing the state of the system before the measurement with the state after the measurement.
\end{itemize}

Suppose we are provided with a system which is in an unknown state
$\rho_n$ randomly drawn from an ensemble
$\{p_n \rho_n \}$ ($p_n$ is the probability of getting the state $\rho_n$);
we want to estimate the parameter $n$
and compare the state after the measurement with the state before the measurement.
The right tool for
 describing  such a process  which  has both a classical (the result of the measurement)
and a quantum (the final state)  output
is a  quantum instrument $\{ \mathcal{T}_{\hat{n}} \}$ (see Section \ref{sec:statechanpovm}).
The quantum instrument $\{ \mathcal{T}_{\hat{n}} \}$ with probability
$p(\hat{n}|n) = \Tr[\mathcal{T}_{\hat{n}}(\rho_n)]$ 
outputs the classical outcome $\hat{n}$ (that is an estimate of $n$)
and the quantum state $\rho_n' = \mathcal{T}_{\hat{n}}(\rho_n)/\Tr[\mathcal{T}_{\hat{n}}(\rho_n)]$:
the closer $\hat{n}$ is to $n$ the greater is the information and
 the closer $\rho_n'$ is to $\rho_n$ the less is the disturbance.

The previous framework can be generalized to the case  of channels.
Consider a quantum network $\mathcal{C}$   that can be linked with a 
single use of an unknown channel $\mathcal{E}_n$ randomly drawn from a set $\{\mathcal{E}_n \}$.
We want the network $\mathcal{C}$ to provide us with an estimate
$\hat{n}$ of $n$,
 but without affecting the output $\mathcal{E}(\rho)$ on the input state $\rho$

\begin{align}\label{eq:tradeofgeneric}
&  \begin{aligned}
    \Qcircuit @C=0.7em @R=1.5em
{
\prepareC{\rho}&\ustick{0}\qw &\multigate{1}{\quad}
&\ustick{1}\qw&\gate{\;\mathcal{E}_n\;}&\ustick{2}\qw&\multigate{1}{\quad}&\ustick{3}\qw\\
&&\pureghost{\quad}&\qw&\ustick{A_1}\qw&\qw&\ghost{\quad}
}
  \end{aligned}
\quad \simeq \quad
\begin{aligned}
    \Qcircuit @C=0.7em @R=1.5em
{
\prepareC{\rho}&\ustick{0}\qw&\gate{\;\mathcal{E}_n\;}&\ustick{3}\qw
}
\end{aligned} \\
&\qquad\quad \underbrace{\qquad\qquad \qquad \qquad \qquad} \nonumber\\
&\qquad\qquad\qquad \;\;\;\;\;\;  \mathcal{C} \rightarrow \hat{n}  \nonumber
\end{align}

We notice that the resulting map $\mathcal{C} \star \mathcal{E}_n$
behaves like a quantum instrument; since $\mathcal{E}_n$
is a channel (i.e. a deterministic map)
we have that $\mathcal{C}$ is actually a generalized instrument $\{\mathcal{C}_{\hat{n}}\}$
(see \ref{def:geninst}).

Obviously, if we are interested only in gathering
 information on the unknown channel, the optimal device
is the one suggested by \emph{channel estimation} \cite{optimalestimunit1}:
 we apply locally the channel to
 the best (according to some prior information)
bi-partite state $\sigma$ and then we perform a suitable measurement $P_i$.
In this case we neglect the action of the channel on the input state of the circuit $\mathcal{E}(\rho)$.
On the other hand, if we are not interested at all in gathering information about the channel $\mathcal{E}$,
the best circuit board simply consists in applying  $\mathcal{E}$ to  $\rho$.
Between these two extremal situations one can ask 
what is the maximum amount of information that is possible to gather
without violating a disturbance threshold.

In this chapter we review Ref. \cite{unitradeoff}
derived the best generalized instrument which achieves this task 
 when the unknown channel is a unitary transformation, for any possible information-disturbance rate.

\subsection{Optimization of the tradeoff}

We  now address the information-disturbance problem in the unitary case.
Suppose we are provided with an unknown unitary gate
 $\mathcal{V} \in SU(d)$ picked randomly according to the 
Haar distribution; we now look for the
best generalized instrument 
 $\{ \mathcal{R}_V \in \mathcal{L}(\mathcal{L}(\hilb{H}_{02})\bigotimes_{i=0}^{4}\hilb{H}_i) \}$,
$\int \d V \mathcal{R}_V = \mathcal{R}_\Omega$
 ($V\in SU(d)$) which  performs the best 
estimation of the group parameter $V$ without affecting too much the performance of the unknown gate.
 
We now introduce two figures of merit  in order to quantify the disturbance and the information gain.
Minimization of the disturbance can  be expressed by maximizing 
the channel fidelity $\mathcal{F}$ (defined in Eq. \ref{eq:channelfidelity})
 between the average resulting channel $\int \d V \mathcal{R}_V\star\mathcal{U}=\mathcal{R}_\Omega \star \mathcal{U}$
 and the input unitary $\mathcal{U}$:
\begin{align}\label{eq:tradefigmer}
\mathcal{F}(\mathcal{R}_\Omega \star \mathcal{U},\mathcal{U})
 = \frac{1}{d^2} \Bra{U}_{03}\Bra{U^*}_{12} R_{\Omega} \Ket{U}_{03}\Ket{U^*}_{12}.
\end{align}
A reasonable choice for the  figure of merit 
is the group average of the fidelity (\ref{eq:tradefigmer}):
\begin{align}\label{eq:tradefigmer2}
F({R}_\Omega) := \int \d U\mathcal{F}(\mathcal{R}_\Omega \star \mathcal{U},\mathcal{U})
 = \frac{1}{d^2} \int \d U \Bra{U}_{03}\Bra{U^*}_{12} R_{\Omega} \Ket{U}_{03}\Ket{U^*}_{12}.
\end{align}

Now we need an expression to evaluate
 the amount of information gathered.
The probability of outcome $V$ when the input state of the network  is
 $\rho \in \mathcal{B}(\hilb{H}_0)$ has the following expression
\begin{align}\label{probabilityofV}
p(V|U,\rho)=\Tr_{3}[\mathcal{R}_V \star \mathcal{U} (\rho_{0})].
\end{align}

In our derivation we assume $\rho_{\mathcal{E}}  = d^{-1}I_{0}$ since this condition
arises in two relevant cases:
\begin{itemize}
\item when the input system is prepared in a maximally entangled state with some 
ancillary system; this is the scenario in the protocols of Ref. \cite{bostromprotocol}
\item when the input system is prepared at random in one of the states
of an ensemble $(p_i, \rho_i)$, with the property $\rho_{\mathcal{E}}  = \sum_ip_i\rho_i$.
This is the case  of the protocol  of Ref. \cite{lucamariniprotocol}
\end{itemize}
With this assumption
  Eq.(\ref{probabilityofV}) becomes
\begin{align}\label{probabilityofV2}
p(V|U) = \frac1d \Tr_{3}[(\mathcal{R}_V * \mathcal{U}) I_{0}] = \frac1d \Tr_{03}[\Bra{U^*}R_{V} \Ket{U^*}].
\end{align}
Now we need a payoff function $c(U,V)$ which quantifies
 the error of estimating $V$ when the unknown unitary is $U$: taking inspiration from the previous
definition of disturbance, a good choice is again
the channel fidelity, that is 
\begin{align}
c(U,V) := \mathcal{F}(\mathcal{U}\mathcal{V}) =  \frac{1}{d^2}\BraKet{U}{V} =  \frac{1}{d^2}|\Tr[UV^\dagger]|^2.
\end{align}
Then the information gain is given by
\begin{align}\label{eq:tradeinfogain}
G(R_{V}) &:= \int \!\! dU \! dV p(V|U)c(U,V) =\nonumber \\
&= \frac 1{d^3} \int \!\! dU \! dV \Tr_{03}[\Bra{U^*}_{12} R_{V} \Ket{U^*}_{12}] |\BraKet{U}{V}|^2
\end{align}

The following lemma allows us to restrict to a specific class of generalized instruments.
\begin{lemma}\label{lem:tradeofcovinst}
  For any generalized instrument $\{ \mathcal{R}_{V'} \}, \int \d V' \mathcal{R}_{V'} = \mathcal{R}_{\Omega'}$
there exists another generalized instrument $\{ \mathcal{R}_{{V}}\}, \int \d V \mathcal{R}_{V} = \mathcal{R}_{\Omega} $
such that
\begin{align}\label{def R_V}
R_V &= (V_{0} \otimes V_{1}^* \otimes I_{23}) R_I (V_{0} \otimes V_{1}^* \otimes I_{23})^\dagger \nonumber \\
&=
(I_{01} \otimes V_{2}^{T} \otimes V_{3}^{\dagger} ) R_I ( I_{01} \otimes V_{2}^{T} \otimes V_{3}^{\dagger} )^{\dagger}. \\
&F(R_{\Omega'}) = F(\mathcal{R}_{V}) \label{eq:defrv2}\\
&G(R_{V'}) = G(\mathcal{R}_{V}) \label{eq:defrv3}
\end{align}  
\end{lemma}
\begin{Proof}
This result is a straightforward application of the averaging argument for covariant POVMs \cite{holevo}.
Let $R_V'$ be optimal; then let us consider
\begin{align}
  R_V := \int \d W (W_0 \otimes W_1^* \otimes I_{23})R_{W^\dagger V}(W_0^\dagger \otimes W_1^T \otimes I_{23})
\end{align}
exploiting the properties of the Haar measure $\d W$
it is easy to verify that $R_V$ enjoys the properties (\ref{def R_V}),~(\ref{eq:defrv2}) and (\ref{eq:defrv3}).
\qed
\end{Proof}

Since $R_{\Omega} = \int  \! \! dV R_V$, and reminding Eq. (\ref{def R_V})
we have
\begin{align}\label{eq:romega}
  R_{\Omega} = \int  \! \! dV (V_{0} \otimes V_{1}^* \otimes I_{23}) R_I (V_{0} \otimes V_{1}^* \otimes I_{23})^\dagger.
\end{align}
Applying Theorem \ref{th:groupaverage} we get
$[R_{\Omega}, W_{0} \otimes W_{1}^* \otimes V_{2} \otimes V_{3}^* ] = 0 $
and the normalization conditions $\Tr_{3}[R_\Omega] = R^{(1)}_{01} \otimes I_{2}$,  
$\Tr_1[R^{(1)}] = I_{0}$
become trivially 
\begin{align}\label{eq:tradenormalization2}
\Tr[R_I]= d^2.
\end{align}


Theorem \ref{th:groupaverage} and decomposition (\ref{eq:decompoUU*})
allow us to  rewrite the two figures of merit in the following way:
\begin{align}
&F = \Tr[R_FR_I],  \qquad G =\Tr[R_GR_I] \\
 &R_F = \frac{1}{d^2(d^2-1)}(I_{0123} + d^2 P^p_{01} \otimes P^p_{23} - P^p_{01} \otimes I_{23}- I_{01} \otimes P^p_{23}) \nonumber \\
 &R_G= \frac{1}{d^2(d^2-1)}
 \left( 
 \left( 
 1-\frac{2}{d^2} 
 \right) 
 I_{03} \otimes I_{12} +  I_{03} \otimes P^p_{12} \right) \nonumber
\end{align}
where $P^p_{ij}=d^{-1}\KetBra{I}{I}_{ij}$ is the projector on the one-dimensional invariant
subspace of $V_i\otimes V^*_j$.
Clearly we cannot independently optimize the two figures of merit.
What we can do is to fix a value of $G$ and then maximize $F$.
We now prove that this approach is equivalent to
 fixing a
disturbance-gain rate $0 \leq p \leq 1$ and maximize the convex combination:
\begin{align}\label{eq:convexcomb}
pG + (1-p)F = \Tr[(pR_G+(1-p)R_F) R_I]
\end{align}
Let fix the value $G = \overline{G}$;
now let us suppose that $R_I(p')$ maximize the combination
  $p'G + (1-p')F$ with $p'$  such that 
$p'\Tr[R_I(p')G] = \overline{G}$.
Clearly $R_I(p')$ achieves the maximum value of $F$ since any other
greater value of $F$ would increase $pG + (1-p)F$.
This explain why the optimal information disturbance tradeoff can be 
obtained by maximizing Eq. (\ref{eq:convexcomb}).

Since the only restrictions on $R_I$ are positivity and the normalization given by Eq. (\ref{eq:tradenormalization2}), 
the optimal choice for the operator $R_I$ is to take it proportional to the projector on the eigenspace of
 $pR_G+(1-p)R_F$ corresponding to the maximum eigenvalue; this projector can be shown \cite{tradeoffmax} to be
\begin{align}\label{eq:optimalri}
R_I = \ketbra{\chi}{\chi}\\
 \ket{\chi} = x \Ket{I}_{03}\Ket{I}_{12} + y \Ket{I}_{01}\Ket{I}_{23} 
\quad x,y \in \mathbb{R}^+ \nonumber
\end{align}
Reminding Eq. (\ref{def R_V}) we get
$R_V=\ketbra{\chi_V}{\chi_V}$ with $\ket{\chi_V} = x \Ket{V}_{03}\Ket{V^*}_{12} + y \Ket{I}_{01}\Ket{I}_{23}.$
Normalization condition (\ref{eq:tradenormalization2}) implies that $x$ and $y$   obey
\begin{align}\label{xandy}
d^2x^2+d^2y^2+ 2xyd = d^2
\end{align}
We notice that we correctly have just one free parameter which will depend on the tradeoff ratio $p$.
Fidelity and gain can be calculated in terms of the parameters $x$ and $y$, getting the following expressions
\begin{align}
F  = 1- \frac{d^2-2}{d^2}x^2 \qquad G = \frac{2-y^2}{d^2}
\end{align}
We note that when $x=0, y=1$, we have
$R_V = \Ket{I}\Bra{I}_{01} \otimes \Ket{I}\Bra{I}_{23}$ for all $V$, that is the generalized instrument
is the identity map. 
In this case the performance of the unknown unitary is not affected at all and the channel fidelity reaches its maximum
 $F=1$. On the other hand the information gain 
takes its minimum value $G = \frac{1}{d^2}$ which corresponds to random guessing $U$.
The opposite case $x=1, y=0$ clearly gives the minimum value 
 $F=\frac{2}{d^2}$ and the maximum  $G = \frac{2}{d^2}$, 
which is the same given by the optimal estimation.

\begin{figure}[tb]
\begin{center}
\includegraphics[width=10cm]{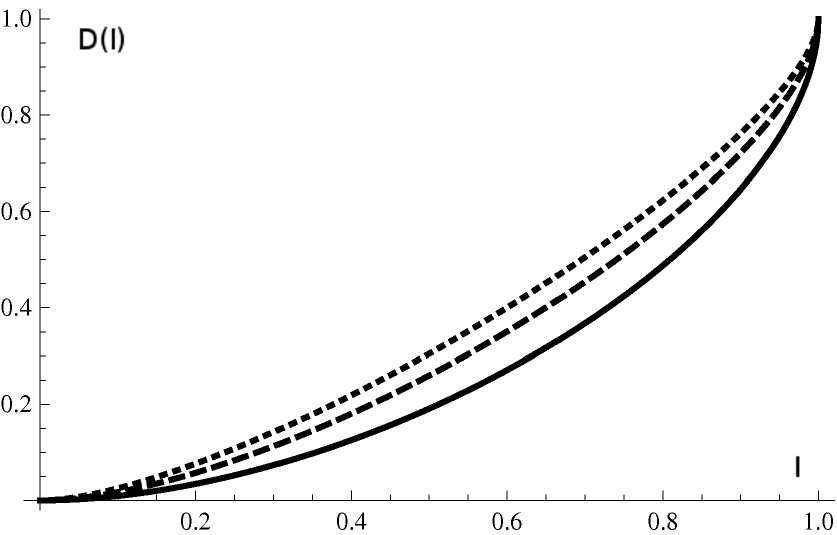}
\caption{\label{fig:tradeoff}
Plot of the lower bound $D(I)$ of the disturbance, corresponding to 
 Eq. (\ref{tradeoff2}), for various value of $d$: solid line, $d=2$; dashed line, $d=3$;dotted line ,  $d=4$. }
\label{fig1}
\end{center}
\end{figure}

Using Eq. (\ref{xandy}), we can easily express $G$ as 
a function of $x$; then, upon eliminating $x$, we can express $F$ as a function of $G$:
\begin{equation}\label{tradeoff1}
\sqrt{(d^2-2)(2- d^2 G)} = \sqrt{(d^2 -1)F -1} -\sqrt{1-F}.
\end{equation}
It seems useful to introduce the variables $ 0 \leq I,D \leq 1 $: 
\begin{eqnarray}
I=\frac{G - G_{min}}{G_{max}-G_{min}} \qquad D= \frac{F_{max}-F}{F_{max}-F_{min}}
\end{eqnarray}
where $G_{max}=2d^{-2}$, $G_{min}=d^{-2}$, $F_{min}= 2d^{-2}$ and $F_{max}=1$.
Expression (\ref{tradeoff1}) can be rewritten in terms of $D$ and $I$:
\begin{eqnarray}\label{tradeoff2}
d^2(D-I)^2 -4D (1-I)=0;
\end{eqnarray}
the plot of Eq.  (\ref{tradeoff2}) is reported in Figure \ref{fig1}.

\subsection{Realization scheme for the optimal network}

We now inspect the structure of the optimal network.
Theorem \ref{th:realgeninst} tells us that the generalized instrument can be realized by
\begin{itemize}
\item a deterministic network
  $\mathcal{S}:\mathcal{B}(\hilb{H}_{02}) \rightarrow \mathcal{B}(\hilb{H}_{13}\otimes \hilb{H}_{A_2}) $;
\item  a POVM
$\{ P_V = R_\Omega^{-\frac12} R_V  R_\Omega^{-\frac12}\} $ on the ancilla space $\hilb{H}_{0'1'2'3'}$.
\end{itemize}
The deterministic network $\mathcal{S}$ 
 can be realized,
 according to Theorem \ref{th:realinet}, as a product of two isometries
$W^{(1)}:\hilb{H}_{0} \rightarrow \hilb{H}_{0\,A_1}$ and $W^{(2)}:\hilb{H}_{2\,A_1} \rightarrow \hilb{H}_{3\,A_2}$,
$\mathcal{S} = Z \cdot Z^\dagger$, $Z = W^{(2)}W^{(1)}$.

Inserting Eq. (\ref{eq:optimalri}) into Eq. (\ref{eq:romega})
we have
\begin{align}
  R_\Omega  = A \, P^p_{01} \otimes P^p_{23} + B \,P^q_{01} \otimes P^q_{23} , 
\qquad \frac1d\Tr_{23}[R_\Omega]=R^{(1)} = aP^p_{01} + bP^q_{01} \\
A = x^2+d^2y^2+2dxy = d^2-(d^2-1)x^2 \qquad B=\frac{x^2}{(d^2-1)}\\
a = \frac{A}{d}, b = \frac{d^2-1}{d}B \qquad a+(d^2-1)b=d  \nonumber\\
\end{align}

The explicit expression of $W^{(1)}$ is given by specializing Eq. (\ref{eq:realisometries})
\begin{align}
W^{(1)} = (I_1 \otimes R^{(1)\frac12 *}_{1'0'})(\Ket{I}_{11'}\otimes T_{0 \rightarrow 0'} ) = \\
= \frac{1}{\sqrt{d}}\left( y\Ket{I}_{1'0'} \otimes T_{0 \rightarrow 1} 
+ x\Ket{I}_{11'}\otimes T_{0 \rightarrow 0'}\right)
\end{align}
If we input a pure state $\ket{\psi}$ in the first isometry, we will have as the output the superposition 
$\frac{1}{\sqrt{d}}\left( y\Ket{I}_{1'0'} \otimes \right( \ket{\psi} \left)_{1}
+ x\Ket{I}_{11'}\otimes \ket{\psi}_{0'}\right)$.

The explicit expression for the second isometry is given by:
\begin{align}
W^{(2)}
= I_3 \otimes {R_\Omega}^{\frac12}_{\;0'1'2'3'}R^{(1)-\frac{1}{2}}_{1'0'} \Ket{I}_{33'} T_{2 \rightarrow 2'}
\end{align} 
Thanks to Eq. (\ref{eq:realpovminst})  the POVM on the ancilla space ($0'1'2'3'$) can be written as
\begin{align}
P_V = \ketbra{\eta_V}{\eta_V} \qquad
\ket{\eta_V} = R_\Omega^{-\frac12}\ket{\chi_V} 
\end{align}
Isometry ${W^{(2)}}$ together with the POVM $\{ \ketbra{\eta_V}{\eta_V} \}$
can be rewritten as a quantum instrument 
 $\{ \mathcal{T}_V : \mathcal{B}(\hilb{H}_{21'}) \rightarrow \mathcal{B}(\hilb{H}_{3})$ where
 the maps $\{ \mathcal{T}_V \}$ are defined as 
 $\mathcal{T}_V(\rho) \hspace*{-0.07cm}=\hspace*{-0.06cm}  \bra{\eta_V}W^{(2)} \rho W^{(2)\dagger}
\ket{\eta_V}\}$.
Explicit calculation gives:
\begin{align}\label{instrument}
\bra{\eta_V}W^{(2)}  = \sqrt{d} V_{0' \rightarrow 3} \Bra{V}_{21'}
\end{align}
we notice that the final instrument $\{ \mathcal{T}_V \}$ does not depend on the parameters $x$ and $y$.

Summarizing, the quantum network realizing the optimal information disturbance tradeoff
in estimating a unitary transformation is as follows:
\begin{align}
  \begin{aligned}
    \Qcircuit @C=1em @R=1.5em
{
&\ustick{0}\qw&\multigate{2}{\mathcal{W}^{(1)}}&\ustick{1}\qw
&\gate{\;\;\mathcal{U}\;\;}&\ustick{2}\qw&\multimeasureD{1}{d \KetBra{V}{V}}&&&\ustick{3}\qw&
\qw\\
&&\pureghost{\mathcal{W}^{(1)}}&\qw&\ustick{0'}\qw&\qw&\ghost{d \KetBra{V}{V}}&&\qwx&\\
&&\pureghost{\mathcal{W}^{(1)}}&\qw&\ustick{1'}\qw&\qw&\gate{\;\;\mathcal{V}\;\;}&\qw&\qw \qwx&
}
  \end{aligned}
\end{align}
\begin{itemize}
\item  The first isometry $W^{(1)}$ prepares a coherent superposition \\
 $\frac{1}{\sqrt{d}}\left( y\Ket{I}_{1'0'} \otimes \right( \ket{\psi} \left)_{1}
+ x\Ket{I}_{11'}\otimes \ket{\psi}_{0'}\right)$ which is tuned by the parameters $x$ and $y$
 (that is by $p$ in Eq. (\ref{eq:convexcomb} ));
\item the unitary $U$ acts locally on system $1$;
\item at the end the instrument $\{ \mathcal{T}_V \}$ is applied: $\{ \mathcal{T}_V \}$
can realize either an estimate-and-reprapare strategy, or a teleportation protocol.
\end{itemize}

We now give a look to the complete action of the optimal circuit when the input is a pure state
$\ket{\psi}$:
\begin{align}
\ket{\psi} 
&\rightarrow  
\frac{1}{\sqrt{d}}\left( y\Ket{I}_{1'0'} \otimes \right( \ket{\psi} \left)_{1}
+ x\Ket{I}_{11'}\otimes \ket{\psi}_{0'}\right) 
\rightarrow \nonumber \\
&\rightarrow
\frac{1}{\sqrt{d}}\left( y\Ket{I}_{1'0'} \otimes \right( U\ket{\psi} \left)_{2}
+ x\Ket{U}_{21'}\otimes \ket{\psi}_{0'}\right)
 \rightarrow \nonumber \\
&\rightarrow
y U (\ket{\psi})_{3} + x \Tr[V^\dagger U] V (\ket{\psi})_{3}.
\end{align}

We remark that the optimal device essentially combines two strategies:
\begin{enumerate}
\item applying the unknown unitary $U$ to the state $\Ket{I}$, measuring the state $\Ket{U}$, and then
performing the estimated transformation $V$ on the input
 state $\psi$. This is a measure and re-prepare strategy which is optimal if 
$y=0,x=1$ (that is we are interested only in the information gain)

\item  Applying $U$ on the input state and then outputting $U\ket{\psi}$ (in our scheme this last step involves a teleportation protocol).
This is clearly an optimal strategy if $x=0,y=1$, that is if we are interested only in leaving the action of $U$ unaffected.
\end{enumerate}

Surprisingly, the analytical expression of the tradeoff curve 
given in Eq. (\ref{tradeoff2}) is the same as the one for the
estimation of a maximally entangled state \cite{tradeoffmax}.
It is worth noting that this is not a trivial consequence of the isomorphism
$2^{-\frac12}\Ket{U} \leftrightarrow U$; indeed, this mathematical correspondence
cannot be implemented by a physical invertible map. Once a unitary
$U$ is applied to the maximally entangled state $2^{-\frac12}\Ket{I}$ it is
possible to retrieve the transformation $U$ only probabilistically 
(this is the problem of the quantum learning discussed in chapter \ref{chapter:learning}).
Because of this reason there is no operational relation between the
information disturbance tradeoff for unitary transformation and for maximally entangled states
(the former is not a primitive of the latter and viceversa).

Besides its fundamental relevance, the information disturbance tradeoff for transformations
is interesting as a possible eavesdropping for cryptographic protocol
in which the secret key is encoded into a transformation.
However this is not the case of the protocols
\cite{bostromprotocol, lucamariniprotocol} where orthogonal unitaries are used
and the security of the protocol is not based on the information disturbance tradeoff
studied here.
On the other hand the tradeoff we considered is an effective
attack to the alternative $BB84$ protocol
introduced in chapter \ref{chapter:cloning}.
However, this alternative version of the $BB84$
protocol just involves two nonorthogonal unitaries; in principle,
the tradeoff curve for a restricted of unitaries could be more favorable 
to the eavesdropper.

\setcounter{equation}{0} \setcounter{figure}{0} \setcounter{table}{0}\newpage
\section{Learning and cloning of a measurement device}\label{chapter:observables}

As we stressed in the introduction the recent trend of quantum information
is to consider transformations as  information carriers.
Unlike what we did  in all the previous chapters,
in the present one we will not deal with unitary transformations
but with measurements.
We will  consider quantum networks that, upon the insertion
of $N$ uses of an undisclosed measurement device,
reproduce $M$ approximate replicas of it.

When a measurement is an intermediate step of a quantum procedure its
outcome can influence the following operations. 
This feed forward of the classical outcome can be conveniently
described 
using a quantum system into which the outcome is encoded into
perfectly distinguishable
 orthogonal states. In this sense a quantum measurement with only
 classical outcomes can be seen as a channel, which first measures the
 input
 system and based on the outcome prepares a state from a fixed orthogonal set.


In order to achieve  this task different scenarios can be considered:

{\bf $N \rightarrow M$ cloning}\footnote{The term cloning of
  observables  has been used in Ref. \cite{Parisobsclon} referring to
 state cloning machines preserving the statistics of a class of
 observables.}:
 The measurement device and the states we want to measure are available at the same
time;
\begin{align}
  \begin{aligned}
    \Qcircuit @C=0.5em @R=0.5em
{
&\ustick{10}&\qw&\multigate{4}{\;\;\;}\\
&\ustick{9} &\qw&\ghost{\;\;\;}\\
&\ustick{8} &\qw&\ghost{\;\;\;}\\
&\ustick{7} &\qw&\ghost{\;\;\;}& \ustick{0}\qw &\measureD{E_i} & \ustick{1}\cw & \pureghost{\;\;\;} \cw& \ustick{2}\qw & \measureD{E_i}& \ustick{3} \cw & \pureghost{\;\;\;} \cw& \ustick{4}\qw & \measureD{E_i}& \ustick{5} \cw & \pureghost{\;\;\;} \cw \\
&\ustick{6} &\qw &\ghost{\;\;\;}& \qw &\qw& \qw &\multigate{-1}{\;\;\;}& \qw &          \qw& \qw &\multigate{-1}{\;\;\;}& \qw & \qw& \qw &\multigate{-1}{\;\;\;}
}
  \end{aligned}
\; \simeq \; 
  \begin{aligned}
    \Qcircuit @C=1em @R=0.5em
{
& \ustick{10} &\measureD{E_i}\\
& \ustick{9} &\measureD{E_i}\\
& \ustick{8} &\measureD{E_i}\\
& \ustick{7} &\measureD{E_i}\\
& \ustick{6} &\measureD{E_i}
}
  \end{aligned}
\end{align}
(the double wire carries the classical outcome of the measurement).

{\bf $N \rightarrow M$ learning}:
we can use the measurement device $N$ times today and
 we want to replicate the same observables on $M$ systems that will be provided
tomorrow
\begin{align}
  \begin{aligned}
    \Qcircuit @C=0.5em @R=0.5em
{
&&&&&&&&&&&&&&& \ustick{10} \qw&\ghost{\;\;\;} \\
&&&&&&&&&&&&&&& \ustick{9} \qw&\ghost{\;\;\;} \\
&&&&&&&&&&&&&&&\ustick{8} \qw&\ghost{\;\;\;} \\
&&&&&&&&&&&&&&&\ustick{7} \qw&\ghost{\;\;\;} \\
&\multiprepareC{1}{\;\;\;}&\ustick{0} \qw&\measureD{E_i}&\ustick{1} \cw& \pureghost{\;\;\;} \cw&\ustick{2} \qw& \measureD{E_i}&\ustick{3} \cw& \pureghost{\;\;\;} \cw&\ustick{4} \qw& \measureD{E_i}&\ustick{5} \cw& \pureghost{\;\;\;} \cw&&
\ustick{6} \qw&\ghost{\;\;\;} \\
&\pureghost{\;\;\;}& \qw&\qw& \qw&\multigate{-1}{\;\;\;}& \qw&          \qw& \qw&\multigate{-1}{\;\;\;}& \qw& \qw& \qw&\multigate{-1}{\;\;\;}& \qw&  \qw&
\multigate{-5}{\;\;\;}\\
}
  \end{aligned}
\; \; \simeq \; \;
  \begin{aligned}
    \Qcircuit @C=0.5em @R=0.5em
{
\ustick{10} &\measureD{E_i}\\
\ustick{9}&\measureD{E_i}\\
\ustick{8}&\measureD{E_i}\\
\ustick{7}&\measureD{E_i}\\
\ustick{6}&\measureD{E_i}
}
  \end{aligned}
\end{align}

{\bf $N \rightarrow M$ hybrid}:
we have to produce the  replicas at different times
\begin{align}
  \begin{aligned}
    \Qcircuit @C=0.5em @R=0.5em
{
&&&&&&&&\ustick{8}&\ghost{\;\;\;}&&&&&&&\ustick{10}&\ghost{\;\;\;} \\
\ustick{7}&\multigate{1}{\;\;\;}&\ustick{0}\qw&\measureD{E_i}&\ustick{1}\cw& \pureghost{\;\;\;} \cw&\ustick{2} \qw& \measureD{E_i}&\ustick{3}\cw& \pureghost{\;\;\;} \cw&\ustick{4}\qw& \measureD{E_i}&\ustick{5}\cw& \pureghost{\;\;\;} \cw&&&\ustick{9}&\ghost{\;\;\;} \\
\ustick{6}&\ghost{\;\;\;}&\qw&\qw&\qw&\multigate{-1}{\;\;\;}&\qw&          \qw&\qw&\multigate{-2}{\;\;\;}&\qw& \qw&\qw&\multigate{-1}{\;\;\;}&\qw&\qw&\qw&
\multigate{-2}{\;\;\;}\\
}
  \end{aligned}
 \; \simeq \;
  \begin{aligned}
    \Qcircuit @C=0.7em @R=0.5em
{
&\ustick{10}&\measureD{E_i}\\
&\ustick{9}&\measureD{E_i}\\
&\ustick{8}&\measureD{E_i}\\
&\ustick{7}&\measureD{E_i}\\
&\ustick{6}&\measureD{E_i}
}
  \end{aligned}
\end{align}

In the following we will consider some specific scenarios and compare their performances.

\subsection{Formulation of the problem}

In the following we will restrict ourselves to von Neumann measurement, i.e. sharp non degenerate POVMs:
\begin{align}
E_i = \ketbra{i}{i}
\end{align}
where $\{ \ket{i} \}_{i=1}^d$ is an o.n.b. of the Hilbert space $\mathcal{H}$.
We notice that all the POVMs of this kind can be generated
by rotating  a reference POVM $\{ \ketbra{i}{i}  \}_{i=1}^d$
by arbitrary elements of the $\group{SU}(d)$ group as follows
\begin{align}
  E_i^{(U)} = U\ketbra{i}{i}U^\dagger \qquad U \in \group{SU}(d).
\end{align}
 The classical outcome $i$ of the POVM  will be encoded into a quantum system
by preparing the state $\ket{i}$ from a fixed orthonormal basis, which is the same for each 
POVM $\{ E_i^{(U)} \}$.
Within this framework the measurement device is modeled as the following measure-and-prepare
quantum channel $\mathcal{E}^{(U)}:\mathcal{L}(\hilb{H})\rightarrow \mathcal{L}(\hilb{H})$
\begin{align}
  \mathcal{E}^{(U)}(\rho) = \sum_i\Tr[E_i^{(U)}\rho]\ketbra{i}{i}
\end{align}
that measure the POVM $\{ E_i^{(U)} \}$ on its input state 
and outputs the state $\ketbra{i}{i}$ if the outcome is $i$.
The channel $\mathcal{E}^{(U)}$ is represented by its Choi operator
\begin{align}\label{eq:obsdepolachan}
  {E}^{(U)} = \sum_i {E_i^{(U)}}^T \otimes \ketbra{i}{i} = \sum_i U^* \ketbra{i}{i} U^T \otimes \ketbra{i}{i} 
\end{align}
The $N$ uses of the measurement device are then represented by the tensor product
$E^{(U)}_{01} \otimes  \cdots \otimes E^{(U)}_{2N-2 \; 2N-1}$
where 
the input and the output space of the $k$-th  use
of the measurement device
are denoted by $2k-2$ and $2k-1$ respectively.
We introduce the following notation:
\begin{align}
\hilb{H}_{\defset{or}} := \bigotimes_{k=1}^N \hilb{H}_{2k-2},
\qquad
\hilb{H}_{\defset{cl}} := \bigotimes_{k=1}^N \hilb{H}_{2k-1} .  
\end{align}

Since we want   the replicating network 
$\mathcal{R} $
to behave as $M$ copies of the POVM $\{ E^{(U)}_{i}  \}$ upon insertion of the $N$ uses 
$\mathcal{E}^{(U)}$, 
we have that $\mathcal{R}$ is actually a generalized instrument $\{ \mathcal{R}_{\vec i} \}$
where $\vec{i}$ is the $M$-tuple of outcomes $(i_1, \dots, i_M)$.
The overall resulting POVM is then
 \begin{align}\label{eq:obsfinalpovm}
   G^{(U)}_{\vec i} = (R_{\vec i}* {E}_{01}^{(U)} * \cdots {E}_{2N-2 \; 2N-1}^{(U)})^T \\
\nonumber {R}_{\vec i} =  \mathcal{L}(\hilb{H}_{\defset{or}} \otimes \hilb{H}_{\defset{cl}} \otimes \hilb{H}_{\defset{re}}) 
\quad \hilb{H}_{\defset{re}} = \bigotimes_{k=1}^{M}\hilb{H}_{2N+k-1}
 \end{align}
where  $\hilb{H}_{2N+k-1}$
denotes
the input space of the
$k$-th replica.

Our task is to find the network $\mathcal{R}_{\vec i}$
such that $G^{(U)}_{\vec i}$ is as close as possible to
to $M$ uses of $\{ E_i^{(U)} \}$, i.e
\begin{align}
  \{ G^{(U)}_{\vec i} \} \simeq \{ E_{i_1}^{(U)} \otimes E_{i_2}^{(U)}\otimes \cdots \otimes E_{i_M}^{(U)} \} := \{E_{\vec i}^{(U)}\}.
\end{align}
In order to quantify the performances of the replicating
network,
we need to introduce a criterion which quantify the closeness between two POVMs.
the following lemma provides such a tool:
\begin{lemma}[distance criterion for POVM]
Let $\Sigma := \{ 1,\dots ,d \}$ be a finite  set of events and
 $\{ P_i \in \mathcal{L}(\hilb{H}) \}$ and $\{ Q_j \in \mathcal{L}(\hilb{H}) \}$ be two POVMs.
Consider now the quantity
 \begin{align}
   \mathscr{F} := \frac1d \sum_i \Tr[P_iQ_i]
 \end{align}
and suppose that either  $\{ P_i  \}$ or $\{ Q_j \}$
is a von Neumann measurement.
Then $\mathscr{F}=1 \Leftrightarrow P_i = Q_i \forall i$
\end{lemma}
\begin{Proof}
If $\{ P_i \}$
is a von Neumann measurement 
we have $P_i= \ketbra{i}{i}$
where $\ket{i}$ is an orthonormal basis of $\hilb{H}$.
Then we have $Q_i = P_i \Rightarrow Q_i = \ketbra{i}{i}$
and 
\begin{align}
   \mathscr{F}= \frac1d \sum_i \Tr[P_iQ_i] = \frac1d \sum_i \Tr[\ketbra{i}{i}] = 1
 \end{align}
On the other hand if $\mathscr{F}=1$ we have
\begin{align}
 & d= \sum_i \Tr[P_iQ_i] = \sum_i \bra{i}Q_i\ket{i} = \sum_{ij} \bra{i}Q_j\ket{i} - \sum_{i\neq j} \bra{i}Q_j\ket{i}= \nonumber \\
& = \Tr\left[ \sum_j Q_j \right] - \sum_{i\neq j} \bra{i}Q_j\ket{i} = d  - \sum_{i\neq j} \bra{i}Q_j\ket{i} \Rightarrow
\sum_{i\neq j} \bra{i}Q_j\ket{i} = 0 \nonumber
\end{align}
Since $Q_j \geq 0$ 
$\sum_{i\neq j} \bra{i}Q_j\ket{i} = 0 \Rightarrow \bra{i}Q_j\ket{i} \forall i \neq j$
which implies $Q_j = \alpha_j \ketbra{j}{j}$ with $\alpha_j \geq 0$.
Finally  the condition $\sum_j \alpha_j \ketbra{j}{j} = I$ implies $\alpha_j = 1$ and 
thus $Q_j = P_j$.
\qed \end{Proof}

Assuming that the unknown POVM $ \{E_i^{(U)}\}$
is randomly drawn according to the Haar distribution,
we choose the quantity:
\begin{align}\label{eq:obsfigmer}
F :=    \int \d U \mathscr{F}( \{ G^{(U)}_{\vec i} \} \{E_{\vec i}^{(U)}\})
\end{align}
as a figure of merit.

After fixing one of the possible scenarios ($N \rightarrow M$ cloning, learning or  hybrid)
our task is to find the optimal generalized instrument  $\mathcal{R}_{\vec i}$
 maximizing the quantity $F \hspace*{-0.09cm}:=\hspace*{-0.12cm}    \int\hspace*{-0.08cm} \d U
\mathscr{F}( \{ G^{(U)}_{\vec i} \} \{E_{\vec i}^{(U)}\})$.

\subsection{Symmetries of the replicating network}\label{sec:obssymm-repl-netw}

Here we exploit the symmetries of the figure of merit 
(\ref{eq:obsfigmer}) to simplify the optimization problem.
The first simplification relies on the fact that some wires of the network
carry only classical information, representing the outcome of the measurement.

\begin{lemma}[Restriction to diagonal networks]
\hspace*{-0.2cm} The  optimal generalized instrument $\{ \mathcal{R}_{\vec i}\}$, \hspace*{-0.08cm}
with $\sum_{\vec{i}}
\mathcal{R}_{\vec i} = \mathcal{R}_{\Omega}$ maximizing Eq. (\ref{eq:obsfigmer}),
can be chosen to satisfy:
\begin{align}\label{eq:obsdiagcomb}
  R_{\vec i} = \sum_{\vec j}  R'_{\vec{i},\vec{j} } \otimes \ketbra{\vec{j}}{\vec{j}},
\end{align}
where $\vec{j} = (j_1, \dots, j_N)$, $\ket{\vec{j}} := \ket{j_1}_1\otimes \cdots \otimes \ket{j_N}_{2N-1} \in \hilb{H}_{\defset{cl}}$,
$0 \leq R'_{\vec{i},\vec{j} } \in   \mathcal{L}
( \hilb{H}_{\defset{or}} \otimes \hilb{H}_{\defset{re}})$
and $\sum_{\vec j}$ is a shorthand for $\sum_{j_1, \dots , j_N=1}^d$.
\end{lemma}
\begin{Proof}
  Let $\{R_{\vec i}\}$ be a generalized instrument.
Let us define $\{\tilde{R}_{\vec i}\}$ as
\begin{align}
  \tilde{R}_{\vec i} :=  \sum_{\vec j} \bra{\vec{j}} R_{\vec i}\ket{\vec{j}} \otimes \ketbra{\vec{j}}{\vec{j}}
\qquad 
\ket{j_1}_1\otimes \cdots \otimes \ket{j_N}_{2N-1}.
\end{align}
We now prove that $\{\tilde{R}_{\vec i}\}$
is a generalized instrument:
reminding Eq. (\ref{eq:obsdepolachan}),  we have
\begin{align}\label{eq:obsdigonalnet}
&\sum_{\vec{i}}   \tilde{R}_{\vec i} =
\sum_{\vec{i}}    \sum_{\vec j} \bra{\vec{j}} R_{\vec i}\ket{\vec{j}} \otimes \ketbra{\vec{j}}{\vec{j}} = 
\sum_{\vec j} \bra{\vec{j}} R_{\Omega}\ket{\vec{j}} \otimes \ketbra{\vec{j}}{\vec{j}}= \nonumber \\
& = R_\Omega * \left(\sum_{j_1} \ketbra{j_1}{j_1}\otimes \ketbra{j_1}{j_1} \right)
* \cdots * 
\left(
\sum_{j_1} \ketbra{j_N}{j_N}\otimes \ketbra{j_N}{j_N}
\right) = \nonumber\\
&= 
 R_\Omega *  E^{(I)}
* \cdots * 
E^{(I)}
\end{align}
where the link is performed on the space $\hilb{H}_{\defset{cl}}$.
The operator in Eq. (\ref{eq:obsdigonalnet}) is the Choi-Jamio\l kowsky of a deterministic 
quantum network with the same normalization of 
$R_\Omega$.
Finally we show that $\{{R}_{\vec i}\}$ and $\{\tilde{R}_{\vec i}\}$
when linked with the $N$ uses of $E^{(U)}$
produce the same replicas $\{ G^{(U)}_{\vec{i}}\}$:
\begin{align}
  G^{(U)}_{\vec{i}} &= {(R_{\vec i}* {E}_{01}^{(U)} * \cdots {E}_{2N-2 \; 2N-1}^{(U)})}^T = \nonumber \\
&= {(  \sum_{\vec{j}}  (\bra{\vec{j}}_{\defset{or}} {U^\dagger}^{\otimes N}   \bra{\vec{j}}_{\defset{cl}})
  R_{\vec i} (U^{\otimes N}  \ket{\vec{j}}_{\defset{or}} \ket{\vec{j}}_{\defset{cl}})}^T = \nonumber \\
&=  {(  \sum_{\vec{j}}  (\bra{\vec{j}}_{\defset{or}} {U^\dagger}^{\otimes N}   \bra{\vec{j}}_{\defset{cl}})
  \tilde{R}_{\vec i} (U^{\otimes N}  \ket{\vec{j}}_{\defset{or}} \ket{\vec{j}}_{\defset{cl}})}^T =  \nonumber \\
& = {(\tilde{R}_{\vec i}* {E}_{01}^{(U)} * \cdots {E}_{2N-2 \; 2N-1}^{(U)})}^T.
\end{align}
\qed \end{Proof}
We now exploit the form of Eq. (\ref{eq:obsdiagcomb}) to 
simplify the expression of the fidelity in  Eq. (\ref{eq:obsfigmer}) as follows:
\begin{align}\label{eq:obsfigmer2}
  F &:=    \int \d U \mathscr{F}( \{ G^{(U)}_{\vec i} \} \{E_{\vec i}^{(U)}\})  =  \nonumber \\
& = \frac{1}{d^M}\int \d U \sum_{\vec{i},\vec{j}}
\bra{\vec{i}}_{\defset{re}}{U^T}^{\otimes N} \bra{\vec{j}}_{\defset{or}}{U^\dagger}^{\otimes N}
R'_{{\vec{i},\vec{j} }}
{U^*}^{\otimes N}\ket{\vec{i}}_{\defset{re}} {U}^{\otimes N} \ket{\vec{j}}_{\defset{or}}.
\end{align}

The following lemma exploits the symmetry properties of Eq. (\ref{eq:obsfigmer2})
and simplifies the structure of the $R'_{{\vec{i},\vec{j} }}$:
\begin{lemma}[Restriction to covariant networks]
  The operators  $R'_{{\vec{i},\vec{j} }}$ that maximize Eq. (\ref{eq:obsfigmer2})
can be chosen to satisfy the commutation relation
\begin{align}\label{eq:obscommutr}
  [R'_{{\vec{i},\vec{j} }}, U^{\otimes N}_{\defset{or}} \otimes {U^*}^{\otimes M}_{\defset{re}}] = 0
\end{align}
\end{lemma}
\begin{Proof}
  The proof consists in the same averaging argument
we used in proving lemmas \ref{Lemma} , \ref{lem:learncov} and \ref{lem:tradeofcovinst}
\qed \end{Proof}
The commutation relation (\ref{eq:obscommutr})
allows us to rewrite the figure of merit has:
\begin{align}\label{eq:obsfigmer3}
  F  = \frac{1}{d^M}\int \d U \sum_{\vec{i},\vec{j}}
\bra{\vec{i}}_{\defset{re}}\bra{\vec{j}}_{\defset{or}}
R'_{{\vec{i},\vec{j} }}
\ket{\vec{i}}_{\defset{re}} \ket{\vec{j}}_{\defset{or}}
\end{align}

Another symmetry of our figure of merit is related to the possibility
of relabeling the outcomes of a POVM. We shall denote by $\sigma$ the
element of $\group{S}_d$, the group of permutations of $d$ elements as well
as the linear operator that permutes the elements of basis
$\{\ket{i}\}$ according to this permutation
($\sigma\ket{i}\equiv\ket{\sigma(i)}$).

\begin{lemma}[Relabeling symmetry]\label{lem-obspermlemma}
  Without loss of generality we can assume 
that the operators  $R'_{{\vec{i},\vec{j} }}$ that maximize Eq. (\ref{eq:obsfigmer2})
satisfy the relation
\begin{align}\label{eq:obsperminvproperty}
  R'_{{\vec{i},\vec{j} }} =  R'_{\sigma(\vec{i}),\sigma(\vec{j})}
\end{align}
 where we shortened
$\sigma(\vec{i})\equiv(\sigma(i_1),\ldots,\sigma(i_M))$,
$\sigma(\vec{j})\equiv(\sigma(j_1),\ldots, \sigma(j_N))$.
\end{lemma}

\begin{Proof}
Without loss of generality we can suppose that the $R'_{\sigma(\vec{i}),\sigma(\vec{j})}$'s satisfy
Eq.~(\ref{eq:obscommutr}).    Let us then define 
\begin{eqnarray}
 \widetilde{R'}_{\vec{i},\vec{j}}=\frac{1}{d!}
\sum_{\sigma\in \group{S}_d} R'_{\sigma(\vec{i}),\sigma(\vec{j})}
 \label{eq:obsdefPermSym}
\end{eqnarray}
 This corresponds to a valid instrument
$\{\widetilde{R'}_{\vec{i}}\}$, because it is a convex combination of
instruments obtained from $R_{\sigma(\vec{i}),\sigma(\vec{j})}$ by
relabeling the outcomes of the inserted and replicated measurements by
permutation $\sigma$.
 Let us now evaluate the figure of merit for this new
instrument:
\begin{align}
&F(\widetilde{R'}_{\vec{i},\vec{j}})=\frac{1}{d^M}\sum_{\vec{i},\vec{j}}
\bra{\vec{i}}\bra{\vec{j}}
\widetilde{R'}_{\vec{i},\vec{j}}
\ket{\vec{i}}\ket{\vec{j}}  =  \frac{1}{d^M d! }
\sum_{\vec{i},\vec{j}}
\bra{\vec{i}}\bra{\vec{j}}
\left(
\sum_{\sigma\in \group{S}_d}
R^{\sigma(\vec{i}),\sigma(\vec{j})}
\right)
\ket{\vec{i}}\ket{\vec{j}} = \nonumber\\
& =  \frac{1}{d^M d! }
\sum_{\vec{i},\vec{j}}
\bra{\vec{i}}\bra{\vec{j}}
\left(
\sum_{\sigma\in \group{S}_d}
\sigma^{\otimes N} \otimes \sigma^{\otimes M}
R'_{\sigma(\vec{i}),\sigma(\vec{j})}
\sigma^{\otimes N} \otimes \sigma^{\otimes M}
\right)
\ket{\vec{i}}\ket{\vec{j}} = \label{eq:obscompatsym} \\
&= \sum_{\sigma\in \group{S}_d}
\sum_{\vec{i},\vec{j}}
\bra{\sigma(\vec{i})}\bra{\sigma(\vec{j})}
R'_{\sigma(\vec{i}),\sigma(\vec{j})}
\ket{\sigma(\vec{i})}\ket{\sigma(\vec{j})} = F({R'}_{\vec{i},\vec{j}})
\end{align}
where the identity (\ref{eq:obscompatsym})
follows from  the commutation relation
(\ref{eq:obscommutr}) with
$ U = U^* = \sigma$.
It is easy to prove that 
$\widetilde{R'}_{\vec{i},\vec{j}}$ satisfies
Eq. (\ref{eq:obsperminvproperty}).

\qed \end{Proof}

\begin{remark}
  It is worth notice that the properties
(\ref{eq:obsdiagcomb}), (\ref{eq:obscommutr}) and (\ref{eq:obsperminvproperty})
induce the following structure of the replicated POVMs:
\begin{align}
  G^{(U)}_{\sigma(\vec{i})} = {(U \sigma)}^{\otimes M} G^{(I)}_{\vec{i}} {(\sigma U^\dagger)}^{\otimes M}
\end{align}
\end{remark}

The advantage of using the above symmetry is in the reduction the number of
independent parts of the  generalized instrument.
Let us define the equivalence relation between strings
$\vec i$ and $\vec i'$ as
\begin{equation}
  \vec i\sim\vec i'\quad\Leftrightarrow\quad \vec i=\sigma(\vec i'),
\end{equation}
for some permutation $\sigma$. Thanks to Eq.
(\ref{eq:obsperminvproperty}) there are only as many independent
$R_{\vec{i},\vec{j}}$ as there are equivalence classes among sequences
$\vec{i},\vec{j}$. For the simplest case $M=N=1$ and arbitrary
dimension $d\geq 2$, there are only two classes, which we denote by
$xx$ and $xy$. The reason is that for any couple $i',j'$ there is a
permutation $\sigma$ such that $\sigma(1)=i'$ and $\sigma(2)=j'$, thus
the classes are defined by the conditions $i=j$ or $i\neq j$,
respectively. For all the cases where $M+N=3$ (e.g. $N=1, M=2$ or
$N=2, M=1$), the vectors $\vec i$ and $\vec j$ have three components.
Then, there are four or five equivalence classes depending on the
dimension $d$ being two or greater than two, respectively. We denote
these equivalence classes by $xxx, xxy, xyx, xyy, xyz$ and the set of
these elements by $\Rel{3}$.  In the general case, it is clear that
the cardinality of classes is given by the number of disjoint
partitions of a set with cardinality $M+N$, with number $p$ of parts
$p\leq d$.  For $M+N\geq d$, this number is known as Bell number
$B_{M+N}$, and is recursively defined as follows
\begin{equation}
  B_{k+1}:=\sum_{j=0}^{k}{{k}\choose{j}} B_j.
\end{equation}
In the case $M+N<d$ the solution is provided by the sum for
$k=1,\dots,d$ of numbers of disjoint partitions of a set with $N+M$
elements into $k$ subsets, which is the sum of Stirling numbers of the
second kind $S(M+N,k)$. The Stirling numbers are given by the
following formula
\begin{equation}
  S(n,k):=\frac1{k!}\sum_{j=0}^k(-1)^j{k\choose j}(k-j)^n,
\end{equation}
thus providing the following expression for the cardinality of classes
$\Rel{M+N}$
\begin{equation}
  \Rel{M+N}=\sum_{k=1}^d\frac1{k!}\sum_{j=0}^k(-1)^j{k\choose j}(k-j)^n.
\end{equation}
Exploiting Lemma \ref{lem-obspermlemma}   we can write the optimal
generalized instrument as follows
\begin{equation}
  S_{\vec x,\vec y}:=R'_{\vec i,\vec j}=R'_{\sigma(\vec i),\sigma(\vec j)},
\end{equation}
where $(\vec x,\vec y)$ is a couple of strings of indices that
represents one equivalence class. We will denote by $\defset L$ the
set of equivalence classes $\defset L:=\{(\vec x,\vec y)\}$. The
figure of merit can finally be written as follows
\begin{equation}
  F=\frac1{d^M}\sum_{(\vec x,\vec y)\in\defset L}n(\vec x,\vec y)\<S_{\vec x,\vec y}\>,
  \label{eq:finfigm}
\end{equation}
where $n(\vec x,\vec y)$ is the cardinality of the equivalence class
denoted by the couple $(\vec x,\vec y)$, and $\<S_{\vec x,\vec
  y}\>=\bra {\vec i}\bra{\vec j}R'_{\vec i,\vec j}\ket{\vec i}\ket{\vec
  j}$ for any string $\vec i,\vec j$ in the equivalence class denoted
by $(\vec x,\vec y)$.
As a consequence of Schur's lemmas, the
condition of Eq.~\eqref{eq:obscommutr} implies the following structure
for the operators $S_{\vec x,\vec y}$ (see Appendix \ref{appendicegruppi} for
the details)
\begin{equation}
  S_{\vec x,\vec y}=\bigoplus_\nu P^\nu \otimes r^\nu_{\vec x,\vec y},
\end{equation}
where $\nu$ labels the irreducible representations in the
Clebsch-Gordan series of $U^{\otimes M}_{\defset{out}}\otimes
{U^*}^{\otimes N}_{\defset{in}}$, and $P^\nu$ acts as the identity on
the invariant subspaces of the representations $\nu$, while
$r^\nu_{\vec x,\vec y}$ acts on the multiplicity space of the same
representation. In the simplest case $M+N=2$ we have
\begin{equation}
  R_{a,b}=P^p r_{a,b}^p  + P^q  r_{a,b}^q,
  \label{eq:rxxxy}
\end{equation}
where
 $ P^p$ and $ P^q $
are defined in Eq. (\ref{eq:decompoUU*}).
and $r_{a,b}^p$ and $r_{a,b}^q$ are non-negative numbers.
 In the case
$M+N=3$, with $M,N\neq0$ we have two different decompositions,
depending whether $d>2$ or $d=2$.
When $d>2$,  we have (see Eq. (\ref{decomposition}))
\begin{align}\label{eq:obsdecompor21}
  R_{\vec x,\vec y} = P^\alpha \otimes r_{\vec x,\vec y}^\alpha+
  P^\beta r_{\vec x,\vec y}^\beta + P^\gamma r_{\vec x,\vec y}^\gamma.
\end{align}
When $d=2$ we have that $\dim(\hilb{H}_{\gamma,-})=0$ and the decomposition
  (\ref{eq:obsdecompor21}) becomes 
\begin{align}\label{eq:obsdecompor21dim2}
  R_{\vec x,\vec y} = P^\alpha\otimes r^\alpha_{\vec x,\vec y} +
  P^\beta r_{\vec x,\vec y}^\beta.
\end{align}

\subsection{Optimal learning}
In this section we derive the optimal quantum learning
of a von Neumann measurement; 
we will consider the following scenarios:
\begin{itemize}
\item $1 \rightarrow 1$ learning
\item $2 \rightarrow 1$ learning
\item $3 \rightarrow 1$ learning
\item $1 \rightarrow 2$ learning
\end{itemize}

\subsubsection{$1 \rightarrow 1$ case}
Consider the case in which today we are provided with a single use of
a measurement device, and we need a replica to measure a state that
will be prepared tomorrow; this scenario is described by the following
scheme
\begin{align}\label{eq:obs11learn}
  \begin{aligned}
    \Qcircuit @C=0.7em @R=1em {
      &&&\ustick{2}& \ghost{\;\;\;}\\
      \multiprepareC{1}{\;\;\;} &\ustick{0} \qw & \measureD{E^{(U)}} & \ustick{1} \cw &\pureghost{\;\;\;} \cw\\
      \pureghost{\;\;\;}& \qw& \qw&\qw & \multigate{-2}{\;\;\;}\\
    }
  \end{aligned}
\end{align}
Using the labeling as in Eq. (\ref{eq:obs11learn}) and exploiting the
results of Section \ref{sec:obssymm-repl-netw} for the case $M+N=2$,
we have
\begin{align}
  &\defset L=\{(x,x),(x,y)\},\nonumber\\
  &R_{{i}_{210}}=\ket i\bra i_{1}\otimes R_{{x,x}_{20}}+(I-\ket i\bra i)_1\otimes R_{{x,y}_{20}} \label{eq:obsrifromrij}\nonumber\\
  & R_{a,b}=P^p r_{a,b}^p + P^q r_{a,b}^q,\quad (a,b)\in \defset L
\end{align}
Exploiting the identity $\bra i\bra j P^p\ket i\ket j=\delta_{ij}1/d$,
and considering that $n(x,x)=d$ and $n(x,y)=d(d-1)$, the figure of
merit in Eq.~\eqref{eq:finfigm} for the can be rewritten as
\begin{align}
  F=& \<R_{x,x}\>+(d-1)\<R_{x,y}\>=
  \nonumber\\
  &\sum_{\nu \in \{p, q \}}\left(
  r_{x,x}^{\nu} \Delta_{x,x}^{\nu} + (d-1) r_{x,y}^{\nu} \Delta_{x,y}^{\nu}\right),
\label{eq:obsfigMeritL11}
\end{align}
where $\Delta_{x,x}^{p}=\frac1d$, $\Delta_{x,y}^{p}=0$, and
$\Delta^q_{a,b}=1-\Delta^p_{a,b}$.  Let us now write the normalization
conditions for the generalized instrument in terms of operators
$R_{i,j}$.  We have that that $R_\Omega := \sum_i R_i$ has to be the
Choi operator of a deterministic quantum network and
must satisfy Eq.  (\ref{eq:recnorm2}), that is
\begin{align}\label{eq:obsnorm11lear}
  R_\Omega = I_2 \otimes I_1 \otimes \rho \qquad \Tr[\rho]=1, \quad
  \rho\geq 0.
\end{align}
The commutation relation (\ref{eq:obscommutr}) implies $[\rho,U^*] = 0$
that by Schur's lemmas gives
\begin{align}\label{eq:obsinputstate}
  \rho = \frac{I}{d}.
\end{align}
Now, exploiting Eqs. (\ref{eq:obsrifromrij}) and
(\ref{eq:obsinputstate}), Eq.  (\ref{eq:obsnorm11lear}) becomes
\begin{equation}
  I_1\otimes R_{x,x}+(d-1)I_1\otimes R_{x,y}=\frac Id
  \label{eq:obsnormL11}
\end{equation}
Substituting the expression of Eq.~\eqref{eq:rxxxy} in Eq.
(\ref{eq:obsnormL11}), we obtain
\begin{align}
  &r_{x,x}^p+(d-1)r^p_{x,y} = r_{x,x}^q+(d-1)r^q_{x,y} = \frac1d.
  \label{eq:obsnormcond11final}
\end{align}
From the constraint (\ref{eq:obsnormcond11final}) the following bound follows
\begin{align}\label{eq:obsmaxfidelity}
  F =& \sum_{\nu}\left(r_{x,x}^{\nu} \Delta_{x,x}^{\nu} +
  (d-1) r_{x,y}^{\nu} \Delta_{x,y}^{\nu}\right) \le\nonumber\\
  &\sum_{\nu \in \{p, q \}} \overline{\Delta}^{\nu} \left(
    r_{x,x}^{\nu} + (d-1) r_{x,y}^{\nu} \right) = \frac{d+1}{d^2},
\end{align}
where $\overline{\Delta}^{\nu} := \max_{ij} \Delta_{i,j}^{\nu}$. The
bound (\ref{eq:obsmaxfidelity}) is achieved by
\begin{align}
  r_{x,x}^q=r_{x,y}^{p}=0, \quad
  r_{x,x}^p=\frac{1}{d},\quad
  r_{x,y}^{q}=\frac{1}{d(d-1)},\nonumber
\end{align}
which corresponds to generalized instrument
\begin{align}
  R_i&=\ketbra ii_1\otimes\frac1d P^p +(I-\ketbra ii)_{1}\otimes \frac1{d(d-1)}P^{q},
\end{align}
that replicates the original Von Neuman measurement as follows
\begin{align}
  &G^{(U)}_i=R^{i}*{E^{(U)}_{10}}^T=\nonumber\\
  &\frac{1}{d(d-1)} U\ketbra{i}{i}_{1}U^{\dagger} +
  \frac{d^2-d-1}{d^2(d-1)}I.
\end{align}
The optimal learning strategy can be realized by the following network
\begin{align}
  \begin{aligned}
    \Qcircuit @C=1em @R=1em {
      &                                   &              &             &\ustick{2}&\multimeasureD{2}{d R^i}\\
      &\multiprepareC{1}{\frac{1}{d}|I\kk \bb I|}&\ustick{0}\qw& \gate{E^{(U)}} & \ustick{1} \cw& \pureghost{ d R_i} \cw\\
      &\pureghost{\frac{1}{d}|I\kk \bb I|}& \qw & \ustick{A_1} \qw & \qw&
      \ghost{d R_i} }
  \end{aligned}
\end{align}

\subsubsection{$2 \rightarrow 1$ case}

We now consider the case in which we have two uses of $E^{(U)}$ at our
disposal
\begin{align}\label{eq:obs21learn}
  \begin{aligned}
    \Qcircuit @C=0.7em @R=1em { &&&&&&& \ustick{4}& \ghost{\;\;\;}\\
      \multiprepareC{1}{\;\;\;} &\ustick{0} \qw & \measureD{E^{(U)}} &
      \ustick{1} \cw & \pureghost{\;\;\;} \cw & \ustick{2} \qw &
      \measureD{E^{(U)}} & \ustick{3} \cw &
      \pureghost{\;\;\;} \cw \\
      \pureghost{\;\;\;}& \qw& \qw&\qw & \multigate{-1}{\;\;\;}&\qw &
      \qw&\qw &
      \multigate{-2}{\;\;\;} \\
    }
  \end{aligned}
\end{align}
Exploiting the symmetries introduced in Section
\ref{sec:obssymm-repl-netw} we have
\begin{align}
  &\defset L=\{(x,xx),(x,xy),(x,yx),(x,yy),(x,yz)\}\nonumber\\
  &R_i = \sum_{j,k} \ketbra{j}{j}_3 \otimes \ketbra{k}{k}_1 \otimes R'_{i,jk}  \label{eq:obslear21strucr}\\
  &[R'_{i,jk}, U_4 \otimes U_2^* \otimes U_0^*] = 0 \label{eq:obsler21comr}\\
  &R'_{i,jk} = \left\{\begin{array}{lcl}
      R_{x,xx} & {\rm if}& i=j=k \\
      R_{x,xy} & {\rm if}& i=j\neq k \\
      R_{x,yx} & {\rm if}& i=k\neq j \\
      R_{x,yy} & {\rm if}& j=k\neq i \\
      R_{x,yz} & {\rm if}& i\neq j \neq k \neq i.
    \end{array}
  \right.
  \label{eq:obslear21perminv}
\end{align}
The figure of merit (\ref{eq:obsfigmer3}) becomes
\begin{align}
  F = \frac1d\sum_{(a,bc)\in\defset L}n(a,bc)\<R_{a,bc}\>.
\end{align}
Let us now consider the normalization condition of the following
generalized instrument $\vec R$
\begin{align}
  \sum_iR_i = I_4 \otimes I_3 \otimes S_{210} \qquad \Tr_{2}[S]=
  I_1 \otimes \rho_0.
\end{align}
Exploiting Eq. (\ref{eq:obslear21strucr}) we have
\begin{align}
  & \sum_iR_i = \sum_{i, j,k} \ketbra{j}{j}_3 \otimes \ketbra{k}{k}_1
  \otimes R'_{i,jk} =
  I_4 \otimes I_3 \otimes S_{210} \nonumber \\
  &\sum_{i, k} \ketbra{k}{k}_1 \otimes R'_{i,jk} = I_4 \otimes
  S_{210},
  \quad \forall j \nonumber \\
  &\sum_{i} R'_{i,jk} = I_4 \otimes \bra{k}S_{210}\ket{k}_1,\quad
  \forall j,k
\end{align}
Exploiting the property (\ref{eq:obsperminvproperty}) we have
\begin{align}
  &I_4  \otimes \bra{k}S_{210}\ket{k}_1 =   \sum_{i} R'_{i,jk} = \sum_{i} R'_{\sigma(i),\sigma(j)\sigma(k)} = \nonumber\\
  &=I_4  \otimes (\bra{k} \sigma) S_{210} (\sigma\ket{k}_1)\quad \forall j,k.
\end{align}
This finally implies
\begin{align}
  \sum_{i} R'_{i,jk} = I_4 \otimes T_{20} \;\; \forall j,k \qquad
  \Tr_{20}[T] = 1. \label{eq:obsnorminner}
\end{align}
Eq. (\ref{eq:obsnorminner}) implies that the optimal strategy can be parallelized
\begin{align}\label{eq:obs21learnpar}
  \begin{aligned}
    \Qcircuit @C=0.7em @R=1em {
      &&&\ustick{4}&\ghost{\;\;\;}\\
      \multiprepareC{2}{\;\;\;}&\ustick{0} \qw&\measureD{E^{(U)}} &\ustick{1}\cw& \pureghost{\;\;\;}\cw\\
      \pureghost{\;\;\;}&\ustick{2}\qw&\measureD{E^{(U)}} &\ustick{3}\cw& \pureghost{\;\;\;}\cw\\
      \pureghost{\;\;\;}&\qw&\qw&\qw&\multigate{-3}{\;\;\;} }
  \end{aligned}
\end{align}
Eq. (\ref{eq:obs21learnpar}) induces a further symmetry of the
problem:
\begin{lemma}\label{lem:obs21perminv1}
  The operator $R'_{i,jk}$ in Eq. (\ref{eq:obslear21strucr}) can be
  chosen to satisfy:
  \begin{align}\label{eq:obspermsymmetry}
    R'_{i,jk} = \mathsf{S}R'_{i,kj}\mathsf{S} \quad \forall k,j
  \end{align}
  where $\mathsf{S}$ is the swap operator
  $\mathsf{S}\ket{k}_2\ket{j}_0 = \ket{j}_2\ket{k}_0$.
\end{lemma}
\begin{Proof}
  The proof consists in the standard averaging argument.  let us
  define $\overline{R}_{i,jk}:=\frac12 (R'_{i,jk} + \mathsf{S} R'_{i,kj}
  \mathsf{S})$.  It is easy to prove that $\{ \overline{R}_{i,jk} \}$
  satisfies the normalization (\ref{eq:obsnorminner}) and that gives
  the same value of $F$ as $R'_{i,kj}$.\qed
\end{Proof}
Eq. (\ref{eq:obspermsymmetry}) together with the decomposition
(\ref{eq:obsdecompor21}) gives
\begin{align}\label{eq:obssimplifiedr}
  &\sigma_z r_{a,bc}^{\alpha} \sigma_z = r_{a,cb}^{\alpha} \quad
  r_{a,bc}^{\beta} = r_{a,cb}^{\beta} \quad
  r_{a,bc}^{\gamma}=r_{a,cb}^{\gamma}
\end{align}

where $\sigma_z = \begin{pmatrix}1 & 0 \\0& -1\end{pmatrix}$ and we
used the property (\ref{eq:groupcambiasegnoswap}).

Considering that $n(x,xx)=d$, $n(x,xy)=n(x,yx)=n(x,yy)=d(d-1)$, and
$n(x,yz)=d(d-1)(d-2)$, and that $\mathsf SR_{x,xy}\mathsf S=R_{x,yx}$, the figure of
merit in Eq.~\eqref{eq:finfigm} can be written as
\begin{align}\label{eq:obsfigmer21redux}
  F =&\<R_{x,xx}\>+(d-1)\<R_{x,yy}\> + 2(d-1)\<R_{x,xy}\> + \nonumber\\
  &(d-1)(d-2) \<R_{x,yz}\> =\nonumber \\
  & = \sum_{\nu} \Tr[\Delta_{x,xx}^\nu r_{x,xx}^\nu +(d-1)\Delta_{x,yy}^\nu r_{x,yy}^\nu +\nonumber\\
  & 2(d-1) \Delta_{x,xy}^\nu r_{x,xy}^\nu +(d-1)(d-2)
  \Delta_{x,yz}^\nu r_{x,yz}^\nu]
\end{align}
where
\begin{equation}
  \Delta_{a,bc}^\nu := \Tr_{\hilb{H}_\nu}[\ketbra{ijk}{ijk}],
\end{equation}
and $i,jk$ is any triple of indices in the class denoted by $a,bc$.
Notice that in the case $d=2$ the last term in the sum of
Eq.~\eqref{eq:obsfigmer21redux} is 0. In particular, by direct
calculation we have
\begin{align} \label{eq:obsdeltas}
  &\Delta_{x,xx}^\alpha=
  \begin{pmatrix}
    \frac{2}{d+1}& 0 \nonumber\\
    0 & 0
  \end{pmatrix},
 \quad\Delta_{x,xy}^\alpha= \frac12
  \begin{pmatrix}
    \frac{1}{d+1} & \frac{1}{\sqrt{d^2-1}} \nonumber\\
    \frac{1}{\sqrt{d^2-1}} & \frac{1}{d-1}
  \end{pmatrix},
 \quad\Delta_{x,yx}^\alpha= 
\sigma_z\Delta_{x,xy}^\alpha \sigma_z
\nonumber\\
  &\Delta_{x,yy}^\alpha= \Delta_{x,yz}^\alpha= 0,\nonumber\\
  &\Delta_{x,xx}^\beta = \frac{d-1}{d+1},\quad \Delta_{x,xy}^\beta =
  \frac{d}{2(d+1)},\nonumber\\
  & \Delta_{x,yy}^\beta = 1,\quad \Delta_{x,yz}^\beta = \frac12, \nonumber\\
  &\Delta_{x,xx}^\gamma =   \Delta_{x,yy}^\gamma =  0, \quad
  \Delta_{x,xy}^\gamma = \frac{d-2}{2(d-1)},\quad
  \Delta_{x,yz}^\gamma = \frac12.
\end{align}
The commutation relation  (\ref{eq:obscommutr}) implies
$  [I_4 \otimes  T_{20}, U^*_4 \otimes U_2 \otimes U_0] = 0$
and taking the trace on  $\hilb{H}_4$ we get
\begin{align}\label{eq:obscommutt}
  [T_{20}, U_0 \otimes U_2] = 0,
\end{align}
which by theorem \ref{th:characterizationcommutant} and the
decomposition (\ref{eq:UUdecomposition}) implies $T_{20} = t_+P^+ +
t_-P^- $. The normalization $\Tr_{20}[T]=1$ becomes $d_+t_++d_-t_- =
1$ and Eq.  (\ref{eq:obsnorminner}) becomes
\begin{align}
\sum_{(a,bc)\in \defset L}&\frac{n(a,bc)}{d^2} \left(r_{a,bc}^\alpha \otimes P^\alpha + r_{a,bc}^\beta P^\beta + r_{a,bc}^\gamma P^\gamma\right) = \nonumber \\
  &I_4 \otimes (t_+P^+ + t_-P^-) =\nonumber\\
  &t_+(\ketbra ++\otimes P^{\alpha} + P^\beta) + t_-(\ketbra --\otimes
  P^{\alpha} + P^\gamma),
\end{align}
independently of $j,k$. This in turn implies that
\begin{align}
  &t_+ = \sum_{(a,bc)\in \defset L}\frac{n(a,bc)}{d^2}\bra +r_{a,bc}^{\alpha}\ket += \sum_{(a,bc)\in  \defset L}\frac{n(a,bc)}{d^2}r_{a,bc}^{\beta} \nonumber\\
  &t_- = \sum_{(a,bc)\in \defset L} \frac{n(a,bc)}{d^2}\bra
  -r_{a,bc}^{\alpha}\ket-
  =\sum_{(a,bc)\in \defset L}\frac{n(a,bc)}{d^2}r_{a,bc}^{\gamma} \nonumber \\
  &0 = \sum_{(a,bc)\in \defset L}
  \frac{n(a,bc)}{d^2}
\bra{\pm}
r_{a,bc}^{\alpha}
\ket{\mp}
  \label{eq:obsnorm21final}
\end{align}
where we exploited the decomposition (\ref{eq:UUU*decomposition}). Let
us now introduce the notation
\begin{align}\label{eq:obs21defnizdirridot}
  &s_{x,xx}^{\nu} := r_{x,xx}^{\nu}&&s_{x,xy}^{\nu} :=  (d-1)r_{x,xy}^{\nu} \nonumber\\
  &s_{x,yx}^{\nu} := (d-1)r_{x,yx}^{\nu} & &s_{x,yy}^{\nu} :=
  (d-1)r_{x,yy}^{\nu}\nonumber\\
  & s_{x,yz}^{\nu} := (d-2)(d-1)r_{x,yz}^{\nu}.
\end{align}
Exploiting Eq. (\ref{eq:obslear21perminv}) and Eq.
(\ref{eq:obssimplifiedr}) the constraint (\ref{eq:obsnorm21final})
becomes
\begin{align}\label{eq:obsnorm21ultrafinal}
  & s_{x,xx}^{\alpha} + s_{x,yy}^{\alpha} =
  \begin{pmatrix}
    t_+ & 0 \\
    0 & t_-
  \end{pmatrix}
  \nonumber\\
  &s_{x,xx}^{\beta} +   s_{x,yy}^{\beta} = t_+\nonumber\\
  &s_{x,xx}^{\gamma} +   s_{x,yy}^{\gamma} = t_-  \nonumber \\
  & s_{x,xy}^{\alpha}+ \sigma_zs_{x,xy}^{\alpha}\sigma_z + s_{x,yz}^{\alpha} =
  \begin{pmatrix}
    (d-1)t_+ & 0 \\
    0 & (d-1)t_-
  \end{pmatrix}
  \nonumber \\
  &2s_{x,xy}^{\beta}+ s_{x,yz}^{\beta} = (d-1)t_+ \nonumber \\
  &2s_{x,xy}^{\gamma}+s_{x,yz}^{\gamma} = (d-1)t_-
\end{align}
and the figure of merit (\ref{eq:obsfigmer21redux}) becomes
\begin{align}\label{eq:obsfigmer21ultraridotta}
  F = \sum_{\nu}\sum_{(a,bc)\in \defset L} \Tr[\Delta_{a,bc}^\nu s_{a,bc}^\nu]
\end{align}
We are now ready to derive the optimal learning network;
we will proceed as follows: i) first we will maximize the value of $F$
for a fixed value of $t_+$ (remember that $t_- = (1-d_+t_+ )/d_-$) and then ii) we will find
the value of $t_+$ that maximize $F$.
The figure of merit can be rewritten as:
\begin{align}
  F = F_\alpha + F_\beta + F_\gamma
\end{align}
where
\begin{align}
  F_\nu =& \sum_{(a,bc)\in \defset L}\Tr[\Delta_{a,bc}^\nu s_{a,bc}^\nu].
\end{align}
We now maximize $F_\beta$ and $F_\gamma$
for the case $d \geq 3$.
Reminding the expressions (\ref{eq:obsdeltas}) for the $\Delta_{i,jk}^\nu$ we have:
\begin{align}\label{eq:obsboundfbeta}
  F_\beta =& \sum_{(a,bc)\in \defset L}\Tr[ \Delta_{a,bc}^\beta s_{a,bc}^\beta]\leq \nonumber\\
  &\max(\Delta_{x,xx}^\beta, \Delta_{x,yy}^\beta)t_+ +
  \max(\Delta_{x,xy}^\beta, \Delta_{x,yz}^\beta)(d-1)t_+ =\nonumber \\
  &\Delta_{x,yy}^\beta t_+ + \Delta_{x,yz}^\beta(d-1)t_+= \nonumber \\
  & t_+ + \frac{d-1}{2}t_+= \frac{d+1}{2} t_+
\end{align}
and
\begin{align}\label{eq:obsboundfgamma}
  F_\gamma =& \sum_{(a,bc)\in \defset L}\Tr[ \Delta_{a,bc}^\gamma s_{a,bc}^\gamma]\leq\nonumber\\
  &\max(\Delta_{x,xx}^\gamma, \Delta_{x,yy}^\gamma)t_- +
  \max(\Delta_{x,xy}^\gamma, \Delta_{x,yz}^\gamma)(d-1)t_- =\nonumber\\
  &\Delta_{x,yz}^\gamma \frac{d-1}{2} t_- = \frac{d-1}{2} t_- .
\end{align}
where we used the normalizations constraints~(\ref{eq:obsfigmer21ultraridotta}).
The upper bounds (\ref{eq:obsboundfbeta}) and ~(\ref{eq:obsboundfgamma})
can be achieved by taking
\begin{align}
&s_{x,xx}^{\beta}=s_{x,xy}^{\beta}=s_{x,yx}^{\beta}=s_{x,xx}^{\gamma}=s_{x,xy}^{\gamma}=s_{x,yx}^{\gamma}=0, \nonumber \\
&s_{x,yy}^{\beta}=t_+, \quad s_{x,yz}^{\beta}=(d-1)t_+, \nonumber \\
&s_{x,yy}^{\gamma}=t_-, \quad s_{x,yz}^{\gamma}=(d-1)t_-. \nonumber
\end{align}
For $d=2$ the irreducible representation denoted by $\gamma$
and the $x,yz$ class do not exist and the optimization
yields $s_{x,xy}^\beta = \frac{d-1}{2}t_+ $.

Let us now consider $F_\alpha$
(in this case there is no difference between $d\geq 3$ and $d=2$);
reminding the expression of the $\Delta_{i,jk}^\alpha$ we have:
\begin{align}
  F_\alpha =& \sum_{(a,bc)\in \defset L}\Tr[\Delta_{a,bc}^\alpha s_{a,bc}^\alpha]= \nonumber \\
  & \Tr[\Delta_{x,xx}^\alpha s_{x,xx}^\alpha] + \Tr[
  2\Delta_{x,xy}^\alpha s_{x,xy}^\alpha] =
  \nonumber \\
  & \Tr\left[ \begin{pmatrix}
      \frac2{d+1} & 0 \\
      0 & 0
  \end{pmatrix}
s_{x,xx}^\alpha +
\begin{pmatrix}
    \frac{1}{d+1} &\frac{1}{\sqrt{d^2-1}}  \\
    \frac{1}{\sqrt{d^2-1}} &  \frac{1}{d-1}
  \end{pmatrix}
 s_{x,xy}^\alpha \right] \leq \nonumber \\
&
    \frac2{d+1}
t_+
 +
\Tr \left[
  \begin{pmatrix}
    \frac{1}{d+1} &\frac{1}{\sqrt{d^2-1}}  \\
    \frac{1}{\sqrt{d^2-1}} &  \frac{1}{d-1}
  \end{pmatrix}
 s_{x,xy}^\alpha \right],
\end{align}
the bound can be achieved by taking
\begin{align}
  s_{x,xx}^\alpha =
  \begin{pmatrix}
    t_+ & 0 \\
    0 & t_-
  \end{pmatrix}.
\end{align}

\begin{figure}[tb]
  \begin{center}
    \includegraphics[width=10cm ]{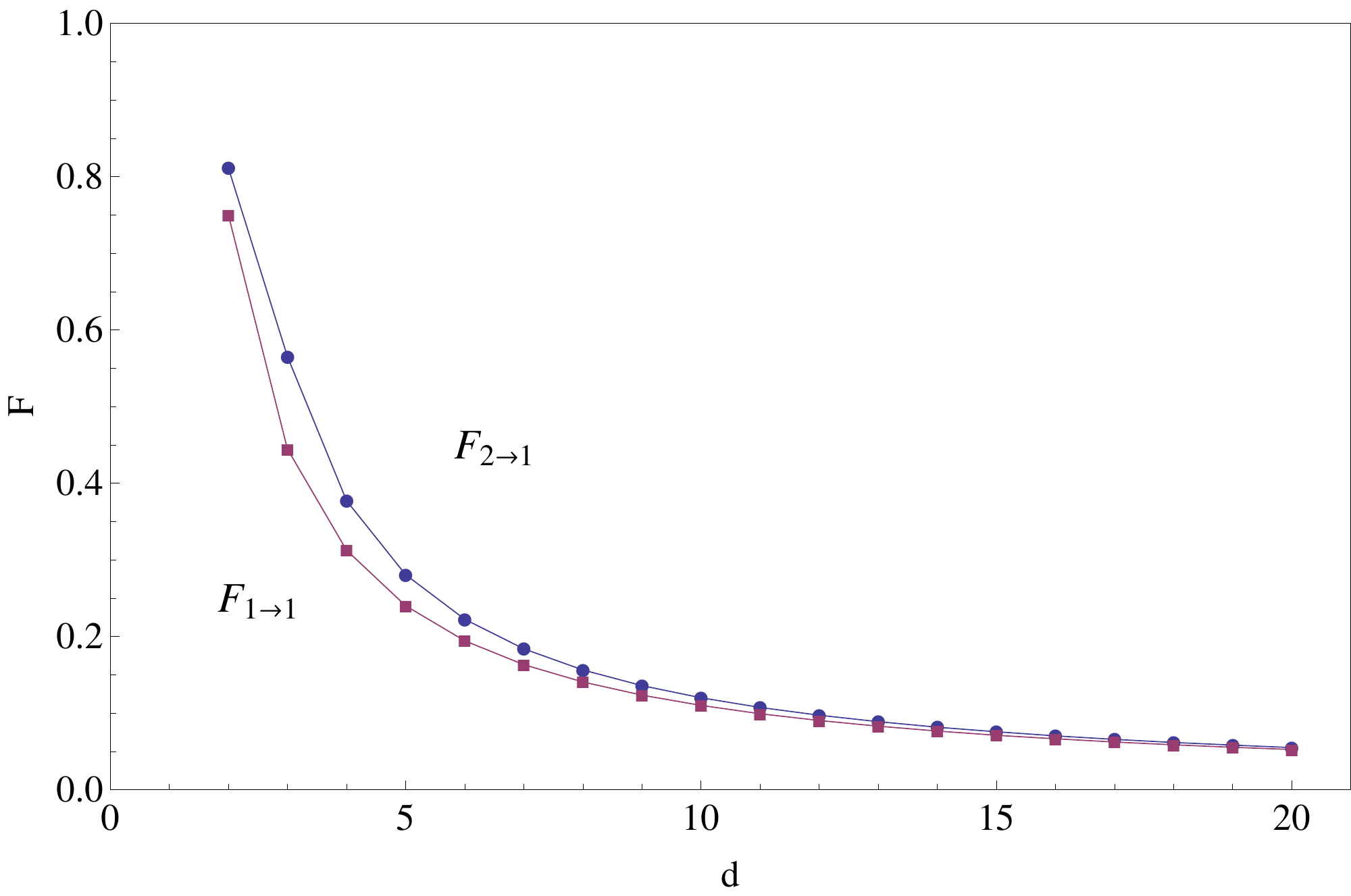}
    \caption{\label{fig:learnobs} Optimal learning of a measurement
      device: we present the value of $F$ for different values of the
      dimension $d$.  The squared dots represent the optimal learning
      from a single use ($1 \rightarrow 1$ learning) while the round
      dots represent the optimal learning from two uses ($2
      \rightarrow 1$ learning).  }
  \end{center}
\end{figure}

Let us now focus on the expression $\Tr[  \Delta_{x,xy}^\alpha s_{x,xy}^\alpha]$.
The normalization constraint (\ref{eq:obsnorm21ultrafinal})
  for the operator $s_{x,xy}^\alpha$ can be rewritten as:
\begin{align}
  &\begin{pmatrix}
   2s_{x,xy}^{\alpha,+,+}+s_{x,yz}^{\alpha,+,+}  &s_{x,yz}^{\alpha,+,-}\\
    s_{x,yz}^{\alpha,-,+}& 2s_{x,xy}^{\alpha,-,-}+s_{x,yz}^{\alpha,-,-}
 \end{pmatrix}=(d-1)
 \begin{pmatrix}
   t_+  & 0 \\
   0 & t_-
 \end{pmatrix}
\end{align}
which implies
\begin{align}\label{eq:obs21normforalpha}
  &  s_{x,yz}^{\alpha,+,-} = s_{x,yz}^{\alpha,-,+} = 0 \nonumber \\
  & s_{x,yz}^{\alpha,+,+}+  2s_{x,xy}^{\alpha,+,+} =    (d-1)t_+ \nonumber \\
  & s_{x,yz}^{\alpha,-,-}+ 2s_{x,xy}^{\alpha,-,-} = (d-1)t_-.
\end{align}
Then we have
\begin{align}
  \Tr[ \Delta_{x,xy}^\alpha s_{x,xy}^\alpha] &=
  \frac{s_{x,xy}^{\alpha,+,+}}{d+1} +
  \frac{s_{x,xy}^{\alpha,+,-}}{\sqrt{d^2-1}}+\nonumber\\
  &\frac{s_{x,xy}^{\alpha,-,+}}{\sqrt{d^2-1}}+
  \frac{s_{x,xy}^{\alpha,-,-}}{d-1} \leq  \nonumber \\
  &\frac{ s_{x,xy}^{\alpha,+,+}}{d+1}+
  2\frac{\sqrt{s_{x,xy}^{\alpha,+,+}
      s_{x,xy}^{\alpha,-,-}}}{\sqrt{d^2-1}} +
  \frac{s_{x,xy}^{\alpha,-,-}}{d-1} \leq   \label{eq:obs21usolapos} \\
  &\frac{(d-1)t_+}{2(d+1)} +
  \frac{\sqrt{(d-1)t_+t_-}}{\sqrt{d+1}}+ \frac{t_-}{2}
  \label{eq:obs21usoilconst}
\end{align}
where we used the positivity of the operator $s_{x,xy}^\alpha$ for the
inequality (\ref{eq:obs21usolapos}) and the normalization
(\ref{eq:obs21normforalpha}) for the second inequality
(\ref{eq:obs21usoilconst}). The upper bound in Eq.
(\ref{eq:obs21usoilconst}) can be achieved by taking
\begin{align}
  s_{x,xy}^\alpha =
  \frac{(d-1)}2   \begin{pmatrix}
    t_+ & \sqrt{t_+t_-}\\
    \sqrt{t_+t_-} & t_-
  \end{pmatrix}
\end{align}
We can now write the figure of merit as:
\begin{align}\label{eq:obs21figmerasfunctionoft}
  F &= F_\alpha + F_\beta + F_\gamma = \nonumber \\
  &= \frac{(d-1)t_+}{2(d+1)} +
  \frac{\sqrt{(d-1)t_+t_-}}{\sqrt{d+1}}+ \frac{t_-}{2}+ 
\frac{d+1}{2} t_++ 
\frac{d-1}{2} t_- 
= \nonumber \\
&=
\frac{d^2 + 3d}{2(d+1)}t_+ +   \frac{\sqrt{(d-1)t_+t_-}}{\sqrt{d+1}} + \frac{d}2 t_-
\end{align}
The last step of the optimization can be easily done by making the
substitution $t_- = d_-^{-1}(1-d_+t_+)$ in Eq.
(\ref{eq:obs21figmerasfunctionoft}) and then maximizing $F = F(t_+)$.
We will omit the details of the derivation and we rather show a plot
(Fig.~\ref{fig:learnobs}) representing the value of $F$ depending on
the dimension
With the optimal learning network
the replicated POVM has the following form:
\begin{align}
  G_i^{(U)} = \frac{dF-1}{d-1}U \ketbra{i}{i}U^\dagger +\frac{1-F}{d-1}
\end{align}
\subsubsection{$3 \rightarrow 1$ case}
In this section we consider a learning network
exploiting $3$ uses of the measurement device
and  produces a  single replica:
\begin{align}\label{eq:obslera31generalschem}
  \begin{aligned}
    \Qcircuit @C=0.5em @R=1em {
      &&&&&&&&&&& \ustick{6}& \ghost{\;\;\;}\\
      \multiprepareC{1}{\;\;\;} &\ustick{0} \qw & \measureD{E^{(U)}} &
      \ustick{1} \cw &\pureghost{\;\;\;} \cw & \ustick{2} \qw &
      \measureD{E^{(U)}} & \ustick{3} \cw &\pureghost{\;\;\;} \cw &
      \ustick{4} \qw &
      \measureD{E^{(U)}} & \ustick{5} \cw &\pureghost{\;\;\;} \cw  \\
      \pureghost{\;\;\;}& \qw & \qw&\qw &\multigate{-1}{\;\;\;}&\qw &
      \qw&\qw &\multigate{-1}{\;\;\;}&\qw & \qw&\qw &
      \multigate{-2}{\;\;\;}}
  \end{aligned}\ .
\end{align}
In order to simplify the problem we restricy ourselves to the qubit
case, that is we set $d=2$. The derivation of the optimal learning
network turns out to be very cumbersome althogh it follows the same
lines as for the $2 \rightarrow 1$ case. The $3 \rightarrow 1$
scenario deserves interest because the optimal strategy does not allow
for a strategy using the $3$ uses of the measurement device in
parallel.

Let us consider the normalization condition for the generalizd
instrument $\{ R_i \}$:
\begin{align}
  &\sum_{ijkl}\ketbra{jkl}{jkl}_{531} \otimes R'_{i,jkl} = I_{65}
  \otimes S_{43210} \nonumber\\
  &\Tr_{4}[S]= I_3 \otimes T_{210}
  \label{eq:norm3to1}
\end{align}
This implies
\begin{align}
  &\sum_{i} R'_{i,jkl} = I_{6}\otimes
  \bra{kl}S_{43210}\ket{kl}_{31}\quad \forall j,\nonumber\\
  &\bra{kl}\Tr_4[S]\ket{kl}= \bra{l}T\ket{l}_1\quad\forall k.
\end{align}
From the relabeling symmetry $R'_{i,jkl} \hspace*{-0.1cm}=\hspace*{-0.1cm}
R'_{\sigma(i),\sigma(j)\sigma(k)\sigma(l)}$ we have $\bra{kl}S\ket{kl}
\hspace*{-0.1cm}=\hspace*{-0.1cm} \bra{\sigma(k)\sigma(l)}S|\sigma(k)$ $\sigma(l)\rangle$, and consequently
\begin{align}
  \bra{kl}\Tr_4[S]\ket{kl}_{31} = \frac1{d^2}\Tr_{431}[S]=:\widetilde{T}_{20},\quad\forall k,l.
\end{align}
This fact along with Eq.~\eqref{eq:norm3to1} allows us to conclude
that
\begin{align}
  \Tr_4[S] =& \Tr_4\left[\sum_{kl}\ketbra{kl}{kl}_{31}\otimes \bra{kl}S_{43210}\ket{kl} \right] = \nonumber \\
  &\sum_{kl}\ketbra{kl}{kl}_{31}\otimes \widetilde{T}_{20} = I_{31} \otimes \widetilde{T}_{20}
\end{align}
that implies that we can exploit the first two uses in parallel.  We
notice that in general $\bra{kl}S\ket{kl}$ $=
\bra{\sigma(k)\sigma(l)}S\ket{\sigma(k)\sigma(l)}$ does not imply that
$\bra{ kl}S\ket{kl} = \widetilde{S}$ is independent of $k,l$, but only
that $\bra{ kl}S\ket{kl} = \widetilde{S}_{ab}$, where $a,b$ denotes
the equivalence class of the couple $(k,l)$. Consequently, we cannot
in general assume that all the examples can be used in parallel. In
fact, the optimal learning network has the following causal structure
\begin{align}\label{eq:obs31learnseq}
  \begin{aligned}
    \Qcircuit @C=0.7em @R=1em { \multiprepareC{2}{\;\;\;}&\ustick{0}
      \qw&\measureD{E^{(U)}} &\ustick{1}\cw& \pureghost{\;\;\;} \cw&
      &&\ustick{6}&
      \ghost{\;\;\;}\\
      \pureghost{\;\;\;}&\ustick{2}\qw&\measureD{E^{(U)}}
      &\ustick{3}\cw& \pureghost{\;\;\;} \cw& \ustick{4}
      \qw&\measureD{E^{(U)}} &\ustick{5}\cw&
      \pureghost{\;\;\;}\cw\\
      \pureghost{\;\;\;}&\qw&\qw&\qw& \multigate{-2}{\;\;\;}& \qw&\qw
      &\qw& \multigate{-2}{\;\;\;} }
  \end{aligned}\ .
\end{align}
where the state of system $4$ depends on the classical outcome on
system $3$ and $1$. The optimal fidelity achieves the value $F \simeq
0,87$ (we remind that for the $1 \rightarrow 1$ case we had $F = 0,75$
while for the $2 \rightarrow 1$ case we had $F= 0,81$).
\begin{remark}
  One can wonder whether without assuming any symmetry it is possible
  to find a non-symmetric parallel strategy $\{R_i\}$ that achieves
  the optimal value of $F$.  However we remind that for any strategy
  $\{R_i\}$ we can build a symmetric one with the same normalization,
  that is without spoiling the parallelism, and giving the same
  fidelity. Since the optimal symmetric network cannot be parallel,
  we have that any other optimal network has to be sequential as well.
\end{remark}
As we pointed out in Remark \ref{parallelgruppi}
the optimality of the parallel strategy is a common feature of the tasks involving
group transformation. On the other hand, if the set of transformation considered
is covariant under a group representation but does not form a group, the parallelism cannot be proven:
the set of channels  in Eq. (\ref{eq:obsdepolachan})
falls in the latter case.
A similar situation arises in the Grover 
algorithm \cite{Grover},  that can be rephrased as the
estimation  of an unknown unitary from the set  $\{ U_{n} = I- 2\ketbra{n}{n} \}$;
also in this case the unitaries $\{ U_{n}  \}$ 
do  not a group and the optimal algorithm, as it was proved in Ref. \cite{Zalka},
cannot be parallelized.

Quantum channel discrimination is a typical example of a task
in which the optimality of sequential strategies easily arises.
In Ref. \cite{memoryeffects} it was found that discrimination of unitary channels
can be optimally performed in parallel, but as shown in Refs.
\cite{watrousdiscrimi, wangyngdiscri}, there exist examples of non-unitary channels
that can be better discriminate by sequential strategies.

\subsubsection{$1 \rightarrow 2$ case}\label{sec:obslear12}

Our goal in this scenario is to create two replicas of the measurement after it was used once
($1 \rightarrow 2$ learning).
\begin{align}\label{eq:obs12learn}
  \begin{aligned}
    \Qcircuit @C=0.7em @R=1em { & &
      & \ustick{3}& \ghost{\;\;\;}\\
      &&& \ustick{2}& \ghost{\;\;\;}\\
      \multiprepareC{1}{\;\;\;} &\ustick{0} \qw & \measureD{E^{(U)}} & \ustick{1} \cw &\pureghost{\;\;\;} \cw\\
      \pureghost{\;\;\;}& \qw& \qw&\qw & \multigate{-3}{\;\;\;}\\
    }
  \end{aligned}
\end{align}
Using the symmetries we introduced in Section \ref{sec:obssymm-repl-netw}
we have
\begin{align}
  &\defset L=\{(xx,x),(xx,y),(xy,x),(xy,y),(xy,z)\}\nonumber\\
  &  R_{ij}= \sum_k \ketbra{k}{k}_1 \otimes R'_{ij,k}  \label{eq:obslear12defr}\\
  & [R'_{ij,k}, U_3 \otimes U_2 \otimes U^*_0]=0  \label{eq:obslear12coomr}\\
  &R'_{i,jk}= \left\{\begin{array}{lcl}
      R_{xx,x} & {\rm if}& i=j=k \\
      R_{xx,y} & {\rm if}& i=j\neq k \\
      R_{xy,x} & {\rm if}& i=k\neq j \\
      R_{xy,y} & {\rm if}& j=k\neq i \\
      R_{xy,z} & {\rm if}& i\neq j \neq k \neq i.
    \end{array}
  \right.
\end{align}
and the figure of merit becomes
\begin{align}
  F =\frac1{d^2} \sum_{(ab,c)\in \defset L}n(ab,c)\<R_{ab,c}\>
\end{align}
The commutation relations of $R_{ab,c}$ with $U_{3}\otimes
U_{2}\otimes U^{*}_{0} $ is very similar to the one in Eq.
(\ref{eq:obsler21comr}) for the $2 \rightarrow 1$ case, because
$U^*\otimes U\otimes U$ has same invariant subspaces as
$U\otimes U^{*}\otimes U^*$.  This enables us to write
\begin{align}\label{eq:obsdecompor12}
  R_{ab,c} = P^\alpha \otimes r_{ab,c}^\alpha + P^\beta r_{ab,c}^\beta
  + P^\gamma r_{ab,c}^\gamma
\end{align}
The following lemma introduces an additional symmetry property of the
generalized instrument $\{ R_{ij} \}$.
\begin{lemma}\label{lem:obs12perminv2}
  The operators $R_{ab,c}$ in Eq. (\ref{eq:obsdecompor12}) can be chosen to be satisfy
\begin{align}\label{eq:obs12permsymmetry}
  R_{ab,c} = \mathsf{S}R_{ba,c}\mathsf{S} \quad \forall a,b,c
\end{align}
where $\mathsf{S}$ is the swap operator $\mathsf{S}\ket{k}_2\ket{j}_3 = \ket{j}_2\ket{k}_3$.
\end{lemma}
\begin{Proof}
See Lemma \ref{lem:obs21perminv1}.
  \qed
\end{Proof}
\begin{remark}
The symmetry  (\ref{eq:obs12permsymmetry}) translates
the possibility to exchange the inputs of the two replicas (Hilbert spaces $\hilb{H}_2$
and $\hilb{H}_3$) together with exchanging the measurement outcomes corresponding to these two replicas.
\end{remark}
Inserting the decomposition (\ref{eq:obsdecompor12}) into Eq.
(\ref{eq:obs12permsymmetry}) and reminding Eq.
(\ref{eq:groupcambiasegnoswap}) we have
\begin{align}\label{eq:obs12permsymbasta}
  r_{ab,c}^\nu &=  r_{ba,c}^\nu \qquad \mbox{if } \nu = \beta ,\gamma  \nonumber \\
  r_{ab,c}^\alpha &= \sigma_z r_{ba,c}^\alpha \sigma_z \
\end{align}
Let us now consider the normalization constraint for the generalized
instrument $R_{ij}$; since $\sum_{i,j}R_{ij}$ has to be a
deterministic network we have
\begin{align}
  \sum_{ij} R_{ij}= I_{321}\otimes \rho_0,\quad \Tr[\rho]=1 \label{eq:obsnormL12}
\end{align}
where $\rho$ has to be a positive operator. The commutation relation
(\ref{eq:obslear12coomr}) implies $[\rho,U]=0$ and so we have
$\rho=\frac{1}{d}I_0$.  Writing $I_{3210}$ as $\sum_k\ket{k}\bra{k}_1
\otimes ( I_{m_\alpha}\otimes P^{\alpha}+P^{\beta}+P^{\gamma})$ we can
rewrite the normalization conditions as follows
\begin{align}
\sum_{ij} {R'}_{ij,k} ^{\nu} =
 \frac{1}{d} I_{m_{\nu}} .
\label{eq:obsnormL21final}
\end{align}
If we  use the following definitions
\begin{align}\label{eq:obs12defnizdirridottiss}
  \begin{array}{ll}
    s_{xx,x}^{\nu} := r_{xx,x}^{\nu}
    &s_{xx,y}^{\nu} := (d-1) r_{xx,y}^{\nu} \\
    s_{xy,x}^{\nu} :=  (d-1)r_{xy,x}^{\nu}  &
     s_{xy,z}^{\nu} :=  (d-1)(d-2)r_{xy,z}^{\nu} 
  \end{array}
\end{align}
the normalization becomes
\begin{align}
  s_{xx,x}^{\nu} + s_{xx,y}^{\nu} + 2 s_{xy,x}^{\nu} +  s_{xy,z}^{\nu} = \frac1d, 
\quad \mbox{if }\nu = \beta , \gamma \nonumber \\
    s_{xx,x}^{\alpha} + s_{xx,y}^{\alpha} + s_{xy,x}^{\alpha} + 
\sigma_z s_{xy,x}^{\alpha} \sigma_z
 +  s_{xy,z}^{\alpha} = \frac1d I_{m_\alpha} \label{eq:obslear12normcons}
\end{align}
where we used the relabeling symmetry.
Let us now express the figure of merit in terms of the
$s_{ab,c}^\nu$:
\begin{align}
  F =& F_\alpha + F_\beta + F_\gamma \label{eq:obs12fnu} \\
  F_\nu =\frac1d& \Tr[\Delta^\nu_{xx,x}s_{xx,x}^{\nu} + 
2\Delta^\nu_{xy,x}s_{xy,x}^{\nu}+
\Delta^\nu_{xx,y}s_{xx,y}^{\nu}+
\Delta^\nu_{xy,z}s_{xy,z}^{\nu}]
] \nonumber
\end{align}
where $\Delta_{ab,c}^\nu$ are the same as the $\Delta_{a,bc}^\nu$ in Eq.  (\ref{eq:obsdeltas})
taking into account the change of Hilbert space labelling from
$\hilb{H}_0, \hilb{H}_2, \hilb{H}_4$ to $\hilb{H}_2, \hilb{H}_3, \hilb{H}_0$.
The maximization of $F_\beta$ and $F_\gamma$
is  simple and yelds
\begin{align}
&  F_\beta = \frac1{d^2} \qquad
 F_\gamma = \frac1{2d^2}\\
 &s_{xx,x}^{\beta}=s_{xy,x}^{\beta}=s_{xy,y}^{\beta}=s_{xy,z}^{\beta}=0 \nonumber \\
 &s_{xx,x}^{\gamma}=s_{xx,y}^{\gamma}=s_{xy,x}^{\gamma}=s_{xy,y}^{\gamma}=0, \nonumber \\
 &s_{xx,y}^{\beta}=s_{xy,z}^{\gamma}=\frac{1}{d^2}. \nonumber \\
\end{align}
Let us now consider the maximization of $F_\alpha$.
Inserting the explicit expression of the $\Delta_{ab,c}^\alpha$ into
Eq. (\ref{eq:obs12fnu}) we have
\begin{align}
 d F_\alpha =& \Tr \left[\begin{pmatrix} s_{xx,x}^{\alpha, +, +} & s_{xx,x}^{\alpha, +, -}\\
  s_{xx,x}^{\alpha, -, +} & s_{xx,x}^{\alpha, -, -}
\end{pmatrix}
\begin{pmatrix}
  \frac2{d+1}&0\\
  0&0
\end{pmatrix}\right]+\nonumber\\
&\Tr\left[\begin{pmatrix}
  s_{xy,x}^{\alpha, +, +} & s_{xy,x}^{\alpha, +, -}\\
  s_{xy,x}^{\alpha, -, +} & s_{xy,x}^{\alpha, -, -}
\end{pmatrix}
\begin{pmatrix}
  \frac1{d+1}&\frac1{\sqrt{d^2-1}}\\
  \frac1{\sqrt{d^2-1}}&\frac1{d-1}
\end{pmatrix}
\right] = \nonumber \\
&= \frac{2s_{xx,x}^{\alpha, +, +}}{d+1}+
\frac{s_{xy,x}^{\alpha, +, +}}{d+1}+\frac{s_{xy,x}^{\alpha, -, -}}{d-1}+\frac{2s_{xy,x}^{\alpha, +, -}}{\sqrt{d^2-1}} \leq \nonumber \\
&\leq\frac2{(d+1)}\left( \frac1d - 2s_{xy,x}^{\alpha, +, +}  \right)+
\frac{s_{xy,x}^{\alpha, +, +}}{d+1}+\frac{s_{xy,x}^{\alpha, -, -}}{d-1}+
2\sqrt{\frac{s_{xy,x}^{\alpha, +, +}s_{xy,x}^{\alpha, -, -}}{d^2-1}}\leq \nonumber \\
& \leq
\frac{5d-3}{2d(d^2-1)}
-\frac{3s_{xy,x}^{\alpha, +, +}}{d+1}+
2\sqrt{\frac{s_{xy,x}^{\alpha, +, +}}{2d(d^2-1)}}
\label{eq:obs12upperboundFa}
\end{align}
where in the derivation of the bound (\ref{eq:obs12upperboundFa}) we
used the positivity of $s_{xy,x}^{\alpha}$ and the constraints  (\ref{eq:obslear12normcons}).
   The upper bound
(\ref{eq:obs12upperboundFa}) can be achieved by taking
\begin{align}
  &s_{xx,x}^\alpha=
  \begin{pmatrix}
    \frac1{d} - 2a &0\\
    0&0
  \end{pmatrix}\quad
  s_{xy,x}^\alpha=
  \begin{pmatrix}
    a&\sqrt{ \frac{1}{2d} a}\\
    \sqrt{ \frac{1}{2d} a}&\frac{1}{2d}
  \end{pmatrix}, \nonumber\\
  & s_{xy,z}^\alpha=s_{xx,y}^\alpha=0
\end{align}
where we defined $a := s_{xy,x}^{\alpha,+,+}$.
Eq. (\ref{eq:obs12upperboundFa}) gives the value of $F_\alpha$
as a function of $a$; the maximization of $F_\alpha(a)$ with the constraint
$0\leq a \leq \frac1{d} $ is easy and gives
\begin{align}
  F_\alpha = 
 \frac{4(2d-1)}{3d^2(d^2-1)} \qquad \mbox{for }  a = \frac{d+1}{18d(d-1)} .
\end{align}
and then for $d \geq 3$ we have
\begin{align}\label{eq:maxfidelear12}
  F = F_\alpha + F_\beta + F_\gamma =
  \frac{3d^2+4d+4\sqrt{d^2-1}-3}{2d^2(d^2-1)}\sim\frac3{2d^2} .
\end{align}
For $d=2$ the invariant subspace $\hilb{H}_\gamma$
does not appear and the fidelity becomes $F = F_\alpha +F_\beta = \frac{7+2\sqrt3}{12}$.

In the next section we consider a different scenario which is less
restrictive than the learning scheme we have considered up to now.
Similarly to what we had when comparing the optimal cloning and the
optimal learning of a unitary, relaxing the constraints of the network
allows to achieve better pefomances

\subsection{Optimal cloning}\label{clon}

In this section we turn our attention to the cloning scenario. As we
previously discussed, this scheme is less restrictive than the learning
one, since we allow both the $M$ states to be measured and the $N$
uses of the measurement device to be available at the same time.

We consider the case in which we are provided with a single use of the
measurement device and we want to produce two replicas:
\begin{align}
  \begin{aligned}
    \Qcircuit @C=1em @R=1em {
      &\ustick{0}&\multigate{1}{\;\;\;}&\ustick{2}\qw&\measureD{E^{U}}& \ustick{3} \cw & \pureghost{\;\;\;} \cw& \\
      &\ustick{1}&\ghost{\;\;\;} &\qw &\qw &\qw
      &\multigate{-1}{\;\;\;}& }
  \end{aligned}
\end{align}
We can require for the optimal $1 \rightarrow 2$ cloning network the
same symmetries we had for the $1 \rightarrow 2$ learning network. The
set $\defset L$ in this case is $\defset
L=\{(x,xx),(x,xy),(x,yx),(x,yy),(x,yz)\}$. Then the figure of merit
becomes
\begin{align}\label{eq:obsclonfigmerfinall}
  F = & \sum_\nu \sum_{(a,bc)\in \defset L}\Tr[\Delta_{a,bc}^\nu
  s_{a,bc}^\nu]
\end{align}
where $\Delta_{a,bc}^\nu$, $ s_{a,bc}^\nu$ are the same as in section
\ref{sec:obslear12}. The normalization condition for the $1
\rightarrow 2$ cloning scenario is different from the $1 \rightarrow
2$ learning. Instead of Eq.  (\ref{eq:obsnormL21final}) we have
\begin{align}
  \sum_{i,jk} \ketbra{i}{i}_3 \otimes R'_{i,jk} = I_3 \otimes S_{210}
  \qquad \Tr_2[S] = I_{10}
\end{align}
which implies the following
\begin{align}
  I_{10}=&d\Tr_2[R_{x,xx}+R_{x,yy}]+\nonumber\\
  &d(d-1)\Tr_2[R_{x,xy}+R_{x,yx}+R_{x,yz}].
  \label{eq:normR}
\end{align}
From the commutation $[R_{a,bc},U^*_2\otimes U_1\otimes U_0]$ it
follows that $[\Tr_2[R_{a,bc}], U_1\otimes U_0]$ and then, exploiting
the decomposition (\ref{eq:UUdecomposition}) we have
\begin{align}\label{eq:obslearnormconst2}
  t_+ P^+ + t_- P^-=&d\Tr_2[R_{x,xx}+R_{x,yy}]+\nonumber\\
  &d(d-1)\Tr_2[R_{x,xy}+R_{x,yx}+R_{x,yz}].
\end{align}
and finally by Eq.~\eqref{eq:normR} $t_+ = t_-=1$. Exploiting the
decomposition $R_{a,bc} = \sum_\nu P^\nu \otimes r_{a,bc}^\nu$ along
with Eq.  (\ref{eq:groupdecompoprojector}), the normalization
constraint (\ref{eq:obslearnormconst2}) becomes
\begin{align}
  P^\pm=& P^\pm\sum_\nu\sum_{(a,bc)\in \defset L}\Tr_2[P^\nu \otimes s_{a,bc}^\nu]=\nonumber\\
  &\frac1{d_\pm}\sum_{(a,bc)\in \defset L}(d_{\alpha} s_{a,bc}^{\alpha,\pm,\pm} +  d_{\delta_\pm} s_{a,bc}^{\delta_\pm})P^\pm,
\end{align}
where $\delta_+=\beta$ and $\delta_-=\gamma$. Exploiting the
relabeling symmetry (\ref{eq:obsperminvproperty}) and the permutation
symmetry (\ref{eq:obs12permsymmetry}) we have
\begin{align}
  d_+ &= d_\alpha \sum_{(a,bc)\in \defset L}s_{a,bc}^{\alpha,+,+} +
  d_{\beta}\sum_{(a,bc)\in \defset L} s_{a,bc}^{\beta},\label{eq:obsclonnormconst4} \\
  d_- &= d_\alpha \sum_{(a,bc)\in \defset L} s_{a,bc}^{\alpha,-,-} +
  d_{\gamma} \sum_{(a,bc)\in \defset L} s_{a,bc}^{\gamma}. \label{eq:obsclonnormconst3}
\end{align}
If we  introduce the notation
\begin{align}
  & s_{a,bc}^{\beta} :=
  \begin{pmatrix}
    s_{a,bc}^{\beta}&0\\
    0&0
  \end{pmatrix}
\qquad
  s_{a,bc}^{\gamma} :=
  \begin{pmatrix}
    0&0\\
    0&s_{a,bc}^{\gamma}
  \end{pmatrix}
 \nonumber \\
& \Pi^+ =
  \begin{pmatrix}
    1&0\\
    0&0
  \end{pmatrix}
  \qquad \Pi^- =
  \begin{pmatrix}
    0&0\\
    0&1
  \end{pmatrix}
\end{align}
the normalization constraints (\ref{eq:obsclonnormconst4}) and (\ref{eq:obsclonnormconst3})
can be rewritten as
\begin{align}\label{eq:obsclonfinalconst}
  &\Pi^+ \left( \sum_{\nu,(a,bc)\in \defset L} d_\nu s_{a,bc}^{\nu}
  \right) \Pi^+ = d_+ \nonumber\\
  &\Pi^- \left( \sum_{\nu,(a,bc)\in \defset L} d_\nu s_{(a,bc)}^{\nu}
  \right) \Pi^- = d_-.
\end{align}
In order to solve the optimization problem we have to find the set ${
  \bf \mathsf{r} } := \{ r_{\ell}^{\nu}, \ell:= (a,bc) \in \defset{L}, \nu \in
\{\alpha, \beta \gamma \} \}$, $r_{\ell}^{\nu} \in
\mathcal{L}(\mathbb{C}^2), r_{\ell}^{\nu} \geq 0$ subjected to the
constraint (\ref{eq:obsclonfinalconst}) that maximizes the figure of
merit (\ref{eq:obsclonfigmerfinall}); we will denote as ${ \bf
  \mathsf{M} }$ the set of all the ${ \bf \mathsf{r} }$ satisfying Eq.
(\ref{eq:obsclonfinalconst}).  Since the figure of merit
(\ref{eq:obsclonfigmerfinall}) is linear and the set ${ \bf \mathsf{M}
}$ is convex, a trivial result of convex analysis states that the
maximum of a convex function over a convex set is achieved at an
extremal point of the convex set.  We now give two necessary
conditions for a given ${ \bf \mathsf{r} }$ to be an extremal point of
${ \bf \mathsf{M} }$.  Let us start with the following
\begin{Def}[Perturbation]
  Let ${ \bf \mathsf{s} }$ be an element of ${ \bf \mathsf{M} }$.  A
  set of hermitian operators ${ \bf \mathsf{z} }:= \{ z_{\ell}^{\nu}
  \}$ is a \emph{perturbation} of ${ \bf \mathsf{s} }$ if there exists
  $\epsilon \geq 0$ such that
  \begin{align}
    { \bf \mathsf{s} } + h { \bf \mathsf{z} } \in { \bf \mathsf{M} }
    \qquad\forall h \in [-\epsilon, \epsilon ]
  \end{align}
  where we defined ${ \bf \mathsf{s} } + h { \bf \mathsf{z} } := \{
  s_{\ell}^{\nu} + h z_{\ell}^{\nu}|h\in[-\epsilon,\epsilon]\}$.
\end{Def}
By the definition of perturbation it is easy to prove that an element
${ \bf \mathsf{s} }$ of ${ \bf \mathsf{M} }$ is extremal if and only
if it admits only the trivial perturbation $z_{\ell}^{\nu}=0$ $\forall
\ell, \nu $.  We now exploit this definition to prove two necessary
conditions for extremality.
\begin{lemma}\label{lem:secondextremalcond}
Let  ${ \bf \mathsf{s} }$ be an extremal element of ${ \bf \mathsf{M} }$.
Then $s_{\ell}^{\nu}$ has to be rank one for all $\ell, \nu$.
\end{lemma}
\begin{Proof}
  Suppose that there is a $s_{\ell '}^{\nu '} = \left(
   \begin{array}{cc}
     a & b \\
     c & d
   \end{array}
\right)
 \in { \bf \mathsf{s} }$
which is not rank one;
then there exist $\epsilon$ such that
${ \bf \mathsf{z} } := \{0,\dots,0,z_{\ell '}^{\nu '},0, \dots, 0  \}$,
$ z_{\ell '}^{\nu '} =
\left(
   \begin{array}{cc}
     0 & 1 \\
     1 & 0
   \end{array}
\right)$ is an admissible perturbation. \qed
\end{Proof}

This lemma tells us that w.l.o.g.  we can assume the optimal ${ \bf
  \mathsf{s} }$ to be a set of rank one matrices.  Let us now consider
a set ${ \bf \mathsf{s} }$ such that $s_{\ell}^{\nu}$ is rank one for
all $\ell, \nu$; any admissible perturbation ${ \bf \mathsf{z} }$ of
${ \bf \mathsf{s} }$ must satisfy
\begin{align}
& z_{\ell}^{\nu} = c_\ell^\nu s_{\ell}^{\nu} \qquad c_\ell^\nu \in  \mathbb{R} \label{eq:perturbprespos}\\
&
  \Pi^+
\left(
 \sum_{\nu,\ell} d_\nu
c_\ell^\nu s_{\ell}^{\nu}
\right)
\Pi^+ =
 \Pi^-
 \left(
\sum_\nu
 d_\nu
c_\ell^\nu s_{\ell}^{\nu}
 \right)
\Pi^- = 0.
\label{eq:perturbpreservnorm}
\end{align}
where the constraint (\ref{eq:perturbprespos})
is required in order to
have $s_{\ell}^{\nu} + h  z_{\ell}^{\nu} \geq 0$,
 while Eq.  (\ref{eq:perturbpreservnorm})
tells us that
${ \bf \mathsf{s} } + h { \bf \mathsf{z} }$ satisfies
the normalization (\ref{eq:obsclonfinalconst}).
Let us now consider the map
\begin{align}
&f: \mathcal{L}(\mathbb{C}^2) \rightarrow \mathbb{C}^2 \qquad
f(A) :=
\left(
 \begin{array}{c}
\Pi^+   A \Pi^+\\
\Pi^-  A  \Pi^-
 \end{array}
\right)
 \nonumber \\
&f
\left(
 \begin{array}{cc}
 a&b\\
 c&d
 \end{array}
\right)
=
\left(
 \begin{array}{c}
a\\
d
 \end{array}
\right) \nonumber
  \end{align}
exploiting this definition
Eq. (\ref{eq:perturbpreservnorm})
becomes
\begin{align}
 \sum_{\nu, \ell} c_\ell^\nu f(s^\nu_\ell) =
\left(
 \begin{array}{c}
   0 \\
   0
 \end{array}
\right).
\end{align}
Suppose now that
the set $ \overline{{ \bf \mathsf{r}} }$
has $N \geq 3$  elements;
then $\{ f(\overline{r}_\ell^\nu) \}$ is a set
of $N \geq 3$ vectors of $\mathbb{C}^2$
that cannot be linearly independent.
That being so, there exists a set of
coefficients $\{ c_\ell^\nu \}$ such that
$ \sum_{\nu, \ell} c_\ell^\nu f(s^\nu_\ell) =0$
and then
$ z_{\ell}^{\nu} = c_\ell^\nu \overline{s}_{\ell}^{\nu}$
is a perturbation of $ \overline{{ \bf \mathsf{r}} }$.
We have then proved the following lemma
\begin{lemma}\label{lem:seconextremalitycond}
  Let $ { \bf \mathsf{s}} $
be an extremal element of $\defset{M}$.
Then  $ { \bf \mathsf{s}} $ cannot have
more than $2$ elements.
\end{lemma}
Lemma \ref{lem:secondextremalcond} and Lemma
\ref{lem:seconextremalitycond} provide two sufficient conditions for extremality
that allow us to restrict the search
of the optimal $ { \bf \mathsf{s}} $ among the ones that satisfy
\begin{align}
{ \bf \mathsf{s}} = \{s_{\ell'}^{\nu'}, s_{\ell''}^{\nu''}  \}
\qquad \rank(s_{\ell'}^{\nu'})=\rank(s_{\ell''}^{\nu''})=1 \nonumber\\
 \Pi^i
\left(
 \sum_{\nu,\ell} d_\nu
s_{\ell}^{\nu}
\right)
\Pi^i =
 d_i \quad  i = +,-
\end{align}
The set of the admissible $ { \bf \mathsf{s}} $
is small enough to allow us to  compute
the value of $F$ for all the possible cases.
It turns out that the best choice is to take
\begin{align}
{ \bf \mathsf{s}} = \{    s^\alpha_{xx,x},  s^\alpha_{xy,x}  \} \nonumber \\
   s^\alpha_{xx,x} =
\left(
 \begin{array}{cc}
\frac{9d_+ -1}{9d} &0\\
0&0
 \end{array}
\right)
\qquad
   s^\alpha_{xy,x} =
\left(
 \begin{array}{cc}
\frac{1}{9d}&\frac{\sqrt{d_-}}{3d}\\
\frac{\sqrt{d_-}}{3d}&\frac{d_-}{d}
 \end{array}
\right); \nonumber
\end{align}
the corresponding value of $F$ is
\begin{align}\label{eq:obsclonoptimalfide}
  F = \frac{4}{3d}
\end{align}
which is much higher
then the maximum value  (\ref{eq:maxfidelear12})
achieved by the $1 \rightarrow 2$
learning scheme.

\setcounter{equation}{0} \setcounter{figure}{0} \setcounter{table}{0}\newpage
\section{Conclusion}
The aim of this work was
twofold.
The first part was devoted to   present a unified
description of Quantum Networks in terms of their Choi
operators.
The core result of this approach
are Theorem \ref{th:normcond}  and Theorem \ref{th:realinet}
that prove the isomorphism between the set of deterministic Quantum Networks
and a set of suitably normalized positive operators.
This result can be then generalized to probabilistic Networks.
The second key ingredient of the theory
is the notion of link product
(see Definition \ref{def:link}) that allows us to express the composition of quantum 
networks in terms of their Choi operators (Theorem \ref{th:linknet}).

In the second part of the work,
we made use of this formalism to solve some relevant optimization problems.
The  representation of Quantum Networks as positive operators
is extremely efficient in 
handling tasks that involve manipulation of transformations
like  process tomography (Chapter \ref{chaptertomo}) and 
cloning, learning and inversion of transformations
(Chapters \ref{chapter:cloning}, \ref{chapter:learning} and \ref{chapter:inversion}).

Even if the tools provided by the general theory of Quantum Networks
simplify a lot many scenarios,
it is also true that in order to analytically carry on the optimization
we had to make 
 a clever use of the symmetries of the various problems.
The full power of the general theory reveals itself when combined
with the techniques provided by the group representation theory
(Appendix \ref{appendicegruppi}):
this happy marriage lies at the core of the results
achieved in the optimization
problems involving unitary transformations.

However, in many problems in quantum information theory
like for example in channel discrimination \cite{maxdiscrimi, memoryeffects, watrousdiscrimi},
we cannot exploit such  strong symmetry properties;
the general theory of quantum network
is still powerful \cite{memoryeffects}
but the results from group theory cannot be applied.
A possible way out (in some cases the only one)
is the numerical approach.
The set of the admissible  Choi
operators  of  Quantum network with fixed causal structure,
is a convex set of positive operator.
It is then possible to implement
computer routines \cite{CVX, watrousroutine}
that solve the semidefinite program 
corresponding to the optimization problem in exam.

\addcontentsline{toc}{section}{Acknowledgment}
\section*{Acknowledgements}
We thank M. Sedlak for his productive collaboration on Quantum Networks theory,
which provided some of the results reviewed here.
This work is supported by Italian Ministry of Education through grant PRIN 2008 and the EC through project COQUIT. Research at Perimeter Institute for Theoretical Physics is supported in part by the Government of Canada through NSERC and by the Province of Ontario through MRI.



\appendix



\setcounter{equation}{0} \setcounter{figure}{0} \setcounter{table}{0}\newpage
\section{Channel Fidelity}\label{chap:channelfidelity}
This short appendix has the purpose to introduce the 
\emph{channel fidelity} as 
a notion of distance between
quantum channels.
This definition was introduced in \cite{raginskyfide} and discussed in \cite{belavkinrag}
In the following we will review  the definition of channel fidelity
and some of its most relevant properties.

\begin{Def}[Channel Fidelity]\label{def:channelfidelity}
  Let $\mathcal{C}\in\mathcal{L}(\mathcal{L}(\hilb{H})_a,\mathcal{L}(\hilb{H})_b)$
 and $\mathcal{D}\in\mathcal{L}(\mathcal{L}(\hilb{H})_a,\mathcal{L}(\hilb{H})_b)$ be two 
quantum channels
and $C$ and $D$ be their Choi-Jamio\l kowsky operators.
We  call \emph{channel fidelity}
the following expression
\begin{align}\label{eq:channelfidelity}
  \mathcal{F}(\mathcal{C},\mathcal{D}) := f\left(\frac{C}{d_a},\frac{D}{d_a} \right)
\end{align}
where $f$ is the state fidelity $f(\rho, \sigma) := |\Tr[\sqrt{\sigma^{\frac12} \rho \sigma^{\frac12}}]|^2$.
\end{Def}
The channel fidelity enjoys many properties inherited by the state fidelity:
\begin{lemma}[Properties of channel fidelity]
  The channel fidelity $\mathcal{F}$ defined in definition \ref{def:channelfidelity}
enjoys the following properties:
\begin{itemize}

\item $0 \le \mathcal{F}(\mathcal{C},\mathcal{D}) \le 1$, 
and $\mathcal{F}(\mathcal{C},\mathcal{D}) = 1$ if and only if $\mathcal{C}=\mathcal{D}$.

\item $\mathcal{F}(\mathcal{C},\mathcal{D}) = \mathcal{F}(\mathcal{D},\mathcal{C})$ (symmetry).

\item For any two isometric channels $\mathcal{V}$ and $\mathcal{W}$ 
(i.e., $\mathcal{V}(\rho)=V\rho V^\dagger$ and $\mathcal{W}(\rho) = W\rho W^\dagger$
 with isometry $V$ and $W$), $\mathcal{F}(\mathcal{V},\mathcal{W}) = (1/d^2)|\Tr{(U^\dagger V)}|^2$.

\item For any $0 < \lambda < 1$, $\mathcal{F}(\mathcal{C},\lambda \mathcal{D}_1 + (1-\lambda)\mathcal{D}_2) \ge \lambda \mathcal{F}(\mathcal{C},\mathcal{D}_1) + (1-\lambda)\mathcal{F}(\mathcal{C},\mathcal{D}_2)$ (concavity).

\item $\mathcal{F}({\mathcal{C}_1}\otimes{\mathcal{C}_2},{\mathcal{D}_1}\otimes{\mathcal{D}_2}) = \mathcal{F}(\mathcal{C}_1,\mathcal{D}_1)\mathcal{F}(\mathcal{C}_2,\mathcal{D}_2)$ (multiplicativity with respect to tensoring).

\item $\mathcal{F}$ is invariant under composition with unitary  channels, i.e., for any unitary channel $\mathcal{U}$, $\mathcal{F}(\mathcal{U} \star \mathcal{C},\mathcal{U} \star \mathcal{D}) = \mathcal{F}(\mathcal{C},\mathcal{D})$.

\item $\mathcal{F}$ does not decrease under composition with arbitrary channels, i.e., for any channel $\mathcal{R}$,
 $\mathcal{F}(\mathcal{R} \star \mathcal{C}, \mathcal{R} \star \mathcal{D}) \ge \mathcal{F}(\mathcal{C},\mathcal{D})$.

\end{itemize}
\end{lemma}
\begin{Proof}
  See Ref. \cite{raginskyfide}
\qed \end{Proof}
The following lemma provides a physical interpretation of the channel fidelity
$\mathcal{F}(\mathcal{A},\mathcal{B})$
between two channels $\mathcal{A}$ and $\mathcal{B}$
(one of them unitary)
as the fidelity between the output states of $\mathcal{A}$ and $\mathcal{B}$
uniformly averaged over all input pure states.
\begin{lemma}\label{lem:equivchanfidstatefid}
Let $\mathcal{A}\in\mathcal{L}(\mathcal{L}(\hilb{H})_a,\mathcal{L}(\hilb{H})_b)$
 and $\mathcal{B}\in\mathcal{L}(\mathcal{L}(\hilb{H})_a,\mathcal{L}(\hilb{H})_b)$ be two channels
and let us define $d = \dim(\hilb{H}_a)$.
If either $\mathcal{A}$ or $\mathcal{B}$
is a unitary channel we have
\begin{align}\label{eq:equivchanfid}
  \mathsf{F} := \int \d \varphi f(\mathcal{A}(\ketbra{\varphi}{\varphi}), 
\mathcal{B}(\ketbra{\varphi}{\varphi})) 
=
\frac{d}{d+1}\mathcal{F}(\mathcal{A},\mathcal{B})
+
\frac1{d+1}
\end{align}
where $\ket{\varphi}\in \hilb{H}_a$,
$\d \varphi$ is the normalized ($\int \d \varphi = 1$)  Haar measure over the set of pure states
and
 $f$ is the state fidelity.
\end{lemma}
\begin{Proof}
First we notice that   
 we can parametrize each vector $\ket{\psi} \in \hilb{H}_{a}$ as
$U\ket{0}$ where $\ket{0}$ is a fixed vector and $U$ is a unitary operator on $\hilb{H}_a$;
with this parametrization the measure $\d \varphi$
becomes the usual Haar measure $\d U$ of $\group{SU}(d)$.
The left hand side of Eq. (\ref{eq:equivchanfid}) now becomes:
\begin{align}\label{eq:equivchanfid2}
    \mathsf{F} = \int \d U 
f(\mathcal{A}(U\ketbra{0}{0}U^\dagger),\mathcal{B}(U\ketbra{0}{0}U^\dagger)).
\end{align}
Now suppose that $\mathcal{B}$
is a unitary channel $\mathcal{B} = V \cdot V^\dagger$.
Eq. (\ref{eq:equivchanfid2}) becomes:
\begin{align}
\mathsf{F} 
&= 
\int \d U 
f(\mathcal{A}(U\ketbra{0}{0}U^\dagger),\mathcal{B}(U\ketbra{0}{0}U^\dagger)) = \nonumber \\
&= \int \d U 
f(\mathcal{A}(U\ketbra{0}{0}U^\dagger),VU\ketbra{0}{0}U^\dagger V^\dagger)
=\nonumber\\
&=\int \d U 
\bra{0}U^\dagger V^\dagger( \ I \otimes \bra{0} U^T) A ( I \otimes   U^*\ket{0} )VU\ket{0}
=\nonumber\\
&=\Tr\left[ (  V^\dagger \otimes I  ) A (V \otimes I )\left(\int \d U U \otimes U^* 
 (\ket{0}\ket{0}\bra{0}\bra{0}) U^\dagger \bra{0} U^T \right)  \right] 
\end{align}
Reminding Theorem \ref{th:groupaverage}
and the  decomposition (\ref{eq:decompoUU*}) we have
\begin{align}
  \int \d U U \otimes U^*  (\ket{0}\ket{0}\bra{0}\bra{0}) U^\dagger \bra{0} U^T  = \frac1{d(d+1)}\KetBra{I}{I}+
\frac1{d(d+1)}I 
\end{align}
that leads to 
\begin{align}
\mathsf{F} &= 
\frac1{d(d+1)}\Tr\left[  (   V^\dagger \otimes I) A (V \otimes I )\KetBra{I}{I}  \right] 
+
\frac1{d(d+1)}\Tr\left[  (  V^\dagger \otimes I) A (V \otimes I )  \right] = \nonumber \\
&=
\frac1{d(d+1)}\Tr\left[  A \KetBra{V}{V}  \right] 
+
\frac1{d+1}=
 \frac d{d+1}\mathcal{F}(\mathcal{A},\mathcal{B}) + \frac1{d+1}
\end{align}\qed
\end{Proof}

\setcounter{equation}{0} \setcounter{figure}{0} \setcounter{table}{0}\newpage
\section{Elements of Group Representation Theory}
\label{appendicegruppi}

This Appendix is an introduction to the basic tools of group representation theory
that are needed in this work. 
The key results of the appendix are the Schur's lemma \ref{lemmadischur}
and the Theorem \ref{th:characterizationcommutant} that allow us to decompose
an operator that commutes  with a unitary representation of a group.
The last section of this appendix is devoted to the decomposition
of some relevant tensor product representations.
All the results in this appendix are presented without proofs;
a more exhaustive presentation  can be found for example in 
\cite{fultonharris, jones, fulton2, barut}.

\subsection{Basic definitions}

\begin{Def}[Group]
  A group $\group{G}$ is a set of elements with a law of composition
that assigns each ordered couple of elements $g_1,g_2 \in \group{G}$
another element $g_1g_2$ of $\group{G}$.
This composition law has to satisfy the following requirements:
\begin{align}
g_1(g_2g_3) = (g_1g_2)g_3  \quad \forall g_1,g_2,g_3 \in \group{G}\\
\exists e \in \group{G} : ge=eg=g \quad \forall g \in \group{G} \\
\forall g \in \group{G} \exists g^{-1} \in \group{G}: gg^{-1} =g^{-1}g =e.
\end{align}
If $\group{G}$ has a finite number of elements we say that $\group{G}$ is a finite group.
\end{Def}
Typical examples of groups are 
\begin{itemize}
\item $\group{GL}(n,\mathbb{R})$: the set of $n \times n$ real invertible matrices with matrix multiplication;
\item $\group{S}_n$: the group of permutation of $n$ objects
 (the composition is the successive operation of permutations);
\item $\group{U}(1)$: the set $1 \times 1$ unitary matrices with matrix multiplication;
\item $\group{SU}(d)$: the set of $d \times d$ unitary matrices with determinant $1$ with matrix multiplication;
\end{itemize}
A relevant class of groups are Lie groups
\begin{Def}[Lie Group]
A group $\group{G}$ which is a differentiable manifold and such that the maps
\begin{align}
(g_1, g_2)\rightarrow g_1g_2, \qquad g_1 \rightarrow g_1^{-1}
\end{align}
are smooth, is a Lie group. If $\group{G}$ (as a manifold) is compact, we say that
$\group{G}$ is a compact Lie group.
\end{Def}
 $\group{GL}(n, \mathbb{R})$, $\group{U}(1)$,  $\group{SU}(d)$ are Lie groups but only
$\group{U}(1)$ and  $\group{SU}(d)$ are compact. From now on we restrict to the case of finite group 
and compact Lie groups.

\begin{Def}[Unitary Representation]
Let $\group{G}$ be a group and $\hilb{H}$ a Hilbert space.
A unitary representation of $\group{G}$ on  $\hilb{H}$ is a map
$g \rightarrow U_g$ from $\group{G}$ to set of bounded linear operator $\mathcal{B}(\hilb{H})$
such that:
\begin{align}
U_g \mbox{ is unitary } \forall g \in \group{G} \\
U_gU_h = U_{gh} \forall g,h \in \group{G} \\
U_e = I.
\end{align}
\end{Def}

\begin{Def}[Equivalent Representation]\label{equivalentrepresentation}
Let $\{U_g \;|\;g \in \group{G} \}$ be a unitary representation of $\group{G}$ on $\hilb{H}$ and
$\{V_g \;|\;g \in \group{G} \}$be  a  unitary representation of $\group{G}$ on $\hilb{K}$.
We say that $\{ U_g  \}$ is equivalent to $\{ V_g \} $ if there exists
an isomorphism
$T : \hilb{H} \rightarrow \hilb{K}$ such that
\begin{align}
TU_g = V_gT \qquad \forall g \in \group{G} \\
T^\dagger T = I_{\hilb{H}} \qquad TT^\dagger = I_{\hilb{K}}
\end{align}
The isomorphism $T$ is often called intertwiner.
\end{Def}

\begin{remark}
The notion of representation  
 makes a bridge between group theory and quantum physics.
Indeed, the action of a group on an Hilbert space  induces a transformation
on the set of quantum states $\mathcal{S}(\hilb{H})$
\begin{align}
\rho \rightarrow U_g\rho U_g^{\dagger} \qquad \rho \in \mathcal{S}(\hilb{H}).
\end{align}
\end{remark}

\begin{Def}[Invariant Subspace]
Let $\{U_g \;|\;g \in \group{G} \}$ a unitary representation of $\group{G}$ on $\hilb{H}$ and
let $\hilb{K} \subseteq \hilb{H}$,  be a subspace of $\hilb{H}$.
We say that $\hilb{K}$ is invariant with respect to $\group{G}$ if
\begin{align}
U_g(\hilb{K}) \subseteq \hilb{H} \qquad \forall g \in \group{G}
\end{align}
\end{Def}

\begin{Def}[Irreducible Representation]
Let $\{U_g \;|\;g \in \group{G} \}$ a unitary representation of $\group{G}$ on $\hilb{H}$ and
let $\hilb{K} \subseteq \hilb{H}$,  be an invariant subspace.
We say that $\{ U_g \}$ is irreducible in $\hilb{K}$ 
if there exists no proper subspace $\hilb{V}$ of $\hilb{K}$
that is invariant 
 with respect to $\group{G}$.
A subspace carrying an irreducible representation is called irreducible subspace.
\end{Def}

\begin{lemma}[Subrepresentation]
Let  $\{ U_g  \}$ be a unitary representation of  $\group{G}$ on $\hilb{H}$
and $\hilb{K}$ be an invariant subspace of  $\hilb{H}$.
The restriction $\{U_g|_{\hilb{K}}\}$ of $\{ U_g  \}$ on  $\hilb{K}$ is still a representation
and it is called a subrepresentation of $\{ U_g \}$.
\end{lemma}

Finite groups and compact lie groups share a very relevant feature that is called
complete reducibility, that is, any representation can be decomposed 
as a discrete sum of irreducible representations. 

\begin{theorem}[Complete Reducibility] \label{completereducibility}
Let $\group{G}$ be a finite group or a compact Lie group and $\{U_g\}$ a unitary representation
of $\group{G}$ on a Hilbert space $\hilb{H}$. Then there exists a discrete set of irreducible unitary 
subrepresentations $\{ U_g|_{\hilb{H}_k}  \}$ such that
\begin{align}
U_g = \bigoplus_{k}U_g|_{\hilb{H}_k}, \qquad \bigoplus_k \hilb{H}_k = \hilb{H} \label{eq:completereducibility}
\end{align}
\end{theorem}

Let $\{ U_g \}$ be a reducible, as opposed to irreducible, representation of a group $\group{G}$
on a Hilbert space $\hilb{H}$.
Suppose now that there are only two invariant subspaces $\hilb{H}_1$ and $\hilb{H}_2$
($\hilb{H} = \hilb{H}_1 \oplus \hilb{H}_2 $) with dimensions $n$ and $m$ respectively.
Then Theorem \ref{completereducibility} says that for all $g\in\group{G}$, $U_g$ can be written in a block
diagonal form 
\begin{align}
U_g = 
\left(
\begin{array}{cc}
U_g^{(1)} & 0 \\
0  & U_g^{(2)}
\end{array}
\right)
\end{align}
where $U_g^{(1)}$ is a $n \times n$ submatrix 
and $U_g^{(2)}$ is a $m \times m$ submatrix.

It can happen that in the decomposition $U = \bigoplus_kU_k$ (we omit the index of the group element)
$U_k$ is equivalent to $U_l$ for some $k \neq l$;
that being so, it is usual to rewrite the decomposition in this way:
\begin{align}
  U= \bigoplus_{\mu \in \rm{irrepS}(U)}\bigoplus_{i=1}^{m_\mu}U_{\mu,i}
\end{align}
where irrepS$(U)$ represents the set of equivalence classes of 
irreducible representations contained in the decomposition of $(U)$
and $i$ labels different representations in the same class; $m_\mu$ is the number of different equivalent irreducible
representations in the same class and it is called multiplicity.
Likewise we write:
\begin{align}
  \bigoplus_k \hilb{H}_k = \hilb{H}  \qquad  \bigoplus_{\mu \in \rm{irrepS}(U)}\bigoplus_{i=1}^{m_\mu}\hilb{H}_{\mu,i}
\end{align}
There is an isomorphism between the spaces $\bigoplus_{i=1}^{m_\mu}\hilb{H}_{\mu,i}$
and $\hilb{H}_\mu \otimes \mathbb{C}^{m_\mu}$
where $\hilb{H}_\mu$ is an abstract Hilbert space of dimension
 $d_\mu$ ($\rm{dim}(\hilb{H}_{\mu,i}) = \rm{dim}(\hilb{H}_{\mu,j})$ for all $i$ and $j$).
If we denote with $T_{ij}^\mu$ the intertwiner connecting the equivalent representation 
$U_{\mu,i}$ and $U_{\mu,j}$ it can be written in the simple form $T_{ij}^\mu = I_{d_\mu} \otimes \ketbra{i}{j}$
where $\{ \ket{i} \}$ is an o.n.b. for the space $\mathbb{C}^{m_\mu}$ and $I_{d_\mu}$ is
the identity on the abstract space $\hilb{H}_\mu$.
Thanks to this isomorphism it is possible to rewrite the decomposition 
\ref{eq:completereducibility} in this way
\begin{align}
  U_g = 
\bigoplus_{\mu \in \rm{irrepS}(U)}
 U_g^{\mu} \otimes I_{m_\mu},
 \qquad \hilb{H} =\bigoplus_{\mu \in \rm{irrepS}(U)}  \hilb{H}_{\mu} \otimes \mathbb{C}^{m_\mu} 
\label{eq:completereducibility2}
\end{align}
It is customary to call $\hilb{H}_{\mu}$ representation space and 
$\mathbb{C}^{m_\mu}$ multiplicity space.


\subsection{Schur lemma and its applications}
\begin{lemma}[Schur]\label{lemmadischur}
  Let $\{ U_g  \}$ and $\{ V_g \}$ two irreducible representations
of the same  group $\group{G}$ on Hilbert spaces $\hilb{H}$ and $\hilb{K}$ respectively.
Let $O: \hilb{H} \rightarrow \hilb{K}$ an operator  such that
 such that $O U_g = V_g O$ for all $g \in \group{G}$.
If $\{ U_g  \}$ and $\{ V_g \}$
are equivalent then  $O = \lambda T$, where $T$  is the isomorphism defined in Definition \ref{equivalentrepresentation}
and $\lambda \in \mathbb{C}$.
If $\{ U_g  \}$ and $\{ V_g \}$ are not equivalent, then $O=0$
\end{lemma}
The Schur lemma is a powerful tool for inspecting
the structure of operators commuting with a group representation.
\begin{theorem}[Characterization of the Commutant] \label{th:characterizationcommutant}
  Let $\{ U_g \}$ be a unitary representation of a group $\group{G}$
and $O \in \mathcal{B}(\mathcal{H})$ an operator such that
$[O,U_g]=0$ for all $g \in \group{G}$.
Then 
\begin{align}
  O = \bigoplus_{\mu \in \rm{irrepS}(U)}I_{d_{\mu}} \otimes O_\mu
\end{align}
\end{theorem}
A typical example of operator in the commutant of a representation
is the group average of an operator.
Suppose that $\{U_g\}$ is a unitary representation
of a finite group $\group{G}$ on an Hilbert space $\hilb{H}$ and $O \in \mathcal{B}(\hilb{H})$.
Then we can define 
\begin{align}
  \overline{O} = \frac{1}{|\group{G}|} \sum_{g \in\group{G}}U_g O U_g^\dagger \label{eq:averagefinite}
\end{align}
where   $|\group{G}|$ is the cardinality of $\group{G}$.
Eq. (\ref{eq:averagefinite}) can be generalized to the case of Lie groups but
to do this  we need a preliminary definition
\begin{Def}[Invariant measure]\label{def:haarmeasure}
  Let $\group{G}$ be a Lie group. A measure $\mu_L(dg)$  on $\group{G}$ is called left invariant if
$\mu_L(gB) = \mu_L(B)$ for any $g \in \group{G}$ and any region $B \subseteq \group{G}$.
A measure $\mu_r(dg)$  on $\group{G}$ is called right invariant if
$\mu_L(Bg) = \mu_L(B)$ for any $g \in \group{G}$ and any region $B \subseteq \group{G}$.
\end{Def}
Any Lie group can be endowed with a right invariant measure and a left invariant measure.
When this to measures coincide the group is called unimodular; in this work we consider only
unimodular group so we can talk about invariant measure without any misunderstanding.
When the Lie group is compact (as it is always the case in this presentation)
the invariant measure can be normalized in this way
\begin{align}
  \int_{\group{G}}dg = 1.
\end{align}
Now we can define the group average for the case of (compact unimodular) Lie groups:
\begin{align}
  \overline{O} = \int_{\group{G}}dg U_g O U_g^\dagger \label{eq:averagelie}
\end{align}
As a consequence of Theorem \ref{th:characterizationcommutant}
we have
\begin{theorem}[Group average of an operator]\label{th:groupaverage}
Let $\{ U_g \}$ be a unitary representation of a finite (compact) group on an Hilbert space
$\hilb{H}$. Let $O$ be an operator in $\mathcal{B}(\hilb{H})$ and $\overline{O}$
its group average (as defined in Eq. \ref{eq:averagefinite} for the finite case and in 
Eq. \ref{eq:averagelie} for the compact case).
Then we have
\begin{align}
  [\overline{O}, U_g]=0 \qquad \forall g \in \group{G} \\
\overline{O} = \bigoplus I_{d_\mu} \otimes \frac{\Tr_{\hilb{H}_\mu}[P^{\mu} O P^{\mu} ]}{d_\mu}
\end{align}
where $P^{\mu}$ is the projector on $\hilb{H}_\mu \otimes \mathbb{C}^{m_\mu}$
and $\Tr_{\hilb{H}_\mu}$ denotes the partial trace over $\hilb{H}_\mu$.
\end{theorem}

\subsection{Relevant decompositions}
In this section we will give some results about
the decomposition into irreducible representations for 

\subsubsection{The symmetric group $\group{S}_n$}
$\group{S}_n$ is the group of permutation of $n$ objects. 
It can be proved that the number
of inequivalent irreducible representation of $\group{S}_n$ is given by the number of 
partition of $n$.\footnote{A partition of an integer $n$ is a way of writing $n$ as a sum of positive integers.}
It is useful to associate each partition $\nu = (\nu_{\rm i})$ of $n$  with a \emph{Young diagram}.
A Young diagram is a collection of boxed arranged in left aligned rows, the row lengths not increasing 
from the top to the bottom; as an example consider the partition
\begin{align}
  n=11 \qquad \nu = (4,3,3,1)
\end{align}
The corresponding Young diagram has the following shape
\begin{align}
\begin{Young} 
  & & & \cr
&& \cr
&&\cr
 \cr
\end{Young}  
\end{align}
The usefulness of this pictorial representation will be more evident in the following section

\subsubsection{Decomposition of $\group{SU}(d)^{\otimes n}$}
At the beginning of this chapter $\group{SU}(d)$ was defined as the group of $d \times d$ unitary matrices $U$
with determinant equal to $1$. This definition identifies $\group{SU}(d)$ with
its smallest-dimensional faithful irreducible representation:
this representation is usually called
the \emph{defining representation}. .
Then  $\group{SU}(d)^{\otimes n}$ will denote the unitary representation $\{ U^{\otimes n} \}$
over the Hilbert space $\hilb{H}^{\otimes n}$ where $\dim(\hilb{H})=d$.
In this section we will use both $\group{SU}(d)^{\otimes n}$ and $U^{\otimes n}$ with the same meaning.

Let now consider the action of  $\group{S}_n$ on factorized vectors:
\begin{align}
  s\cdot(\ket{\psi}_1 \otimes \dots \otimes \ket{\psi}_n) = 
\ket{\psi}_{s^{-1}(1)} \otimes \dots \otimes \ket{\psi}_{s^{-1}(n)}
\qquad
s \in \group{S}_n;
\end{align}
 this action can be  extended by linearity to the whole $\hilb{H}^{\otimes n}$
leading to a representation of $\group{S}_n$ over $\hilb{H}^{\otimes n}$.
This representation of $\group{S}_n$ commutes with the representation
$\group{SU}(d)^{\otimes n}$ and 
it can be proved\footnote{This result  is  the \emph{Schur-Weyl duality}. The aim of  this section 
is to introduce (without claiming to be rigorous) some consequence of this theorem that  are exploited for proving 
many results of Quantum Information Theory.}
that the irreducible subspaces of these two representations are the same.
Each irreducible representation $U_\nu$ in the decomposition 
$ U^{\otimes n} = \bigoplus_\nu U_\nu \otimes I_{m_\nu}$
is then in correspondence with a Young diagram $\nu$.

From a Young diagram one can obtain a \emph{Young tableaux}
filling the empty boxes with the integers numbers from $1$ to $n$; a \emph{standard Young tableau}
is a tableau in which the numbers in each row grow from left to right and the numbers in each
column grow from top to bottom e.g.
\begin{align}
\begin{Young} 
 1 &2 &5 &7 \cr
3&4&8 \cr
6&9&11\cr
10 \cr
\end{Young}
\qquad  
\begin{Young} 
 1 &3 &4 &7 \cr
2&5&6 \cr
8&10&11\cr
9 \cr
\end{Young}  \nonumber
\end{align}
Given an irreducible representation $U_\nu$ in
 the decomposition of $ \group{SU}(d)^{\otimes n}$, the dimension $m_\nu$ of the corresponding
multiplicity space is given by the number of admissible standard Young tableaux
associated to the Young diagram corresponding to $U_\nu$.

The following combinatorial procedure gives the dimension of $\hilb{H}_\nu$:
\begin{itemize}
\item Given the Young diagram $\nu$  number the rows and the columns with integer numbers
$1,2,\dots,n$ from top to bottom for the rows and $1,2,\dots,m$ from left to right for the columns;
\item
associate  each box $\rm{b}$ of $\nu$ with the 
expression $\frac{l_{\rm{b}}}{h_{\rm{b}}}$ where:

\begin{align}
l_{\rm{b}}=d+j-i  
\end{align}
$d = \dim(\hilb{H})$, $j$ is the column which the box $\rm{b}$ belongs to
and $i$ is the row which $i$ belongs to;

\begin{align}
h_{\rm{b}}=1 + r+s
\end{align}
$r$ is the number of boxes to the right of $\rm{b}$ in the same row and $s$
is the number of boxes below it in the same column;

\item
Finally we have 
\begin{align}
  \dim(\hilb{H}_\nu) = \prod_{\rm{b}} \frac {l_{\rm{b}}}{h_{\rm{b}}}
\end{align}
\end{itemize}
We notice that when the number of rows is greater than $n$ there will be at least
one box ${\rm{b}}$ for which we have $h_{\rm{b}}=0$;
such diagrams 
correspond to the mapping $g \rightarrow 0 $ for all $g \in \group{SU}(d)$
and can be discarded in the decomposition of $\group{SU}(d)^{\otimes n}$.

Since each irreducible representation of $SU(d)$ appears in the decomposition
of $U^{\otimes n}$ for some $n$, then it is possible to establish the correspondence
\begin{align}
  \mbox{Irreducible representations of $\group{SU}(d)$ } \leftrightarrow &
  \mbox{ Young diagrams} \nonumber\\ 
&\mbox{ with at most $d-1$ columns} \nonumber
\end{align}
A relevant example is the defining representation which corresponds to the Young diagram 
made of a single box 
$  \begin{Young}
    \cr
  \end{Young}$

This $1$ to $1$ correspondence allows to deal with decomposition of tensor product of irreducible
representation of $\group{SU}(d)$ in a diagrammatic way.
If $\{ U_\alpha \}$ and $\{ U_\beta  \}$  are two irreducible representations
of $\group{SU}(d)$ we associate their tensor product representation
$\{ U_\alpha \otimes U_\beta  \}$ with the product $\alpha \times \beta$ 
of the corresponding Young diagrams $\alpha$ and $\beta$.
The following procedure provides the expansion of the product of two Young diagrams as a sum of
Young diagrams.

\vspace{0.2cm}
\noindent{\bf Expansion algorithm for Young diagram}
\begin{itemize}
\item Write the product of two Young diagrams $\alpha$ and $\beta$ 
labelling successive rows of $\beta $ with indexes $a,b,\dots$ 
as follows:
  \begin{align}
    \begin{array}{c}
  \begin{Young}
    &&& \cr
\cr
  \end{Young}    
    \end{array}
\quad
\times
\quad 
\begin{array}{c}
 \begin{Young}
    a&a&a \cr
b \cr
  \end{Young} 
\end{array}
  \end{align}
\item  
At each stage add boxes 
$
\begin{Young}
  a\cr
\end{Young} \dots 
\begin{Young}
  a\cr
\end{Young}$ ,
$
\begin{Young}
  b\cr
\end{Young} \dots 
\begin{Young}
  b\cr
\end{Young}$
from $\beta$ to $\alpha$ one at a time
checking that:
\begin{itemize}
\item the created diagrams $\nu$ have no more than $d$ columns
\item boxes with the same label must not appear in the same column
\item when the the created tableaux is read from left to right
and from top to bottom the sequence of letters $a,b,\dots$ must be such that
any any point of the sequence the number of $b$'s occurred is not bigger
than the number of $a$'s, the number of $c$'s occurred is not bigger
than the number of $b$'s etc.
\end{itemize}
\item Two diagrams $\nu, \mu$ of the shame shape are considered different only if the labeling is different.
\end{itemize}
Finally  we can  write the expansion
  \begin{align}
    \alpha \times \beta = \sum_{\nu}\sum_i \nu_i \label{eq:youngproduct}
  \end{align}
where $\nu$ labels diagram with different shape and $i$ labels different diagrams
with the same shape.
It is worth noting that the product of  Young diagrams, as defined by means of the previous expansion,
enjoys the following properties:
\begin{align}
   \alpha \times \beta  = \beta \times \alpha
\qquad
(\alpha \times \beta) \times \gamma = \alpha \times (\beta \times \gamma)
\end{align}

Each diagram $\nu_i$ in the expansion (\ref{eq:youngproduct}) corresponds
to an irreducible representation in the decomposition $U_\alpha \otimes U_\beta  = \sum_{\nu}\sum_i U_{\nu,i}$;
 diagrams with the same shape represent equivalent representations
 and the number of diagrams
with the same shape but with different labeling gives the dimension $m_\nu$ of the multiplicity space.
Finally we have the following correspondence 
\begin{align}
  \alpha \times \beta = \sum_{\nu}\sum_i \nu_i
 \quad
\leftrightarrow
\quad
U_\alpha \otimes U_\beta  = \sum_{\nu} U_{\nu} \otimes \mathbb{C}^{m_\nu}
\end{align}


The following examples will clarify the previous discussion

\subsubsection{$\bf{U \otimes U}$}
The admissible Young diagrams for $\group{SU}(d)^{\otimes 2}$ are
\begin{align}
\nu_+ =
\begin{array}{c}
   \begin{Young}
    & \cr 
  \end{Young}
\end{array}
\qquad
\nu_ = 
\begin{array}{c}
\begin{Young}
     \cr
\cr
  \end{Young}  
\end{array}
\nonumber
\end{align}
with $\dim(\hilb{H}_+)= \frac{d(d+1)}{2}$ and $\dim(\hilb{H}_-)= \frac{d(d-1)}{2}$.
The admissible standard Young tableaux are
\begin{align}
\nu_+ =
\begin{array}{c}
   \begin{Young}
   1 &2 \cr 
  \end{Young}
\end{array}
\qquad
\nu_- = 
\begin{array}{c}
\begin{Young}
1     \cr
2 \cr
  \end{Young}  
\end{array}
\nonumber
\end{align}
thus we have $m_+ = m_- =1$ and the decomposition 
becomes:
\begin{align}
  U \otimes U = U_+ \otimes U_- \label{eq:UUdecomposition}
\qquad
U_+ \in \mathcal{B}(\hilb{H}_+),\;\;U_- \in \mathcal{B}(\hilb{H}_-).
\end{align}
$\hilb{H}_+$ and $\hilb{H}_-$ can be proved to be the symmetric  and the anti-symmetric subspace of
$\hilb{H} \otimes \hilb{H}$ respectively.
If $\{\ket{i} \; i=1,\dots,d\}$ is a basis for $\hilb{H}$ it is possible to find a basis for $\hilb{H}_+$
 and $\hilb{H}_-$; we have
 \begin{align}
   \hilb{H}_+  =  
\Span \left\{
\ket{n_+} :=
\frac{1}{\sqrt{2}}(\ket{i}\ket{j} + \ket{j}\ket{i}),\; i,j=1 \dots d\right \} \label{eq:symmbasis} \\
   \hilb{H}_-  =  
\Span \left\{
\ket{n_-} :=
\frac{1}{\sqrt{2}}(\ket{i}\ket{j} - \ket{j}\ket{i}),\; i,j=1 \dots d\right \}. \label{eq:antisymmbasis}
 \end{align}
Exploiting Eqs. (\ref{eq:symmbasis},  \ref{eq:antisymmbasis})
it is easy to check that $\hilb{H}_+$ and $\hilb{H}_-$ are invariant subspaces of $\group{SU}(d)^{\otimes 2}$.
We introduce
\begin{align}
P^+  =  \sum_n \ket{n_+} \bra{n_+} \qquad P^-  =  \sum_n \ket{n_-} \bra{n_-}
\end{align}
$P^+$ is the projector on the symmetric subspace and 
$P^-$ is the projector on the antisymmetric subspace.

We notice that the expansion of the product  
$
  \begin{Young}
    \cr
  \end{Young}
\times
 \begin{Young}
    \cr
  \end{Young}
$
would lead to the same decomposition for $U\otimes U$.

\subsubsection{$\bf{U \otimes U \otimes U}$}
The admissible Young diagrams for $\group{SU}(d)^{\otimes 3}$ are
\begin{align}
\alpha =
\begin{array}{c}
   \begin{Young}
    & \cr
\cr 
  \end{Young}
\end{array}
\qquad
\beta = 
\begin{array}{c}
\begin{Young}
&&     \cr
  \end{Young}  
\end{array}
\qquad
\gamma =
\begin{array}{c}
 \begin{Young}
     \cr
\cr 
\cr
  \end{Young} 
\end{array}
.\nonumber
\end{align}
with $\dim(\hilb{H}_\alpha)= \frac{d(d+1)(d-1)}{3}$, $\dim(\hilb{H}_\beta)= \frac{d(d+1)(d+2)}{6}$
$\dim(\hilb{H}_\gamma)= \frac{d(d-1)(d-2)}{6}$.
We notice that for $d=2$ $\dim(\hilb{H}_\gamma)=0$ and the representation labelled by $\gamma$
does not appear in the decomposition.
The admissible standard Young tableaux are
\begin{align}
\alpha_1 =
\begin{array}{c}
   \begin{Young}
   1 &2 \cr 
3 \cr
  \end{Young}
\end{array}
\qquad
\alpha_2 =
\begin{array}{c}
   \begin{Young}
   1 &3 \cr 
2 \cr
  \end{Young}
\end{array}
\qquad
\beta = 
\begin{array}{c}
\begin{Young}
1&2&3     \cr
  \end{Young}
\end{array}
\qquad
\gamma = 
\begin{array}{c}
\begin{Young}
1     \cr
2 \cr
3 \cr
  \end{Young}
\end{array}
\nonumber
\end{align}
thus we have $m_\alpha  =2$,  $m_\beta  =m_\gamma  =1$ and the decomposition 
becomes:
\begin{align}
&  U \otimes U  \otimes U = U_\alpha \otimes I_{m_\alpha} \oplus U_\beta \oplus U_\gamma &\nonumber \\
&U_\alpha \in \mathcal{B}(\hilb{H}_\alpha),\;\;
U_\beta \in \mathcal{B}(\hilb{H}_\beta),\;\;
U_\gamma \in \mathcal{B}(\hilb{H}_\gamma),\;\;
I_{{m_\alpha}} \in \mathbb{C}^{m_\alpha}= \mathbb{C}^{2}.&
\end{align}

An equivalent way to decompose $U \otimes U \otimes U $
is through the expansion of the product
$
  \begin{Young}
    \cr
  \end{Young}
\times
 \begin{Young}
    \cr
  \end{Young}
\times
 \begin{Young}
    \cr
  \end{Young}
$

\subsubsection{$\bf{U \otimes U^*}$}
\label{subsec:uu*}
Let us start with a preliminary definition
\begin{Def}[conjugate representation]
  Let $\{ U_g  \}$ be a unitary representation of a group $\group{G}$.
Then it is possible to define its conjugate representation $\{ U_g^* \}$ in this way:
\begin{align}
U_g^* =  U_{g^{-1}}^T  \qquad \forall g \in \group{G}
\end{align}
\end{Def}
It is straightforward to notice that
the conjugate of the defining representation $U$ of $\group{SU}(d)$ is the one formed by
the complex conjugate matrices $U^*$.
The Young diagram corresponding to the representation $U^*$
is the one corresponding to a column of $d-1$ boxes
\begin{align}
U^*  
\;\;\leftrightarrow\;\;
\left.
\begin{array}{c}
\begin{Young}
    \cr
\cr
  \end{Young}\\
\vdots \\
\begin{Young}
\cr
  \end{Young}
\end{array}
\right\}
\mbox{ $d-1$ boxes}
\end{align}
It is worth noting that for $d=2$ both $U$ and $U^*$ are represented
by the  Young diagram made of a single box.
This  agree with the fact that the defining representation $U$ of $\group{SU}(2)$
and its conjugate $U^*$ are equivalent; Indeed for all $U \in \group{SU}(2)$
we have
\begin{align}
  CU = U^*C \qquad
C = \left(
\begin{array}{cc}
  0&1 \\
-1 & 0
\end{array}
\right) \label{eq:SU2duality}
\end{align}
 
The easiest way to the decompose of $U\otimes U^*$
is exploiting the Young diagrams formalism and the expansion algorithm:
\begin{align}
U \otimes U^* 
\quad
\leftrightarrow
\quad
  \begin{array}{c}
\begin{Young}
    \cr
  \end{Young}
\end{array}
\;
\times
\;
\begin{array}{c}
\begin{Young}
    \cr
\cr
  \end{Young}\\
\vdots \\
\begin{Young}
\cr
  \end{Young}
\end{array}
\quad
=
\quad
\begin{array}{c}
\begin{Young}
    \cr
\cr
\cr
  \end{Young}\\
\vdots \\
\begin{Young}
\cr
  \end{Young}
\end{array}
\;
\oplus
\;
\begin{array}{l}
\begin{Young}
  &  \cr 
\cr
  \end{Young}\\
\begin{array}{c}
\vdots
\end{array}\\
\begin{Young}
\cr
  \end{Young}
\end{array}
\quad
\leftrightarrow
\quad
U_p 
\oplus
U_q
\label{eq:UU*Youngdecomposition}
\end{align}
where $U_p \in \mathcal{B}(\hilb{H}_p)$ and $U_q \in \mathcal{B}(\hilb{H}_q)$
$\dim(\hilb{H}_p)=1$, $\dim(\hilb{H}_q)=d^2-1$.
An explicit form for the projectors on $\hilb{H}_p$ and $\hilb{H}_q$ can be given:
\begin{align}\label{eq:decompoUU*}
  P^p = d^{-1}\KetBra{I}{I} \qquad P^q = I \otimes I-P^p.
\end{align}

\subsubsection{$\bf{U \otimes U \otimes U^*}$}
We can decompose the representation $U \otimes U \otimes U$
as follows. First, as we  showed previously, $U \otimes U$
can be  decomposed as $U_+ \oplus U_-$ and so 
we have $U \otimes U \otimes U^* = (U_+ \oplus U_-) \otimes U^* = (U_+ \otimes U^*) \oplus ( U_- \oplus U^*)$.
We now further decompose $U_+ \otimes U^*$  and  $U_- \oplus U^*$:
\begin{align}
U_+ \otimes U^* 
\;
\leftrightarrow
\;
  \begin{array}{c}
\begin{Young}
  &  \cr
  \end{Young}
\end{array}
\;
\times
\;
\begin{array}{c}
\begin{Young}
    \cr
\cr
  \end{Young}\\
\vdots \\
\begin{Young}
\cr
  \end{Young}
\end{array}
\;
=
\;
\begin{array}{l}
\begin{Young}
  & & \cr 
\cr
  \end{Young}\\
\begin{array}{c}
\vdots
\end{array}\\
\begin{Young}
\cr
  \end{Young}
\end{array}
\;
\oplus
\;
\begin{array}{l}
\begin{Young}
  &  \cr 
\cr
\cr
  \end{Young}\\
\begin{array}{c}
\vdots
\end{array}\\
\begin{Young}
\cr
  \end{Young}
\end{array}
\;
\leftrightarrow
\;
U_{\beta,+} 
\oplus
U_{\alpha,+}
 \nonumber \\
U_- \otimes U^* 
\;
\leftrightarrow
\;
  \begin{array}{c}
\begin{Young}
    \cr
\cr
  \end{Young}
\end{array}
\;
\times
\;
\begin{array}{c}
\begin{Young}
    \cr
    \cr
  \end{Young}\\
\vdots \\
\begin{Young}
\cr
  \end{Young}
\end{array}
\;
=
\;
\begin{array}{l}
\begin{Young}
  &  \cr 
& \cr
  \end{Young}\\
\begin{array}{c}
\vdots
\end{array}\\
\begin{Young}
\cr
  \end{Young}
\end{array}
\;
\oplus
\;
\begin{array}{l}
\begin{Young}
  &  \cr 
\cr
\cr
  \end{Young}\\
\begin{array}{c}
\vdots
\end{array}\\
\begin{Young}
\cr
  \end{Young}
\end{array}
\;
\leftrightarrow
\;
U_{\gamma,-} 
\oplus
U_{\alpha,-}
\label{eq:U_+U*Youngdecomposition}
\end{align}

Then the following decomposition holds:
\begin{align}
&  U \otimes U \otimes U^* = U_{\alpha,+} \oplus U_{\alpha,-} \oplus U_{\gamma,-} \oplus U_{\beta,+} &
\label{eq:UUU*decomposition}\\
\nonumber\\
&\dim(\hilb{H}_{\alpha,+}) = \dim(\hilb{H}_{\alpha,-}) = d,& \nonumber\\
& \dim(\hilb{H}_{\beta,+}) = d \frac{d^2+d-2}{2},
\qquad\dim(\hilb{H}_{\gamma,-}) = d \frac{d^2-d-2}{2}&
\nonumber
\end{align}
We notice that for $d=2$ the subspace $\dim(\hilb{H}_{\gamma,-})=0$
Since $U_{\alpha,+}$ and $U_{\alpha,-}$ are equivalent representations the decomposition (\ref{eq:UUU*decomposition})
can be rewritten as
\begin{align}
   U \otimes U \otimes U^* = U_{\alpha}\otimes I_{m\alpha} \oplus U_{\gamma} \oplus U_{\beta}
 \qquad I_{m_\alpha} \in \mathcal{B}(\mathbb{C}^2)
\end{align}
where we relabeled $\hilb{H}_{\beta,+} = \hilb{H}_\beta$, $\hilb{H}_{\gamma,-} = \hilb{H}_\gamma$
and $\hilb{H}_{\alpha,+} \oplus \hilb{H}_{\alpha,-} = \hilb{H}_\alpha \otimes \mathbb{C}^2$.
We now provide two basis for  $\hilb{H}_{\alpha,+}$ and $\hilb{H}_{\alpha,-}$
\begin{align}
  \hilb{H}_{\alpha,+} = \Span \left\{  \ket{k_{\alpha,+}}:=
\frac{1}{\sqrt{2(d+1)}} 
\left(\Ket{I}_{02}\ket{k}_1+\Ket{I}_{12}\ket{k}_0 \right)  \right\} \nonumber\\
  \hilb{H}_{\alpha,-} = \Span \left\{  \ket{k_{\alpha,-}}:=
\frac{1}{\sqrt{2(d-1)}} 
\left(\Ket{I}_{02}\ket{k}_1-\Ket{I}_{12}\ket{k}_0 \right)  \right\} \label{eq:alphabasis}.
\end{align}
where we introduced the labeling $\hilb{H} \otimes \hilb{H} \otimes \hilb{H}
:= \hilb{H}_0 \otimes \hilb{H}_1 \otimes \hilb{H}_2$ and  $\ket{k}_{i}$ means
$\ket{k} \in \hilb{H}_i$.
We notice the properties
\begin{align}\label{eq:groupcambiasegnoswap}
  \mathsf{S}\ket{k_{\alpha,+}} = \ket{k_{\alpha,+}} \qquad   \mathsf{S}\ket{k_{\alpha,-}} = -\ket{k_{\alpha,-}}
\end{align}
where $\mathsf{S}$ is the swap operator $\mathsf{S}\ket{\psi}_0\ket{\phi}_1=\ket{\phi}_0\ket{\psi}_1$
In terms of these two basis the isomorphism  between $\hilb{H}_{\alpha,+}$ and $\hilb{H}_{\alpha,-}$
has the following form:
\begin{align}
  T^{\alpha,+,-}= \sum_k\ket{k_{\alpha,+}}\bra{k_{\alpha,-}}.
\end{align}
From Eqs. (\ref{eq:U_+U*Youngdecomposition}) and (\ref{eq:alphabasis})  we can derive the  expression
for the projectors on $\hilb{H}_{\beta}$ and  $\hilb{H}_{\gamma}$:
\begin{align}
P^{\beta} = P^+\otimes I_2 -  T^{\alpha,+,+} 
\qquad
P^{\gamma} = P^-\otimes I_2 -  T^{\alpha,-,-} \label{eq:groupdecompoprojector}\\ 
  T^{\alpha,+,+}= \sum_k\ket{k_{\alpha,+}}\bra{k_{\alpha,+}} \qquad T^{\alpha,-,-}= \sum_k\ket{k_{\alpha,-}}\bra{k_{\alpha,-}}.
\end{align}
$T^{\alpha,+,+}$ is the projector on $\hilb{H}_{\alpha,+}$ and
$T^{\alpha,-,-}$ is the projector on $\hilb{H}_{\alpha,-}$.

Exploiting Theorem \ref{th:characterizationcommutant}
any operator $O$ satisfying the commutation
$[O, U \otimes U \otimes U^*]$
can be decomposed as
\begin{align}\label{decomposition}
  O = \sum_{\nu \in \defset{S}} \sum_{i,j = \pm} T^{\nu,i,j} o_{\nu}^{i,j} \qquad o_{\nu}^{i,j} \in \mathbb{R}
\end{align}
where 
$\defset{S} = \{\alpha, \beta, \gamma  \} $,
 $T^{\beta,+,-} =  T^{\beta,-,+} =T^{\beta,-,-} = 0$,
$T^{\gamma,+,-} =  T^{\gamma,-,+} =T^{\gamma,+,+} = 0$,
$T^{\beta,+,+} = P^\beta$ and $T^{\beta,+,+} = P^\beta$.

\subsubsection{$\bf{U \otimes U \otimes U \otimes U}$ ($2$-dimensional case)}
Expanding the product 
$\;
  \begin{Young}
    \cr
  \end{Young}
\times
 \begin{Young}
    \cr
  \end{Young}
\times
 \begin{Young}
    \cr
  \end{Young}
\times
 \begin{Young}
    \cr
  \end{Young}
\times
 \begin{Young}
    \cr
  \end{Young}
\;
$ 
leads to the decomposition 
 \begin{align}
  \begin{array}{c}
 \begin{Young}
&&&    \cr 
  \end{Young}   
  \end{array}
\oplus
\begin{array}{c}
 \begin{Young}
&&    \cr 
\cr
  \end{Young} 
\end{array}
\oplus
\begin{array}{c}
 \begin{Young}
&&    \cr 
\cr
  \end{Young} 
\end{array}
\oplus
\begin{array}{c}
 \begin{Young}
&&    \cr 
\cr
  \end{Young} 
\end{array}
\oplus
\begin{array}{c}
 \begin{Young}
&    \cr 
& \cr
  \end{Young} 
\end{array}
\oplus
\begin{array}{c}
 \begin{Young}
&    \cr 
& \cr
  \end{Young} 
\end{array}
\nonumber
\end{align}
\begin{align}
& U^{\otimes 4} = U_a \oplus U_b \otimes I_{m_b} \oplus U_c \otimes I_{m_c} &
\label{eq:U^4decomposition}\\
\nonumber \\
&\dim(\hilb{H}_a) = 5
\quad
\dim(\hilb{H}_b) = 3
\quad
\dim(\hilb{H}_c) = 1& \nonumber\\
& \mathbb{C}^{m_b}=\mathbb{C}^{3}
\quad
\mathbb{C}^{m_c}=\mathbb{C}^{2}& \nonumber
\end{align}
Since we are considering the case $d=2$,
$U$ and $U^*$ are equivalent and
 the decomposition (\ref{eq:U^4decomposition})
can be generalized to the cases in which one or more $U$ is replaced with $U^*$.
For example we have:
\begin{align}
  U^* \otimes U \otimes U \otimes U = 
(C \otimes I^{\otimes 3})\, U_a \oplus U_b \otimes I_{m_b} \oplus U_c \otimes I_{m_c} \, (C \otimes I^{\otimes 3})
\end{align}
where $C$ was defined in Eq. (\ref{eq:SU2duality}).





\addcontentsline{toc}{section}{References}
\begin{thebibliography}{99}


\bibitem{holevo}
A. S. Holevo,
\emph{Probabilistic and Statistical Aspects of Quantum Theory},
North Holland, Amsterdam (1982)


\bibitem{nielsenchuang}
M. A. Nielsen, I. L. Chuang,
\emph{Quantum Computation
and Quantum Information},
Cambridge University Press, Cambridge (2000)


\bibitem{helstrom}
 C. W. Helstrom, 
\emph{Quantum Detection and Estimation Theory},
 Academic Press, New York (1976)

\bibitem{acin}
 A. Ac\'{\i}n,
 Phys. Rev. Lett. {\bf 87}, 177901 (2001)

\bibitem{leaderdiscr}
G. M. D'Ariano, P. Lo Presti, and M. G. A. Paris, 
Phys. Rev. Lett. {\bf 87}, 270404 (2001)

\bibitem{maxdiscrimi}
M. F. Sacchi
 J. Opt. B {\bf 7}, S333 (2005) 


\bibitem{memoryeffects}
 G. Chiribella, G. M. D'Ariano, P. Perinotti,
Phys. Rev. Lett. {\bf 101}, 180501 (2008) 

\bibitem{watrousdiscrimi}
A. W. Harrow, A. Hassidim, D. W. Leung, J. Watrous,
 Phys. Rev. A {\bf 81}, 032339 (2010)
 

\bibitem{noprogramming}
M. A. Nielsen, I. L. Chuang
 Phys. Rev. Lett. {\bf 79}, 321 (1997)

\bibitem{teleport1}
S. F. Huelga, J. A. Vaccaro, A. Chefles, and M. B. Plenio,
Phys. Rev. A {\bf 63}, 042303 (2001)

\bibitem{teleport2}
S. D. Bartlett, W. J. Munro,
 Phys. Rev. Lett. {\bf 90}, 117901 (2003)

\bibitem{teleport3}
 Y.-F. Huang, X.-F. Ren, Y.-S. Zhang, L.-M. Duan, G.-C. Guo,
 Phys. Rev. Lett. {\bf 93}, 240501 (2004)

\bibitem{tomoexp1}
Y. S. Weinstein, T. F. Havel, J. Emerson, N. Boulant, M. Saraceno, S. Lloyd, D. G. Cory,
J. Chem. Phys. {\bf 121(13)}, 6117-6133 (2004)

\bibitem{tomoexp2}
 J. L. O'Brien, G. J. Pryde, A. Gilchrist, D. F. V. James, N. K. Langford, T. C. Ralph, A. G. White,
 Phys. Rev. Lett. {\bf 93}, 080502 (2004)

\bibitem{watrousgame}
 G. Gutoski and J. Watrous,
 Proc. of the 39th Annual
ACM Symposium on Theory of Computation, 565 (2007).

\bibitem{deutschjoz} D. Deutsch, R. Jozsa, Proc. R. Soc. Lond. A {\bf 439}, 553-558 (1992).

\bibitem{Grover}
Grover L.K.
Proceedings of the 28th Annual ACM Symposium on the Theory of Computing, 212 (1996)

\bibitem{shor} P. Shor, SIAM Rev. {\bf 41}, pp. 303-332 (1999).

\bibitem{bitcommitment1}
 H. P. Yuen, quant-ph/0207089.

\bibitem{bitcommitment2}
 G. M. D'Ariano, D. Kretschmann, D. M. Schlingemann, R. F. Werner,
 Phys. Rev. A {\bf 76} 032328 (2007).

\bibitem{pironio} S. Pirandola, S. Mancini, S. Lloyd, and S. L.
  Braunstein, Nature Physics {\bf 4}, 726 - 730 (2008).

%
%
\bibitem{QCA}
G. Chiribella, G. M. D'Ariano,  P. Perinotti,
Phys. Rev. Lett. {\bf 101}, 060401 (2008)

\bibitem{comblong}
G. Chiribella, G. M. D'Ariano,  P. Perinotti,
Phys. Rev. A {\bf 80}, 022339 (2009)


\bibitem{optimaltomo}
A. Bisio, G. Chiribella, G. M. D'Ariano, S. Facchini, and P. Perinotti,
Phys. Rev. Lett. {\bf 102}, 010404 (2009).


\bibitem{tomoieee}
A. Bisio, G. Chiribella, G. M. D'Ariano, S. Facchini, and P. Perinotti,
IEEE Journal of Selected Topics in Quantum Electronics {\bf 15} 1646 (2009)


\bibitem{cloningunit}
G. Chiribella, G. M. D'Ariano, P. Perinotti,
Phys. Rev. Lett. {\bf 101}, 180504 (2008) 


\bibitem{optimallearning}
A. Bisio, G. Chiribella, G. M. D'Ariano, S. Facchini, P. Perinotti
Phys. Rev. A {\bf 81}, 032324 (2010) 



\bibitem{algorithm}
A. Bisio, G. Chiribella, G. M. D'Ariano,  P. Perinotti,
Phys. Rev. A {\bf 83}, 022325 (2011)


\bibitem{unitradeoff}
     A. Bisio, G. Chiribella, G. M. D'Ariano, P. Perinotti,
Phys. Rev. A {\bf 82}, 062305 (2010).


\bibitem{learnobs}
 A. Bisio, G. M. D'Ariano, P. Perinotti,  M. Sedl\'ak,
Physics Letters A {\bf 375}, 3425-3434
(2011).

\bibitem{clonobs} A. Bisio, G. M. D'Ariano, P. Perinotti,
  M. Sedl\'ak, 
(accepted in Phys. Rev. A).
arXiv:1103.5709

\bibitem{raginskyfide}
M. Raginsky
Phys. Lett. A {\bf 290}, 11 (2001)


\bibitem{depillis}
J. de Pillis,
 Linear Transformations Which Preserve Hermitian and Positive Semidefinite Operators, 
Pacific J. of Math. {\bf 23}, 129 (1967)


\bibitem{choiisom}
M.-D. Choi,
Lin. Alg. and Appl. {\bf 10}, 285 (1975)

\bibitem{jamioisom}
A. Jamio\l kowski
Rep. Mod. Phys. {\bf 3}, 275 (1972)


\bibitem{stinespring}
W. F. Stinespring,
Proc. Amer. Math. Soc. {\bf 6}, 211 (1955)

\bibitem{JMP}
G. Chiribella, G. M. D'Ariano, P. Perinotti
 J. Math. Phys. {\bf 50}, 042101 (2009)

\bibitem{ozawadilationtheorem}
M. Ozawa.
 J. Math. Phys. {\bf 25}, 79  (1984)

 
\bibitem{semicausal1}
T. Eggeling, D. Schlingemann, R. F. Werner, 
Europhys. Lett. {\bf 57}, 782-788 (2002).



\bibitem{semicausal2}
M. Piani, M. Horodecki, P. Horodecki, R. Horodecki
Phys. Rev. A {\bf 74}, 012305 (2006)

\bibitem{PPOVMziman}
M. Ziman,
Phys. Rev. A {\bf 77}, 062112 (2008).


%
%

\bibitem{tomoptics}
D. T. Smithey, M. Beck, M. G. Raymer, and A. Faridani,
 Phys. Rev. Lett. {\bf 70}, 1244 (1993).

\bibitem{tomoptics2}
K. Vogel and H. Risken,
Phys. Rev. A {\bf 40}, 2847 (1989).

\bibitem{tomoptics3}
G. M. D'Ariano, C. Macchiavello, and M. G. A. Paris,
Phys. Rev. A {\bf 50}, 4298 (1994)


\bibitem{busch}
P. Busch,
Int. J. Theor. Phys. {\bf 30}, 1217 (1991).

\bibitem{optdataproc}
G. M. D'Ariano and P. Perinotti, 
Phys. Rev. Lett. {\bf 98}, 020403 (2007).


\bibitem{scott1}
A. J. Scott,
Phys. A {\bf 39}, 13507 (2006).

\bibitem{AAPT1}
G. M. D'Ariano, P. Lo Presti,
Phys. Rev. Lett. {\bf 86}, 4195 (2001).

\bibitem{AAPT2}
 W. D\"{u}r  and J. I. Cirac,
 Phys. Rev. A {\bf 64}, 012317 (2001).

\bibitem{FRAMES1}
R. J. Duffin, A. C. Schaeffer, 
Trans. Am. Math. Soc. {\bf 72}, 341 (1952).

\bibitem{FRAMES2}
P. G. Casazza,
Taiw. J. Math. {\bf 4}, 129 (2000)

\bibitem{statistica}
G .Casella, R. L.  Berger,
\emph{Statistical Inference},
 Duxbury Press (2001).


\bibitem{covpovmnorm1}
G. M. D'Ariano. P. Perinotti, M. F. Sacchi,
J. Opt.B: Quantum and Semicl. Optics {\bf 6}, S487 (2004)

\bibitem{scott2}
A. J. Scott,
J. Phys. A {\bf 39}, 13507 (2006)

\bibitem{contmeasurdiscrete}
G. Chiribella, G. M.D'Ariano, D. M. Schlingemann,
Phys. Rev. Lett. {\bf 98}, 020403 (2007)

\bibitem{zeilingerwalt}
P. Walther, A. Zeilinger,
Phys. Rev. A {\bf 72}, 010302(R) (2005)

%
%



\bibitem{nocloning}
W. K. Wootters, W.H.Zurek,
Nature {\bf 299}, 802 (1982)

\bibitem{optclonvlado}
V. Buzek, M. Hillery,
Physics World {\bf 14},  25 (2001).

\bibitem{optclonwerner}
R. Werner,
Phys. Rev. A {\bf 58}, 1827 (1998)

\bibitem{asymcloningfiura}
J. Fiurasek, R. Filip, N. J. Cerf
Quant. Inform. Comp. {\bf 5}, 583 (2005). 

\bibitem{cloningreview}
V. Scarani, S. Iblisdir, N. Gisin, and A. Ac\'{i}n,
Rev. Mod. Phys. {\bf 77}, 1225 (2005)



\bibitem{BB84}
C. H. Bennett, G. Brassard,
Proceedings IEEE Int. Conf. on Computers, Systems and Signal Processing,
Bangalore, India (IEEE New York, 1984), pp. 175-179





%
%

\bibitem{quantmemory1}
R. Zhao, Y. O. Dudin, S. D. Jenkins, C. J. Campbell, D. N. Matsukevich, T. A. B. Kennedy, A. Kuzmich,  
Nature Physics {\bf 5}, 100 (2009) 

\bibitem{quantmemory2}
A. I. Lvovsky, B. C. Sanders, W. Tittel
Nature Photonics {\bf 3}, 706 - 714 (2009)

\bibitem{quantmemory3}
B. Julsgaard, J. Sherson, J. I. Cirac, J. Fiurasek, E. S. Polzik
   Nature {\bf 432}, 482 - 486 (2004)



\bibitem{programming1}
  G. Vidal, L. Masanes, J. I. Cirac
     Phys. Rev. Lett. {\bf 88}, 047905 (2002)
    
\bibitem{programming2}
    M. Ziman, V. Buzek
    Phys. Rev. A {\bf 72}, 022343 (2005)
    
\bibitem{programming3}
 G. M. D'Ariano, P. Perinotti
Phys. Rev. Lett. {\bf 94}, 090401 (2005)



\bibitem{programming4}
     M. Micuda, M. Jezek, M. Dusek, J. Fiurasek
    Phys. Rev. A {\bf 78}, 062311 (2008)
    

\bibitem{optimalestimunit1}
G. Chiribella, G. M. D'Ariano, M. F. Sacchi,
Phys. Rev. A {\bf 72} 042338 (2005)


\bibitem{optimalestimunit2}
G. Chiribella, G. M. D'Ariano, P. Perinotti, M. F. Sacchi,
Phys. Rev. Lett. {\bf 93} 18053 (2004)

\bibitem{optimalestmbuzek}
V. Buzek, R. Derka, S. Massar,
Phys. Rev. Lett. {\bf 82}, 2207 (1999)




%
%

\bibitem{spekkensframe}
S. D. Bartlett, T. Rudolph, R. W. Spekkens, P. S. Turner,
New J. Phys. {\bf 11}, 063013 (2009)

\bibitem{maxlikelihood}
G. Chiribella, G. M. D'Ariano, P. Perinotti, M. F. Sacchi,
Phys. Rev. A {\bf 70}, 062105 (2004)



%
%
%



\bibitem{heisenberg} 
W. Heisenberg,
 Zeitsch. Phys. {\bf 43}, 172 (1927).


\bibitem{tradeoffscully} 
M. O. Scully, B.-G. Englert, and H. Walther, 
Nature {\bf 351}, 111 (1991). 

\bibitem{tradeofffuchs-peres}
C. A. Fuchs and  A. Peres,
Phys. Rev. A {\bf 53}, 2038 (1996).


\bibitem{tradeoffbanaszek}
K Banaszek, 
 Phys. Rev. Lett. {\bf 86}, 1366 (2001) .


\bibitem{tradeoffozawa}
M. Ozawa,
Ann. Phys. {\bf 311}, 350 (2004).


\bibitem{tradeoffmax}
M. F. Sacchi,
Phys. Rev. Lett. {\bf 96}, 220502 (2006).

 
 \bibitem{tradeoffdema}
 F. Sciarrino, M. Ricci, F. De Martini, R. Filip, and L. Mi\v sta Jr., Phys. Rev. Lett. {\bf 96}, 020408 (2006). 
 
 \bibitem{tradeoffmacca}   L. Maccone, Phys. Rev. A {\bf 73}, 042307 (2006).

\bibitem{tradeoffwerner}
D. Kretschmann, D. Schlingemann and R. F. Werner, IEEE Trans. Inf. Theory {\bf 4}, 1708 (2008).

\bibitem{tradeoffbusc}
F. Buscemi, M. Hayashi, M. Horodecki,
Phys. Rev. Lett. {\bf 100}, 210504 (2008).

\bibitem{bostromprotocol}
K. Bostr\"{o}m, T. Felbinger,
Phys. Rev. Lett. {\bf 89}, 187902 (2002).

\bibitem{lucamariniprotocol}
M. Lucamarini, S. Mancini,
Phys. Rev. Lett. {\bf 94}, 14051 (2005).



%
%



\bibitem{Parisobsclon}
A. Ferraro, M. Galbiati, M. G. A. Paris,
 J. Phys. A {\bf 39},
L219-L228 (2006).


\bibitem{boydvander}
S. Boyd, L. Vanderberghe,
\emph{Convex Optimization}
Cambridge University Press, Cambridge (2004)





\bibitem{Zalka}
 C. Zalka
 Phys. Rev. A {\bf 60},  2746 (1999)

\bibitem{wangyngdiscri}
G.Wang, M. Ying.
Phys. Rev. A {\bf 73},  042301 (2006)



%
%



\bibitem{belavkinrag}
 V. P. Belavkin, G. M. D'Ariano, M. Raginsky
J. Math. Phys. {\bf 46}, 062106 (2005)


%
%

\bibitem{fultonharris}
W. Fulton and J. Harris,
\emph{Representation theory: a first course},
Springer, (1996)



\bibitem{jones}
H. F. Jones,
\emph{Groups, Representations and Physics}
Taylor and Francis (1990)

\bibitem{fulton2}
W. Fulton, 
\emph{Young tableaux : with applications to representation theory and geometry}
 Cambridge University Press, Cambridge (1997)



\bibitem{barut}
 A. O. Barut, R. Raczka, 
\emph{Theory of group representations and applications}
 World Scientific, Singapore  (1986)


\bibitem{CVX}
M. Grant, S. Boyd, http://cvxr.com/cvx/


\bibitem{watrousroutine}
J. Watrous,
private communication (2010)


\end{thebibliography}


\newpage
\fancyhead[LO]{References}

\end{document}